\global\def\draftcontrol{0}

   \def\versionno{ ks desitter}
\catcode`\@=11

\expandafter\ifx\csname draftcontrol\endcsname\relax\global\def\draftcontrol{0}
\fi

{\count255=\time\divide\count255 by 60
\xdef\hourmin{\number\count255}
\multiply\count255 by-60\advance\count255 by\time
\xdef\hourmin{\hourmin:\ifnum\count255<10 0\fi\the\count255}}
\def\draftdate{\number\month/\number\day/\number\year\ \ \ \hourmin }

\newcommand\makepapertitle{\par
  \begingroup
    \renewcommand\thefootnote{\@fnsymbol\c@footnote}%
    \def\@makefnmark{\rlap{\@textsuperscript{\normalfont\@thefnmark}}}%
    \long\def\@makefntext##1{\parindent 1em\noindent
            \hb@xt@1.8em{%
                \hss\@textsuperscript{\normalfont\@thefnmark}}##1}%
     \newpage
     \global\@topnum\z@   % Prevents figures from going at top of page.
     \@makepapertitle
     \thispagestyle{empty}\@thanks
  \endgroup
  \setcounter{footnote}{0}%
  \global\let\thanks\relax
  \global\let\makepapertitle\relax
  \global\let\@makepapertitle\relax
  \global\let\@thanks\@empty
  \global\let\@author\@empty
  \global\let\@date\@empty
  \global\let\@title\@empty
  \global\let\title\relax
  \global\let\author\relax
  \global\let\date\relax
  \global\let\and\relax
  \def\version{\let\version\@version\@gobble}
}
\def\@makepapertitle{%
  \newpage
   \ifnum\draftcontrol=1 {}
   \version\versionno
   \vskip 3em%
   \else
   \hfill\hbox to 3cm {\parbox{4cm}{\@pubnum}\hss}%
   \vskip 3em%
   \fi
   \begin{center}%
   \let \footnote \thanks
     {\LARGE {\@title}}%
     \vskip 1.5em%
     {\normalsize%\large
       \lineskip .5em%
       \begin{tabular}[t]{c}%
         \@author
       \end{tabular}\par}%
     \vskip 1.5em%
     {\@bstract}%
     \end{center}%
     \vskip 1.5em
     \@date%
   \par
}

\gdef\@pubnum{}
\def\pubnum#1{%
  \gdef\@pubnum{#1}}

\gdef\@bstract{}
\def\Abstract#1{%
  \gdef\@bstract{%
   \parbox{\textwidth-0pc}{%
   \centerline{\bf Abstract}\penalty1000%
\kern.2cm%
\noindent%\abstractfont \baselineskip=12pt
\renewcommand\baselinestretch{1.0}%
{#1}}}
}

\def\ps@paper{\let\@mkboth\@gobbletwo%
     \ifnum\draftcontrol=1
    \def\@oddfoot{\hbox to \textwidth{\tiny \versionno \hfil\tiny\draftdate}%
    \hskip -\textwidth \hbox to \textwidth{\hfil\rm\thepage\hfil}}%
     \else\def\@oddfoot{\hbox to \textwidth{\hfil\rm\thepage\hfil}}
     \fi
     \let\@evenfoot\@oddfoot
}
\def\body{\clearpage
          \pagestyle{paper}
    }
\def\@version#1{\ifnum\draftcontrol=1
\typeout{}\typeout{#1}\typeout{}
\vskip3mm\centerline{\hbox{\fbox{\normalsize{\tt DRAFT -- #1 -- }
                   {\draftdate}}}}\vskip3mm
\fi}
\let\version\@version
\long\def\eqlabel#1{\ifnum\draftcontrol=1
                    \tag@false  % there are some problems with multline without this
                    \tag*{(\theequation) \hbox to -0.2cm{\hspace{0cm}\small{#1}\hss}}
                    \refstepcounter{equation}
                    \edef\@currentlabel{\theequation}
                    \ltx@label{#1}          % use old LaTeX \label instead of new definition
                    \else
                    \label{#1}
                    \fi
                    }
\let\st@bibitem\@bibitem
\let\st@lbibitem\@lbibitem
\ifnum\draftcontrol=1
  \def\@bibitem#1{%
    \st@bibitem{#1}\a@@label{#1}\ignorespaces}
  \def\@lbibitem[#1]#2{%
    \st@lbibitem[#1]{#2}\a@@label{#2}\ignorespaces}
  \def\a@@label#1{%
    \gdef\a@lab{\smash{\normalfont\small#1}}
    \ifvmode
      \if@inlabel
        \global\setbox\@labels\hbox{%
          \llap{\a@lab\let\a@lab\relax
                \kern\@totalleftmargin\kern\marginparsep}%
          \box\@labels}%
      \fi
    \fi}
\fi
\documentclass[12pt,letterpaper]{article}
\usepackage{amsmath,amssymb,array,calc,epsfig,rotating,bm,xcolor}
\usepackage[sort]{cite}
\usepackage{graphicx,esint,float}
\usepackage{psfrag,verbatim}
\usepackage[makeroom]{cancel}
\usepackage{xcolor}
\ifnum\draftcontrol=1
\fi
\renewcommand\baselinestretch{1.25}
\renewcommand\section{\@startsection {section}{1}{\z@}%
                                   {-3.5ex \@plus -1ex \@minus -.2ex}%
                                   {2.3ex \@plus.2ex}%
                                   {\normalfont\large\bfseries}}
\renewcommand\subsection{\@startsection{subsection}{2}{\z@}%
                                   {-3.25ex\@plus -1ex \@minus -.2ex}%
                                   {1.5ex \@plus .2ex}%
                                   {\normalfont\normalsize\bfseries}}
\renewcommand\subsubsection{\@startsection{subsubsection}{3}{\z@}%
                                   {-3.25ex\@plus -1ex \@minus -.2ex}%
                                   {1.5ex \@plus .2ex}%
                                   {\normalfont\normalsize\it}}
\renewcommand\paragraph{\@startsection{paragraph}{4}{\z@}%
                                   {-3.25ex\@plus -1ex \@minus -.2ex}%
                                   {1.5ex \@plus .2ex}%
                                   {\normalfont\normalsize\bf}}

\numberwithin{equation}{section}

\def\revise#1       {\raisebox{-0em}{\rule{3pt}{1em}}%
                     \marginpar{\raisebox{.5em}{\vrule width3pt\
                     \vrule width0pt height 0pt depth0.5em
                     \hbox to 0cm{\hspace{0cm}{%
                     \parbox[t]{4em}{\raggedright\footnotesize{#1}}}\hss}}}}

\newcommand\nxt[1]  {\\\fnxt#1}
\newcommand{\ie}{{\it i.e.,}\ }
\newcommand{\eg}{{\it e.g.,}\ }

\def\cala         {{\cal A}}

\def\cale         {{\cal E}}
\def\calf         {{\cal F}}

\def\call         {{\cal L}}
\def\calm         {{\cal M}}
\def\caln         {{\cal N}}
\def\calo         {{\cal O}}
\def\calp         {{\cal P}}
\def\calq         {{\cal Q}}
\def\calr         {{\cal R}}
\def\cals         {{\cal S}}

\def\reals        {{\mathbb R}}
\def\zet          {{\mathbb Z}}

\def\del          {\partial}

\def\tr           {\mathop{\rm Tr}}

\def\sqr#1#2{{\vcenter{\vbox{\hrule height.#2pt
 \hbox{\vrule width.#2pt height#1pt \kern#1pt
 \vrule width.#2pt}\hrule height.#2pt}}}}

\def\a{\alpha}
\def\b{\beta}
\def\w{\omega}
\def\r{\rho}
\def\dd{\delta}

\def\e{\epsilon}
\def\c{\chi}

\def\g{\gamma}

\def\hh{\hat{h}}
\def\hk{\hat{k}}
\def\hs{\hat{s}}
\def\hf{\hat{f}}

\def\hK{\hat{K}}
\def\aa1{\phi}
\def\cc1{\psi}
\def\hh{\hat{h}}

\def\l{\lambda}

\def\Om{\Omega}
\def\om{\Omega}

\def\hr{\hat{\rho}}
\def\hf{\hat{f}}
\def\hK{\hat{K}}

\def\ha{\hat{a}}

\def\csb{{\chi\rm{SB}}}

\def\t{\tau}
\def\s{\sigma}

\def\hw{\hat{\omega}}

\def\ha{\hat{a}}

\def\as{{\rm TypeA}_{s}}
\def\ab{{\rm TypeA}_{b}}

\catcode`\@=12

\begin{document}

%%%
%%%%%% text starts here
%%%%%%%%%

\title{\bf $\chi\rm{SB}$ of cascading gauge theory  in de Sitter}

\date{December 7, 2019}
%\date\today

\author{
Alex Buchel\\[0.4cm]
\it $ $Department of Applied Mathematics\\
\it $ $Department of Physics and Astronomy\\ 
\it University of Western Ontario\\
\it London, Ontario N6A 5B7, Canada\\
\it $ $Perimeter Institute for Theoretical Physics\\
\it Waterloo, Ontario N2J 2W9, Canada
}

\Abstract{${\cal N}=1$ supersymmetric $SU(N)\times SU(N+M)$ cascading gauge
theory of Klebanov et.al \cite{Klebanov:2000nc,Klebanov:2000hb}
spontaneously breaks chiral symmetry in Minkowski space-time. We
demonstrate that in de Sitter space-time the chiral symmetry breaking
occurs for the values of the Hubble constant $H\lesssim 0.7 \Lambda$,
as well as in the narrow window $0.92(1) \Lambda \le H\le
0.92(5) \Lambda $.  We give a precise definition of the strong
coupling scale $\Lambda$ of the cascading gauge theory, which is related
to the glueball mass scale in the theory $m_{glueball}$ and the
asymptotic string coupling $g_s$ as $\Lambda\sim g_s^{1/2}
m_{glueball}$.
}

\makepapertitle

\body

\version\versionno
\tableofcontents

\section{Introduction and summary}\label{intro}
Consider $\caln=1$ supersymmetric $SU(N+M)\times SU(N)$  
gauge theory with two chiral superfields $A_1$, $A_2$ in the $(N+M,\overline{N})$ representation,
and two chiral superfields $B_1$, $B_2$ in the $(\overline{N+M},{N})$ representation, in four dimensional Minkowski
space-time $\reals^{3,1}$. This theory has two
gauge couplings $g_1$, $g_2$ associated with the two gauge group factors, and a quartic superpotential
\begin{equation}
W\ \sim\ \tr\left(A_iB_jA_kB_\ell\right)\e^{ik}\e^{j\ell}\,.
\eqlabel{defw}
\end{equation}
When $M=0$, both gauge couplings are exactly marginal, and the theory flows to a strongly coupled
superconformal fixed point --- the  Klebanov-Witten (KW) theory \cite{Klebanov:1998hh}.
KW infrared (IR) fixed point global symmetry
\begin{equation}
G:\qquad \underbrace{SU(2)\times SU(2)}_{\rm flavour}\ \times \underbrace{U(1)}_{R-{\rm symmetry}}\,,
\end{equation}
together with the superconformal invariance implies non-perturbatively large
anomalous dimensions for the chiral superfields:
\begin{equation}
\g(A_i)=\g(B_j)=-\frac 14\,.
\eqlabel{irdim}
\end{equation}
When $M\ne 0$, conformal invariance of $SU(N+M)\times SU(N)$ gauge theory is broken: while the sum of the
gauge coupling remains exactly marginal \cite{Klebanov:2000hb},
\begin{equation}
\frac{4\pi^2}{g_1^2}+\frac{4\pi^2}{g_2^2}=\frac{\pi}{g_s}={\rm const}\,,
\eqlabel{sum}
\end{equation}
where $g_s$ is the asymptotic string coupling of the gravitational dual \cite{Herzog:2001xk},
the perturbative $\b$-function of the difference of the couplings is nonzero \cite{Herzog:2001xk}:
\begin{equation}
\frac{8\pi^2}{g_1^2}-\frac{8\pi^2}{g_2^2}=M \ln\frac{\Lambda}{\mu}\biggl(3+2(1-\g(\tr(A_iB_j)))\biggr)
=6 M \ln\frac{\Lambda}{\mu}\left(1+\calo\left(\frac{M^2}{N^2}\right)\right)\,.
\eqlabel{difference}
\end{equation}
$\Lambda$ is the strong coupling scale of the theory. Given \eqref{sum} and \eqref{difference},
the effective weakly coupled description of $SU(N+M)\times SU(N)$ gauge theory exists only in a
finite-width energy band centered about $\Lambda$  --- one encounters Landau poles both in the
IR
\begin{equation}
g_2^2\to \infty\qquad {\rm as}\qquad \mu\to \mu_{IR}\equiv \Lambda e^{-\frac{\pi}{3g_s M}}\,,
\eqlabel{irsingularity}
\end{equation}
and the ultraviolet (UV),   
\begin{equation}
g_1^2\to \infty\qquad {\rm as}\qquad \mu\to \mu_{UV}\equiv \Lambda e^{+\frac{\pi}{3g_s M}}\,,
\eqlabel{uvsingularity}
\end{equation}
to leading order in $M^2/N^2$.
As explained in \cite{Klebanov:2000hb}, to extend the theory past the strong coupling regions one must
perform the self-similar transformations (Seiberg dualities \cite{Seiberg:1994pq}):
$N\to N-M$ for $\mu\lesssim \mu_{IR}$ and $N\to N+M$ for $\mu\gtrsim \mu_{UV}$. Thus, extension
of the effective $SU(N+M)\times SU(N)$ description to all energy scales involves an infinite sequence
--- a {\it cascade} --- of Seiberg dualities with the renormalization group flow of the effective rank
\cite{Buchel:2000ch,Krasnitz:2002ct,Aharony:2006ce}
\begin{equation}
N=N(\mu)\ \sim\ g_s M^2\ln \frac {\mu}{\Lambda}\,.
\eqlabel{neff}
\end{equation}
Although there are infinitely many duality steps in the UV, there is only a finite
number of the duality transformations as one flows to the IR --- when $N$ is an integer
multiple of $M$ (plus 1) one ends up in the IR with the $SU(M+1)$ gauge theory.  
The latter theory confined in the IR with a spontaneous breaking of the
$U(1)_R$ (chiral symmetry),
\begin{equation}
U(1)_R\to \zet_2\,.
\eqlabel{z2breaking}
\end{equation}
The IR properties of the cascading gauge theories were reviewed in \cite{Herzog:2001xk}
(see also \cite{Dymarsky:2005xt}); an important
feature of the theory is the characteristic scale in the glueball mass spectrum:
\begin{equation}
m_{glueball}\equiv \frac{\e^{2/3}}{Mg_s\a'}\,,
\eqlabel{defmglue}
\end{equation}
where $\e$ is a conifold deformation parameter of the holographic dual \cite{Klebanov:2000hb},
and $\a'=\ell_s^2$ is the string scale.

Previous studies focused on the fate of the chiral symmetry and  the confinement in
the cascading gauge theory at finite temperature.
At finite temperature, there are three different spatially homogeneous and isotropic phases of the theory.
We classify them as follows: 
\begin{itemize}
\item PhaseA$_s$ --- the deconfined phase with the unbroken chiral symmetry, \ie $U(1)$, 
see \cite{Buchel:2000ch,Buchel:2001gw,Gubser:2001ri,Aharony:2007vg}; 
\item PhaseA$_b$ --- the deconfined phase with the broken chiral symmetry, \ie $\zet_2$,
see \cite{Buchel:2010wp,Buchel:2018bzp};
\item PhaseB --- the confined phase with the broken chiral symmetry, \ie $\zet_2$,
see \cite{Klebanov:2000hb}.
\end{itemize}
Notice that confinement triggers the spontaneous breaking of the chiral symmetry \cite{Klebanov:2000hb}: there is no
spatially homogeneous and isotropic phase which is confined with $U(1)$ chiral symmetry. 
It will be instructive to have a geometrical classification of these phases, in the
warped-deformed conifold holographic dual of the theory \cite{Klebanov:2000hb,Aharony:2005zr,Buchel:2010wp}.
To this end, consider analytical continuation along the time direction $t\to t_E\equiv i t$.
Euclidean time $t_E$ is then periodically identified as
\begin{equation}
t_E\qquad \sim\qquad t_E+\frac 1T\,,
\eqlabel{te}
\end{equation}
where $T$ is the equilibrium temperature of the phase. Topologically, the compact directions of the
holographic dual are
\begin{equation}
\begin{split}
&{\rm unbroken\ chiral\ symmetry:}\qquad \underbrace{S^1}_{\rm thermal\ circle}\qquad\times\qquad
\underbrace{S^1\times S^2\times S^2}_{U(1)-{\rm symmetric}\ T^{1,1}}\,;\\
&{\rm broken\ chiral\ symmetry:}\qquad \underbrace{S^1}_{\rm thermal\ circle}\qquad\times\qquad
\underbrace{S^2\times S^3}_{\zet_2-{\rm symmetric}\ T^{1,1}}\,.
\end{split}
\eqlabel{themtop}
\end{equation}
We can thus geometrically characterize different phases depending on which cycle shrinks to
zero size in the interior of the ten-dimensional Euclidean type IIB supergravity dual: 
\begin{equation}
\begin{split}
&{\rm PhaseA}_s:\qquad \underbrace{S^1}_{\rm thermal\ circle}\to 0 \qquad \& \qquad {S^1\times S^2\times S^2}\ {\rm is\ finite}\,;\\
&{\rm PhaseA}_b:\qquad \underbrace{S^1}_{\rm thermal\ circle}\to 0 \qquad \& \qquad {S^2\times S^3}\ {\rm is\ finite}\,;\\
&{\rm PhaseB}:\qquad \underbrace{S^1}_{\rm thermal\ circle}\ {\rm is\ finite} \qquad \& \qquad S^2\to 0\qquad \&\qquad S^3\ {\rm is\ finite}\,.
\end{split}
\eqlabel{thermaltopology}
\end{equation}
According to \cite{Aharony:2007vg} there is the first-order confinement/deconfinement
phase transition between ${\rm PhaseA}_s$ and PhaseB at\footnote{The precise expression
for $\Lambda_{thermal}$ was reported in \cite{Bena:2019sxm}.}
\begin{equation}
T_{c}=0.614(1)\ \frac{\Lambda_{thermal}}{Pg_s^{1/2}}=0.614(1)\ \frac{3^{1/2}e^{1/3}}{2^{7/12}}\
\frac{\e^{2/3}}{P g_{s}^{1/2}}=0.220(2)\ g_s^{1/2} m_{glueball}\,,
\eqlabel{confdeconf}
\end{equation}
where the relation between $P$ and $M$ is given by \eqref{defpm} and $m_{glueball}$ is defined as in
\eqref{defmglue}. At temperature $T<T_c$ the phase ${\rm PhaseA}_s$ is metastable ---
it becomes perturbatively unstable  below  $T_\csb<T_c$ \cite{Buchel:2010wp},
\begin{equation}
T_{\csb}=0.542(0)\ \frac{\Lambda_{thermal}}{Pg_s^{1/2}}=0.194(3)\ g_s^{1/2} m_{glueball}\,.
\eqlabel{csb}
\end{equation}
The symmetry broken deconfined phase PhaseA$_b$ exists only
for  $T\ge T_{\csb}$ or for energy densities $\cale\le \cale_{\csb}$ \cite{Buchel:2018bzp},
\begin{equation}
\begin{split}
\cale_\csb =&1.270(1)\ \frac{\Lambda_{thermal}^4}{16\pi G_5}=1.270(1)\
\frac{2^{2/3}e^{4/3}}{192\pi^4}\ (Mg_s)^4\ m_{glueball}^4\\
&=4.089(6)\times 10^{-4}\ \times\ (Mg_s)^4\  m_{glueball}^4\,,
\end{split}
\eqlabel{ecrit}
\end{equation}
where $G_5$ is given by \eqref{g5deff}.
PhaseA$_b$ has larger thermal free energy density than that of the chirally symmetric deconfined phase
PhaseA$_s$ at the corresponding temperature,
and thus it does not dominate the canonical ensemble. On the other hand, PhaseA$_b$
is entropically favored over PhaseA$_s$ at the corresponding energy density,
and thus is the dominant phase in the microcanonical ensemble. According to
 \cite{Buchel:2018bzp} the phase PhaseA$_b$ is thermodynamically unstable, and thus it
 is dynamically (perturbatively) unstable towards developing spatial inhomogeneities
\cite{Buchel:2005nt}.

In this paper we would like to understand vacua of the cascading gauge theories in de Sitter space-time
(flat or closed spatial slicing\footnote{There is no difference between them at late times
as the curvature effects are diluted as $\propto\exp(-2 H t)$.})
\begin{equation}
ds_4^2=-dt^2+e^{2 H t} d\boldsymbol{x}^2\,,\qquad {\rm or}\qquad
ds_4^2=-dt^2 + \frac{1}{H^2}\cosh^2(Ht)\left(dS^3\right)^2\,,
\eqlabel{ds4metric}
\end{equation}
where $H$ is a Hubble constant. Specifically, we would like to provide
the classification of late-time states of the cascading gauge theory akin to
spatially homogeneous and isotropic thermal phases $\{$PhaseA$_s$, PhaseA$_b$, PhaseB$\}$
reviewed above. Of course there are crucial differences between the thermal equilibrium physics
and the late time de Sitter dynamics:
\nxt Thermodynamics can be studied in canonical or microcanonical
ensembles\footnote{As we emphasized above the thermal equilibrium phase structure is different in the two ensembles
of the cascading gauge theory.}. The latter one is
suitable to study the dynamics of the equilibration process. The de Sitter evolution of the gauge theory
states is eternally sourced by the space-time accelerated expansion and thus is (loosely)
equivalent to the microcanonical ensemble; there is no correspondence to the canonical ensemble.
\nxt Insisting on spatial homogeneity and isotropy, an initial state
typically\footnote{Not all strongly interacting systems
equilibrate. See \cite{Balasubramanian:2014cja} for a holographic example.} relaxes to a thermal
equilibrium configuration, which can be assigned a thermal (time-independent) entropy density. 
The holographic dynamics of the conformal gauge theories with a simple scale transformation
can be mapped to an evolution in Minkowski space-time \cite{Buchel:2017pto} ---
here the late-time de Sitter vacua are conformally equivalent to the equilibrium states
of the microcanonical ensemble. There is no equilibration of non-conformal gauge theories
at late-times in de Sitter \cite{Buchel:2017pto}\footnote{See also \cite{Buchel:2019qcq}
for a detailed recent analysis.}: the comoving entropy density production rate is nonzero.  
In \cite{Buchel:2017qwd} it was pointed out that the comoving entropy production rate $\calr$
can be attribute entirely to the spatial expansion
\[
{\rm volume}\bigg|_{physical}\ =\ e^{3 H t}\ {\rm volume}\bigg|_{comoving}\,,
\]
while the physical entropy density $s$ approaches a constant (time-independent)
entanglement entropy $s_{ent}$:
\begin{equation}
\lim_{t\to\infty} s\equiv s_{ent}= H^3\ \calr\,.
\eqlabel{limsent}
\end{equation}
In holography, the non-equilibrium entropy density $s=s(t)$ is associated with the
Bekenstein entropy of the dynamical apparent horizon (AH) \cite{Booth:2005qc,Figueras:2009iu}.
In \cite{Buchel:2017lhu} an example of a fully nonlinear holographic evolution from initially
homogeneous and isotropic state in de Sitter
was presented where the late-time dynamics approaches de Sitter vacuum  with entanglement
entropy \eqref{limsent}.

Implementing de Sitter holographic dynamics as in \cite{Buchel:2017lhu} for the cascading gauge theories
is outside the scope of this paper. Rather, as in \cite{Buchel:2017pto} and
\cite{Buchel:2019qcq}, we assume that we specify a well-defined spatially homogeneous and
isotropic initial state\footnote{We believe that restriction to
homogeneity and isotropy is not relevant for the late-time dynamics, given the
accelerated background space-time expansion.} (a well-defined
initial condition for the gravitation evolution)
in a holographic dual. This would correspond to some coarse grained state in the gauge theory
specified with the density matrix $\r$. We identify the von Neumann entropy $\cals$
\[
\cals=-\tr(\r\ln\r)\,,
\]
with the Bekenstein entropy of the AH in the holographic
dual\footnote{This procedure is implicit in all examples of holographic
evolutions in Chesler-Yaffe framework \cite{Chesler:2013lia}. Besides 'holographic quenches'
of background space-time \cite{Chesler:2008hg} (similar to de Sitter 'quenches' of interest here)
it was successfully applied to quenches of the coupling
constants of relevant operators in \cite{Buchel:2012gw,Buchel:2013lla}.}.
Partial differential equation of the
gravitational dual at late times reduce to system of ordinary differential equations \cite{Buchel:2017lhu}
which we analyze in details here. Inequivalent de Sitter vacua of the cascading gauge theory
are characterized with different values of the entanglement entropy density $s_{ent}$.
The true (dominant) vacuum is the one which results in the largest $s_{ent}$ for a fixed
Hubble constant $H$ and a fixed strong coupling scale of the theory $\Lambda$, see \eqref{deflambdaf},
\begin{equation}
\Lambda
=\frac{2^{1/6}e^{1/3}g_s^{1/2}}{3^{3/2}}\ m_{glueball} \ \approx\ 0.3 g_s^{1/2} m_{glueball} \,.
\eqlabel{lambdds}
\end{equation}

Parallel to classification of the thermal equilibrium states, we now explain topological/symmetry
considerations to classify de Sitter vacua of cascading gauge theory --- the discussion is more intuitive
for the closed spatial slicing in \eqref{ds4metric}.
To access AH (and thus to evaluate $s_{ent}$),
the dual gravitational bulk must be described in Eddington-Finkelstein (EF) coordinates. 
Fefferman-Graham (FG) coordinates cover only a patch of the former, which is
outside of the EF frame AH \cite{Buchel:2017lhu},
and thus is not suitable for the computation of the vacuum entanglement
entropy. Still, FG frame is useful to implement analytical continuation to
Euclidean (Bunch-Davies) vacuum 
\begin{equation}
-d\t^2 + \frac{1}{H^2}\cosh^2(H\t)\left(dS^3\right)^2\ \underbrace{\longrightarrow}_
{\t\to i\frac{\theta+\pi/2}{H}}\ \frac{1}{H^2}\biggl((d\theta)^2
+\sin^2(\theta)\ \left(dS^3\right)^2\biggr)=\frac{1}{H^2}\
\left(dS^4\right)^2\,.
\eqlabel{fgn1int}
\end{equation}
Topologically, the compact directions of the Euclidean FG frame holographic dual are
(compare with \eqref{themtop})
\begin{equation}
\begin{split}
&{\rm unbroken\ chiral\ symmetry:}\qquad \underbrace{S^4}_{dS_4^{\rm Euclidean}}\qquad\times\qquad
\underbrace{S^1\times S^2\times S^2}_{U(1)-{\rm symmetric}\ T^{1,1}}\,;\\
&{\rm broken\ chiral\ symmetry:}\qquad \underbrace{S^4}_{dS_4^{\rm Euclidean}}\qquad\times\qquad
\underbrace{S^2\times S^3}_{\zet_2-{\rm symmetric}\ T^{1,1}}\,.
\end{split}
\eqlabel{desittertop}
\end{equation}
Parallel to \eqref{thermaltopology}, we can geometrically characterize different 
de Sitter vacua of the cascading gauge theory depending on which cycle
shrinks to zero size in the interior of the ten-dimensional Euclidean FG frame
type IIB supergravity dual:
\begin{equation}
\begin{split}
&{\rm TypeA}_s:\qquad \underbrace{S^4}_{dS_4^{\rm Euclidean}}\to 0 \qquad \& \qquad {S^1\times S^2\times S^2}\ {\rm is\ finite}\,;\\
&{\rm TypeA}_b:\qquad \underbrace{S^4}_{dS_4^{\rm Euclidean}}\to 0 \qquad \& \qquad {S^2\times S^3}\ {\rm is\ finite}\,;\\
&{\rm TypeB}:\qquad \underbrace{S^4}_{dS_4^{\rm Euclidean}}\ {\rm is\ finite} \qquad \& \qquad S^2\to 0\qquad \&\qquad S^3\ {\rm is\ finite}\,.
\end{split}
\eqlabel{dstopology}
\end{equation}
To evaluate $s_{ent}$ we proceed in two steps\footnote{The same two-step procedure was also used in
computation of the de Sitter vacuum entanglement entropy in $\caln=2^*$ gauge theory in
\cite{Buchel:2019qcq}.}:
\begin{itemize}
\item first, we construct the FG frame vacua, subject to the 'boundary conditions'
\eqref{dstopology} (see appendix \ref{apb1} for the technical details);  
\item second, we use coordinate transformation to the EF frame for each of these vacua
(see \cite{Buchel:2017lhu} and appendix \ref{apb2} for the technical details),
and access the corresponding AH.
\end{itemize}

We summarize now our results:
\begin{itemize}
\item ${\rm TypeA}_s$ de Sitter vacua were studied previously in
\cite{Buchel:2001iu,Buchel:2002wf,Buchel:2013dla}. These vacua share resemblance with
the thermal deconfined chirally symmetric states of the cascading gauge theory, \ie PhaseA$_s$.
We find here that
\begin{equation}
s_{ent}(\Lambda,H)\bigg|_{{\rm TypeA}_s}\ne 0\,,
\eqlabel{sas}
\end{equation}
and vanishes as 
\begin{equation}
s_{ent}(\Lambda,H)\bigg|_{{\rm TypeA}_s}\ \propto\ H^3\ \biggl(\ln\frac{H^2}{\Lambda^2}\biggr)^{-3/4}
\qquad {\rm as} \qquad H\gg\Lambda\,,
\eqlabel{vanishconflimit}
\end{equation}
\ie  in the conformal limit.
${\rm TypeA}_s$ de Sitter vacua exist only when
\begin{equation}
H\ \gtrsim\ H_{min}^s\,,\qquad  H_{min}^s=0.7\Lambda\approx 0.2\ g_s^{1/2} m_{glueball}\,.
\eqlabel{hcritltypeas}
\end{equation}
As $\frac{H^2}{\Lambda^2}$ decreases, the Kretschmann scalar at the AH in the holographic dual increases,
making supergravity approximation less reliable. $H_{min}^s$ in \eqref{hcritltypeas} should be interpreted
as the value of the Hubble constant at which the supergravity approximation breaks down.
We identify the rapid growth of the curvature in the gravitational dual to TypeA$_s$ de Sitter vacua
with collapsing of the compact manifold (a deformed $T^{1,1}$) at the location of the apparent
horizon --- as a result, $s_{ent}$ vanishes in this limit as well.
\item ${\rm TypeA}_b$ de Sitter vacua are constructed here for the first
time\footnote{We introduce a novel technique used to identify phases/vacua with
spontaneously broken symmetry.}. These vacua share resemblance with
the thermal deconfined states of the cascading gauge theory with the spontaneously broken chiral symmetry, \ie PhaseA$_b$.
We find here that
\begin{equation}
s_{ent}(\Lambda,H)\bigg|_{{\rm TypeA}_b}\ne 0\,.
\eqlabel{sab}
\end{equation}
${\rm TypeA}_b$ de Sitter vacua exist only when
\begin{equation}
H\ \ge\ H_{min}^b\,,\qquad  H_{min}^b=0.92(1)\Lambda\approx 0.276\ g_s^{1/2} m_{glueball}\,.
\eqlabel{hcritl}
\end{equation}
As $\frac{H^2}{\Lambda^2}$ increases, the Kretschmann scalar at the AH in the holographic dual increases,
making the supergravity approximation less reliable.
\item We find that while
\begin{equation}
s_{ent}(\Lambda,H_{min}^b)\bigg|_{{\rm TypeA}_s}=s_{ent}(\Lambda,H_{min}^b)\bigg|_{{\rm TypeA}_b}\,,
\eqlabel{bsequality}
\end{equation}
de Sitter vacua with the spontaneously broken chiral symmetry are entropically favored within a narrow window 
for the values of the Hubble constant
\begin{equation}
\begin{split}
&s_{ent}(\Lambda,H)\bigg|_{{\rm TypeA}_b}\ \ge\ s_{ent}(\Lambda,H)\bigg|_{{\rm TypeA}_s}\,,\qquad  H_{max}\ge H\ge H_{min}^b\,, 
\end{split}
\eqlabel{bswindow}
\end{equation}
where
\begin{equation}
H_{max}=0.92(5)\Lambda\approx 0.278\ g_s^{1/2} m_{glueball}\,.
\eqlabel{defhm}
\end{equation}
${\rm TypeA}_b$ de Sitter vacua continue to exist for $H>H_{max}$, however they have smaller $s_{ent}$
compare to the corresponding ${\rm TypeA}_s$ de Sitter vacua. 
\item ${\rm TypeB}$ de Sitter vacua were studied previously in
\cite{Buchel:2013dla}.  These vacua share resemblance with
the thermal confined states of the cascading gauge theory with the spontaneously broken chiral symmetry, \ie PhaseB.
We find here that
\begin{equation}
s_{ent}(\Lambda,H)\bigg|_{{\rm TypeB}}=0\,.
\eqlabel{sb}
\end{equation}
We emphasize that \eqref{sb} does not mean that the  coarse grained entropy of the cascading gauge theory
vanishes --- in fact, during de Sitter evolution the  entropy production rate is always positive
(see section \ref{theorem}). What \eqref{sb} states is that the comoving entropy production rate in
TypeB vacuum vanishes at late times (much like it does  in conformal gauge theories \cite{Buchel:2017lhu}).
As a result, TypeB vacuum is never realized as the late-time attractor of a dynamical evolution for
a generic cascading gauge theory state in de Sitter, provided vacua TypeA$_s$ or TypeA$_b$ exist.
Neither of the latter vacua exists for $H\lesssim H_{min}^s$, see \eqref{hcritltypeas},
thus\footnote{While this is likely to be true in general, the statement is strictly precise for
the de Sitter evolution of spatially homogeneous and isotropic states of the cascading gauge theory.}
\begin{equation}
{ TypeB\ de\ Sitter\ vacum\ is\ a\ late-time\ attractor\ provided}\ H\lesssim  H_{min}^s\,.
\eqlabel{typebdom}
\end{equation}
\end{itemize}
Of cause, \eqref{typebdom} implies that TypeB vacua must exist at least for $H> H_{min}^s$; in fact
we find (see section \ref{typebsugra}) that TypeB vacua exist\footnote{This should be understood
in the same sense as existence of TypeA$_s$ vacua: the supergravity approximation
used to construct TypeB vacua is robust against higher-derivative $\a'$ corrections
from the full string theory.} for
\begin{equation}
H\lesssim H_{max}^B\,,\qquad H_{max}^B=0.966(5) \Lambda\ > H_{min}^s=0.7\Lambda\,.
\eqlabel{typebexists}
\end{equation}

Eqs.~\eqref{bswindow} and \eqref{typebdom} represents our main, and
somewhat unexpected result:

\bigskip

\noindent\fbox{%
    \parbox{\textwidth}{%
{\color{red} $SU(N)\times SU(N+M)$ cascading gauge theory with a strong coupling scale $\Lambda$ undergoes spontaneous
chiral symmetry breaking in de Sitter space time with a Hubble constant $H$ provided
\[
H\lesssim  H_{min}^s<H_{min}^b\qquad \&\qquad  H_{min}^b\ \le\ H\ \le H_{max}\,.
\]
The critical values $H_{min}^s$, $H_{min}^b$ and $H_{max}$ are of order the strong coupling scale of the theory $\Lambda$.}
}%
}
\bigskip

The rest of the paper is organized as follows. In section \ref{action} we discuss holographic
dual effective action of cascading gauge theory.  Section \ref{action} contains a guide to
set of Appendices with technical details. Cascading gauge theory de Sitter vacuum
entanglement entropy is identified with the Bekenstein entropy of the AH in the holographic dual
at late times, see section \ref{ah}. In section \ref{ah10} we identify AH in ten dimensional
holographic dual and compute its area density. In section \ref{ah5} we establish that both the
location of the AH and its associated entropy density is invariant upon Kaluza-Klein reduction
on the warped-deformed $T^{1,1}$. In section \ref{theorem} we prove a theorem that as long as
the background geometry of the holographic dual is nonsingular, the area density of the
AH does not decrease with time. In section \ref{sentb} we show that whenever vacua of TypeB
exist, their entanglement entropy vanishes, see \eqref{sb}. Section \ref{typeasv}
devoted to TypeA$_s$ de Sitter vacua. Numerical results are presented in section \ref{typeasnum}:
we construct first the dual holographic backgrounds in the FG frame, transform them to the EF frame,
identify the location of the apparent horizon and compute the vacuum entanglement entropy,
see fig.~\ref{comparesent}. At each step we triple-check the numerical results
by making use of distinct and independent computational schemes, see appendix \ref{apc}.
Comparison of the results from the different
computational schemes in the overlapping regions of the parameter space
is shown in figs.~\ref{figure2},\ref{figure2c},\ref{errorsent}. In section
\ref{typeasc} we make use of the computational SchemeII to discuss the conformal
limit of TypeA$_s$ vacua, \ie $H\gg \Lambda$, and establish \eqref{vanishconflimit}.
The validity of the supergravity approximation of the holographic dual to TypeA$_s$
de Sitter vacua is discussed in section \ref{typeashmin}. We establish a rapid
growth of the Kretschmann scalar of the background geometry \eqref{ef1i}
evaluated at the AH for small values of $\frac{H^2}{\Lambda^2}$,
and associate this growth  with ``collapsing'' of the deformed $T^{1,1}$,
see figs.~\ref{k} and \ref{t11size}. Extrapolating the numerical data,
we estimate the value of the Hubble constant $H_{min}^s$, see \eqref{hcritltypeas},
when the Kretschmann scalar diverges --- we take this value as a limiting value of
$H$ below which TypeA$_s$ vacua stop existing.
We study TypeA$_b$ vacua with the spontaneously broken chiral symmetry in section \ref{typeabv}.
We begin in section \ref{onsetab} with identification of the critical value $H_{min}^b$,
see \eqref{hcritl}, below
which TypeA$_b$ vacua do not exist. This is done computing the linearized chiral symmetry breaking
perturbations on top of TypeA$_s$ vacua with the explicit symmetric breaking parameter --- the gaugino
mass term. At this critical value $H=H_{min}^b$ all the symmetry breaking expectation values
diverge, see fig.~\ref{flucuvir}. We explain how TypeA$_b$ vacua,
with the spontaneous symmetry breaking, can be constructed at values of the
Hubble constant close to $H_{min}^b$ using the linearized perturbations on top of TypeA$_s$ vacua with
the explicit symmetry breaking. Numerical construction of
TypeA$_b$ vacua in section \ref{typeabnum} follows the discussion of section \ref{typeasnum}.
Section \ref{typeabnum} contains the central result of the paper --- fig.~\ref{comparesentasab}:
it establishes that the chiral symmetry breaking of the cascading gauge theory in de Sitter space-time
occurs in a narrow range of values of the Hubble constant, see \eqref{bswindow}.  
The validity of
the supergravity approximation of the holographic dual to TypeA$_b$ de Sitter vacua is
discussed in section \ref{typeabsugra}.
TypeB de Sitter vacua are discussed in section \ref{typebv}. These vacua have vanishing
entanglement entropy \eqref{sb}; however, they exist for arbitrary small $\frac{H}{\Lambda}$,
approaching the extremal Klebanov-Strassler solution \cite{Klebanov:2000hb} as $\frac{H}{\Lambda}\to 0$.
We discuss TypeB vacua, first as a deformation of the extremal KS solution, and followed later
by the numerical construction in two different computational schemes in section \ref{typebnum}. 
In section \ref{typebsugra} we present an indication that TypeB vacua exist only
for $H\lesssim H_{max}^B$ \eqref{typebexists} --- in this limit the 3-cycle of the dual geometry
supporting the RR 3-form flux becomes vanishingly small in string units, making the
supergravity approximation not reliable as indicated by the rapid growth of the Kretschmann scalar of the
background geometry  evaluated at the AH, see fig.~\ref{kTypeB}. Since
both TypeA$_s$ and TypeA$_b$ vacua cease to exist below certain value of the Hubble constant,
specifically for $H\lesssim H_{min}^s$, and $H_{max}^B>H_{min}^s$, TypeB vacua become late-time
attractors of the dynamical evolution of the cascading gauge theory in de Sitter for
$H\lesssim H_{min}^s$. We conclude in section \ref{conclude} highlighting open questions and
future directions.

\section{Dual effective actions of the cascading gauge theory}\label{action}

Consider $SU(2)\times SU(2)\times \zet_2$ invariant states of
the cascading gauge theory on a 4-dimensional manifold
$\calm_4\equiv \del\calm_5$. In the planar limit and at large 't Hooft
coupling, one can consistently truncate the theory to a finite number of
operators \cite{Buchel:2010wp}: a stress-energy tensor $T_{ij}$,
a pair of dimension-3 operators $\calo_3^{\a=\{1,2\}}$
(dual to gaugino condensates for each of the gauge group factors),
a pair of dimension-4 operators $\calo_4^{\b=\{1,2\}}$,
and dimension-6,7,8 operators $\calo_6,\calo_7,\calo_8$. 
Effective gravitational action on a 5-dimensional manifold
$\calm_5$ describing holographic dual of such states was derived
in \cite{Buchel:2010wp}:
\begin{equation}
\begin{split}
&S_5\left[g_{\mu\nu}\leftrightarrow T_{ij},\{\Omega_i,h_i,\Phi\}\leftrightarrow
\{\calo_3^\a,\calo_4^\b,\calo_6,\calo_7,\calo_8\}\right]= \frac{108}{16\pi G_5} 
\int_{\calm_5} {\rm vol}_{\calm_5}\ \Omega_1 \Omega_2^2\om_3^2\ \times\\
&\times \biggl\lbrace 
 R_{10}-\frac 12 \left(\nabla \Phi\right)^2
-\frac 12 e^{-\Phi}\left(\frac{(h_1-h_3)^2}{2\om_1^2\om_2^2\om_3^2}+\frac{1}{\om_3^4}\left(\nabla h_1\right)^2
+\frac{1}{\om_2^4}\left(\nabla h_3\right)^2\right)
\\
&-\frac 12 e^{\Phi}\left(\frac{2}{\om_2^2\om_3^2}\left(\nabla h_2\right)^2
+\frac{1}{\om_1^2\om_2^4}\left(h_2-\frac P9\right)^2
+\frac{1}{\om_1^2\om_3^4} h_2^2\right)
\\
&-\frac {1}{2\Omega_1^2\Omega_2^4\om_3^4}\left(4\Omega_0+ h_2\left(h_3-h_1\right)+\frac 19 P h_1\right)^2
\biggr\rbrace\,,\\
\end{split}
\eqlabel{5action}
\end{equation}
where $\Omega_0$ is a constant in the definition of the
5-form flux\footnote{In the limit of vanishing
3-form fluxes, $\Omega_0=\frac{L^4}{108}$, where $L$ is the asymptotic $AdS_5$ radius.},
see \eqref{fluxes}, $R_{10}$ is given by
\begin{equation}
\begin{split}
R_{10}=R_5&+\left(\frac{1}{2\om_1^2}+\frac{2}{\om_2^2}+\frac{2}{\om_3^2}-\frac{\om_2^2}{4\om_1^2\om_3^2}
-\frac{\om_3^2}{4\om_1^2\om_2^2}-\frac{\om_1^2}{\om_2^2\om_3^2}\right)-2\Box \ln\left(\om_1\om_2^2\om_3^2\right)\\
&-\biggl\{\left(\nabla\ln\om_1\right)^2+2\left(\nabla\ln\om_2\right)^2
+2\left(\nabla\ln\om_3\right)^2+\left(\nabla\ln\left(\om_1\om_2^2\om_3^2\right)\right)^2\biggr\}\,,
\end{split}
\eqlabel{ric5}
\end{equation}
and $R_5$ is the five-dimensional Ricci scalar of the metric 
\begin{equation}
ds_{5}^2 =g_{\mu\nu}(y) dy^{\mu}dy^{\nu}\,,
\eqlabel{5met}
\end{equation}
that forms part of the ten dimensional full metric
\begin{equation}
ds_{10}^2 = ds_{5}^2 + ds^2_{T^{1,1}}\,, \qquad ds^2_{T^{1,1}} = \Omega_1^2(y) g_5^2 + \Omega_2^2(y) (g_3^2 + g_4^2) + \Omega_3^2(y) (g_1^2 + g_2^2).
\eqlabel{10dmetric}
\end{equation}

One-forms $\{g_i\}$ (for $i=1,\cdots,5$) are the usual forms defined in the warped-squashed $T^{1,1}$ and are given as in \cite{Buchel:2010wp}, for coordinates $0 \leq \psi \leq 4 \pi$, $0 \leq \theta_a \leq \pi$ and $0 \leq \phi_a \leq 2 \pi$ ($a=1,2$). All the covariant derivatives $\nabla_\lambda$  are
with respect to the metric \eqref{5met}. Fluxes (and dilaton $\Phi$) are parameterized in such a way that functions $h_1(y), h_2(y), h_3(y)$ appear as
\begin{equation}
\begin{split}
F_5 &= \calf_5+\star \calf_5\,,\\
\calf_5 &=\left(4\Omega_0+h_2(y) (h_3(y)-h_1(y)) +\frac P9 h_1(y)\right)\
g_1\wedge g_2\wedge g_3\wedge g_4\wedge g_5\,,
\\
B_2 & =  h_1(y) g_1 \wedge g_2 + h_3(y) g_3 \wedge g_4, \\
F_3=&\frac 19 P\ g_5\wedge g_3\wedge g_4+h_2(y)\ \left(g_1\wedge g_2-g_3\wedge g_4\right)\wedge g_5
\\
&\qquad +\left(g_1\wedge g_3+g_2\wedge g_4\right)\wedge d\left(h_2(y)\right)\,,\\
\Phi & = \Phi (y),
\end{split}
\eqlabel{fluxes}
\end{equation}
Parameter $P$ must be appropriately quantized \cite{Aharony:2007vg,Herzog:2001xk}:
\begin{equation}
\frac{1}{4\pi^2\a'}\ \int_{\rm 3-cycle:\ \theta_2=\phi_2=0} F_3\qquad =\qquad \frac{2P}{9\a'}\ \in \zet\,,
\eqlabel{pquantization}
\end{equation}
thus
\begin{equation}
P=\frac 92 M\a'\,,
\eqlabel{defpm}
\end{equation}
corresponding to the number $M$ of fractional branes
(the difference of ranks of the cascading gauge theory gauge group factors)
on the conifold. Finally, $G_5$ is the five dimensional effective gravitational constant  
\begin{equation}
G_5\equiv \frac{G_{10}}{{\rm{vol}}_{T^{1,1}}}=\frac{27}{16\pi^3}\ G_{10}\,,
\eqlabel{g5deff}
\end{equation}
where $16 \pi G_{10}=(2\pi)^7(\a')^4$ is  10-dimensional gravitational constant of 
type IIB supergravity.

Chirally symmetric 
states of the cascading gauge theory correspond to enhancement of the
global symmetry\footnote{In the planar limit.} $SU(2)\times SU(2)\times \zet_2\to
SU(2)\times SU(2)\times U(1)$,
and are described by the gravitational configurations of \eqref{5action} 
subject to constraints\footnote{This is a consistent truncation
of the cascading gauge theory to $U(1)$ symmetric
sector constructed in \cite{Aharony:2005zr}.}   
\begin{equation}
h_1=h_3\,,\qquad h_2=\frac{P}{18}\,,\qquad \om_2=\om_3\,,
\eqlabel{cinv}
\end{equation}
or in the boundary QFT language \cite{Buchel:2010wp},
\begin{equation}
\calo_3^\a=0\,,\qquad \calo_7=0\,.
\eqlabel{symmetriclimit}
\end{equation}

We find it convenient to introduce 
\begin{equation}
\begin{split}
h_1=&\frac 1P\left(\frac{K_1}{12}-36\Om_0\right)\,,\qquad h_2=\frac{P}{18}\ K_2\,,\qquad 
h_3=\frac 1P\left(\frac{K_3}{12}-36\Om_0\right)\,,\\
\Om_1=&\frac 13 f_c^{1/2} h^{1/4}\,,\qquad \Om_2=\frac {1}{\sqrt{6}} f_a^{1/2} h^{1/4}\,,\qquad 
\Om_3=\frac {1}{\sqrt{6}} f_b^{1/2} h^{1/4}\,.
\end{split}
\eqlabel{redef}
\end{equation}

The ultimate goal is to compute the entanglement entropy of the cascading gauge theory ---
using the dual holographic picture with the effective gravitational action \eqref{5action} ---
in distinct vacua (see \eqref{dstopology}) in four dimensional de Sitter space-time.
As explained in the introduction, this is done in two steps:
\nxt constructing de Sitter vacua in Fefferman-Graham coordinate frame
\begin{equation}
\begin{split}
&ds_{10}^2 = \frac{1}{h^{1/2}\r^{2}} \left(-d\t^2 + e^{2 H \t}d\boldsymbol{x}^2\right)+
\frac{h^{1/2}}{\r^{2}} \left(d\r\right)^2\\
&\qquad\qquad\qquad\qquad+ \frac {f_c h^{1/2}}{9}  g_5^2
+ \frac{f_a h^{1/2}}{6} (g_3^2 + g_4^2) + \frac{f_b h^{1/2}}{6} (g_1^2 + g_2^2)\,,\\
&h=h(\r)\,,\qquad f_{a,b,c}=f_{a,b,c}(\r)\,,
\end{split}
\eqlabel{fg1i}
\end{equation}
subject to appropriate topological/symmetry restrictions \eqref{dstopology};
\nxt using diffeomorphism transformation to represent the FG frame vacua in
Eddington-Finkelstein coordinate frame 
\begin{equation}
\begin{split}
&ds_{10}^2 = 2dt\ \left(dr-a\ dt\right)+\s^2 e^{2 Ht}\ d\boldsymbol{x}^2 + \frac 19 w_{c2}\ g_5^2
+ \frac16\w_{a2}\ (g_3^2 + g_4^2) + \frac16\w_{b2}\ (g_1^2 + g_2^2)\,,\\
&a=a(r)\,,\qquad \s=\s(r)\,,\qquad \w_{a2,b2,c2}=\w_{a2,b2,c2}(r)\,.
\end{split}
\eqlabel{ef1i}
\end{equation}
It is important to keep in mind that EF frame vacua \eqref{ef1i} are the late-time
limits of the evolution in EF frame:
\begin{equation}
\begin{split}
&ds_{10}^2 = 2dt\ \left(dr-A\ dt\right)+\Sigma^2\ d\boldsymbol{x}^2 + \Omega_1^2 \ g_5^2
+ \Omega_2^2\ (g_3^2 + g_4^2) + \Omega_3^2\ (g_1^2 + g_2^2)\,,\\
&A=A(t,r)\,,\qquad \Sigma=\Sigma(t,r)\,,\qquad \Omega_{1,2,3}
=\Omega_{1,2,3}(t,r)\,.
\end{split}
\eqlabel{ef2i}
\end{equation}
We now summarize technical details delegated to various Appendices.
\begin{itemize}
\item In appendix \ref{efeoms} we derive the equations of motion in the holographic bulk for the
evolution of generic spatially homogeneous and isotropic state of the cascading gauge theory
in de Sitter space-time, see \eqref{ev1}-\eqref{con2}. We explain how to take
the late time limit $t\to \infty$ in \eqref{ef2i} to obtain \eqref{ef1i}.
The EF frame vacuum equations of motion are given by \eqref{efv1}-\eqref{efc2}.
The latter equations of motion have symmetries SEF1-SEF4 \eqref{efsym1}-\eqref{efsym4},
which are used to set up and validate numerics (see appendix \ref{apc}).
\item We begin appendix  \ref{apb} presenting gravitational bulk equations of motion in
FG frame \eqref{kseq2}-\eqref{kseq10}. These equations of motion have
(corresponding to SEF1-SEF4)
symmetries SFG1-SFG4 \eqref{efsym1}-\eqref{efsym4}, which are used to set up and
validate numerics (see appendix \ref{apc}). In appendix \ref{apb1} we explain the near
boundary (UV) $\r\to 0$ and the interior (IR) $\r\to \infty$ asymptotics. UV 
asymptotics are used to classify non-normalizable coefficients (defining parameters of the
cascading gauge theory): the asymptotic string coupling $g_s$ \eqref{sum} and the
strong coupling scale $\Lambda$ of the theory \eqref{lambdds}, and the normalizable coefficients: the
expectation values of boundary gauge theory
operators\footnote{Developing the precise holographic dictionary between these normalizable
coefficients and the corresponding expectation values, while interesting, is not
important for the results presented, and thus is outside the scope of the paper.}:
$\{T_{ij},\calo_3^{\a=\{1,2\}},\calo_4^{\b=\{1,2\}},\calo_6,\calo_7,\calo_8\}$. IR
asymptotics are used to classify the distinct de Sitter vacua of the theory
\eqref{dstopology},  as well to ensure that the bulk geometry is smooth
as the corresponding cycles shrinks to zero size ($S^4$ for TypeA$_s$ and TypeA$_b$, and $S^2$ for TypeB
vacua).
\item TypeA$_s$ vacua enjoy unbroken chiral symmetry; appendix \ref{apb11}
presents the UV and IR asymptotics in FG frame obtained in \cite{Buchel:2013dla} and
translates the coefficients governing the expansion to those used for the characterization of
TypeA$_b$ vacua,
see \eqref{map1}-\eqref{map5}.
\item appendix \ref{apb2} establishes the map between EF and FG frame description
for each type of the vacua:  TypeA$_s$, TypeA$_b$ and  TypeB.
\item In the limit $H\to 0$, TypeB vacuum in FG frame represents the extremal
KS solution \cite{Klebanov:2000hb}. We use this limit in appendix \ref{ksextremal}
to related the strong coupling scale $\Lambda$ of the cascading gauge theory
to the complex structure conifold deformation parameter $\epsilon$
used in  \cite{Klebanov:2000hb}, see \eqref{deflambdaf}.
\item appendix \ref{apc} covers numerical procedures for construction of FG frame dual backgrounds
(see \ref{apcfg}) and EF frame dual backgrounds (see \ref{apcef}). We introduce three
different computational schemes --- SchemeI, SchemeII and SchemeIII \eqref{compschemes} ---
explain how they are related and outline their computational advantages
in accessing different regions of the parameter space of the model. We introduce the
AH location function $\call_{AH}$ \eqref{defcall}, used to identify the apparent horizon. 
\item appendix \ref{apd} presents technical details for construction of TypeA$_s$ de
Sitter vacua in computational scheme SchemeII in the conformal limit, \ie $b\to 0$.
\item appendix \ref{kretschmann} collects the expression for the Kretschmann scalar \eqref{defK}
of the background geometry \eqref{ef1i}.  It is used to test the validity of the supergravity
approximation. 
\item appendix \ref{lincsb} contains equations of motion and the asymptotic expansions
for the chiral symmetry breaking perturbations about FG frame TypeA$_s$ de Sitter vacua with explicit
symmetry breaking parameter --- the gaugino mass term. These perturbations are used to identify
TypeA$_b$ vacua "close'' to TypeA$_s$ vacua.
\end{itemize}

\section{Apparent horizon in de Sitter evolution of the cascading gauge theory}\label{ah}

Apparent horizon\footnote{In general AH
is observer dependent. It is natural to define AH with respect to an observer reflecting the symmetries of the spatial slices --- homogeneity and isotropy in $\boldsymbol{x}$ in \eqref{ef2i}, see \cite{Chesler:2013lia}.
Such an identification correctly reproduces the hydrodynamic
limit \cite{Buchel:2016cbj}
and can be proven to comply with the second law of thermodynamics \cite{Buchel:2017pto,Buchel:2017lhu},
thus serving as a useful definition of the dynamical (nonequilibrium)
entropy.}  in holographic dual
is crucial for identifying the attractor vacuum for the evolution
of generic homogeneous and isotropic states of the cascading gauge theory
in de Sitter: given competing trajectories for the evolution, dynamics proceeds
along trajectory resulting in the maximum entropy at late times.
We identify AH directly in ten-dimensional
EF frame gravitational dual in section \ref{ah10}. We reproduce the same result
in EF gravitational dual of the effective five-dimensional
description in section \ref{ah5}.
Both in ten-dimensions and upon Kaluza-Klein reduction to five
dimensions the area of the AH stays the same. In section \ref{theorem}
we use equations of motion \eqref{ev1}-\eqref{con2} to
prove that the area of the AH is nondecreasing upon evolution.
We identify the (dynamical)
area density of the AH $\cala_{10}(t)$ with the dynamical 
entropy density $s$ of the boundary gauge theory as 
\begin{equation}
a^3s=e^{3Ht} s(t)=\frac{\cala_{10}}{4 G_{10}}=\frac{4\pi}{(2\pi)^7(\a')^4}\ \cala_{10}(t)\,,
\eqlabel{sentgen}
\end{equation}
where $a=e^{Ht}$ is the boundary spatial metric scale factor, see \eqref{ds4metric}.  
The entanglement entropy $s_{ent}$ is related to the late-time limit
of $s$ as
\begin{equation}
\begin{split}
&\lim_{t\to\infty}\ \frac{1}{H^3a^3}\frac{d}{dt}\left(a^3 s\right)\equiv 3 H\times \calr\,,\\
&\lim_{t\to\infty} s(t)\equiv s_{ent}=H^3\calr\,,
\end{split}
\eqlabel{stosent}
\end{equation}
where $\calr$ is the comoving entropy production rate in de Sitter vacuum
first introduced in \cite{Buchel:2017pto}. 
Finally, in section \ref{sentb} we show that
\begin{equation}
\calr\bigg|_{\rm TypeB}=0\qquad\ \Longrightarrow\qquad
s_{ent}\bigg|_{\rm TypeB}=0\,.
\eqlabel{senttypeb}
\end{equation}

\subsection{AH in ten dimensions}\label{ah10}

The apparent horizon  of the bulk gravitational dual to the cascading gauge theory dynamics in
de Sitter is located at the radius $r=r_{AH}$ where the expansion $\theta$ of a congruence
of outward pointing null vectors vanishes (\ie it stops expanding outwards). Working
in the coordinates of equation \eqref{ef2i},  we characterize such a congruence with the null
vector $k=\del_t+A \del_r$. The null vector $k$ points toward the boundary of the space-time
outside of the initial black hole, and points inward inside the initial horizon.

Following \cite{Poisson:2009pwt},
the expansion of a congruence of affine parameterized null vectors $n$ is given by
\begin{equation}
\theta=\nabla_{\a}n^{\a}.
\end{equation}
However, it turns out that $k^{\b}\nabla_{\b}k^{\a}=\del_r A\ k^{\a}$, \ie $k$ is not affine.
To remedy this, we rescale $k$ by $\exp\{\int \del_rA\ d\l\}$, where $\l$ is the parameter along
which the congruence $k$ evolves.  This ensures that the rescaled null vector satisfies the
geodesic equation with $\lambda$ as an affine parameter.
Reference  \cite{Poisson:2009pwt} then gives the expansion of $k$ to be
\begin{equation} \eqlabel{expansion}
\theta = \exp\left[\int \del_rA\ d\l\right]\left(\nabla_{\a}k^{\a}-\del_rA\right).
\end{equation}
Substituting in for $\nabla_{\a}k^{\a}$ computed in the metric  \eqref{ef2i}
\begin{equation}
\nabla_{\a}k^{\a}=\frac{1}{\sqrt{-g}}\del_\a\left(\sqrt{-g} k^\a\right)
=\del_t\ln\left({\Sigma^3\Omega_1\Omega_2^2\Omega_3^2}\right)
+A\ \del_r\ln\left({\Sigma^3\Omega_1\Omega_2^2\Omega_3^2}\right)+\del_rA\,.
\eqlabel{nablak}
\end{equation}
We see that $\theta=0$, when
\begin{equation} \eqlabel{ahloc10}
\del_t\left({\Sigma^3\Omega_1\Omega_2^2\Omega_3^2}\right)
+A\ \del_r\left({\Sigma^3\Omega_1\Omega_2^2\Omega_3^2}\right)\bigg|_{r=r_{AH}}=0\,.
\end{equation}
Eq.~\eqref{ahloc10} determines the location of the AH, \ie $r_{AH}=r_{AH}(t)$.
The area density of the AH $\cala_{10}$ is
\begin{equation}
\cala_{10}=\Sigma^3\Omega_1\Omega_2^2\Omega_3^2\bigg|_{r=r_{AH}}\ \int g_5\wedge g_3\wedge g_4\wedge g_1\wedge g_2=
64\pi^3 \Sigma^3\Omega_1\Omega_2^2\Omega_3^2\bigg|_{r=r_{AH}}\,,
\eqlabel{a1010}
\end{equation}
leading to (see \eqref{sentgen})
\begin{equation}
e^{3Ht}s=\frac{64\pi^3}{4 G_{10}}\ \Sigma^3\Omega_1\Omega_2^2\Omega_3^2\bigg|_{r=r_{AH}}=\frac{1}{4 G_5}\
108\Sigma^3\Omega_1\Omega_2^2\Omega_3^2\bigg|_{r=r_{AH}}\,.
\eqlabel{s10dyn}
\end{equation}

\subsection{AH in Kaluza-Klein reduction to five dimensions}\label{ah5}

We would like to reproduce \eqref{ahloc10} and \eqref{s10dyn} from the five-dimensional perspective.

While the effective action \eqref{5action} is five dimensional, the metric frame used is not Einstein:
\begin{equation}
S_5=\frac{108}{16\pi G_5}\int_{\calm_5}\ {\rm vol}_{\calm_5}\ \Om_1\Om_2^2\Om_3^2\times \biggl\{R_5+\cdots\biggr\}\,.
\eqlabel{eh5}
\end{equation}
This can be fixed with a simple conformal rescaling:
introducing
\begin{equation}
d\tilde{s}_5^2\equiv \tilde{g}_{\mu\nu}dy^\mu dy^\nu=\Om^{10/3}\ ds_5^2
=\Om^{10/3}\ g_{\mu\nu}dy^\mu dy^\nu\,,\qquad \Om^5=\Om_1\Om_2^2\Om_3^2\,,
\eqlabel{ds5t}
\end{equation}
and defining
\begin{equation}
\tilde{G_5}=\frac{G_5}{108}\,,
\eqlabel{tg5}
\end{equation}
the effective action $S_5$ in \eqref{eh5} has now a standard Einstein-Hilbert term with respect to $\tilde{g}$
\begin{equation}
S_5=\frac{1}{16\pi \tilde{G}_5}\int_{\calm_5}\ {\rm \widetilde{vol}}_{\calm_5}\ \times \biggl\{\tilde{R}_5+\cdots\biggr\}\,.
\eqlabel{teh5}
\end{equation}
The new EF frame (compare with \eqref{ef2i}) becomes
\begin{equation}
\begin{split}
&d\tilde{s}_5^2=\Om^{10/3}\biggl[2dt\ \left(dr-A\ dt\right)+\Sigma^2\  d\boldsymbol{x}^2\biggr]
=2dtd\hat{r}-2A\Om^{10/3}\ dt^2+\Om^{10/3}\Sigma^2\  d\boldsymbol{x}^2\,,\\
&d\hat{r}=\Om^{10/3}\ dr\,,
\end{split}
\eqlabel{newef}
\end{equation}
where the second equality defines a new radial coordinate $\hat{r}$.
The congruence of null geodesics is now characterized with
\begin{equation}
\tilde{k}=\del_t+ A\Om^{10/3}\ \del_{\hat{r}}\,,
\eqlabel{tk}
\end{equation}
so that 
\begin{equation}
\tilde{k}^\b \tilde{\nabla}_\b \tilde{k}^\a=\del_{\hat{r}}\left(A\Om^{10/3}\right)\ \tilde{k}^\a\,.
\eqlabel{ntk}
\end{equation}
Since
\begin{equation}
\sqrt{-\tilde{g}}=\Om^5 \Sigma^3\,,
\eqlabel{detgt}
\end{equation}
we have
\begin{equation}
\tilde{\nabla}_\a \tilde{k}^\a=\del_t\ln\left(\Om^5 \Sigma^3\right)+
A\Om^{10/3}\ \del_{\hat{r}}\ln\left(\Om^5 \Sigma^3\right)+\del_{\hat{r}}\left(A\Om^{10/3}\right)\,.
\eqlabel{nablatk}
\end{equation}
For the expansion $\tilde{\theta}$ of the congruence of affine parameterized null vectors we have
(compare with \eqref{expansion})
\begin{equation}
\begin{split}
&\tilde{\theta}\ \propto\ \biggl(\tilde{\nabla}_\a \tilde{k}^\a-\del_{\hat{r}}\left(A\Om^{10/3}\right)\biggr)
=\del_t\ln\left(\Om^5 \Sigma^3\right)+
A\Om^{10/3}\ \del_{\hat{r}}\ln\left(\Om^5 \Sigma^3\right)\\
&=\del_t\ln\left(\Om^5 \Sigma^3\right)+
A\ \del_{r}\ln\left(\Om^5 \Sigma^3\right)=\del_t\ln\left(\Sigma^3\Om_1\Om_2^2\Om_3^2\right)+
A\ \del_{r}\ln\left(\Sigma^3\Om_1\Om_2^2\Om_3^2\right)\,,
\end{split}
\eqlabel{ttheta}
\end{equation}
where in the second line we used the definition of $\hat{r}$ \eqref{newef} and $\Om$ \eqref{ds5t}.
Note that $\tilde{\theta}=0$ in \eqref{ttheta} is equivalent to $\theta=0$ reproducing \eqref{ahloc10}.

The five dimensional area density  $\cala_5$ of the AH in \eqref{newef} is given by
\begin{equation}
\cala_5=\biggl(\Om^{5/3}\Sigma\biggr)^3\bigg|_{r=r_{AH}}=\Sigma^3\Om_1\Om_2^2\Om_3^2\bigg|_{r=r_{AH}}\,,
\eqlabel{a55}
\end{equation}
leading to the dynamical entropy density
\begin{equation}
e^{3Ht}s=\frac{\cala_5}{4\tilde{G_5}}=\frac{1}{4 G_5}\ 108\Sigma^3\Om_1\Om_2^2\Om_3^2\bigg|_{r=r_{AH}}\,,
\eqlabel{s5kk}
\end{equation}
reproducing \eqref{s10dyn}.

\subsection{Area theorem for the AH}\label{theorem}

Following \cite{Buchel:2017pto} and using the equations of motion \eqref{ev1}-\eqref{con2} we prove now that
the dynamical entropy density $s$ defined as in \eqref{s5kk} grows with time $t$, \ie
\begin{equation}
\frac{d\cala_5}{dt}=\frac{d}{dt}\biggl(\Sigma^3\Om_1\Om_2^2\Om_3^2\bigg|_{r=r_{AH}}\biggr)\ \ge\ 0\,.
\eqlabel{dsdt}
\end{equation}

Note that the AH location is determined from (see \eqref{ttheta})
\begin{equation}
0=d_+(\Sigma^3\Om_1\Om_2^2\Om_3^2)\bigg|_{r=r_{AH}}\equiv \del_t(\Sigma^3\Om_1\Om_2^2\Om_3^2)+A\ \del_r(\Sigma^3\Om_1\Om_2^2\Om_3^2)\bigg|_{r=r_{AH}}\,.
\eqlabel{ahlocth}
\end{equation}
Taking $\frac{d}{dt}$ we have
\begin{equation}
\begin{split}
0=&\frac{d}{dt}\biggl(\del_t(\Sigma^3\Om_1\Om_2^2\Om_3^2)+A\ \del_r(\Sigma^3\Om_1\Om_2^2\Om_3^2)\biggr)
\\
=&\biggl\{\del_t+\frac{dr_{AH}}{dt}\times\del_r\biggr\}
\biggl(\del_t(\Sigma^3\Om_1\Om_2^2\Om_3^2)+A\ \del_r(\Sigma^3\Om_1\Om_2^2\Om_3^2)\biggr)\bigg|_{r=r_{AH}}\,,
\end{split}
\eqlabel{th1}
\end{equation}
which is used to algebraically solve for $\frac{dr_{AH}}{dt}\bigg|_{r=r_{AH}}$.
The latter expression is then substituted in
\begin{equation}
\frac{d\cala_5}{dt}=\biggl\{\del_t+\frac{dr_{AH}}{dt}\times\del_r\biggr\}\  \Sigma^3\Om_1\Om_2^2\Om_3^2\bigg|_{r=r_{AH}}\,.
\eqlabel{th2}
\end{equation}
We use equations of motion \eqref{ev1}-\eqref{con2} to eliminate all second order derivative in \eqref{th2};
we further eliminate $\del_t\Sigma$ using \eqref{ahlocth} to arrive at
\begin{equation}
\begin{split}
&\frac{dA_5}{dt}=
 \frac{\del_r(\Sigma^3\Om_1\Om_2^2\Om_3^2)}{\del_r(d_+(\Sigma^3\Om_1\Om_2^2\Om_3^2))}
\times \calf^2\bigg|_{r=r_{AH}}\,,
\end{split}
\eqlabel{da5dtfin1}
\end{equation}
where $\calf^2$ is manifestly positive
\begin{equation}
\begin{split}
&\calf^2=\frac{\Sigma^3}{2592\Omega_2^2 \Omega_3^2\Omega_1 g^2P^2}\times \biggl(
\Omega_1^2 \biggl(8 (d_+K_2)^2 \Omega_2^2 \Omega_3^2 g^3 P^4
+1296 (d_+g)^2 \Omega_2^4 \Omega_3^4 P^2\\
&+9 (d_+K_3)^2 \Omega_3^4 g+9 (d_+K_1)^2 \Omega_2^4 g\biggr)
+1728 \Omega_1^2 \Omega_2^4 \Omega_3^4 g^2 P^2 \biggl(
\left(\frac{2 d_+\Omega_2}{\Omega_2}+\frac{d_+\Omega_3}{\Omega_3}\right)^2\\
&+\left(\frac{d_+\Omega_1}{\Omega_1}+\frac{d_+\Omega_2}{\Omega_2}\right)^2+\left(\frac{d_+\Omega_1}{\Omega_1}
+\frac{d_+\Omega_3}{\Omega_3}\right)^2+\frac{3 (d_+\Omega_3)^2}{\Omega_3^2}
\biggr) \biggr)\,.
\end{split}
\eqlabel{th3}
\end{equation}
Constraint \eqref{con1} can be integrated (once) to obtain
\begin{equation}
\begin{split}
&\del_r\left(\Sigma^3\Om_1\Om_2^2\Om_3^2\right)=\Sigma^3\Om_1\Om_2^2\Om_3^2\ \int^{\infty}_r\
dr\ \calm^2\\
&\calm^2=\frac{2 (\del_r\Om_2)^2}{\Om_2^2}+\frac{2 (\del_r\Om_3)^2}{\Om_3^2}
+\frac{(\del_r\Om_1)^2}{\Om_1^2}+\frac{3 (\del_r\Sigma)^2}{\Sigma^2}
+\frac{(\del_r g)^2}{2g^2}+\frac{g P^2 (\del_r K_2)^2}{324\Om_3^2 \Om_2^2}
\\&+\frac{(\del_r K_3)^2}{288gP^2 \Om_2^4 }
+\frac{(\del_rK_1)^2}{288gP^2 \Om_3^4}\,,
\end{split}
\eqlabel{intconst}
\end{equation}
which implies that
\begin{equation}
\del_r\left(\Sigma^3\Om_1\Om_2^2\Om_3^2\right)\ge 0\,,
\eqlabel{th4}
\end{equation}
provided the integral in \eqref{intconst} is convergent and $\Sigma^3\Om_1\Om_2^2\Om_3^2\ge 0$. 

Note that (see appendix \ref{apb2})
\begin{equation}
\begin{split}
&\lim_{r\to \infty}\ d_+\left(\Sigma^3\Om_1\Om_2^2\Om_3^2\right)=\lim_{r\to \infty}\
A\del_r\left(\Sigma^3\Om_1\Om_2^2\Om_3^2\right)\\
&=\lim_{\r\to 0}\ \underbrace{\frac{1}{2h^{1/2}\r^2}}_{A}\times
\underbrace{(-\r^2)\del_\r}_{\del_r}\left(\underbrace{h^{-3/4}\r^{-3}\exp(3H\int_0^\r h^{1/2}(s)ds )}_{\Sigma^3}\times
\underbrace{\frac{h^{5/4}f_c^{1/2}f_af_b}{108}}_{\Om_1\Om_2^2\Om_3^2}\right)\\
&=\lim_{\rho\to 0}\biggl(\frac{1}{72\r^4}+{\rm subleading}\biggr)\to +\infty\,,
\end{split}
\eqlabel{dpentropy}
\end{equation}
where we transformed first to FG frame and used the boundary asymptotic expansions
\eqref{ksfc}-\eqref{ksh}. Thus,
\begin{equation}
d_+\left(\Sigma^3\Om_1\Om_2^2\Om_3^2\right)>0\,,\qquad r>r_{AH}\qquad \Longrightarrow\qquad
\del_r\left(d_+\left(\Sigma^3\Om_1\Om_2^2\Om_3^2\right)\right)\bigg|_{r=r_{AH}}\ge 0\,,
\eqlabel{dpconstr}
\end{equation}
since the quantity $d_+\left(\Sigma^3\Om_1\Om_2^2\Om_3^2\right)$ changes sign at $r=r_{AH}$, see \eqref{ahlocth}.
Combining \eqref{da5dtfin1}, \eqref{th4} and \eqref{dpconstr} we arrive at \eqref{dsdt}. 

For future reference we present the expressions for the location of the AH and the entanglement entropy density
in de Sitter vacua. Using \eqref{ltlimit1} and \eqref{ltlimit2} we find from \eqref{ahlocth} and \eqref{s10dyn}
\begin{equation}
\begin{split}
&{\rm AH\ location:}\qquad \biggl(3H\ \sigma^3 \w_{c2}^{1/2}\w_{a2}\w_{b2}+a\frac{d}{dr}\biggl\{
\sigma^3 \w_{c2}^{1/2}\w_{a2}\w_{b2}\biggr\}\biggr)\bigg|_{r=r_{AH}}=0\,;\\
&{\rm vacum\ entanglement\ entropy:}\qquad s_{ent}=\frac{1}{4G_5}\ \sigma^3\w_{c2}^{1/2}\w_{a2}\w_{b2}\bigg|_{r=r_{AH}}\,.
\end{split}
\eqlabel{ahsvac}
\end{equation}

\subsection{Entanglement entropy of TypeB de Sitter vacua}\label{sentb}

We demonstrate here that entanglement entropy of TypeB de Sitter vacuum vanishes --- this implies that the
corresponding comoving entropy production rate vanishes. de Sitter comoving entropy production rate
vanishes in conformal field theories as well \cite{Buchel:2019qcq}. In CFTs the reason is simple: de Sitter vacuum is
a conformal transformation of a thermal equilibrium state and 
entropy production is invariant under conformal
transformations \cite{Buchel:2017pto}. We do not understand the physical reason why the same is
true for a de Sitter vacuum in nonconformal gauge theory (TypeB vacuum in the cascading gauge theory).

Using the asymptotic expansion \eqref{typebi} (recall that $z=-r$ \eqref{defzef}) we find for \eqref{ahsvac}
\begin{equation}
\begin{split}
&{\rm AH\ location:}\qquad \frac{3^{3/2}}{2} (h^h_0)^{3/4} (f_{a,0}^h)^{3/2} (s^h_0)^3\  r\ \biggl(
1+3 H (h^h_0)^{1/2} r+\calo(r^2)\biggr)
\bigg|_{r=r_{AH}}=0\\
&\qquad \Longrightarrow\qquad r_{AH}=0\,;\\
&{\rm vacum\ entanglement\ entropy:}\ \ s_{ent}=\frac{1}{4G_5}\ \frac{3^{3/2}}{2} (h^h_0)^{5/4} (f_{a,0}^h)^{3/2}
(s^h_0)^3\ r^2+\calo(r^3)
\bigg|_{r=r_{AH}}\\
&\qquad \Longrightarrow\qquad s_{ent}\bigg|_{\rm TypeB}=0\,.
\end{split}
\eqlabel{ahsvactypeb}
\end{equation}
The result \eqref{ahsvactypeb} stands as long as vacua TypeB exist --- we find
in section \ref{typebsugra} that this is true provided $H\lesssim H_{max}^B$, see \eqref{typebexists}.

\section{TypeA$_s$ de Sitter vacua}\label{typeasv}

TypeA$_s$ vacua in FG frame were discussed in details in \cite{Buchel:2013dla}.
As emphasized in \cite{Buchel:2017pto} and \cite{Buchel:2019qcq} this is not enough
to access vacuum entanglement entropy --- one needs the holographic construction in
EF frame.  In section \ref{typeasnum} we present numerical results for
TypeA$_s$ vacua for generic values of $\frac{H^2}{\Lambda^2}$,
in particular the results for the entanglement entropy, see
fig.~\ref{comparesent}. We discuss TypeA$_s$ in the conformal limit $\Lambda\ll H$ in section
\ref{typeasc}. In section \ref{typeashmin} we estimate $H_{min}^s$
(see \eqref{hcritltypeas}) below which TypeA$_s$ vacua construction in type IIB supergravity
becomes unreliable. We identify the source of breaking of the supergravity approximation.

\subsection{Numerical results: TypeA$_s$}\label{typeasnum}

To begin, we numerically construct TypeA$_s$ de Sitter vacua in FG frame \eqref{fg1i}.
This involves solving ODEs 
\eqref{kseq2}-\eqref{kseq10} in the chirally symmetric limit \eqref{ktks1}, subject to UV asymptotics
(the radial coordinate $\rho\to 0$) \eqref{ktks2}-\eqref{ktks6} and IR asymptotics
(the radial coordinate $\rho\to +\infty$) \eqref{phase2ir1}. There are 8 second order equations
\eqref{kseq2}-\eqref{kseq9} and 1 first order equation \eqref{kseq10}. Imposing
the chirally symmetric limit \eqref{ktks1}, this set of coupled ODEs is reduced to 5 second order equations
for the three metric warp factors $f_2=f_c$, $f_3=f_a=f_b$ and $h$, the single 3-form flux function
$K=K_1=K_3$ ($K_2=1$ in the chiral limit) and the string coupling $g$. The first order equation
\eqref{kseq10} involves (linearly) $f_2'$ and can be used instead of one of the second order equations
(namely, the one involving $f_2''$). Thus, altogether we have a coupled system of 4 second order
ODEs (linear in  $\{f_3'',h'',K'',g''\}$) and a single first order equation (linear in $f_2'$). As a result,
a unique solution must be characterized by $9=2\times 4+1$ parameters; these are the UV/IR parameters 
\begin{equation}
\begin{split}
&{\rm UV:}\qquad \{f_{2,1,0}\,,\, g_{4,0}\,,\, f_{2,4,0}\,,\ f_{2,6,0}\,,\, f_{2,8,0}\}\,;\\
&{\rm IR:}\qquad \{f_{2,0}^h\,,\ f_{3,0}^h\,,\ K_{0}^h\,,\ g_{0}^h\}\,.
\end{split}
\eqlabel{uvirpars}
\end{equation}
The external parameters $\{P,K_0,H,g_s\}$ (the gauge group rank difference $M$ of the
cascading gauge theory \eqref{defpm},
its strong coupling scale $\Lambda$ \eqref{deflambda}, the Hubble constant \eqref{ds4metric},
the  renormalization group flow invariant sum of the gauge couplings \eqref{sum})
labeling the vacuum are fixed with the choice of the
computational scheme \eqref{compschemes}. Of cause, as emphasized in appendix \ref{apcfg},
the results must not depend on which computational scheme is adopted.
We illustrate now that this is indeed the case using the IR parameters in \eqref{uvirpars} as
an example\footnote{The same is true for the UV parameters as well.}.
Comparison of the different computational schemes is done using dimensionless and rescaled quantities:
$\ln \frac {H^{2}}{\Lambda^2}$ (as a vacuum label) \eqref{defks}
and $\{\hf_{2,3,0}^h\,,\, \hK_0^h\,,\, \hat{g}_0^h\}$
\eqref{irtypeascaled}.  Explicitly:
\begin{equation}
\begin{split}
&{\rm SchemeI:}\qquad \ln\frac{H^2}{\Lambda^2}=k_s\,,\ \hf_{2,3,0}^h=f_{2,3,0}^h\,,\ \hK_0^h=K_0^h\,,\ \hat{g}_0^h=g_0^h\,;\\
&{\rm SchemeII:}\qquad \ln\frac{H^2}{\Lambda^2}=\frac 1b+\ln b\,,\ \hf_{2,3,0}^h=\frac{1}{b^{1/2}}f_{2,3,0}^h\,,\
\hK_0^h=\frac 1b K_0^h\,,\ \hat{g}_0^h=g_0^h\,;
\\
&{\rm SchemeIII:}\qquad \ln\frac{H^2}{\Lambda^2}=\frac 14+\ln\a\,,\ \hf_{2,3,0}^h=\frac{1}{\a^{1/2}}f_{2,3,0}^h\,,\
\hK_0^h=K_0^h\,,\ \hat{g}_0^h=g_0^h\,.
\end{split}
\eqlabel{collapsetypeas}
\end{equation}

\begin{figure}[t]
\begin{center}
\psfrag{x}{{$\ln \frac{H^2}{\Lambda^2}$}}
\psfrag{y}{{$\hf_{2,0}^h$}}
\psfrag{z}{{$\hf_{3,0}^h$}}
\psfrag{t}{{$\hK_{0}^h$}}
\psfrag{w}{{$\hat{g}_{0}^h$}}
\includegraphics[width=2.6in]{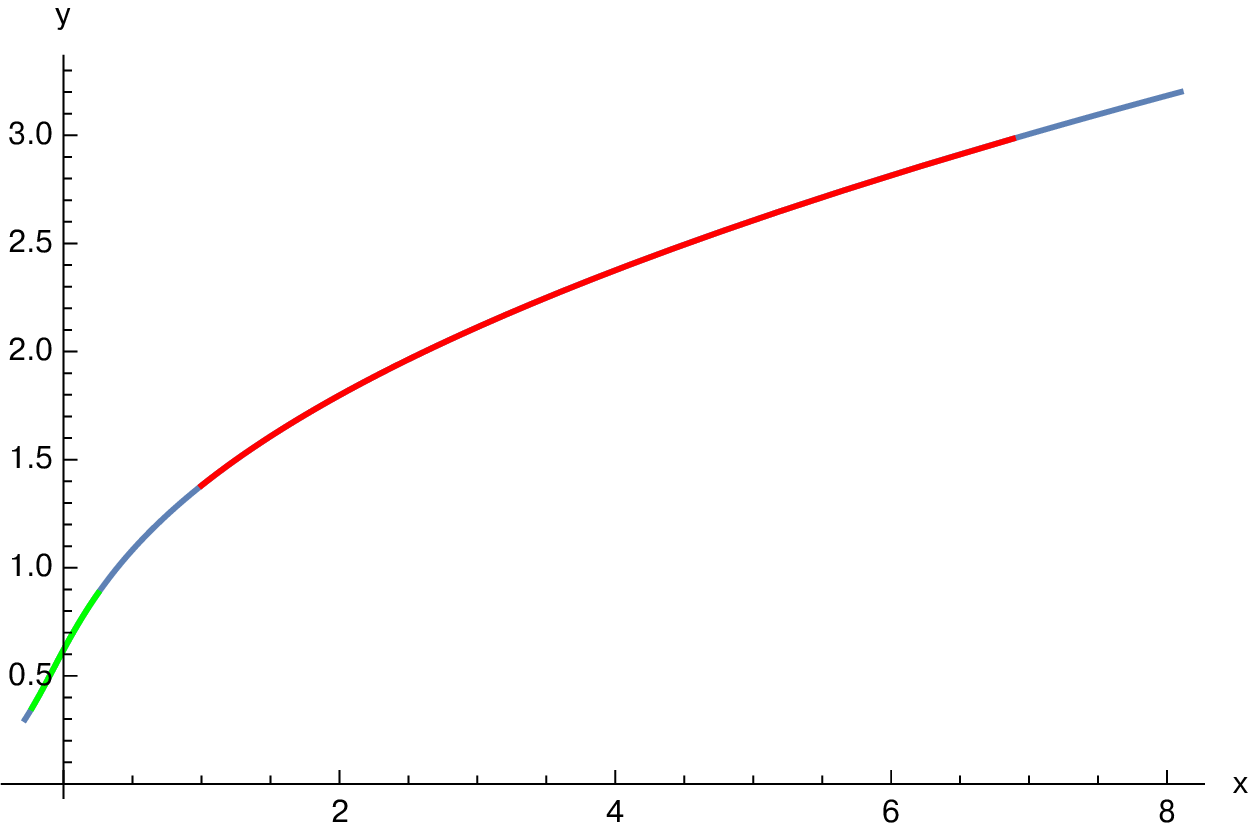}
\qquad \includegraphics[width=2.6in]{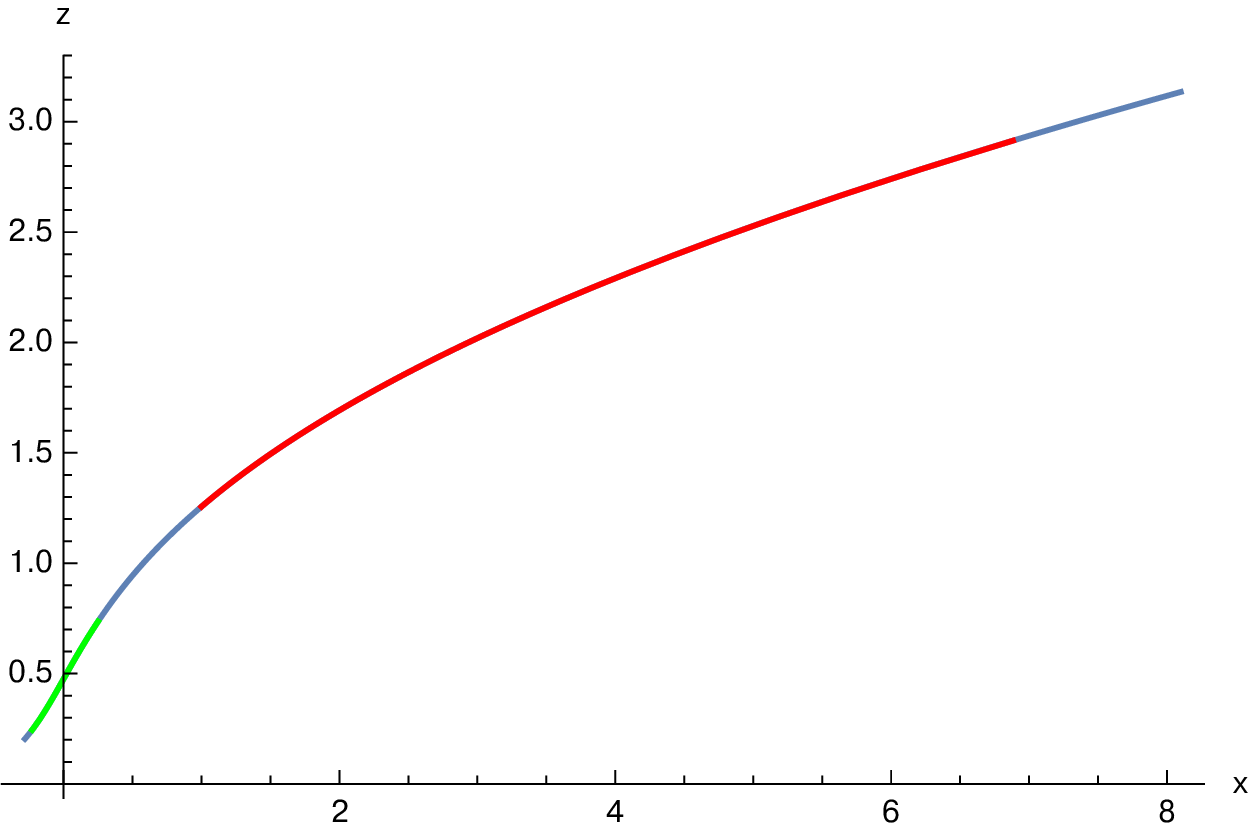}\\
\includegraphics[width=2.6in]{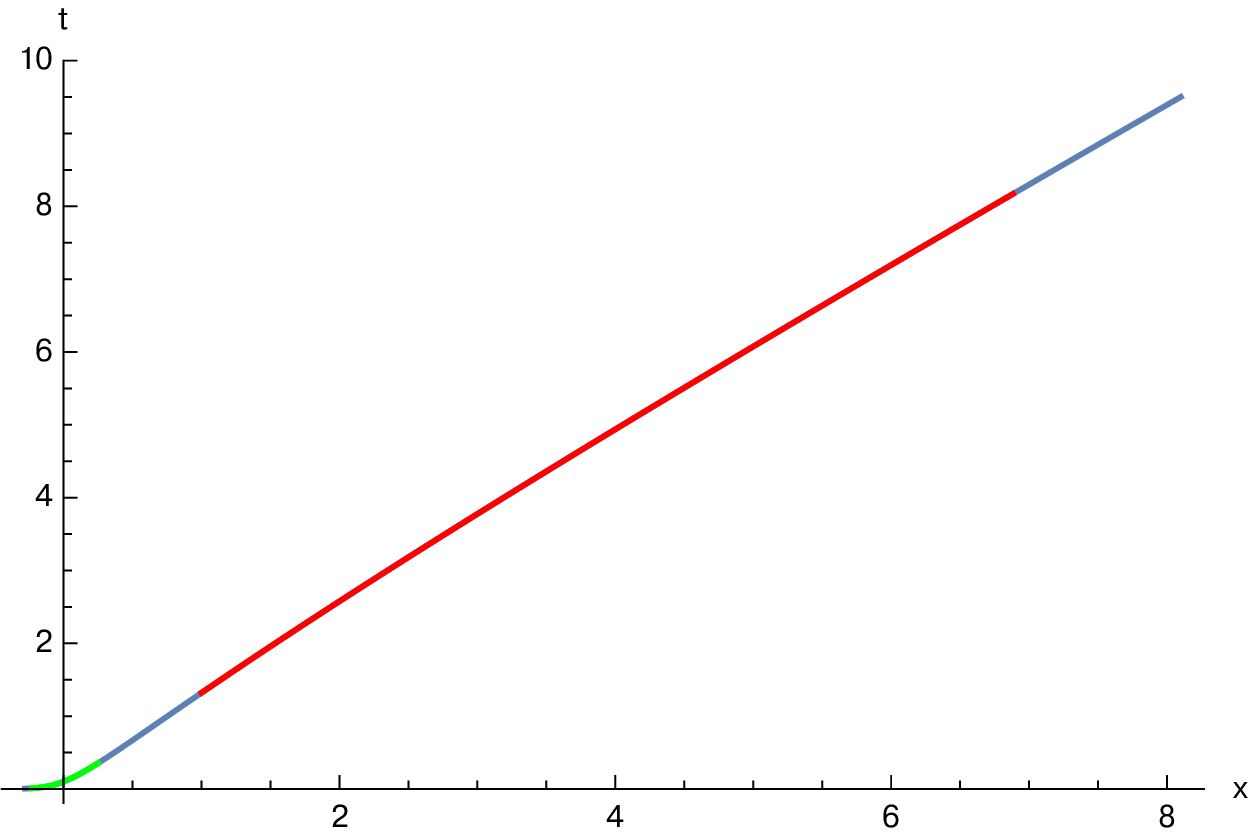}
\qquad
\includegraphics[width=2.6in]{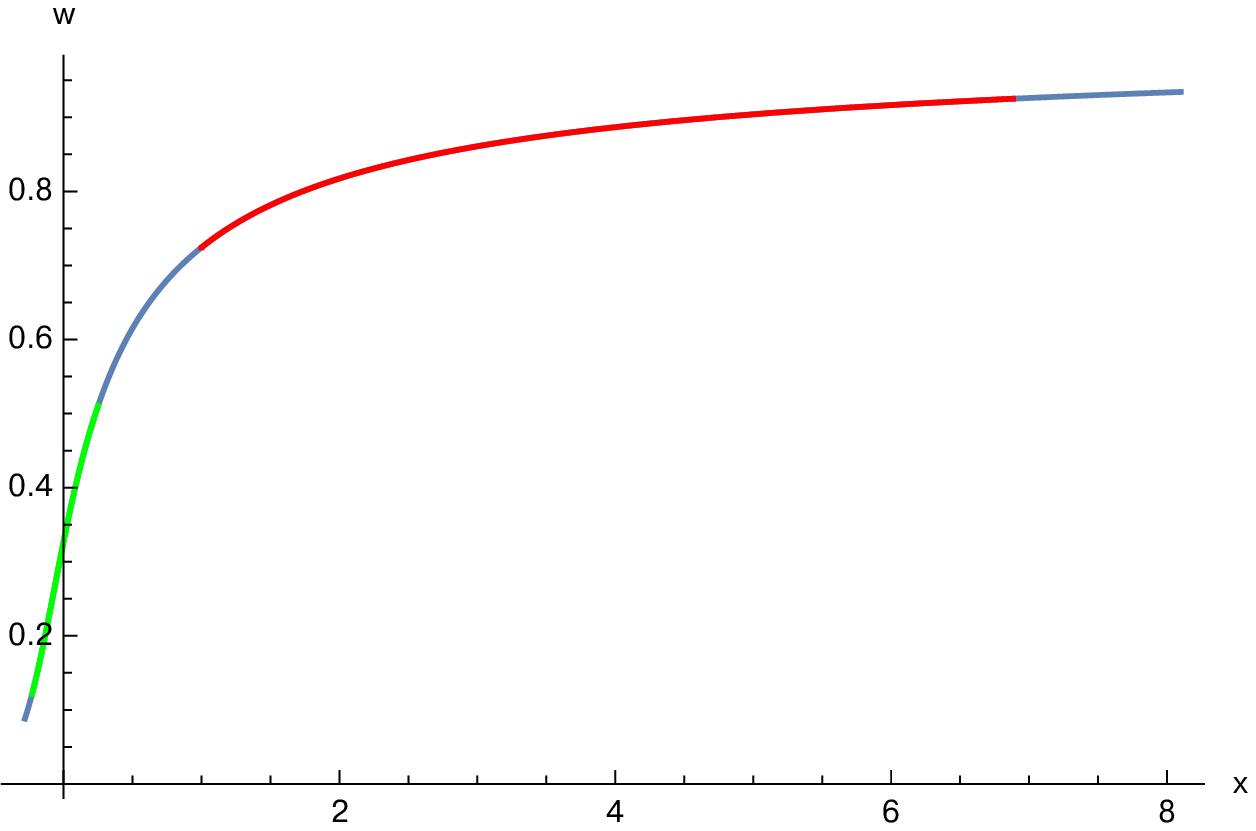}
\end{center}
  \caption{
Infrared parameters $\{\hf_{2,0}^h,\hf_{3,0}^h,\hK_{0}^h,\hat{g}_{0}^h\}$  of the  
Fefferman-Graham coordinate frame of TypeA$_s$ de Sitter vacua of the cascading
gauge theory as functions of $\ln \frac{H^2}{\Lambda^2}$ in different computational
schemes \eqref{compschemes}: SchemeI (blue), SchemeII (red) and Scheme III (green). 
} \label{figure1}
\end{figure}

\begin{figure}[t]
\begin{center}
\psfrag{x}{{$\ln \frac{H^2}{\Lambda^2}$}}
\psfrag{y}{{${\color{red} \hf_{2,0}^h}/{\color{blue} \hf_{2,0}^h}-1$}}
\psfrag{z}{{${\color{green} \hf_{2,0}^h}/{\color{blue} \hf_{2,0}^h}-1$}}
\includegraphics[width=2.6in]{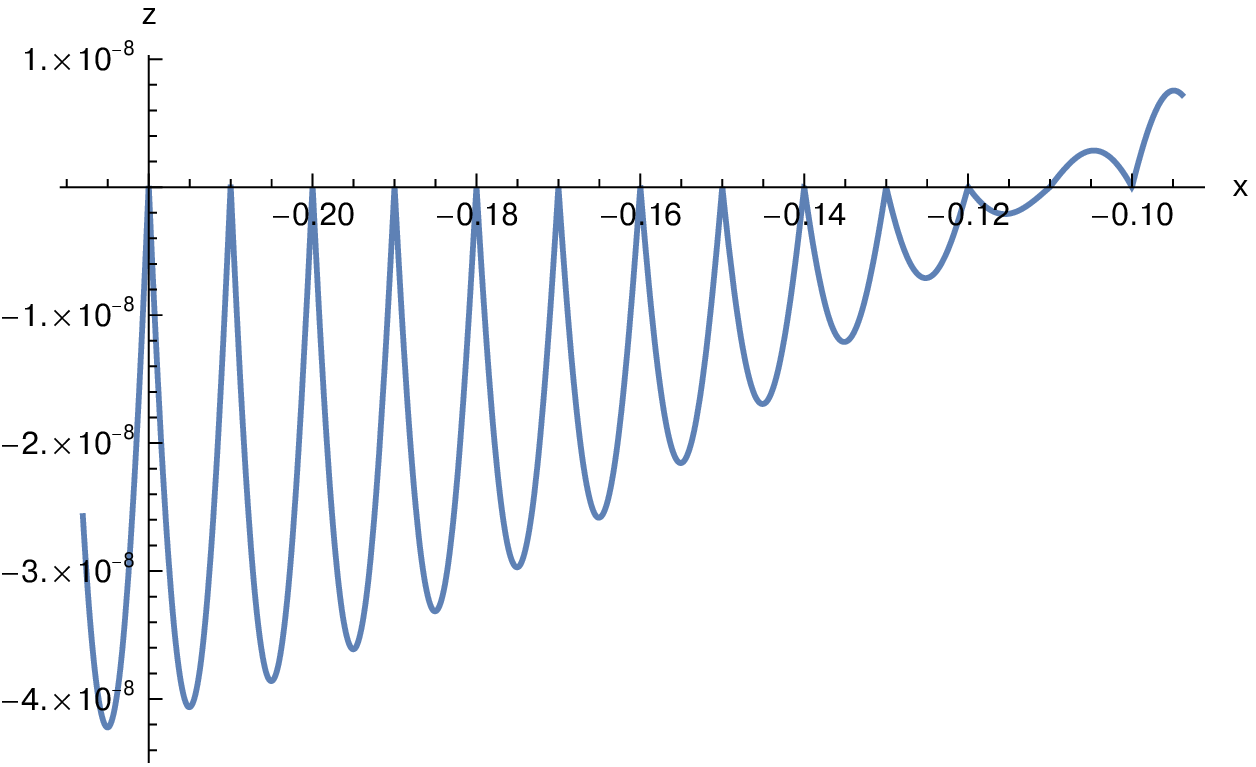}
\qquad \includegraphics[width=2.6in]{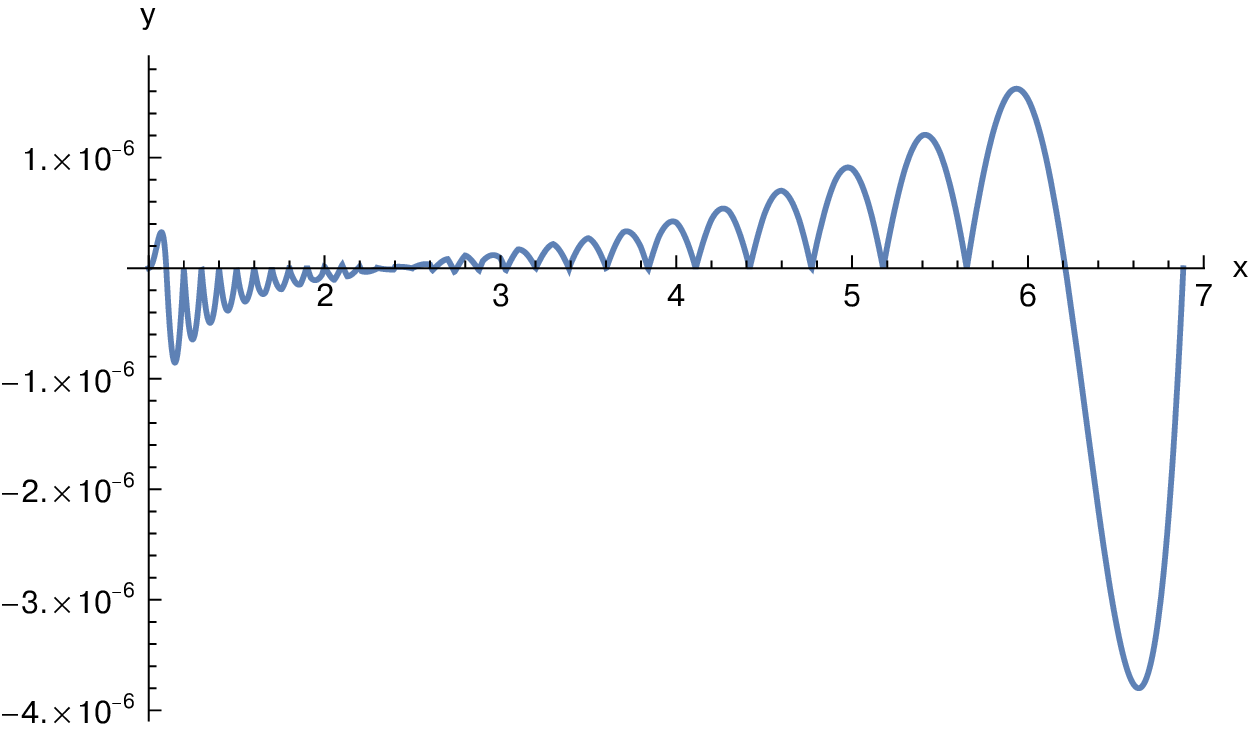}
\end{center}
  \caption{
Left panel: comparison of ${\color{green} \hf_{2,0}^h}$ (the computational scheme SchemeIII) with
${\color{blue} \hf_{2,0}^h}$ (the computational scheme SchemeI). Right panel:
comparison of ${\color{red} \hf_{2,0}^h}$ (the computational scheme SchemeII) with
${\color{blue} \hf_{2,0}^h}$ (the computational scheme SchemeI).  
} \label{figure2}
\end{figure}

Following \eqref{collapsetypeas}, we collect
(subset of the) results of $\{\hf_{2,0}^h,\hf_{3,0}^h,\hK_{0}^h,\hat{g}_{0}^h\}$ as  functions of  
$\ln \frac{H^2}{\Lambda^2}$ in different computational schemes in fig.~\ref{figure1}:
SchemeI (blue curves), SchemeII (red curves) and Scheme III (green curves).
The accuracy of the collapsed results in different schemes is highlighted in fig.~\ref{figure2}
for $\hf_{2,0}^h$ --- the remaining parameters follow the same trend.

Next, FG frame TypeA$_s$ de Sitter vacua have to be reinterpreted in EF frame, see appendix \ref{apb2}.
The diffeomorphism transformation is performed at the radial location
\begin{equation}
\biggl\{{\rm FG:}\qquad \frac 1\r\equiv y=0\biggr\}\qquad \Longleftrightarrow\qquad
\biggl\{{\rm EF:}\qquad r\equiv -z=0\biggr\}\,.
\eqlabel{mapfgef}
\end{equation}
Details of numerical construction of EF frame vacua from FG frame vacua are collected in
appendix \ref{apcef}. An important quantity is the parameter $s_0^h$,  see \eqref{ef1i},
\begin{equation}
s_0^h=\sigma\bigg|_{y=0}^{{\rm FG\ frame}}=\sigma\bigg|_{z=0}^{{\rm EF\ frame}}\,.
\eqlabel{sh0FGEF}
\end{equation}
As with FG frame UV/IR parameters \eqref{uvirpars}, results for $s_0^h$ should not
depend on the choice of the computational scheme, provided we compare properly
dimensionless and rescaled quantities, \ie $\ln \frac{H^2}{\Lambda^2}$ and
$\hat{s}_0^h$ \eqref{finalsh0},
\begin{equation}
\begin{split}
&{\rm SchemeI:}\qquad \ln\frac{H^2}{\Lambda^2}=k_s\,,\qquad  \hs_{0}^h=s_0^h\,;\\
&{\rm SchemeII:}\qquad \ln\frac{H^2}{\Lambda^2}=\frac 1b+\ln b\,,\qquad \hs_{0}^h=\frac{1}{b^{1/4}}s_0^h\,;
\\
&{\rm SchemeIII:}\qquad \ln\frac{H^2}{\Lambda^2}=\frac 14+\ln\a\,,\qquad \hs_{0}^h=\frac{1}{\a^{1/2}}s_0^h\,.
\end{split}
\eqlabel{collapsetypeassigma}
\end{equation}

\begin{figure}[t]
\begin{center}
\psfrag{x}{{$\ln \frac{H^2}{\Lambda^2}$}}
\psfrag{s}{{$\hs_{0}^h$}}
\includegraphics[width=2.6in]{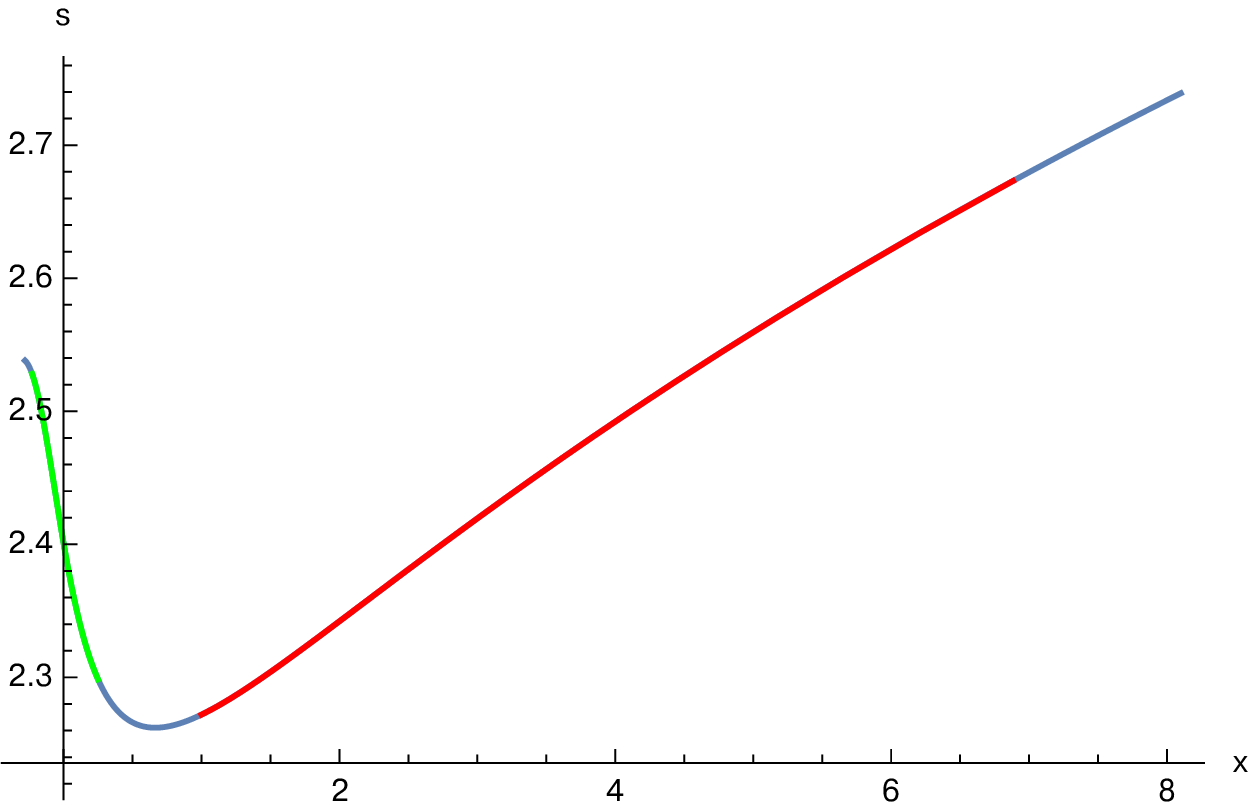}
\end{center}
  \caption{
Parameters $\hs_{0}^h$ of TypeA$_s$ de Sitter vacua of the cascading
gauge theory as functions of $\ln \frac{H^2}{\Lambda^2}$ in different computational
schemes \eqref{compschemes}: SchemeI (blue), SchemeII (red) and Scheme III (green). 
} \label{figure2b}
\end{figure}

\begin{figure}[t]
\begin{center}
\psfrag{x}{{$\ln \frac{H^2}{\Lambda^2}$}}
\psfrag{y}{{${\color{red} \hs_{0}^h}/{\color{blue} \hs_{0}^h}-1$}}
\psfrag{z}{{${\color{green} \hs_{0}^h}/{\color{blue} \hs_{0}^h}-1$}}
\includegraphics[width=2.6in]{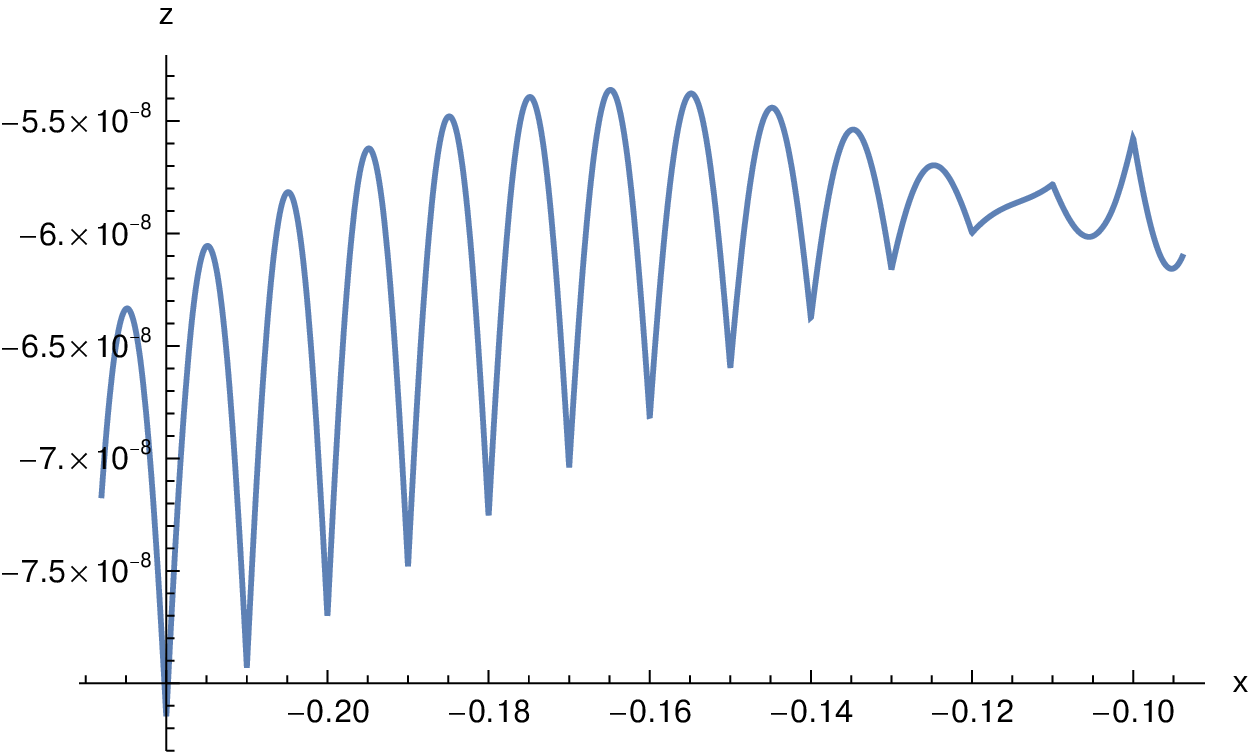}
\qquad \includegraphics[width=2.6in]{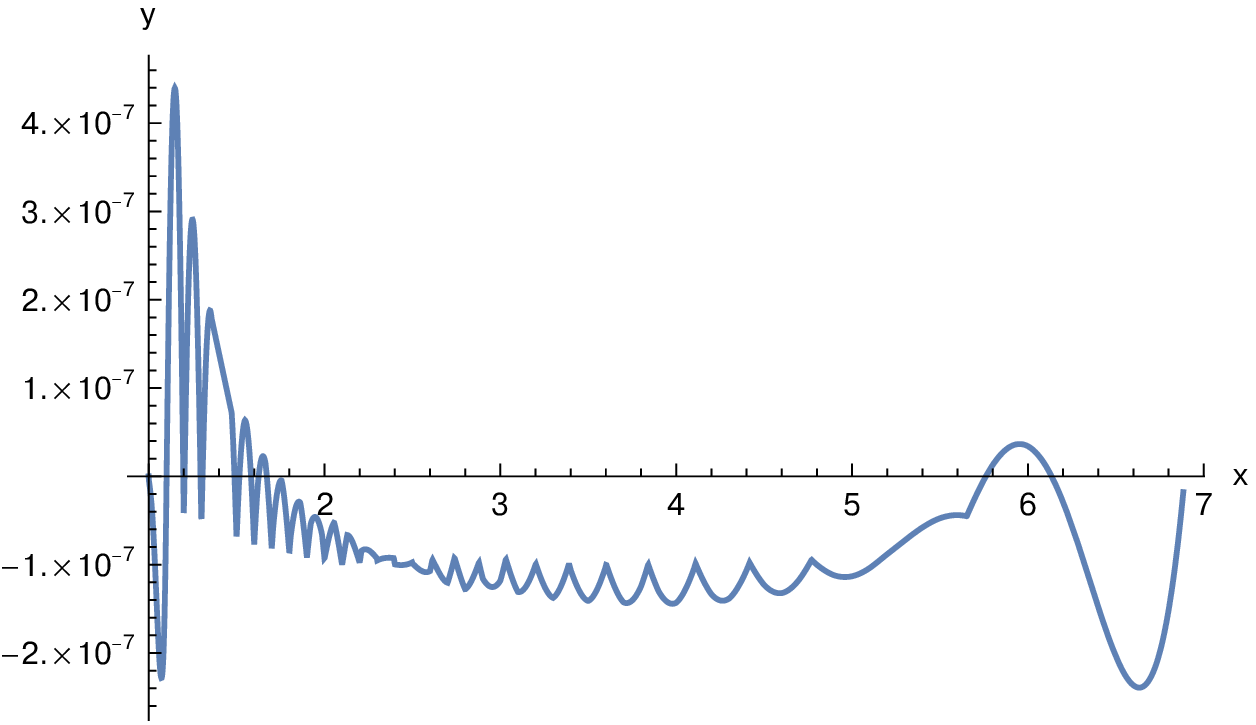}
\end{center}
  \caption{
Left panel: comparison of ${\color{green} \hs_{0}^h}$ (the computational scheme SchemeIII) with
${\color{blue} \hs_{0}^h}$ (the computational scheme SchemeI). Right panel:
comparison of ${\color{red} \hs_{0}^h}$ (the computational scheme SchemeII) with
${\color{blue} \hs_{0}^h}$ (the computational scheme SchemeI).  
} \label{figure2c}
\end{figure}

Following \eqref{collapsetypeassigma}, we collect
(subset of the) results of $\hs_{0}^h$ as functions of  
$\ln \frac{H^2}{\Lambda^2}$ in different computational schemes in fig.~\ref{figure2b}:
SchemeI (blue curve), SchemeII (red curve) and Scheme III (green curve).
The accuracy of the collapsed results in different schemes is highlighted in fig.~\ref{figure2c}.

\begin{figure}[t]
\begin{center}
\psfrag{z}{{$z$}}
\psfrag{l}{{$\call_{AH}$}}
\includegraphics[width=2.6in]{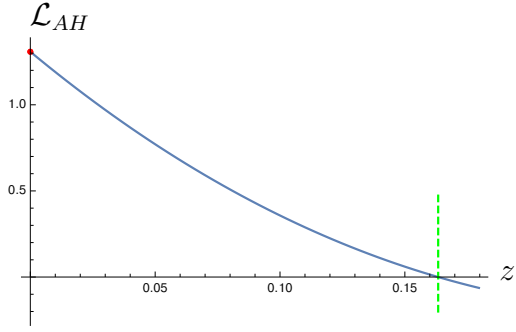}
\end{center}
  \caption{
Apparent horizon location function $\call_{AH}(z)$ in computational scheme SchemeI at $k_s=0$,
\ie at $H=\Lambda$, see \eqref{defcall}. The red dot is $\call_{AH}(0)$, see \eqref{callzsmall}.
Notice that  $\call_{AH}'(0)<0$, see \eqref{callzsmall1}. The vertical green dashed line is the
first zero of $\call_{AH}(z)$: $z_{AH}=0.163346$.
} \label{ahlocplot}
\end{figure}

EF frame equations of motion \eqref{efv2}-\eqref{efc1} are solved subject to the initial conditions
set by the asymptotic expansions \eqref{typeasi} at $z=0$. These equations have to be integrated
on the interval
\begin{equation}
z\in [0,z_{AH}]\,,
\eqlabel{rangez4}
\end{equation}
where $z_{AH}=-r_{AH}$ is the location of the apparent horizon at asymptotically late times, see \eqref{ahsvac}.
To determine the location of the apparent horizon, along with integrating the gravitational
background functions $\{a,\sigma,w_{c2},w_{a2},K_1,g\}$ (remember that $w_{b2}=w_{c2}$, $K_3=K_1$ and
$K_2=1$ when the chiral symmetry is unbroken), we evaluate the AH location function
$\call_{AH}(z)$, see \eqref{defcall}. AH is located at the first zero of this function for $z>0$.
A typical profile of the AH location function is shown in fig.~\ref{ahlocplot}.
Once the AH is identified, TypeA$_s$ vacua entanglement entropy is computed following \eqref{ahsvac}:
\begin{equation}
s_{ent}=\frac{H^3P^4g_s^2}{4G_5}\ \biggl\{
\hat{\sigma}^3 \hat{w}_{c2}^{1/2}\hw_{a2}^2\biggr\}\bigg|_{\hat{z}=\hat{z}_{AH}}
=\frac{3^5M^4g_s^2}{2^5\pi^3}\ H^3\  \biggl\{
\hat{\sigma}^3 \hat{w}_{c2}^{1/2}\hw_{a2}^2\biggr\}\bigg|_{\hat{z}=\hat{z}_{AH}}\,,
\eqlabel{senttypeas}
\end{equation}
where following \eqref{scaleout} we introduced dimensionless and rescaled functions
and the radial coordinate:
\begin{equation}
\begin{split}
&\{z\,,\ a\,,\, \sigma\,,\, w_{c2}\,,\, w_{a2}\,,\, K_{1}\,,\ g \}\qquad
\Longrightarrow\qquad \{\hat{z}\,,\ \ha\,,\, \hat{\sigma}\,,\, \hw_{c2}\,,\, \hw_{a2}\,,\, \hK_{1}\,,\ \hat{g}\}\,;\\
&z={H Pg_s^{1/2}}\ \hat{z}\,,\qquad a=H^2P g_s^{1/2}\ \ha\,,\qquad \sigma=H P^{1/2}g_s^{1/4}\ \hat{\sigma}\,,\\
&w_{c2,a2}=P g_s^{1/2}\ \hw_{c2,a2}\,,\qquad  K_{1}=P^2g_s\ \hK_{1}\,,\qquad g=g_s\ \hat{g}  \,.
\end{split}
\eqlabel{scaleouttypeas}
\end{equation}
In the last equality in \eqref{senttypeas} we used expressions for $G_5$ \eqref{g5deff} and $P$ \eqref{defpm}.
We compute entanglement entropy in different computational schemes; results must agree, provided
we compare dimensionless and rescaled quantities,
\begin{equation}
s_{ent}=H^3 P^4g_s^2\ \hs_{ent}\,.
\eqlabel{defhatsent}
\end{equation}
Explicitly,
\begin{equation}
\begin{split}
&{\rm SchemeI:}\qquad \ln\frac{H^2}{\Lambda^2}=k_s\,,\qquad  \hs_{ent}=s_{ent}\,;\\
&{\rm SchemeII:}\qquad \ln\frac{H^2}{\Lambda^2}=\frac 1b+\ln b\,,\qquad \hs_{ent}=\frac{1}{b^2}s_{ent}\,;
\\
&{\rm SchemeIII:}\qquad \ln\frac{H^2}{\Lambda^2}=\frac 14+\ln\a\,,\qquad \hs_{ent}=\frac{1}{\a^{3/2}}s_{ent}\,.
\end{split}
\eqlabel{collapsetypeasent}
\end{equation}

\begin{figure}[t]
\begin{center}
\psfrag{x}{{$\ln \frac{H^2}{\Lambda^2}$}}
\psfrag{s}{{$4 G_5\ \hat{s}_{ent}$}}
\includegraphics[width=2.6in]{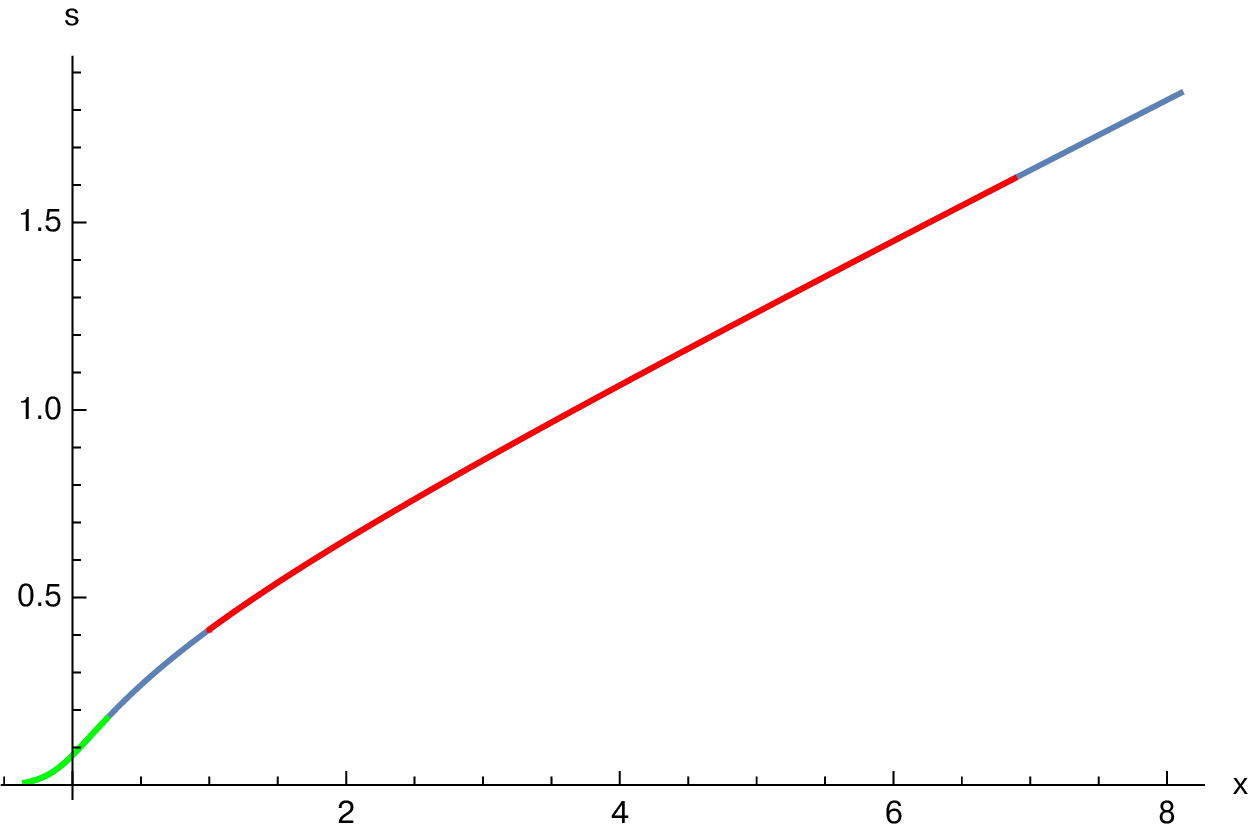}\qquad
\includegraphics[width=2.6in]{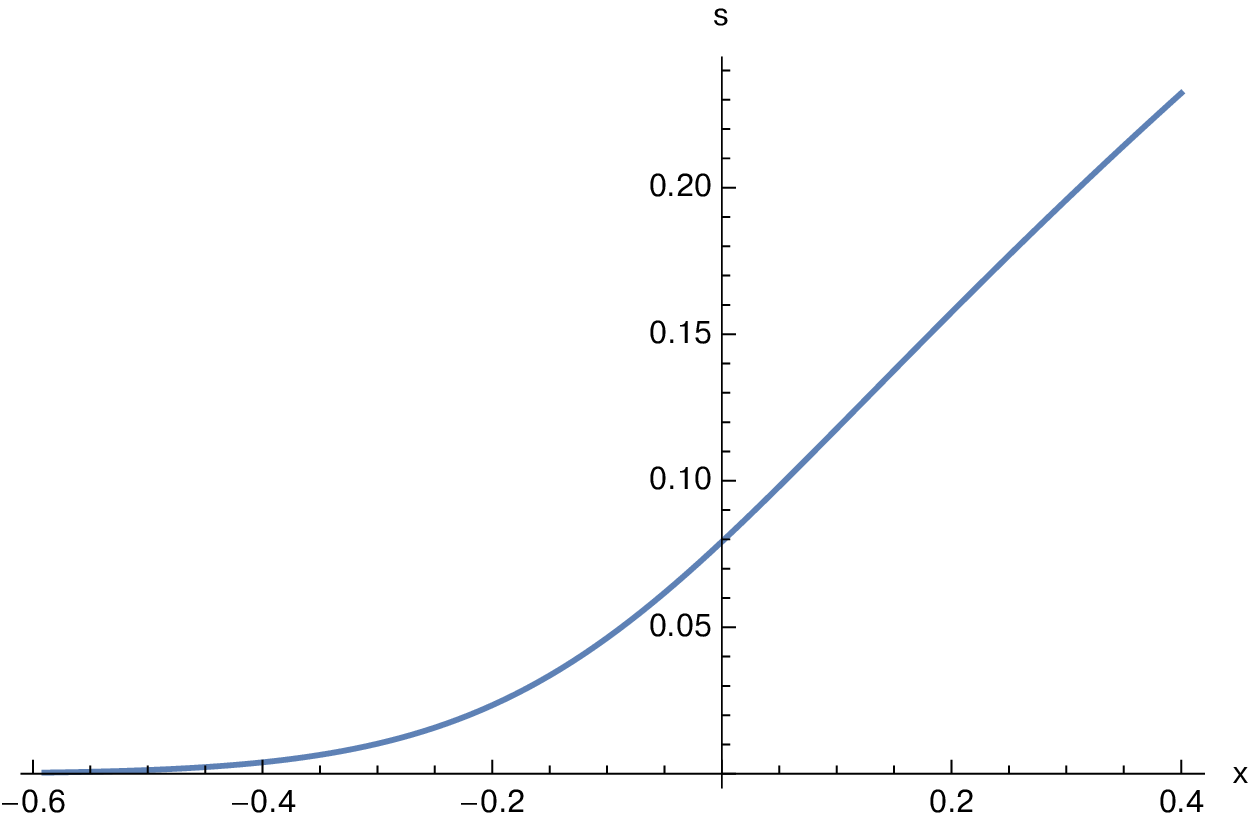}
\end{center}
  \caption{
Left panel: entanglement entropy $\hat{s}_{ent}$  \eqref{defhatsent}
of TypeA$_s$ de Sitter vacua of the cascading
gauge theory as functions of $\ln \frac{H^2}{\Lambda^2}$ in different computational
schemes \eqref{compschemes}: SchemeI (blue), SchemeII (red) and Scheme III (green).
Right panel: entanglement entropy $\hat{s}_{ent}$  \eqref{defhatsent} for small values of
$\frac{H^2}{\Lambda^2}$ --- at the limit of validity of the supergravity approximation,
see section \ref{typeashmin}. 
} \label{comparesent}
\end{figure}

\begin{figure}[t]
\begin{center}
\psfrag{x}{{$\ln \frac{H^2}{\Lambda^2}$}}
\psfrag{y}{{${\color{red} \hs_{ent}}/{\color{blue} \hs_{ent}}-1$}}
\psfrag{z}{{${\color{green} \hs_{ent}}/{\color{blue} \hs_{ent}}-1$}}
\includegraphics[width=2.6in]{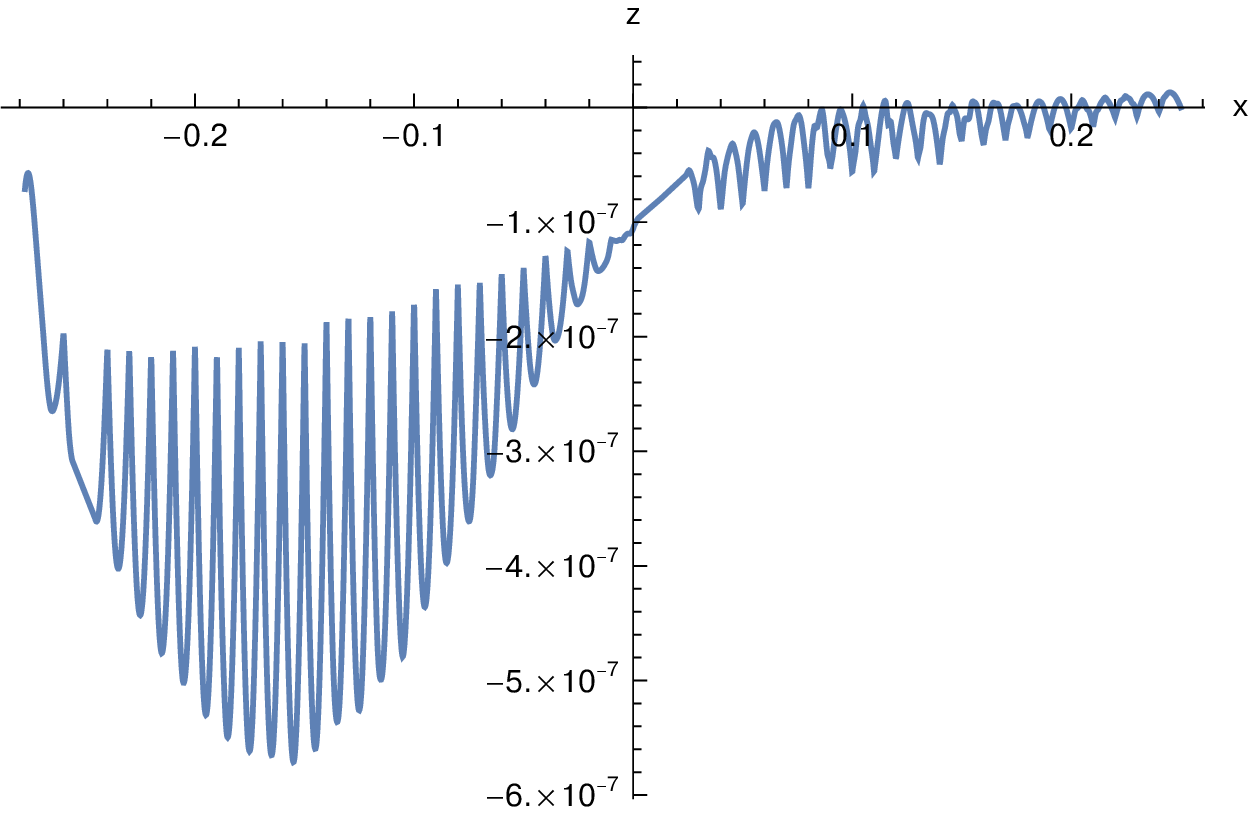}
\qquad \includegraphics[width=2.6in]{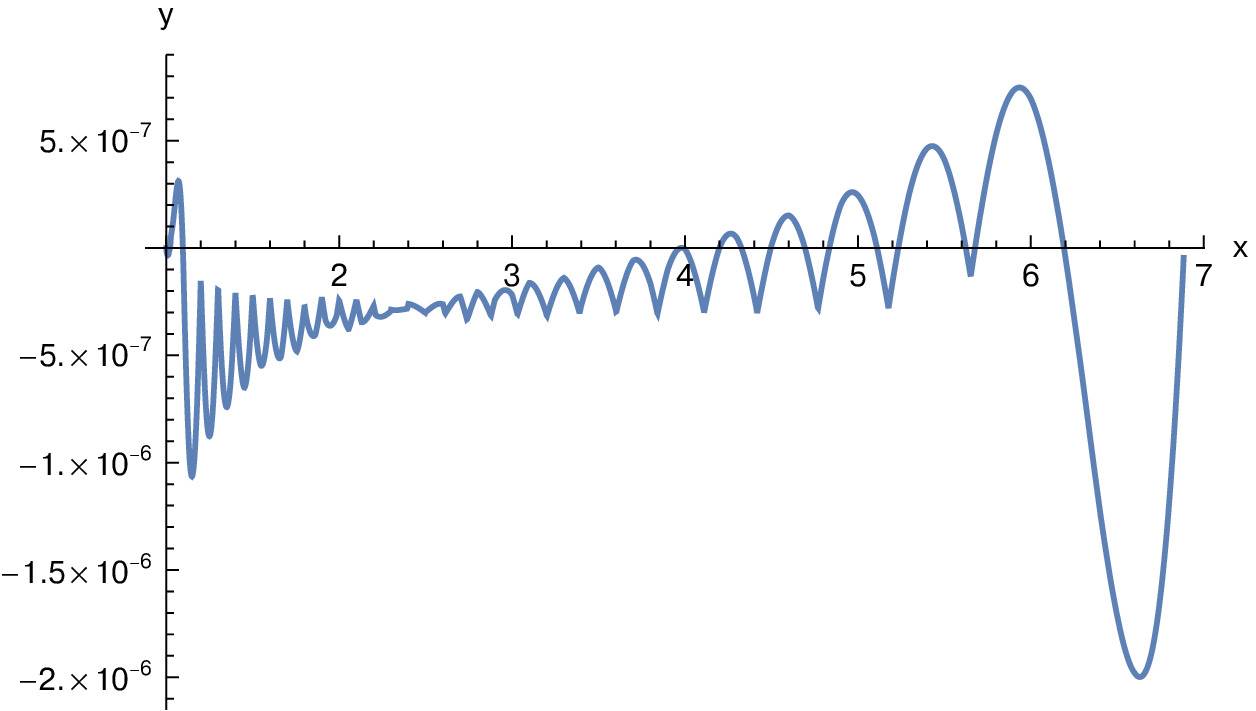}
\end{center}
  \caption{
Left panel: comparison of ${\color{green} \hs_{ent}}$ (the computational scheme SchemeIII) with
${\color{blue} \hs_{ent}}$ (the computational scheme SchemeI). Right panel:
comparison of ${\color{red} \hs_{ent}}$ (the computational scheme SchemeII) with
${\color{blue} \hs_{ent}}$ (the computational scheme SchemeI).  
} \label{errorsent}
\end{figure}

Following \eqref{collapsetypeasent}, we collect
(subset of the) results of $(4G_5\ \hs_{ent})$ as  functions of  
$\ln \frac{H^2}{\Lambda^2}$ in different computational schemes in fig.~\ref{comparesent}:
SchemeI (blue curves), SchemeII (red curves) and Scheme III (green curves).
The accuracy of the collapsed results in different schemes is highlighted in
fig.~\ref{errorsent}.

\subsection{TypeA$_s$ de Sitter vacua in the conformal limit}\label{typeasc} 

The cascading gauge theory is not conformal --- it has a strong coupling scale
$\Lambda$. Thermal states of the cascading gauge theory in Minkowski
space-time at temperature $T\gg \Lambda$ enjoy conformal equation of
state, $\cale=3 \calp$, up to $\calo\left(\frac{1}{\ln (T/\Lambda)}\right)$ corrections,
see \cite{Aharony:2007vg}.
On the gravity side the conformal limit is realized as $P\to 0$ (or Klebanov-Witten \cite{Klebanov:1998hh})
limit. We show here that exactly the same limit on the gravity side of TypeA$ _s$ de Sitter vacua
captures the $H\gg \Lambda$ limit of the cascading gauge theory,
resulting in de Sitter vacuum entanglement entropy density \eqref{vanishconflimit}, vanishing,
as appropriate, for the conformal gauge theory \cite{Buchel:2017pto,Buchel:2019qcq}.

\begin{figure}[t]
\begin{center}
\psfrag{b}{{$b$}}
\psfrag{y}{{$f_{2,0}^h$}}
\psfrag{z}{{$f_{3,0}^h$}}
\psfrag{k}{{$K_{0}^h$}}
\psfrag{g}{{${g}_{0}^h$}}
\includegraphics[width=2.6in]{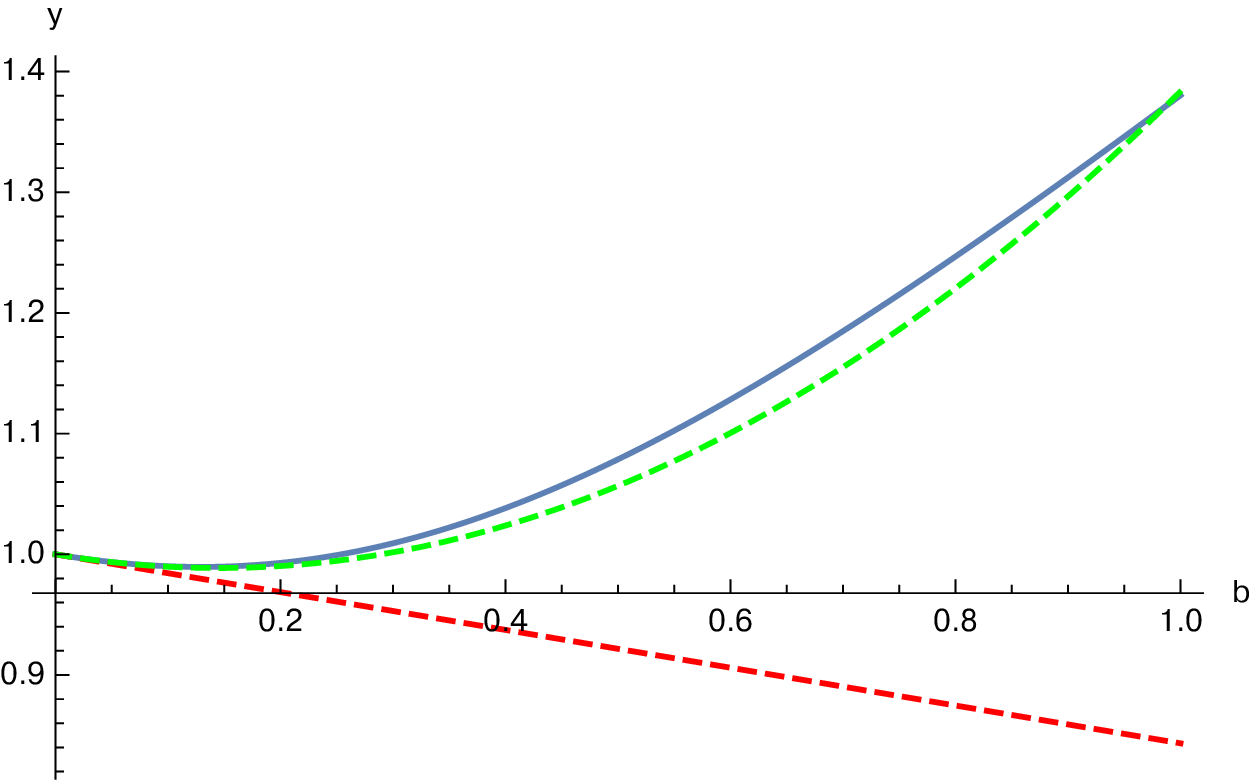}
\qquad \includegraphics[width=2.6in]{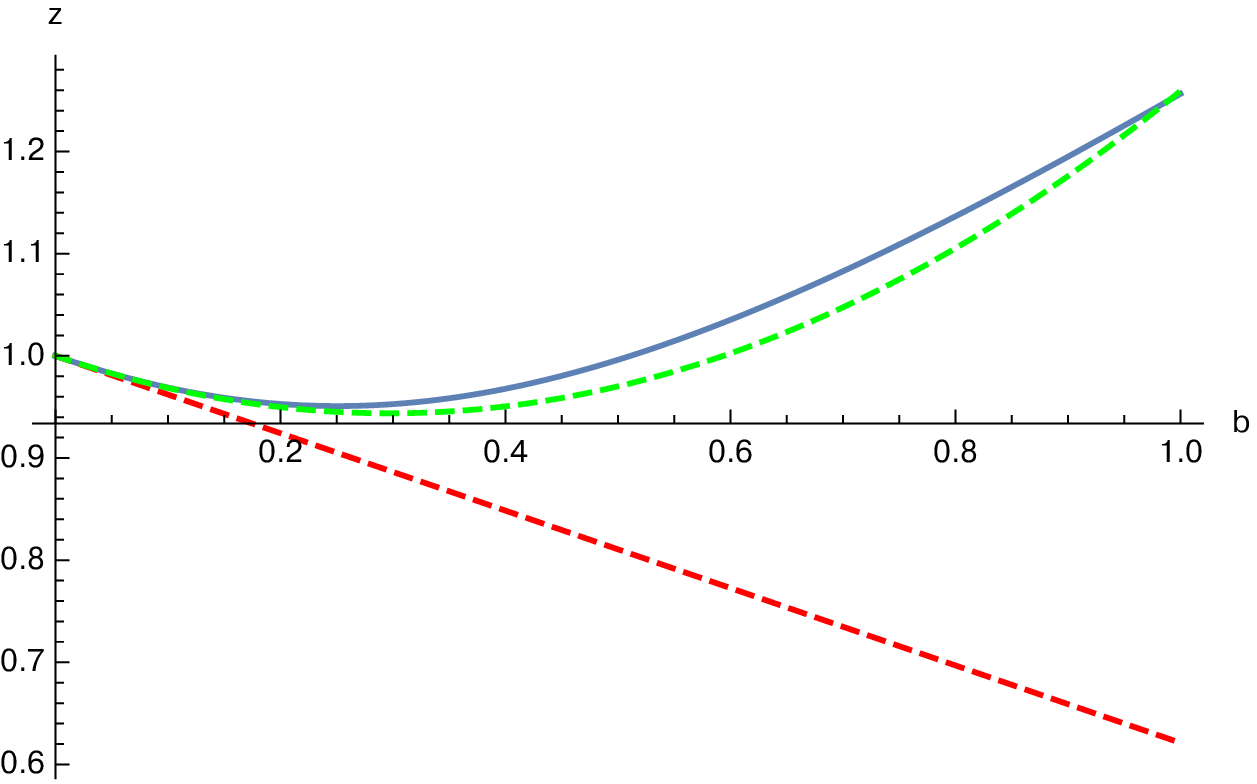}\\
\includegraphics[width=2.6in]{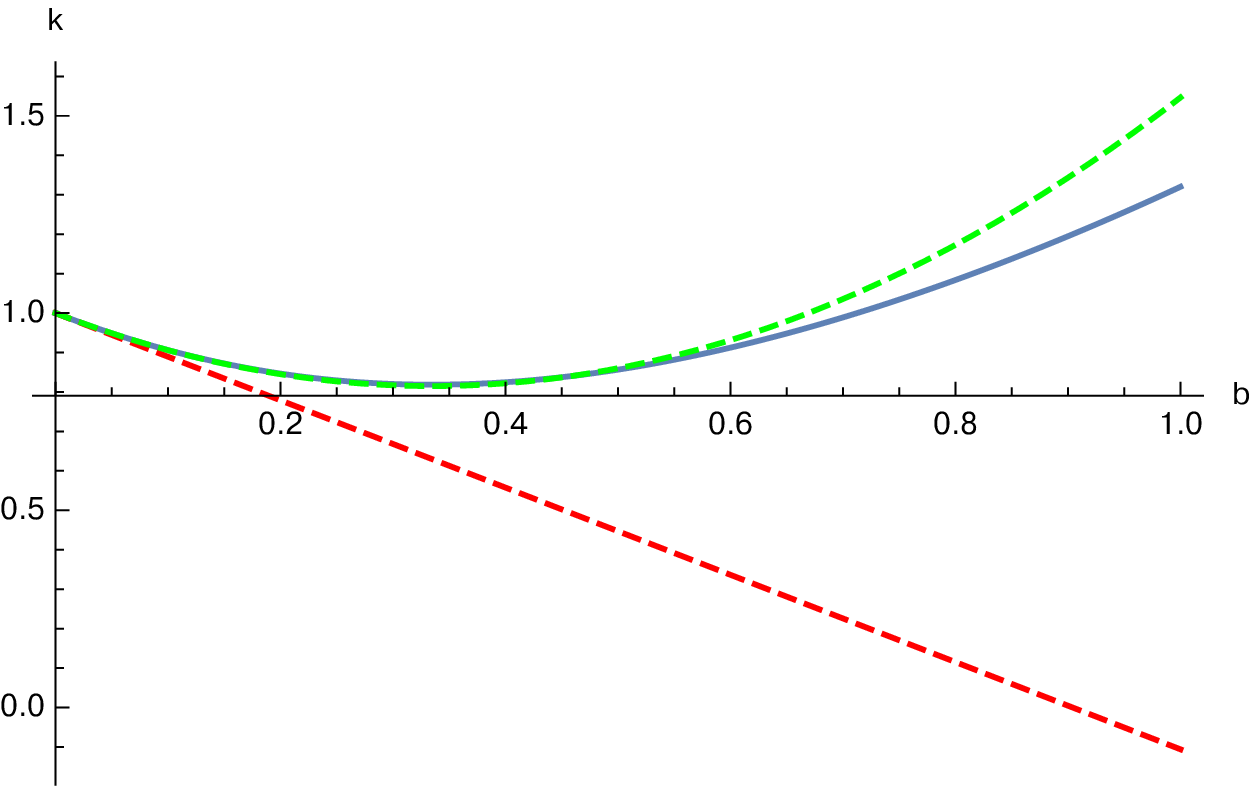}
\qquad
\includegraphics[width=2.6in]{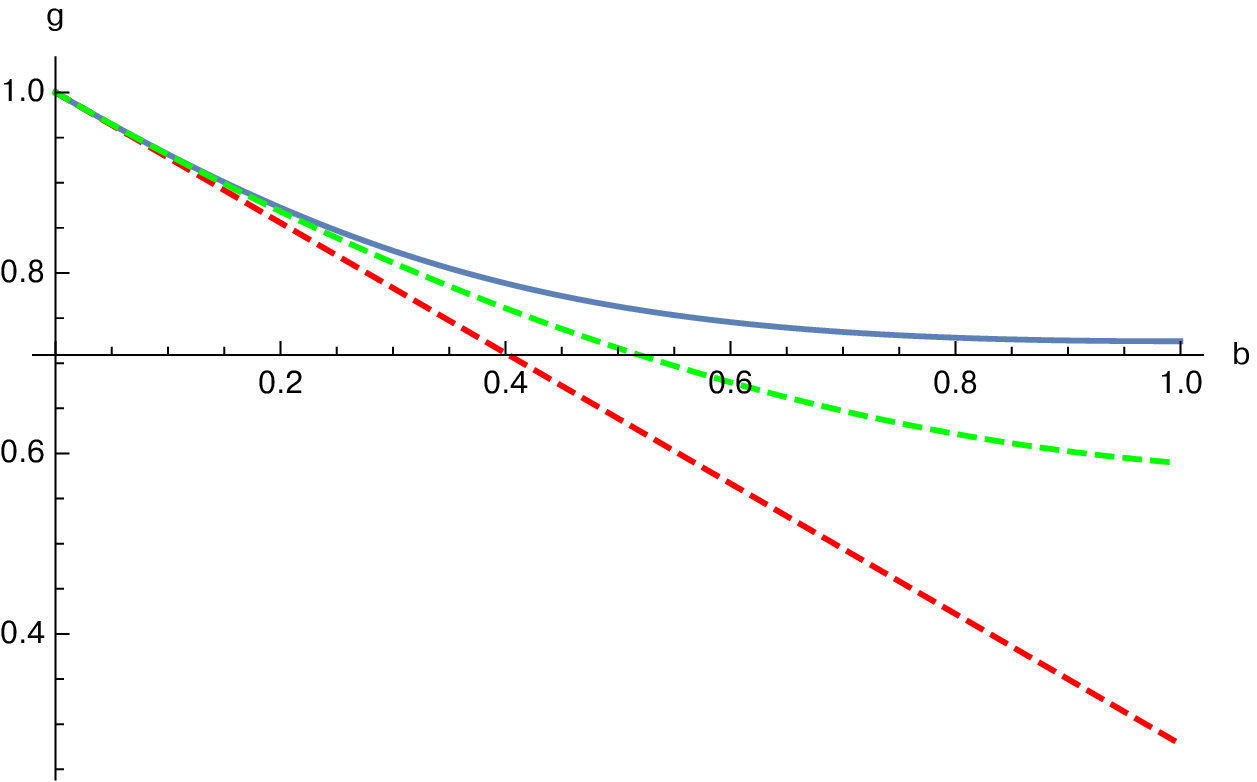}
\end{center}
  \caption{
Infrared parameters $\{f_{2,0}^h,f_{3,0}^h,K_{0}^h,{g}_{0}^h\}$ in the conformal limit $b\to 0$.
Blue curves: results in computational scheme SchemeII; red curves: perturbative approximation to 
order $\calo(b)$;  green curves: perturbative approximation to 
order $\calo(b^2)$; see \eqref{pertcond} with \eqref{numresults}. } \label{figure3}
\end{figure}

\begin{figure}[t]
\begin{center}
\psfrag{b}{{$b$}}
\psfrag{s}{{$s_{0}^h$}}
\includegraphics[width=2.6in]{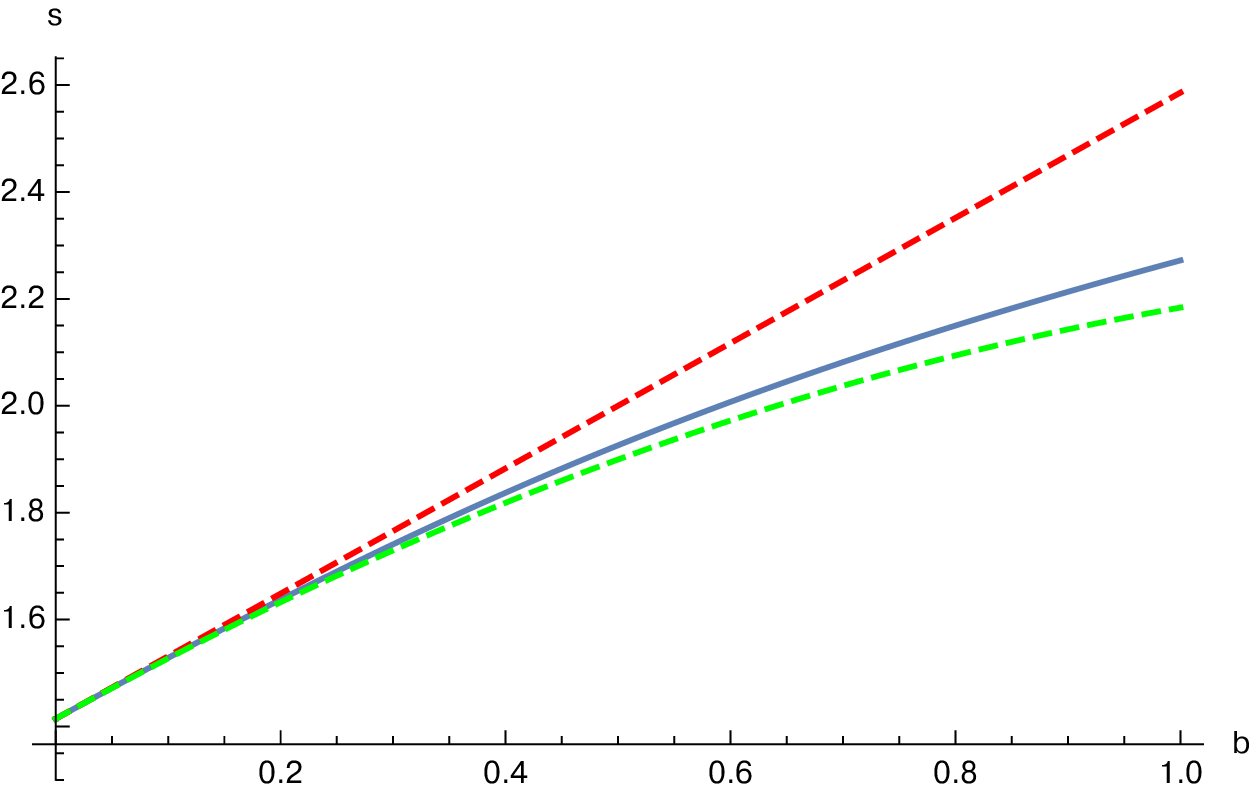}
\end{center}
  \caption{
Infrared parameter ${s}_{0}^h$ in the conformal limit $b\to 0$.
Blue curve: results in computational scheme SchemeII; red curve: perturbative approximation to 
order $\calo(b)$;  green curve: perturbative approximation to 
order $\calo(b^2)$; see \eqref{sh0pert} with \eqref{numsigma}. } \label{figure4}
\end{figure}

To study the conformal limit it is convenient to use the computational scheme SchemeII (see \eqref{compschemes}),
\ie we use the symmetry transformations SFG2-SFG4 of \eqref{kssym1}-\eqref{kssym3} to set
$H=g_s=K_0=1$ and allow $b\equiv P^2$ to vary. The FG frame equations of motion \eqref{kseq2}-\eqref{kseq10}
describing TypeA$_s$ vacua (see also \eqref{ktks1}) can be solved perturbatively as a series expansion in $b$:
\begin{equation}
\begin{split}
&f_2=(1+\r)\ \left(1+\sum_{n=1}^\infty b^n\ f_{2n}(\r)\right)\,,\qquad f_3=(1+\r)\
\left(1+\sum_{n=1}^\infty b^n\ f_{3n}(\r)\right)\,,\\
&h=\frac{1}{4(1+\r)^2}\  \left(1+\sum_{n=1}^\infty b^n\ h_{n}(\r)\right)\,,\ K=1+\sum_{n=1}^\infty b^n\ k_{n}(\r)\,,
\ g=1+\sum_{n=1}^\infty b^n\ g_{n}(\r)\,.
\end{split}
\eqlabel{ktschb1}
\end{equation}
Explicit equations for $\{f_{2n},f_{3n},h_n,k_n,g_n\}$ for $n=1,2$
along with the UV/IR asymptotics
are presented in appendix \ref{apdfg}. Numerically solving these equations we find perturbative
in $b$ predictions for the UV/IR  parameters \eqref{uvirpars}.
As explained in appendix \ref{apcef} we also need the FG frame parameter $s_0^h$, see \eqref{defsh0}.
Given \eqref{ktschb1} we find from \eqref{finalsh0}
\begin{equation}
\begin{split}
s_0^h=&\sqrt{2}\biggl(1+\frac{b}{4}\int_0^\infty ds\ \frac{h_1}{1+s}+\frac{b^2}{32}
\int_0^\infty ds\ \frac{8(1+s) h_2-(1+2s)h_1^2}{(1+s)^2}+\calo(b^3)
\biggr)\\
\equiv&\sqrt{2}\biggl(
1+s^h_{0;1}\ b  +s^h_{0;2}\ b^2+\calo(b^3)\biggr)\,.
\end{split}
\eqlabel{sh0pert}
\end{equation}
Using results of appendix \ref{typeasc} we evaluate the integrals in \eqref{sh0pert} to find
\begin{equation}
s^h_{0;1}=0.828534\,,\qquad s^h_{0;2}=-0.284396\,.
\eqlabel{numsigma}
\end{equation}
Figs.~\ref{figure3}-\ref{figure4} present comparison of the results for the IR parameters
$\{f_{2,0}^h\,,\ f_{3,0}^h\,,\ K_{0}^h\,,\ g_{0}^h\}$ and $s^h_0$ in the computational SchemeII
(blues curves), and independent perturbative $\calo(b)$ (red curves) and $\calo(b^2)$ (green curves)
computations. The agreement is excellent.

Following appendix \ref{apb2} we convert perturbative FG frame construction \eqref{ktschb1}
to EF frame:
\begin{equation}
\begin{split}
&a=-z (1-z)\ \biggl(1+\sum_{n=1}^\infty b^n\ a_n(z)\biggr)\,,\qquad \sigma=\sqrt{2}(1-z)\
\biggl(1+\sum_{n=1}^\infty b^n\ s_n(z)\biggr)\,,\\
&w_{c2}=\frac 12\  \biggl(1+\sum_{n=1}^\infty b^n\ w_{c2n}(z)\biggr)\,,\qquad
w_{a2}=\frac 12\  \biggl(1+\sum_{n=1}^\infty b^n\ w_{a2n}(z)\biggr)\,,\\
&K=1+\sum_{n=1}^\infty b^n\ k_{n}(z)\,,\qquad g=1+\sum_{n=1}^\infty b^n\ g_{n}(z)\,.
\end{split}
\eqlabel{ktefschb1}
\end{equation}
Explicit equations for $\{a_n,s_n,v_n\equiv w_{c2n}+4 w_{a2n},w_{a2n},k_n,g_n\}$ for
$n=1,2$ along with the initial conditions are presented in appendix \ref{apdef}.
The equations for $k_1$ and $g_1$ (\eqref{efn11} and \eqref{efn16} correspondingly)
can be solved analytically; in fact the solutions are just the ${\rm FG}\ \to\ {\rm EF}$ frame
transformations of \eqref{k1anal} and \eqref{analg1}:
\begin{equation}
\begin{split}
&k_1=\frac{z^2-z+1}{4z (z-1)}-\frac14-4 \ln2
+\frac{16 z^3-24 z^2+6 z+1}{4z^{3/2} (1-z)^{3/2}}\ \arctan\sqrt{\frac{z}{1-z}}\,,
\end{split}
\eqlabel{k1z}
\end{equation}
\begin{equation}
\begin{split}
&g_1=-\frac{13 z^4-26 z^3+29 z^2-16 z+1}{32z^2 (1-z)^2}+\frac{13}{32}
-\frac{2z-1}{16z^{5/2} (1-z)^{5/2}}\ \arctan\sqrt{\frac{z}{1-z}}\\
&-\frac{12 z^2-12 z-1}{32z^3 (z-1)^3}\ \arctan^2\sqrt{\frac{z}{1-z}}\,.
\end{split}
\eqlabel{g1z}
\end{equation}

We will show now that the location of the AH $z_{AH}$, as determined from the zero of the AH location
function $\call_{AH}$ \eqref{defcall}, is
\begin{equation}
1-z_{AH}=\calo\left(b^{1/4}\right)\,,
\eqlabel{zahorder}
\end{equation}
and can be determined analytically (in perturbative expansion in $b$) as it is controlled by the singularities
of the EOMs \eqref{efn12}-\eqref{efn15} and \eqref{efn22}-\eqref{efn26} as $u\equiv 1-z\to 0_+$, provided
we use \eqref{k1z} and \eqref{g1z}. From \eqref{k1z}, \eqref{g1z}:
\begin{equation}
k_1=-\frac{\pi}{8} u^{-3/2}-\frac{15\pi}{16} u^{-1/2}+\calo(u^{0/2})\,,\qquad g_1=-\frac{\pi^2}{128} u^{-3}
-\frac{15\pi^2}{128} u^{-2}+\calo(u^{-3/2})\,,
\eqlabel{k1g1}
\end{equation}
leading to\footnote{Subleading terms depend on coefficients
that have to be determined numerically.} (from direct asymptotic analysis of \eqref{efn12}-\eqref{efn15} and \eqref{efn22}-\eqref{efn26})
\begin{equation}
\begin{split}
&v_1=-\frac{3\pi^2}{256}u^{-3}+\frac{51\pi^2}{256}u^{-2}+\calo(u^{-3/2})\,,\qquad
a_1=\frac{3\pi^2}{1024}u^{-3}-\frac{177\pi^2}{1024} u^{-2}+\calo(u^{-3/2})\,,\\
&s_1=\frac{\pi^2}{512}u^{-3}-\frac{33\pi^2}{256} u^{-2}+\calo(u^{-3/2})\,,\qquad
w_{a21}=-\frac{\pi^2}{256}u^{-3}+\frac{9\pi^2}{256} u^{-2}+\calo(u^{-3/2})\,,
\end{split}
\eqlabel{sinn1}
\end{equation}
\begin{equation}
\begin{split}
&k_2=\frac{35\pi^3}{49152} u^{-9/2}-\frac{2985\pi^3}{229376} u^{-7/2}+\calo(u^{-6/2})\,,\\
&g_2=\frac{23\pi^4}{393216}u^{-6}-\frac{571\pi^4}{573440}u^{-5}+\calo(u^{-9/2})\,,\\
&v_2=\frac{21\pi^4}{262144} u^{-6}-\frac{1097\pi^4}{327680} u^{-5}+\calo(u^{-9/2})\,,\\
&a_2=-\frac{13\pi^4}{1310720} u^{-6}+\frac{751\pi^4}{2621440}u^{-5}+\calo(u^{-9/2})\,,
\\
&s_2=-\frac{53\pi^4}{7864320}u^{-6}+\frac{143\pi^4}{524288}u^{-5}+\calo(u^{-9/2})\,,\\
&w_{a22}=\frac{17\pi^4}{786432} u^{-6}-\frac{2599\pi^4}{2293760} u^{-5}+\calo(u^{-9/2})\,.
\end{split}
\eqlabel{sinn2}
\end{equation}
In fact, from the general structure of the perturbative equations we expect 
\begin{equation}
k_n=\calo(u^{-3n+3/2})\,,\qquad \{a,s,v,w_{a2},g\}_n=\calo(u^{-3n})\,, 
\eqlabel{sinnn}
\end{equation}
so that
\begin{equation}
b^n\ k_n\bigg|_{u=u_{AH}=\calo(b^{1/4})}=\calo(b^{n/4+3/8})\,,\qquad b^n\ \{a,s,v,w_{a2},g\}_n\bigg|_{u=u_{AH}=\calo(b^{1/4})}
=\calo(b^{n/4}) \,,
\eqlabel{ordersize}
\end{equation}
rendering successive higher order perturbative corrections in \eqref{ktefschb1} at $z=z_{AH}$ small
despite the singular behavior of $\{a_n,s_n,w_{c2n},w_{a2n},k_n,g_n\}$ in this
limit\footnote{This is similar to the behavior of the phenomenological model \cite{Buchel:2017lhu}
in the conformal limit.}.

\begin{figure}
\begin{center}
\psfrag{b}{{$\ln b$}}
\psfrag{y}{{$\ln(1-z_{AH})$}}
\psfrag{z}{{$\ln(4G_5\ s_{ent})$}}
\includegraphics[width=2.6in]{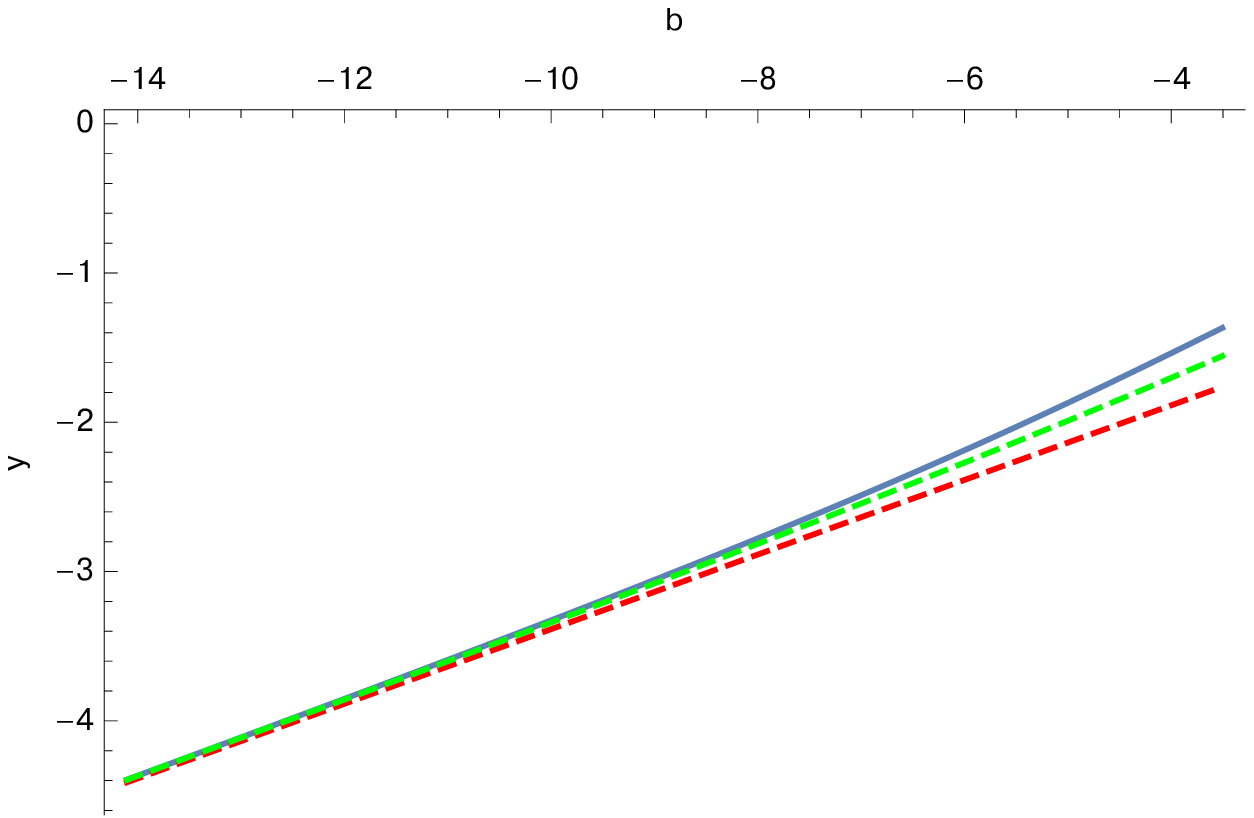}
\qquad
\includegraphics[width=2.6in]{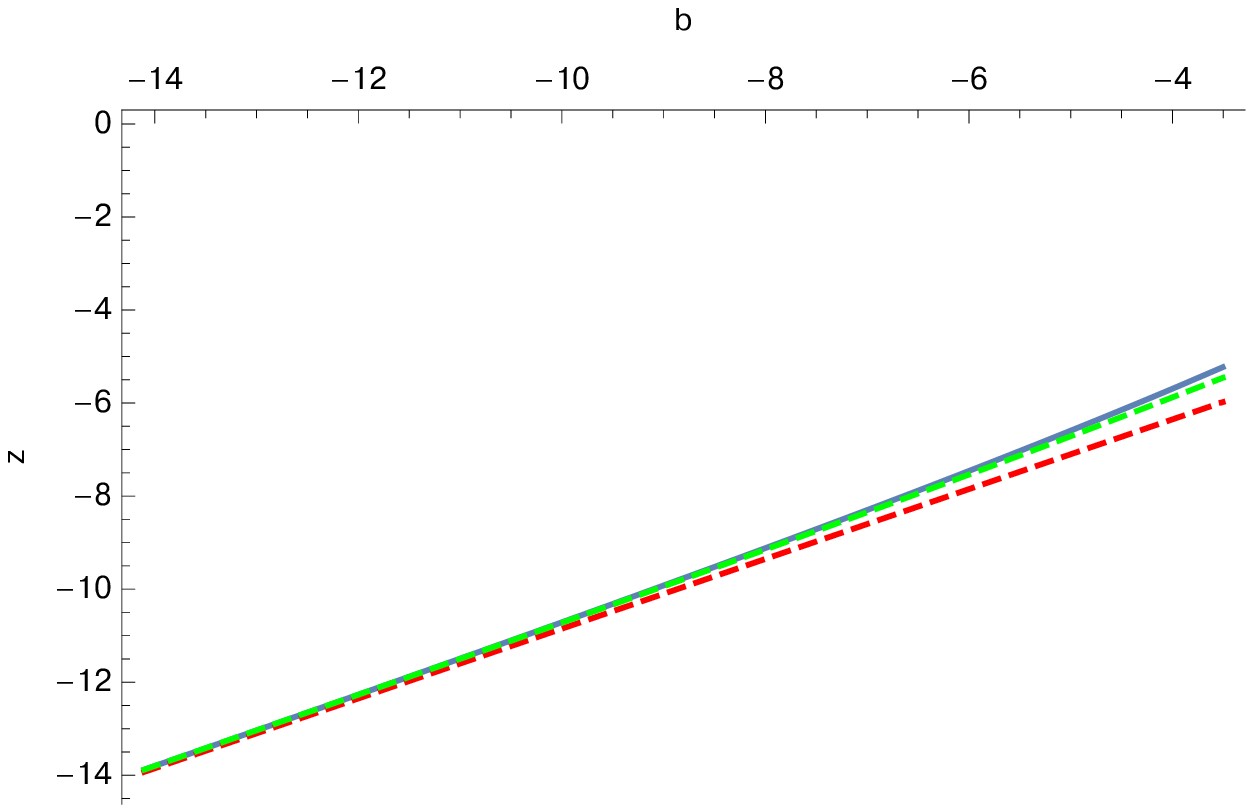}
\end{center}
  \caption{
Location of the apparent horizon $z_{AH}$ (left panel)
and the entanglement entropy $s_{ent}$ (right panel)
of TypeA$_s$ de Sitter vacua in the conformal limit $b\to 0$.
Blue curves: results in computational scheme SchemeII; red curves: leading perturbative approximation;
green curves: next-to-leading perturbative approximation, see \eqref{pertahres} and \eqref{sentconf}.} \label{conflim}
\end{figure}

Given \eqref{sinn1} and \eqref{sinn2} we find from \eqref{defcall}:
\begin{equation}
\begin{split}
&\call_{AH}(u\equiv 1-z)=\frac32 u^3\biggl(
u+b \left(-\frac{3\pi^2}{1024}u^{-3}-\frac{\pi^2}{64}u^{-2}
+\calo(u^{-3/2})\right)\\
&+b^2\left(0\cdot u^{-6}+\frac{349\pi^4}{3932160}u^{-5}
+\calo(u^{-9/2})\right)+\calo\left(b^3\ u^{-9}\right)
\biggr)\,,
\end{split}
\eqlabel{pertcall}
\end{equation}
so that the first zero of the apparent horizon location function occurs at
\begin{equation}
1-z_{AH}=u_{AH}=\frac18 3^{1/4}(2\pi)^{1/2}\ b^{1/4}\ \biggl(
1+\frac16 3^{1/4} (2\pi)^{1/2}\ b^{1/4}+\calo(b^{1/2})\biggr)\,.
\eqlabel{pertahres}
\end{equation}
From \eqref{ahsvac} we find perturbative predictions in the conformal limit for the TypeA$_s$ de Sitter vacua
entanglement entropy:
\begin{equation}
4G_5\ s_{ent}=\frac{1}{1024}3^{3/4} (2\pi)^{3/2}\ b^{3/4}\ \left(1+\frac12 3^{1/4}(2\pi)^{1/2}\ b^{1/4}
+\calo(b^{1/2})
\right)\,.
\eqlabel{sentconf}
\end{equation}
In fig.~\ref{conflim} we compare  numerical results for $z_{AH}$ and $s_{ent}$
in computational scheme SchemeII (blue curves) with the perturbative
predictions \eqref{pertahres} and \eqref{sentconf} at leading (red curves)
and next-to-leading (green curves) orders in the conformal limit: $b\to 0$.  
Restoring dimensional parameters, from \eqref{sentconf},
\begin{equation}
s_{ent}\bigg|_{{\rm TypeA}_s}\propto H^3\ \left(\ln \frac{H^2}{\Lambda^2}\right)^{-3/4}\qquad {\rm as}\qquad
H\gg \Lambda\,.
\eqlabel{sentfinal}
\end{equation}

\subsection{Validity of supergravity approximation for TypeA$_s$ vacua}\label{typeashmin}

\begin{figure}[t]
\begin{center}
\psfrag{x}{{$\ln \frac{H^2}{\Lambda^2}$}}
\psfrag{y}{{$\ln\hat{K}_{AH}$}}
\psfrag{t}{{$1/\hat{K}_{AH}$}}
\includegraphics[width=2.6in]{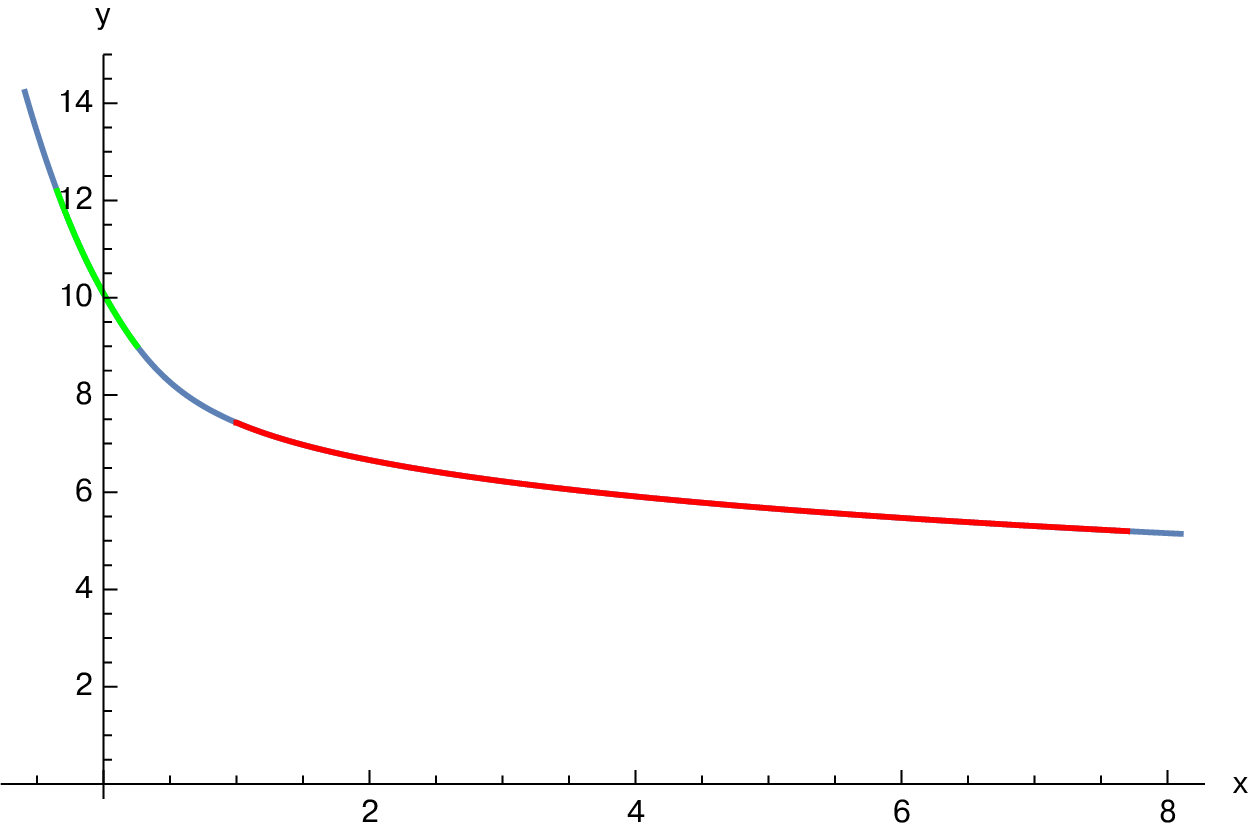}\qquad
\includegraphics[width=2.6in]{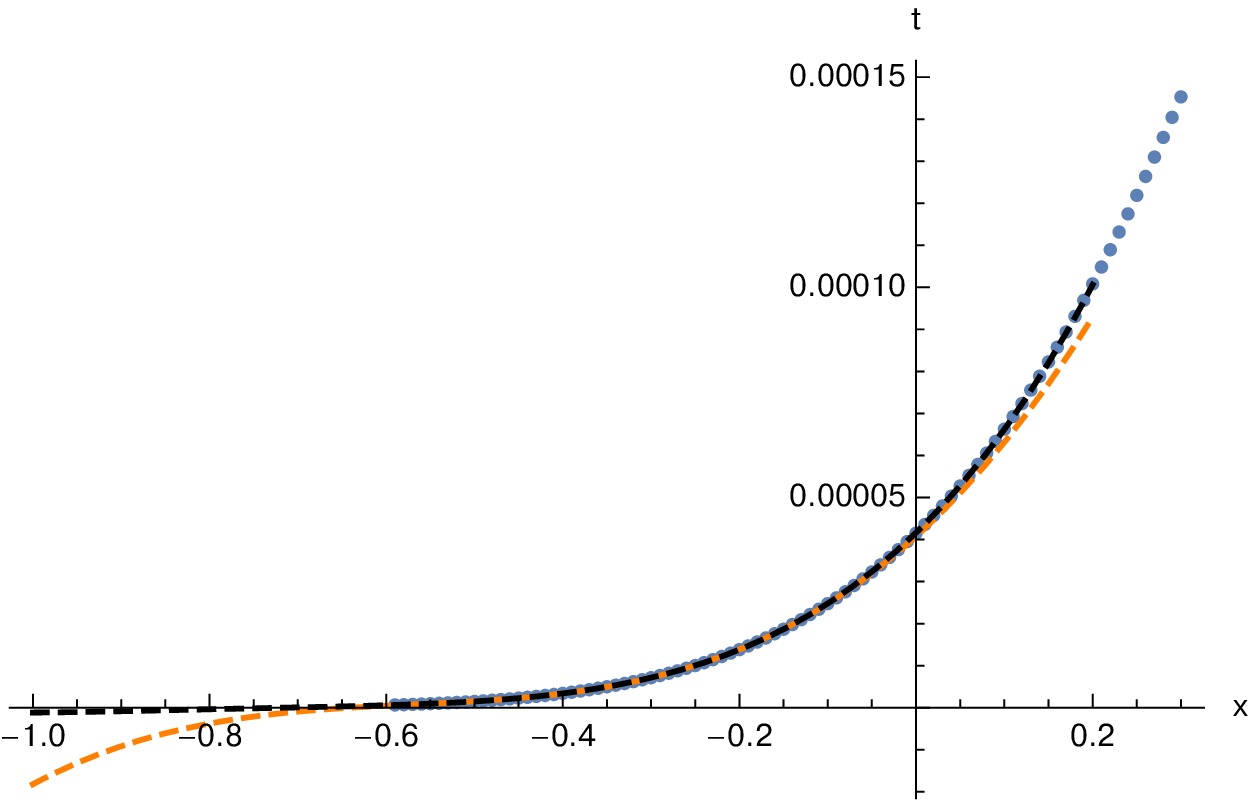}
\end{center}
  \caption{
Left panel: Kretschmann scalar of  \eqref{ef1i} evaluated at the
apparent horizon as  functions of $\ln\frac{H^2}{\Lambda^2}$
in different computation schemes  \eqref{compschemes}:
SchemeI (blue), SchemeII (red) and Scheme III
(green). Right panel: we use order-3 polynomial fit (orange dashed curve)
and order-4 polynomial fit (black dashed curve) to $\frac{1}{\hat{K}_{AH}}$,
see \eqref{zerofits}.
}\label{k}
\end{figure}

\begin{figure}[t]
\begin{center}
\psfrag{x}{{$\ln \frac{H^2}{\Lambda^2}$}}
\psfrag{z}{{$R^2_{T^{1,1}}/(Pg_s^{1/2})$}}
\psfrag{y}{{$\dd_{T^{1,1}}$}}
\includegraphics[width=2.6in]{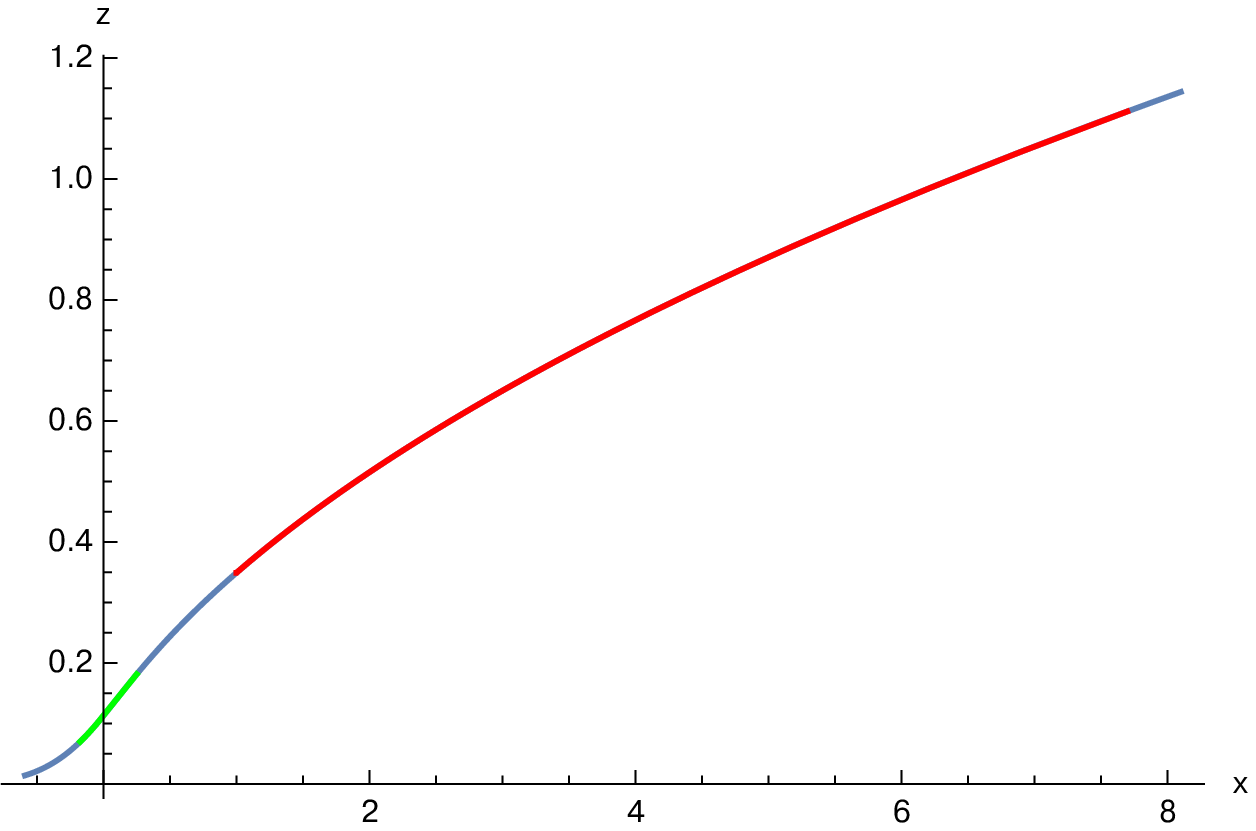}\qquad
\includegraphics[width=2.6in]{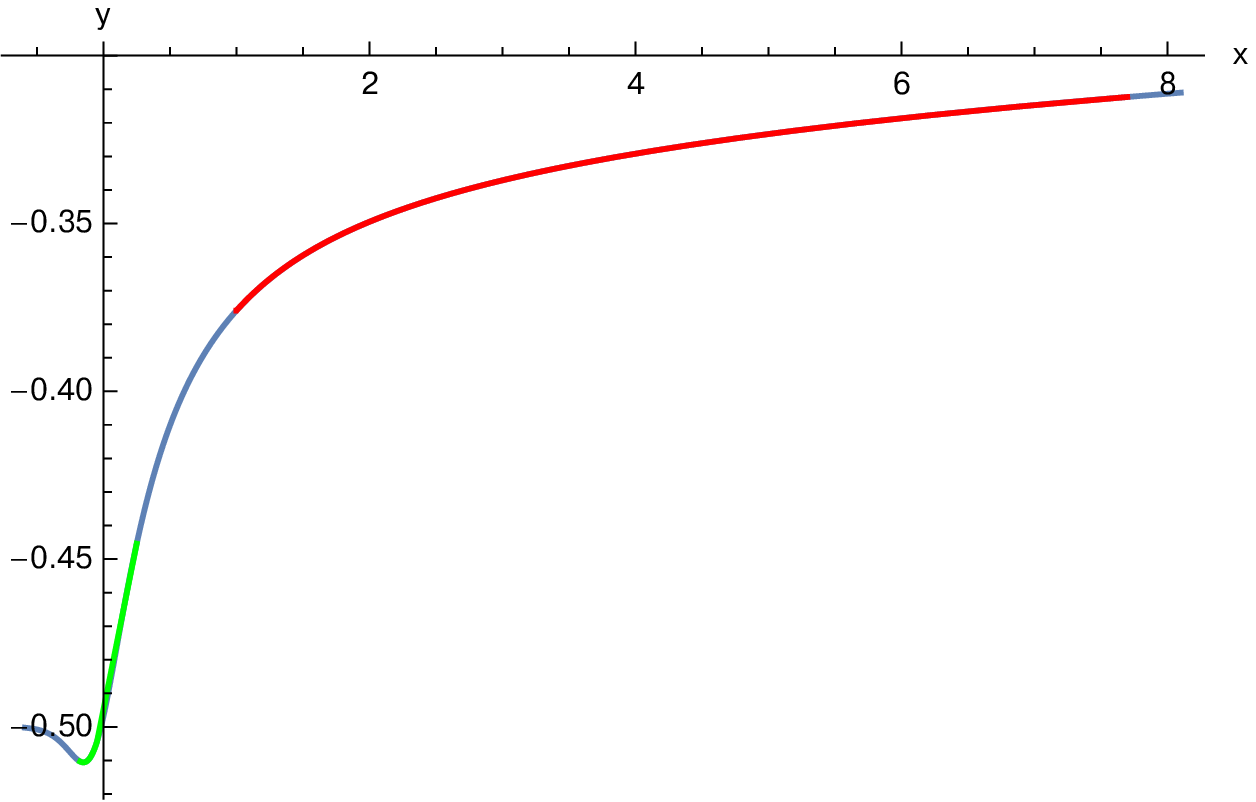}
\end{center}
  \caption{The curvature growth at the apparent horizon
  of the TypeA$_s$ de Sitter vacua gravitational
  dual for small $\frac{H^2}{\Lambda^2}$
  is due to collapsing the compact manifold: the size of deformed $T^{1,1}$,
  see \eqref{sizet11typeas} (left panel). Right panel: the $T^{1,1}$ deformation
  parameter $\dd_{T^{1,1}}$, see \eqref{deft11typeas}.
Results are presented 
in different computation schemes  \eqref{compschemes}:
SchemeI (blue), SchemeII (red) and Scheme III
(green).}\label{t11size}
\end{figure}

Results for the entanglement entropy $s_{ent}$ of TypeA$_s$ de Sitter vacua of the cascading gauge
theory are presented in section \ref{typeasnum}, see fig.~\ref{comparesent}. Notice that it is
a monotonically decreasing function of $\frac{H^2}{\Lambda^2}$. We have been able to obtain
reliable numerical results for
\begin{equation}
\ln\frac{H^2}{\Lambda^2}\ge -0.59\qquad \Longrightarrow\qquad 4G_5\ \hs_{ent}\gtrsim 4.1\times 10^{-4} \,.
\eqlabel{smallhsent}
\end{equation}
Besides numerical (technical) difficulties associated with construction of these vacua, there are conceptual
ones, associated with the breakdown of the supergravity approximation --- the effective action \eqref{5action}
becomes less reliable as the background space-time curvature of \eqref{ef1i} grows.
In fig.~\ref{k} (left panel) we present the Kretschmann scalar of  \eqref{ef1i} evaluated at the
apparent horizon in different computations schemes, see appendix \ref{kretschmann}:
\begin{equation}
\begin{split}
&{\rm SchemeI:}\qquad \ln\frac{H^2}{\Lambda^2}=k_s\,,\qquad \hat{K}=K\,;\\
&{\rm SchemeII:}\qquad \ln\frac{H^2}{\Lambda^2}=\frac 1b+\ln b\,,\qquad \hat{K}=b K\,;
\\
&{\rm SchemeIII:}\qquad \ln\frac{H^2}{\Lambda^2}=\frac 14+\ln\a\,,\qquad \hat{K}=K\,.
\end{split}
\eqlabel{collapsek}
\end{equation}
Notice the fast growth of $K_{AH}$ for small values of $\frac{H^2}{\Lambda^2}$ --- 
in fig.~\ref{k} (right panel) we fit the values of $\frac{1}{\hat{K}_{AH}}$
with order-3 (orange dashed curve) and order-4 (black dashed curve) polynomials.
The fits suggest that the curvature is divergent at
\begin{equation}
\ln\frac{H^2}{\Lambda^2}\bigg|_{orange\ fit}\approx -0.64\,,\qquad
\ln\frac{H^2}{\Lambda^2}\bigg|_{black\ fit}\approx -0.72\,.
\eqlabel{zerofits}
\end{equation}
We take \eqref{zerofits} as an indication that TypeA$_s$ vacua do not
exist\footnote{It would be interesting to rigorously establish this.} for
\begin{equation}
\ln\frac{(H_{min}^s)^2}{\Lambda^2}\ \lesssim\ -0.8\qquad \Longrightarrow\qquad H_{min}^s\ \lesssim\ 0.7 \Lambda\,.
\eqlabel{typeasdne}
\end{equation}
In fig.~\ref{t11size} (left panel) we identify the rapid curvature growth with the fact that the size of
(deformed) $T^{1,1}$, $R_{T^{1,1}}^2$, evaluated at the apparent horizon
\begin{equation}
R_{T^{1,1}}^2\equiv w_{a2}\bigg|_{AH}=Pg_s^{1/2}\ \hw_{a2}\bigg|_{AH}\,,
\eqlabel{sizet11typeas}
\end{equation}
becomes vanishingly small in string units, $P\propto M\a'=M\ \ell_s^2$.
Note that in the limit $R^2_{T^{1,1}}\to 0$ TypeA$_s$ vacua entanglement entropy vanishes, see
\eqref{senttypeas}.
Right panel shows the deformation parameter $\delta_{T^{1,1}}$ of the $T^{1,1}$: the size of the $U(1)$
fiber compare to the $S^2\times S^2$ base,
\begin{equation}
\dd_{T^{1,1}}\equiv 1-\frac{w_{c2}^2}{w_{a2}^2}\bigg|_{AH}=1-\frac{\hw_{c2}^2}{\hw_{a2}^2}\bigg|_{AH}\,.
\eqlabel{deft11typeas}
\end{equation}

\section{TypeA$_b$ de Sitter vacua}\label{typeabv}

TypeA$_b$ vacua have the same topology in Euclidean FG frame as TypeA$_s$ vacua \eqref{dstopology};
they differ in global symmetry: TypeA$_s$ vacua have unbroken $U(1)$ chiral symmetry (in the supergravity approximation),
while the latter symmetry is broken {\it spontaneously} to $\zet_2$ in TypeA$_b$ vacua. The following table
highlights the differences between the dual backgrounds in FG frame  and EF frame:
\begin{center}
\begin{table}[H]
\begin{tabular}{ c | c | c | c| c|}
 &  chiral symmetry & FG frame \eqref{fg1i} & EF frame \eqref{ef1i} & fluxes \eqref{redef}\\
 \hline\hline
 TypeA$_s$&  $U(1)$ & $f_a=f_b$    &  $w_{a2}=w_{b2}$  & $K_1=K_3\ \&\ K_2=1$\\
 \hline
 TypeA$_b$&  $\zet_2$ & $f_a\ne f_b$    &  $w_{a2}\ne w_{b2}$  & $K_1\ne K_3\ \&\ K_2\ne 1$
\end{tabular}
\caption{TypeA de Sitter vacua with broken/unbroken ($ _b\ /\ _s$) chiral symmetry.}
\label{table1}
\end{table}
\end{center}
Unlike TypeA$_s$ vacua, TypeA$_b$ vacua have never been constructed in the literature before ---
morally, they are similar to Klebanov-Strassler black holes, constructed only recently \cite{Buchel:2018bzp}. 
We begin in section \ref{onsetab} with perturbative construction of TypeA$_b$ vacua. Specifically,
we study static linearized perturbations about TypeA$_s$ vacua responsible for the chiral symmetry breaking
$U(1)\to \zet_2$. The symmetry breaking is associated with three operators $\calo_3^{\a=1,2}$ and $\calo_7$
(see section \ref{action})
developing  nonzero expectation values. We break the chiral symmetry {\it explicitly}, by turning on a
non-normalizable component for one of the dim-3
operators\footnote{This was discussed earlier in \cite{Buchel:2010wp}.} (a mass term for one of the gaugino bilinears).
We vary $\frac{H^2}{\Lambda^2}$ keeping the gaugino mass parameter fixed and nonzero --- the signature of the {\it spontaneous}
chiral symmetry breaking is the divergence of all the condensates $\calo_3^{\a=1,2}$ and $\calo_7$ for a particular value
of $\frac{H^2}{\Lambda^2}$, see fig.~\ref{flucuvir}. Once the bifurcation point of TypeA$_b$ vacua off TypeA$_s$
vacua is identified as a function of $\frac{H^2}{\Lambda^2}$, we construct fully nonlinear solution with
spontaneous symmetry breaking slowing increasing the amplitudes of the symmetry breaking expectation values, using
the linearized solution as a seed. Numerical results for TypeA$_b$ vacua are presented in section \ref{typeabnum},
in particular the results for the entanglement entropy  $s_{ent}\bigg|_{{\rm TypeA}_b}$ compare to
the entanglement entropy $s_{ent}\bigg|_{{\rm TypeA}_s}$ at corresponding values of $\frac{H^2}{\Lambda^2}$
are presented in fig.~\ref{comparesentasab}. Validity of supergravity approximation for TypeA$_b$ vacua
is a subject of section \ref{typeabsugra}.

\subsection{TypeA$_b$ vacua from perturbative chiral symmetry breaking of TypeA$_s$ vacua}\label{onsetab}
We will use computational scheme SchemeI \eqref{compschemes}.
Consider static, linearized chiral symmetry breaking fluctuation about TypeA$_s$ in FG frame, see table \ref{table1}:
\begin{equation}
f_a=f_3+\dd f\,,\ f_b=f_3-\dd f\,,\ K_1=K+\dd k_1\,,\ K_2=1+\dd k_2\,,\ K_1=K-\dd k_1\,,
\eqlabel{fldef}
\end{equation}
with the remaining metric functions and the string coupling as in TypeA$_s$ vacua, \ie  $\{f_c=f_2,h,g\}$.
It is straightforward to verify that truncation to $\{\dd f,\dd k_{1,2}\}$ is consistent (at the linearized level).
Equations of motion for the fluctuations and their asymptotic expansions in the UV $(\r\to 0)$ and the IR
($y=\frac 1\r$) are collected in appendix \ref{lincsb}. Once the non-normalizable coefficient
(the explicit chiral symmetry breaking parameter, \ie the gaugino mass term) is fixed to 
$\dd f_{1,0}=1$, the expansions are characterized by 6 UV/IR parameters 
\begin{equation}
\begin{split}
&{\rm UV:}\qquad \{\dd f_{3,0}\,,\, \dd k_{1,3,0}\,,\, \dd f_{7,0}\}\,;\\
&{\rm IR:}\qquad \{\dd f_{0}^h\,,\ \dd k_{1,0}^h\,,\ \dd k_{2,0}^h\}\,,
\end{split}
\eqlabel{uvirparslin}
\end{equation}
which is the correct number of parameters to find a unique solution of 3 second-order differential equations \eqref{fl1}-\eqref{fl3}
for $\{\dd f,\dd k_{1,2}\}$ on the TypeA$_s$ background parameterized by $k_s$.

\begin{figure}[t]
\begin{center}
\psfrag{x}{{$k_s$}}
\psfrag{y}[r]{{${\color{blue} \frac{1}{\dd f_{3,0}}},\frac{1}{\dd k_{1,3,0}},{\color{orange} \frac{1}{\dd f_{7,0}}}\qquad $}}
\psfrag{z}[r]{{${\color{blue} \frac{1}{\dd f_{0}^h}},\frac{1}{\dd k_{1,0}^h},{\color{magenta} \frac{1}{\dd k_{2,0}^h}}$}}
\includegraphics[width=2.6in]{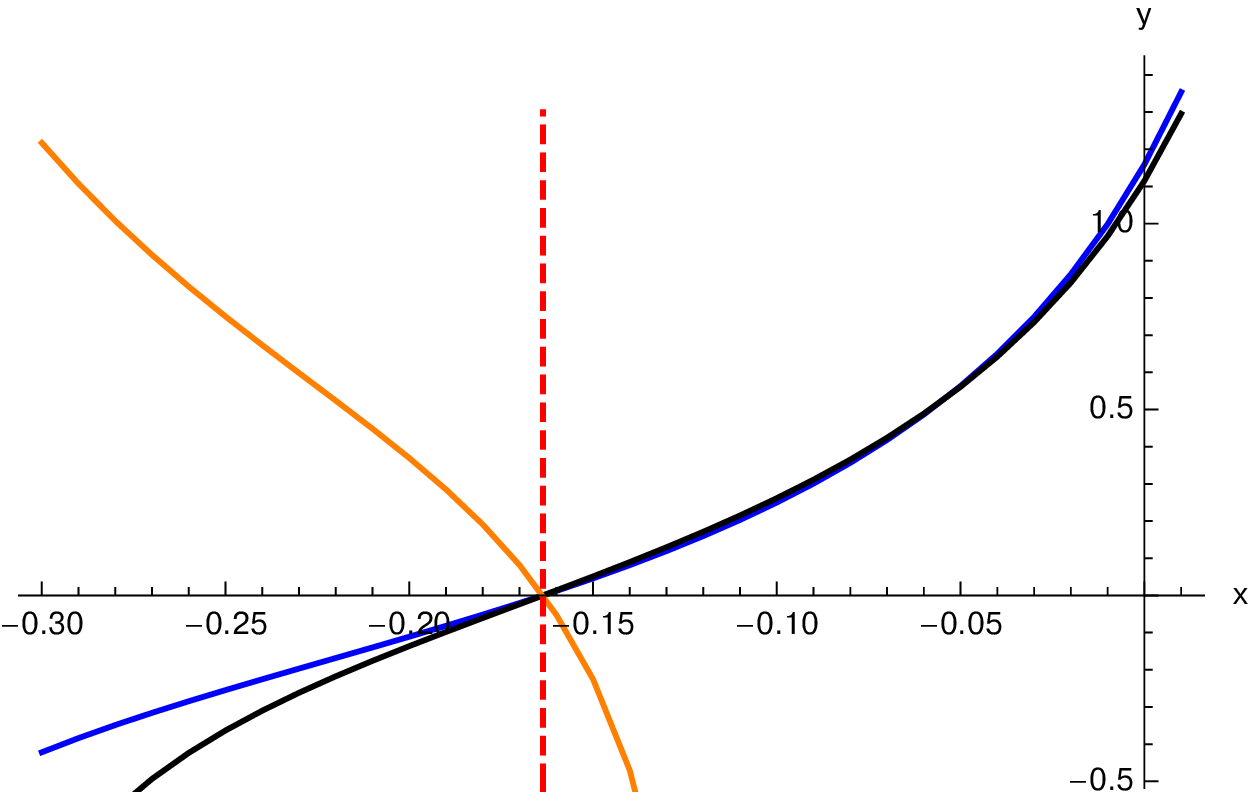}\qquad
\includegraphics[width=2.6in]{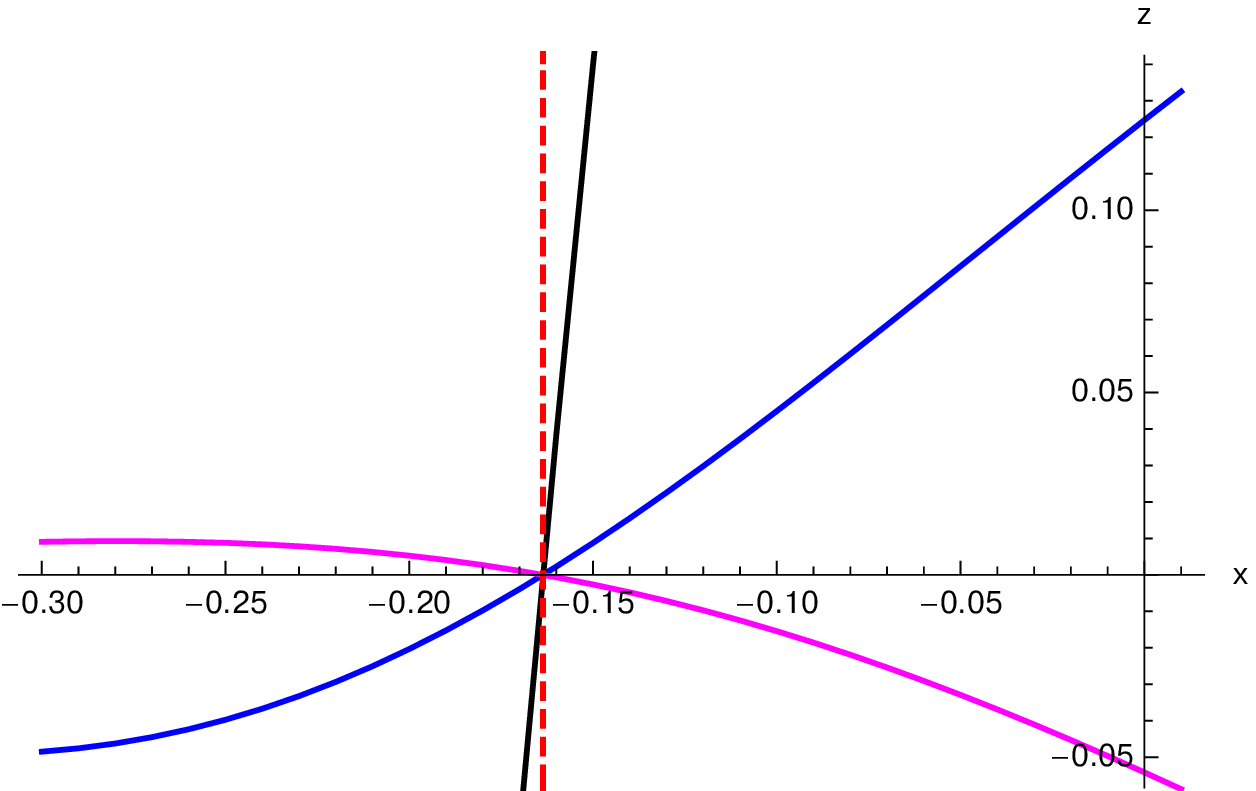}
\end{center}
  \caption{Parameters 
 $\{{\color{blue}\dd f_{3,0}}\,,\, \dd k_{1,3,0}\,,\, {\color{orange}\dd f_{7,0}}\,,\,
 {\color{blue}\dd f_{0}^h}\,,\, \dd k_{1,0}^h\,,\, {\color{magenta}\dd k_{2,0}^h}\}$ 
of the chiral symmetry breaking fluctuations over TypeA$_s$ vacua parameterized by $k_s$,
evaluated at fixed explicit chiral symmetry breaking scale $\dd f_{1,0}=1$, diverge
at $k_s^{crit}$ \eqref{defkcrit}, indicated by a vertical red dashed line. $k_s^{crit}$
identifies the bifurcation point of spontaneous symmetry broken TypeA$_b$ de Sitter vacua off
chirally symmetric TypeA$_s$ de Sitter vacua parameterized by $\ln\frac{H^2}{\Lambda^2}$.
}\label{flucuvir}
\end{figure}

In fig.~\ref{flucuvir} we assemble results for the fluctuation parameters
\eqref{uvirparslin} as $k_s$ label of TypeA$_s$ vacua is varied.
A signature of the spontaneous symmetry breaking is the divergence of all
the parameters, once the scale of the explicit chiral symmetry breaking,
\ie the non-normalizable parameter $\dd f_{1,0}$, is kept fixed. This occurs
at 
\begin{equation}
\ln\frac{(H_{min}^b)^2}{\Lambda^2}=k_s^{crit}=-16363(2)\qquad \Longrightarrow\qquad 
H_{min}^b=0.92(1)\Lambda\,,
\eqlabel{defkcrit}
\end{equation}
represented by vertical dashed red lines. 
We denote the critical value of $H$ corresponding to $k_s^{crit}$ as
$H_{min}^b$ --- we will see in section \ref{typeabnum} that
TypeA$_b$ vacua exist only for $H\ge H_{min}^b$, hence the name.
The value of $k_s^{crit}$ can be computed separately of each of the parameters ---
the fractional differences are of order $\propto 10^{-6}$, excepts for
\begin{equation}
\biggl(\frac{k_s^{crit}\bigg|_{\dd f_{7,0}}}{k_s^{crit}\bigg|_{\dd f_{3,0}}}-1\biggr)\ \propto\ 10^{-4}\,.
\eqlabel{xf70}
\end{equation}

\begin{figure}[t]
\begin{center}
\psfrag{x}{{$k_s$}}
\psfrag{y}{{$\chi_{k_{1,3,0}}$}}
\psfrag{z}{{$\chi_{f_{0}^h}$}}
\includegraphics[width=2.6in]{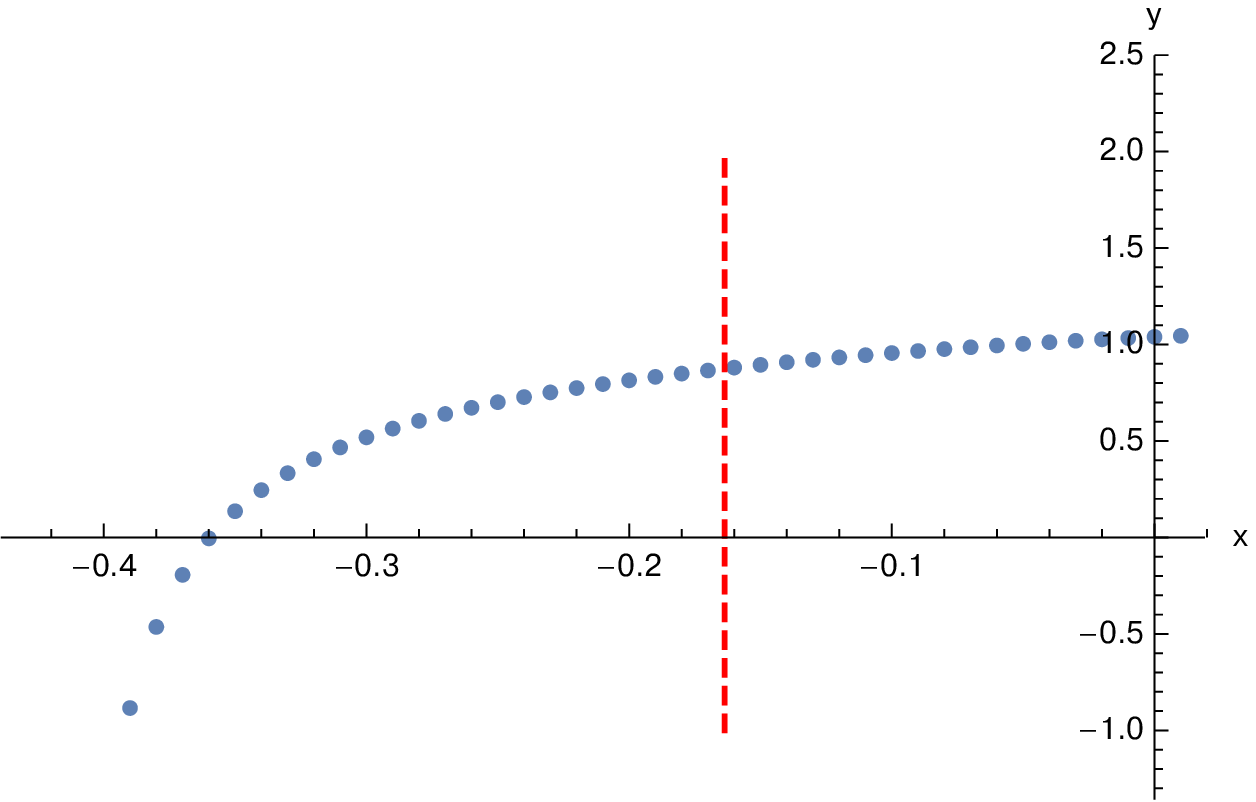}\qquad
\includegraphics[width=2.6in]{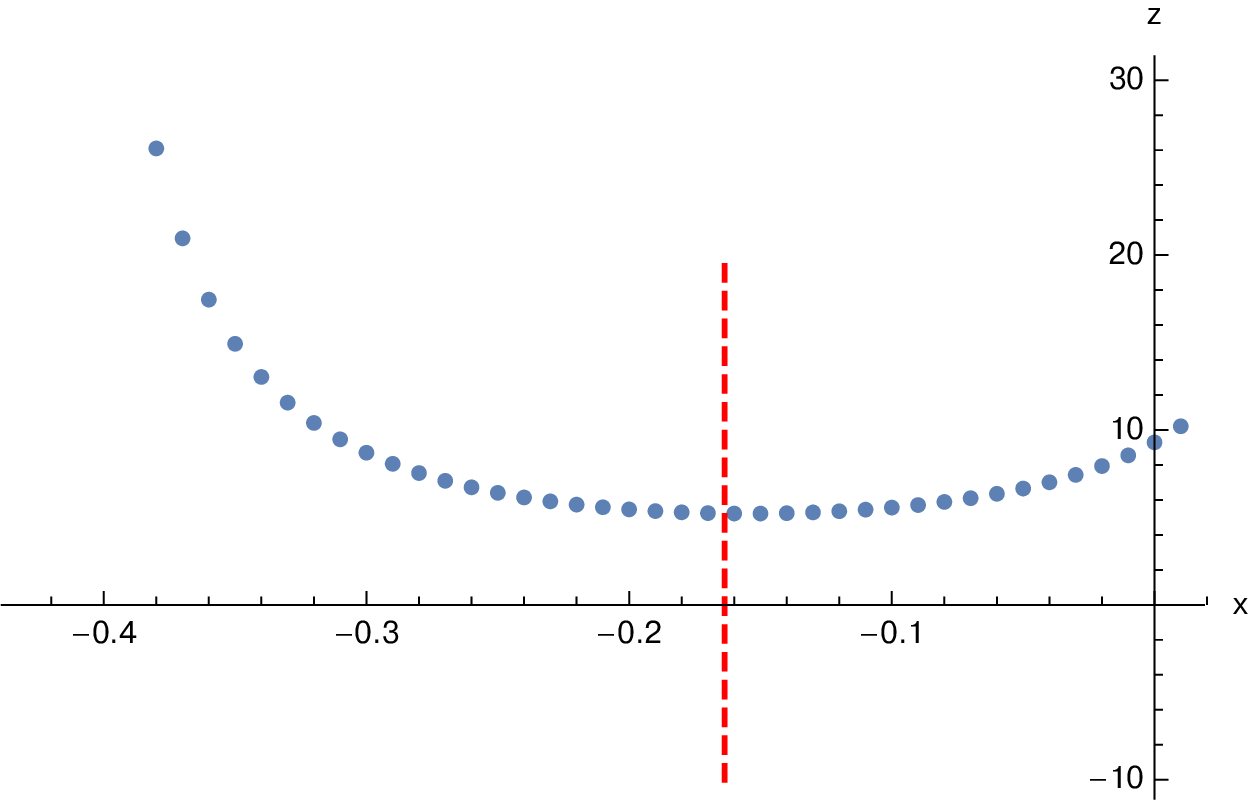}
\end{center}
  \caption{Sample susceptibilities, see \eqref{defkappa}, of the linearized
  chiral symmetry breaking fluctuations. The red dashed vertical line
  denotes $k_s^{crit}$, see \eqref{defkcrit}.
}\label{suceep}
\end{figure}

To use the critical fluctuations as a seed for TypeA$_b$ vacua, we need to know
the 'susceptibilities' 
\begin{equation}
\biggl\{\chi_{k_{1,3,0}}\,,\, \chi_{f_{7,0}}\,,\, \chi_{f_{0}^h}\,,\, \chi_{k_{1,0}^h}\,,\, \chi_{k_{2,0}^h} 
\biggr\}\ \equiv\  \lim_{k_s\to k_s^{crit}}\biggl\{
\frac{\dd k_{1,3,0}}{\dd f_{3,0}}\,,\, \frac{\dd f_{7,0}}{\dd f_{3,0}}\,,\,
\frac{\dd f_{0}^h}{\dd f_{3,0}}\,,\,
\frac{\dd k_{1,0}^h}{\dd f_{3,0}}\,,\,\frac{\dd k_{2,0}^h}{\dd f_{3,0}} 
\biggr\}\,.
\eqlabel{defkappa}
\end{equation}
In fig.~\ref{suceep} we present susceptibilities $\chi_{k_{1,3,0}}$ and  $\chi_{f_{0}^h}$ ---
notice that they are finite at $k_s^{crit}$, represented by vertical dashed red lines.
The other susceptibilities are finite as well; we find:
\begin{equation}
\begin{split}
&\chi_{k_{1,3,0}}=0.8749(7)\,,\qquad \chi_{f_{7,0}}=-0.2373(6)\,,\qquad \chi_{f_{0}^h}=5.230(0)\,,\qquad\\
&\chi_{k_{1,0}^h}=0.3034(2)\,,\qquad \chi_{k_{2,0}^h}=-18.12(6)\,.
\end{split}
\eqlabel{numkappa}
\end{equation}

\begin{figure}[t]
\begin{center}
\psfrag{l}{{$\l$}}
\psfrag{y}{{$k_{2,3,0}$}}
\psfrag{z}{{$f_{a,6,0}$}}
\psfrag{r}{{$f_{a,7,0}$}}
\includegraphics[width=1.7in]{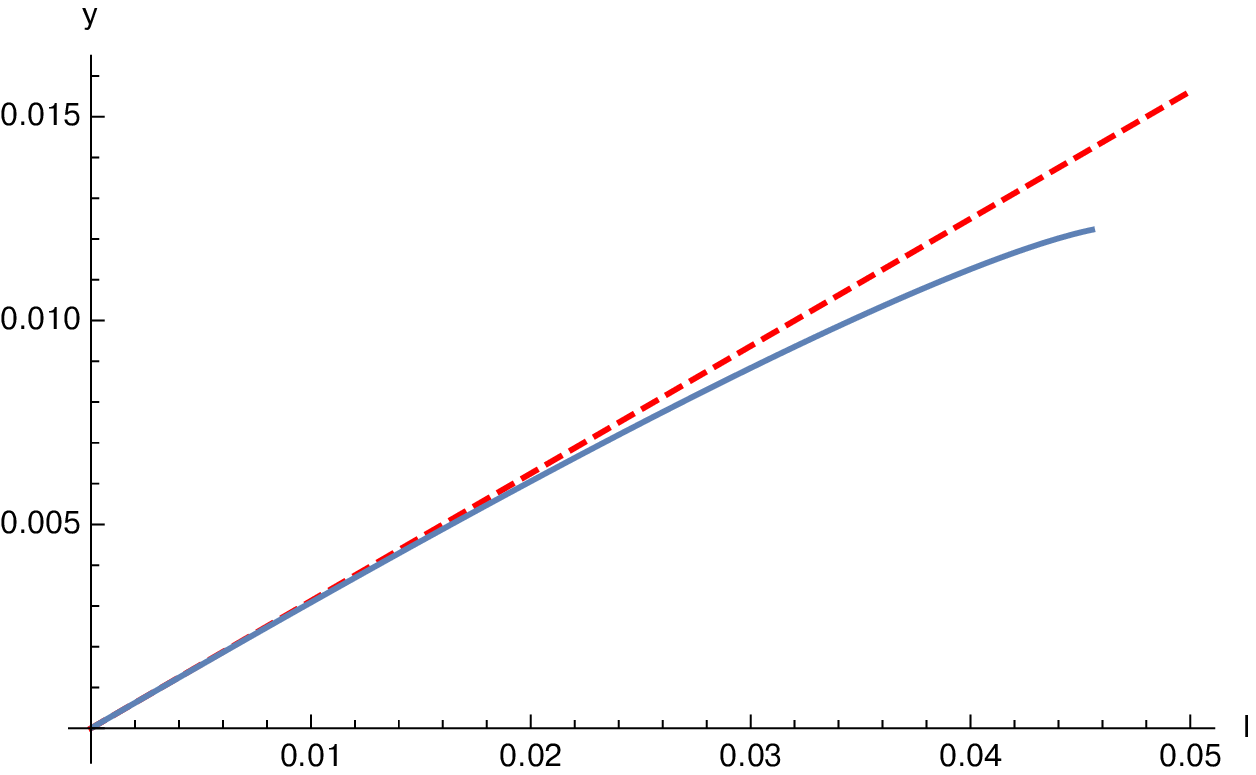}\qquad
\includegraphics[width=1.7in]{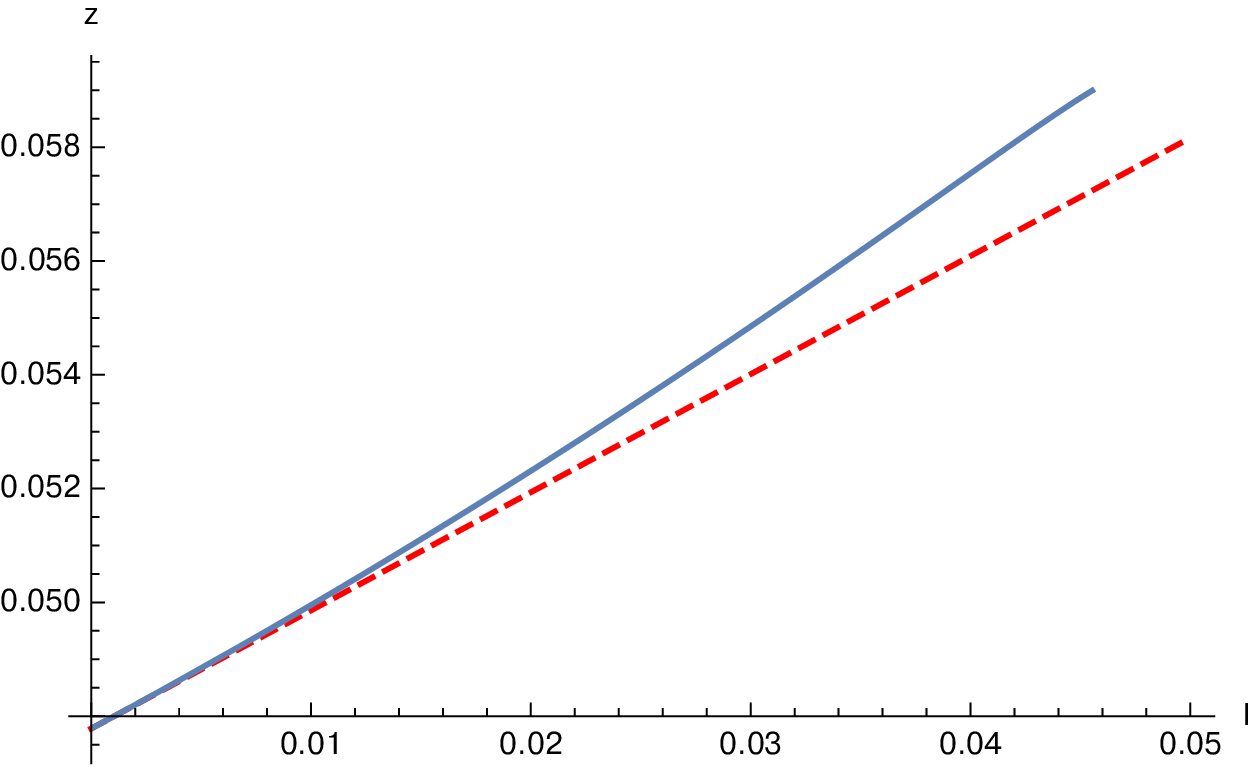}\qquad
\includegraphics[width=1.7in]{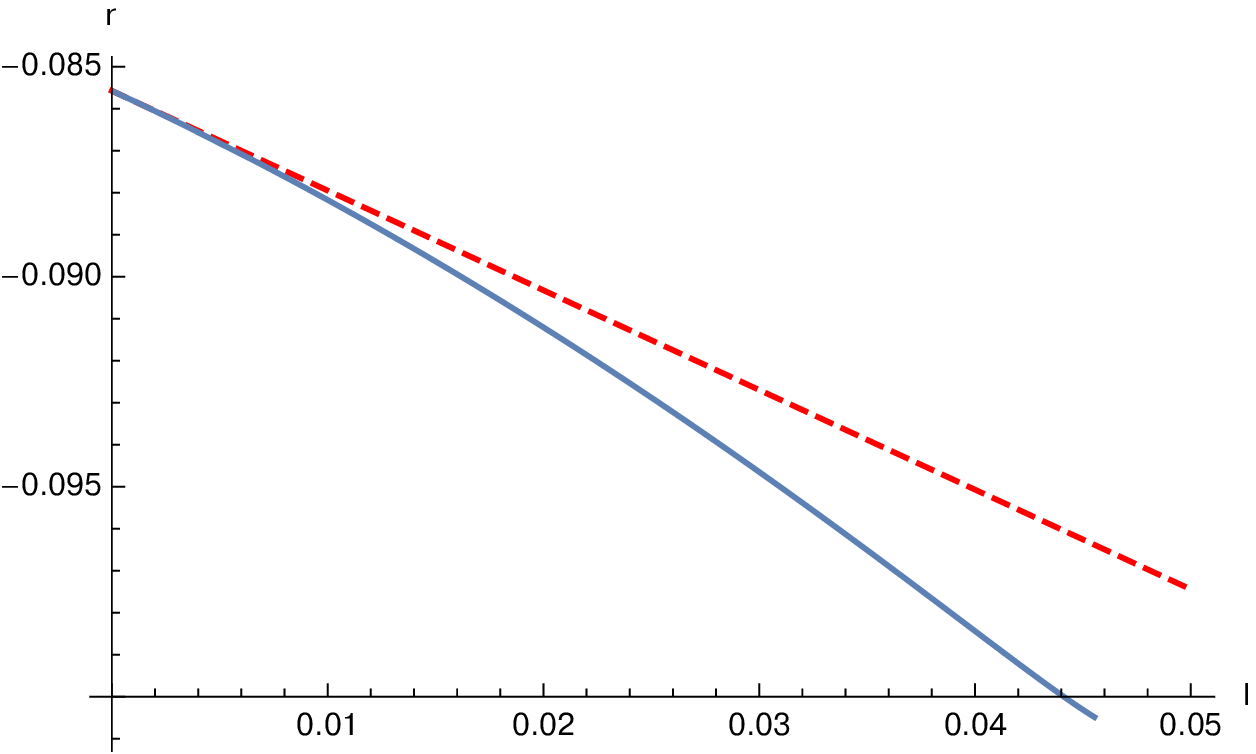}
\end{center}
  \caption{Sample of the UV parameters of TypeA$_b$ de Sitter vacua constructed from
  the 'seed' \eqref{seedtypeab}. The linearized approximations in $\l$ are represented
  by dashed red lines. 
}\label{uvlambda}
\end{figure}

\begin{figure}[t]
\begin{center}
\psfrag{l}{{$\l$}}
\psfrag{u}{{$K_{1,0}^h-K_{3,0}^h$}}
\psfrag{s}{{$f_{a,0}^h-f_{b,0}^h$}}
\psfrag{j}{{$K_{2,0}^h-1$}}
\includegraphics[width=1.7in]{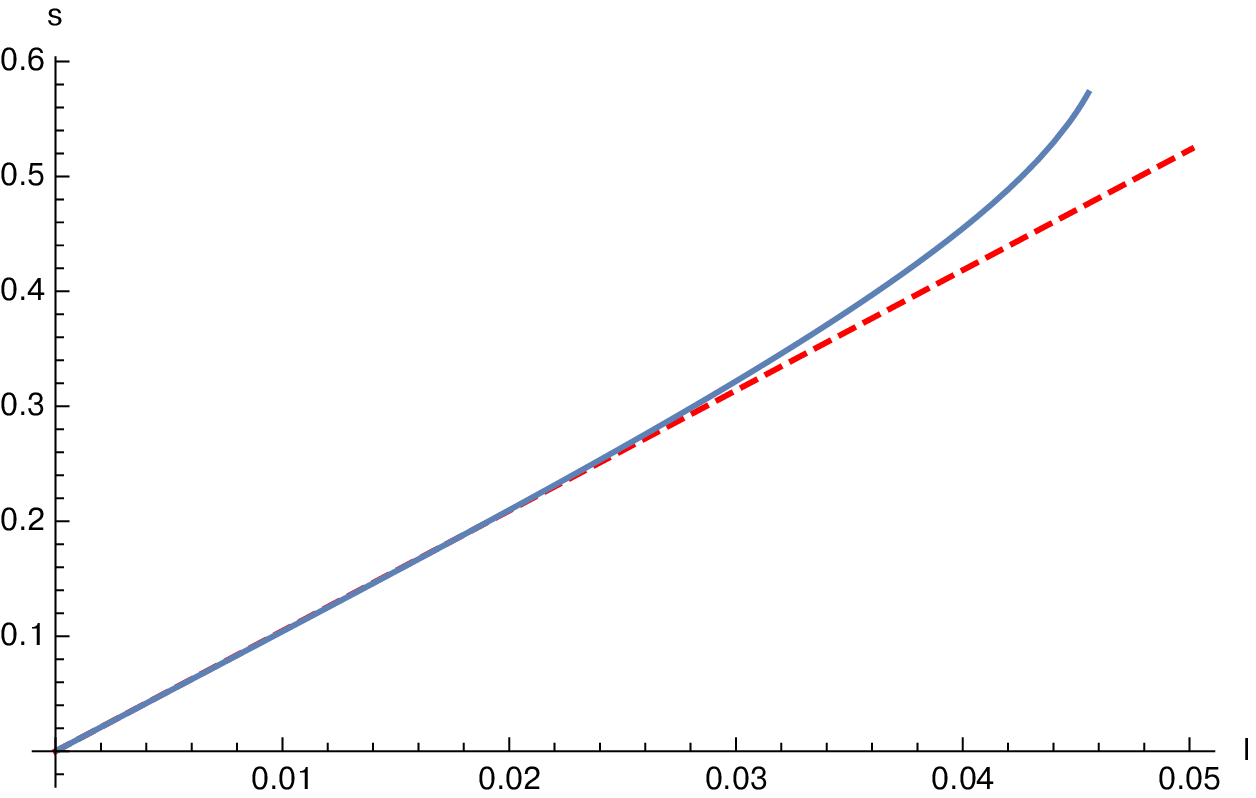}\qquad
\includegraphics[width=1.7in]{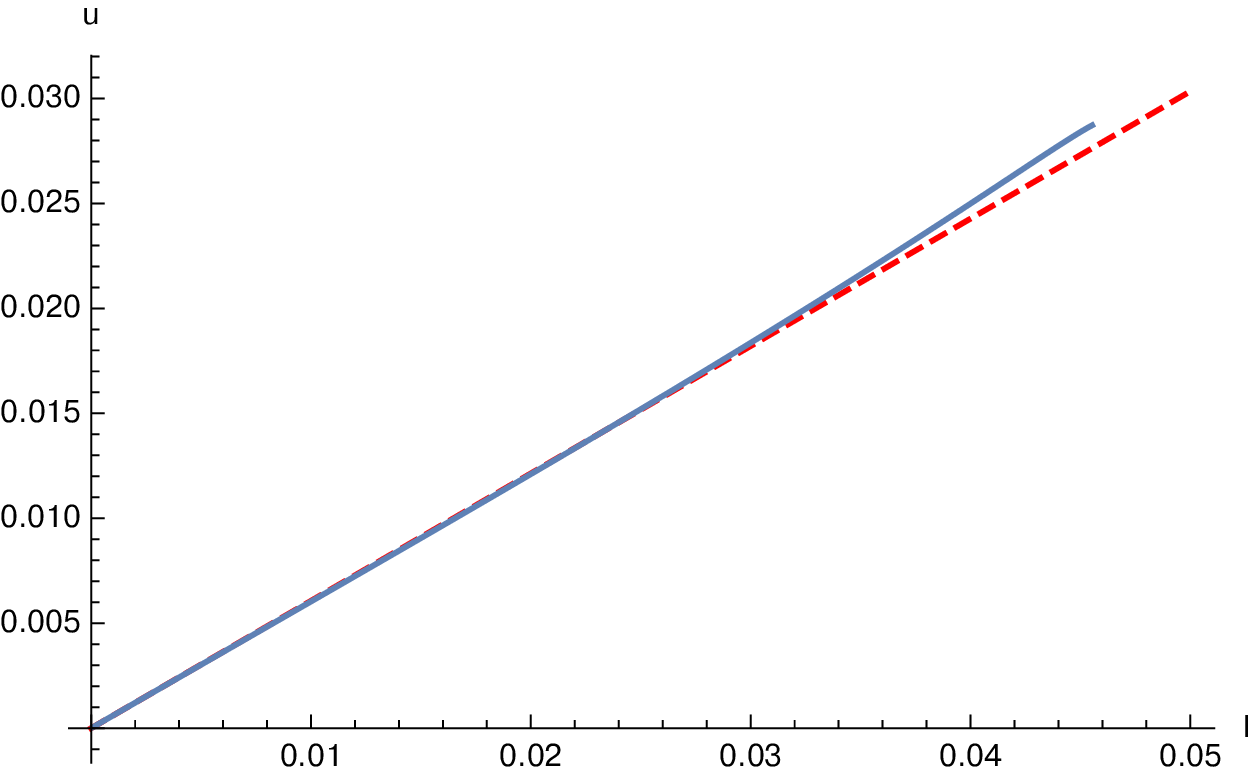}\qquad
\includegraphics[width=1.7in]{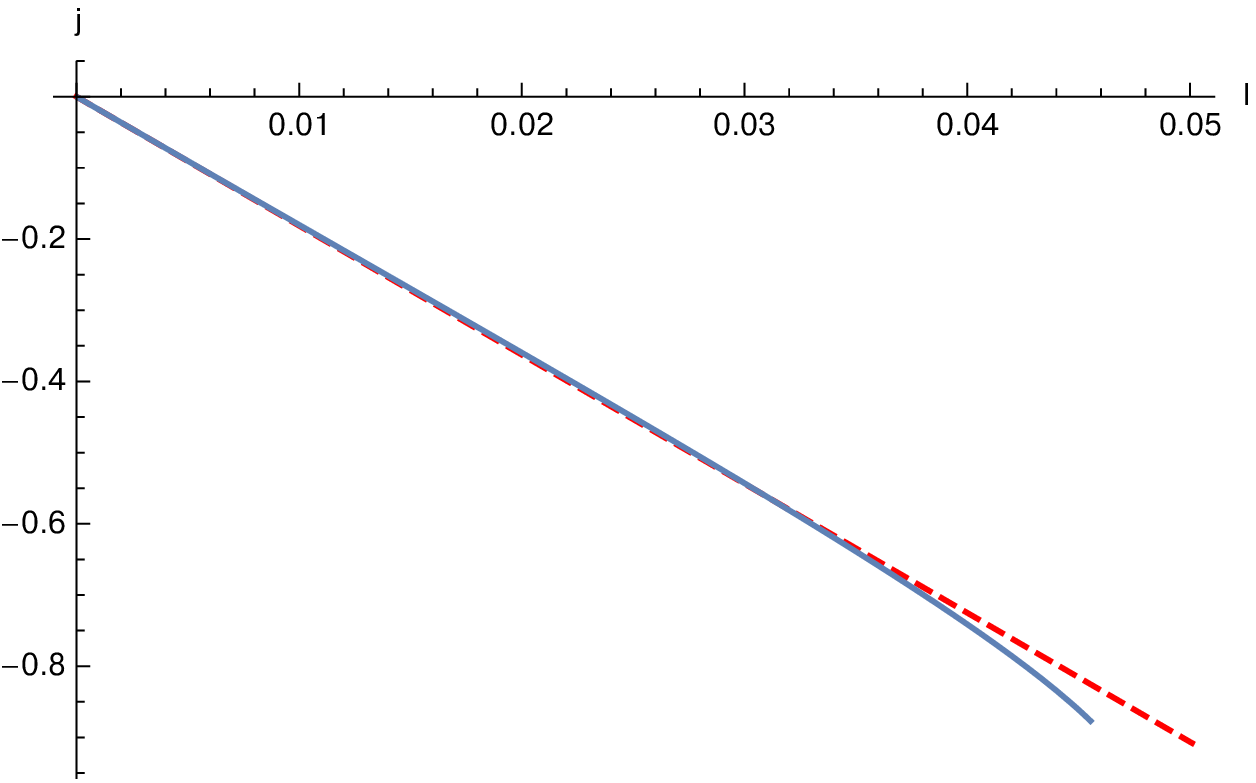}
\end{center}
  \caption{Sample of the IR parameters of TypeA$_b$ de Sitter vacua constructed from
  the 'seed' \eqref{seedtypeab}. The linearized approximations in $\l$ are represented
  by dashed red lines. 
}\label{irlambda}
\end{figure}

Given \eqref{numkappa}, fully nonlinear TypeA$_b$ vacua, with $k_s$ close to $k_s^{crit}$,
can be constructed following the same procedure as the one employed in construction of
Klebanov-Strassler black hole in \cite{Buchel:2018bzp}. We highlight the main steps:
\begin{itemize}
\item We set $k_s=k_s^{crit}$ and compute the corresponding TypeA$_s$ vacuum.
This vacuum is characterized by (see \eqref{uvparkt} and \eqref{irph2par2})
\begin{equation}
\begin{split}
&{\rm UV}:\qquad \{K_0=k_s^{crit}\,,\, H=1\,,\, g_s=1\,,\, f_{2,1,0}^{crit}\,,\,
g_{4,0}^{crit}\,,\, f_{2,4,0}^{crit}\,,\, f_{2,6,0}^{crit}\,,\, f_{2,8,0}^{crit}\}\,;
\\
&{\rm IR}:\qquad \{f_{2,0}^{h,crit}\,,\, f_{3,0}^{h,crit}\,,\, K_{0}^{h,crit}\,,\, g_{0}^{h,crit}\}\,.
\end{split}
\eqlabel{crittypeas}
\end{equation}
Next, we use \eqref{map1}-\eqref{map5} to compute the corresponding 
\begin{equation}
\begin{split}
&{\rm UV}:\qquad \{f_{a,1,0}^{s,crit}\,,\, f_{a,3,0}^{s,crit}\,,\, k_{2,3,0}^{s,crit}
\,,\, g_{4,0}^{s,crit}\,,\, f_{c,4,0}^{s,crit}\,,\, f_{a,6,0}^{s,crit}\,,\, f_{a,7,0}^{s,crit}\,,\,
f_{a,8,0}^{s,crit}\}\,;
\\
&{\rm IR}:\qquad \{f_{a,0}^{h,s,crit}\,,\, f_{b,0}^{h,s,crit}\,,\, f_{c,0}^{h,s,crit}\,,\,
K_{1,0}^{h,s,crit}\,,\, K_{2,0}^{h,s,crit}\,,\,K_{3,0}^{h,s,crit}\,,\, g_{0}^{h,s,crit}\}\,.
\end{split}
\eqlabel{crittypeab0}
\end{equation}
We use superscript $ ^s$ to indicate that UV/IR parameters of TypeA$_b$ vacua
\eqref{uvparks} and \eqref{irph1par} are obtained from the critical TypeA$_s$ vacuum.
\item Let's denote the amplitude of the symmetry breaking condensate (see \eqref{fldef})
\begin{equation}
\dd f_{3,0}\equiv \frac 12 \left(f_{a,3,0}-f_{b,3,0}\right)=\l\,.
\eqlabel{deflambdafluc}
\end{equation}
Then,
\begin{equation}
\left\{\dd k_{1,3,0}\,,\, \dd f_{7,0}\,,\,\dd f_{0}^h\,,\,\dd k_{1,0}^h\,,\,\dd k_{2,0}^h
\right\}=\l\ \{
\chi_{k_{1,3,0}}\,,\, \chi_{f_{7,0}}\,,\, \chi_{f_{0}^h}\,,\, \chi_{k_{1,0}^h}\,,\,
\chi_{k_{2,0}^h}\} +\calo(\l^2)\,.
\eqlabel{allotherveves}
\end{equation}
\item Using \eqref{fldef} and \eqref{uvdf}-\eqref{uvdk2}, \eqref{irdfk1k2},
with $\dd f_{1,0}=0$, in asymptotic expansions  \eqref{ksfc}-\eqref{ksg} and
\eqref{irph1par} we find
\begin{equation}
\begin{split}
&k_s=k_s^{crit}+\calo(\l^2)\,,\qquad f_{a,1,0}=f_{a,1,0}^{s,crit}+\calo(\l^2)\,,\qquad f_{a,3,0}=f_{a,3,0}^{s,crit}+\l+\calo(\l^2)\,,\\
&k_{2,3,0}=k_{2,3,0}^{s,crit}-\l\left(1-\frac 32\chi_{k_{1,3,0}}\right)+\calo(\l^2)\,,\qquad
g_{4,0}=g_{4,0}^{s,crit}+\calo(\l^2)\,,\\
&f_{c,4,0}=f_{c,4,0}^{s,crit}+\calo(\l^2)\,,\\
&f_{a,6,0}=f_{a,6,0}^{s,crit}
-\frac{f_{2,1,0}^{crit}}{64} \biggl(8 (f_{2,1,0}^{crit})^2+18 \chi_{k_{1,3,0}}+12 k_s^{crit}-35\biggr)\ \l
+\calo(\l^2)\,,\\
&f_{a,7,0}=f_{a,7,0}^{s,crit}+\l\ \chi_{f_{7,0}}+\calo(\l^2)\,,\\
&f_{a,8,0}=f_{a,8,0}^{s,crit}-\frac{f_{2,1,0}^{crit}}{1536}\biggl(550-192 (f_{2,1,0}^{crit})^4
-720 \chi_{k_{1,3,0}} (f_{2,1,0}^{crit})^2-480 (f_{2,1,0}^{crit})^2 k_s^{crit}
\\&+36 \chi_{k_{1,3,0}} k_s^{crit}+1184 (f_{2,1,0}^{crit})^2+3840 \chi_{f_{7,0}}-45 \chi_{k_{1,3,0}}
+2304 f_{2,4,0}^{crit}+21 k_s^{crit}
\biggr)\ \l\\
&+\calo(\l^2)\,,\\
&f_{a,0}^h=f_{a,0}^{h,s,crit}+\c_{f_0^h}\ \l+\calo(\l^2)\,,\qquad
f_{b,0}^h=f_{b,0}^{h,s,crit}-\c_{f_0^h}\ \l+\calo(\l^2)\,,\\
&f_{c,0}^h=f_{c,0}^{h,s,crit}+\calo(\l^2)\,,\qquad
K_{1,0}^h=K_{1,0}^{h,s,crit}+\c_{k_{1,0}^h}\ \l+\calo(\l^2)\,,\\
&K_{2,0}^h=K_{2,0}^{h,s,crit}+\c_{k_{2,0}^h}\ \l+\calo(\l^2)\,,\qquad
K_{3,0}^h=K_{3,0}^{h,s,crit}-\c_{k_{1,0}^h}\ \l+\calo(\l^2)\,,\\
&g_{0}^h=g_{0}^{h,s,crit}+\calo(\l^2)\,.
\end{split}
\eqlabel{seedtypeab}
\end{equation}
\item We construct fully nonlinear in $\l$ TypeA$_b$ vacua using
the linearized approximation \eqref{seedtypeab} as a seed. Select UV/IR parameters,
along with the corresponding linearized approximations (dashed red lines)
are shown in figs.~\ref{uvlambda}-\ref{irlambda}.
\end{itemize}

\subsection{Numerical results: TypeA$_b$}\label{typeabnum}

Numerical construction of TypeA$_b$ vacua follows the steps of section \ref{typeasnum}.
In FG frame, there are 8 second order equations
\eqref{kseq2}-\eqref{kseq9} and 1 first order equation \eqref{kseq10}. The first order equation
\eqref{kseq10} involves (linearly) $f_c'$ and can be used instead of one of the second order equations
(namely, the one involving $f_c''$). Thus, altogether we have a coupled system of 7 second order
ODEs (linear in  $\{f_a'',f_b'',h'',K_1'',K_2'',K_3'',g''\}$) and a single first order equation
(linear in $f_c'$). As a result,
a unique solution must be characterized by $15=2\times 7+1$ parameters; these are the UV/IR parameters 
\begin{equation}
\begin{split}
&{\rm UV:}\qquad \{f_{a,1,0}\,,\, f_{a,3,0}\,,\, k_{2,3,0}\,,\, g_{4,0}\,,\, f_{c,4,0}\,,\ f_{a,6,0}\,,\, f_{a,7,0}\,,\, f_{a,8,0}\}\,;\\
&{\rm IR:}\qquad \{f_{a,0}^h\,,\ f_{b,0}^h\,,\ f_{c,0}^h\,,\ K_{1,0}^h\,,\ K_{2,0}^h\,,\ K_{3,0}^h\,,\ g_{0}^h\}\,.
\end{split}
\eqlabel{uvirparstypeab}
\end{equation}
It is rather challenging to find the solutions of the corresponding system of ODEs
in 15-dimensional parameter space by brute force --- fortunately,
we already know some solutions which are close to $k_s^{crit}$, see 
section \ref{onsetab}.

As for the construction of TypeA$_s$ we use three different computation schemes, see appendix
\ref{apcfg}. There are some differences though: both in SchemeII and SchemeIII
we use as a pivot value\footnote{As will be clear from the presented results this is a
convenient value.}
\begin{equation}
K_0^{\star}=-0.161344\,.
\eqlabel{kstar}
\end{equation}
Numerical results must not depend on which computational scheme is adopted.
We illustrate now that this is indeed the case using a sample of  IR parameters in \eqref{uvirparstypeab} as
an example\footnote{The same is true for the rest of IR parameters and the UV parameters as well.}.
Comparison of the different computational schemes is done using dimensionless and rescaled quantities:
$\ln \frac {H^{2}}{\Lambda^2}$ (as a vacuum label) \eqref{defks}
and $\{\hf_{a,b,c,0}^h\,,\, \hK_{1,2,3,0}^h\,,\, \hat{g}_0^h\}$
\eqref{irtypeascaled}.  Explicitly:
\begin{equation}
\begin{split}
&{\rm SchemeI:}\ \ \ln\frac{H^2}{\Lambda^2}=k_s\,,\ \hf_{a,b,c,0}^h=f_{a,b,c,0}^h\,,\
\hK_{1,2,3,0}^h=K_{1,2,3,0}^h\,,\ \hat{g}_0^h=g_0^h\,;\\
&{\rm SchemeII:}\ \ \ln\frac{H^2}{\Lambda^2}=\frac {K_0^{\star}}{b}
+\ln b\,,\ \hf_{a,b,c,0}^h=\frac{1}{b^{1/2}}f_{a,b,c,0}^h\,,\
\hK_{1,2,3,0}^h=\frac 1b K_{1,2,3,0}^h\,,\ \hat{g}_0^h=g_0^h\,;
\\
&{\rm SchemeIII:}\ \ \ln\frac{H^2}{\Lambda^2}=K_0^{\star}+\ln\a\,,\ \hf_{a,b,c,0}^h
=\frac{1}{\a^{1/2}}f_{a,b,c,0}^h\,,\
\hK_{1,2,3,0}^h=K_{1,2,3,0}^h\,,\ \hat{g}_0^h=g_0^h\,.
\end{split}
\eqlabel{collapsetypeab}
\end{equation}

\begin{figure}[t]
\begin{center}
\psfrag{x}{{$\ln \frac{H^2}{\Lambda^2}$}}
\psfrag{f}[r]{{$\hf_{a,0}^h-\hf_{b,0}^h$\qquad  }}
\psfrag{k}[r]{{$\hK_{1,0}^h$\qquad }}
\includegraphics[width=2.6in]{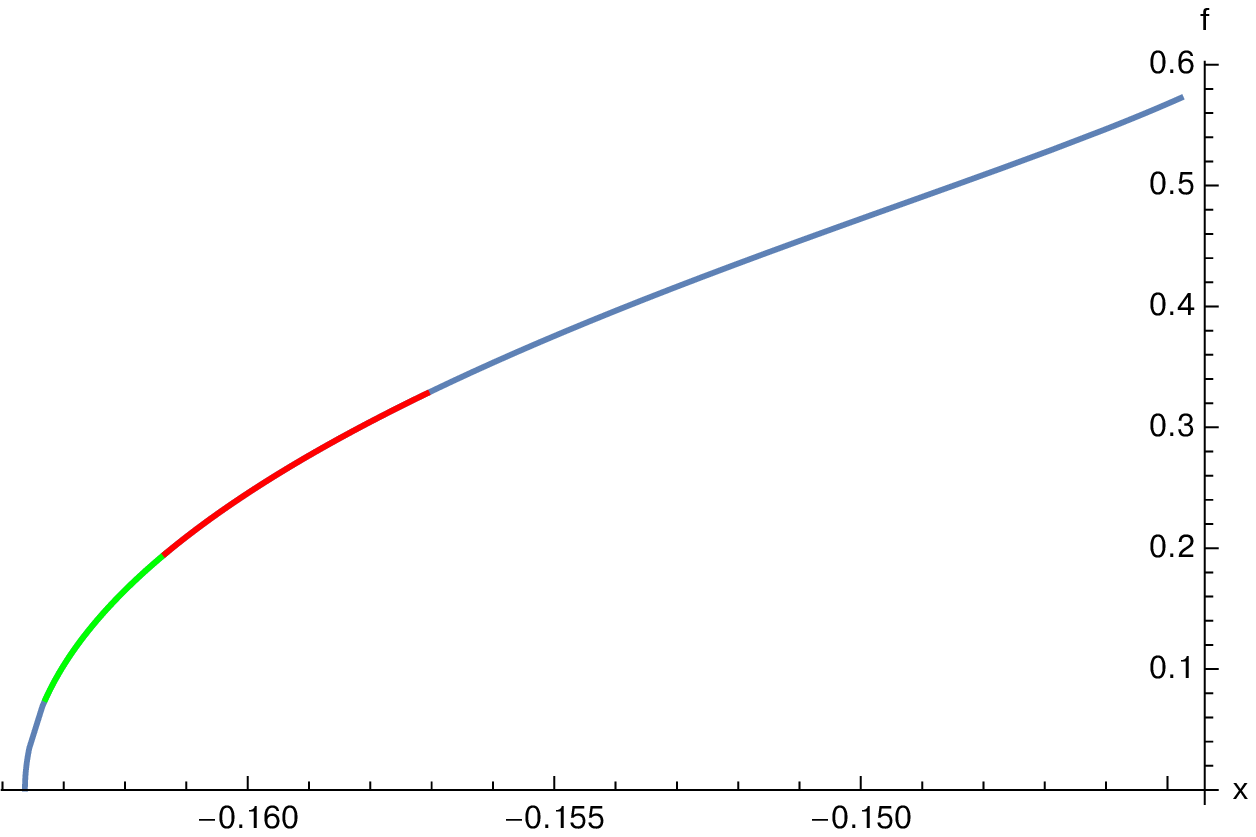}
\qquad \includegraphics[width=2.6in]{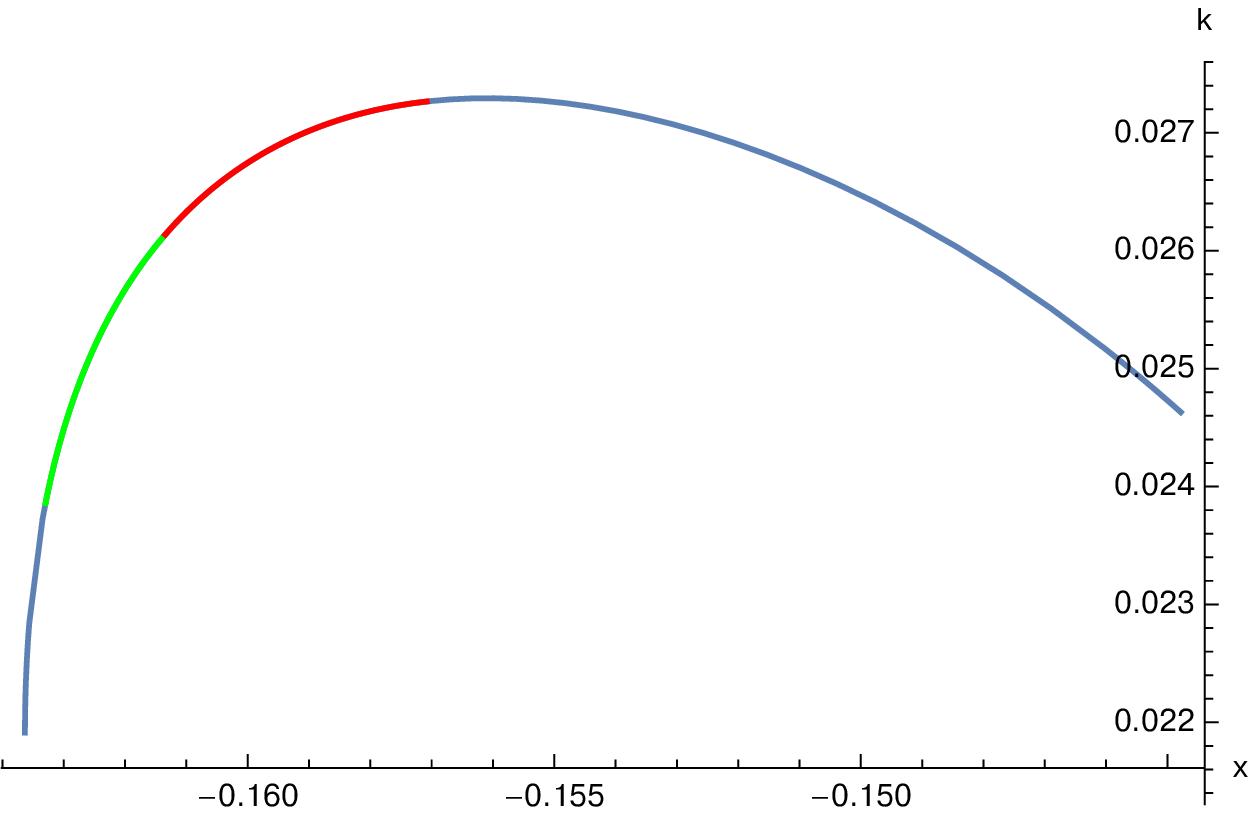}
\end{center}
  \caption{
Infrared parameters $\{\hf_{a,0}^h-\hf_{b,0}^h,\hK_{1,0}^h\}$  of the  
Fefferman-Graham coordinate frame of TypeA$_b$ de Sitter vacua of the cascading
gauge theory as functions of $\ln \frac{H^2}{\Lambda^2}$ in different computational
schemes \eqref{collapsetypeab}: SchemeI (blue), SchemeII (red) and Scheme III (green). 
} \label{plottypeabfg1}
\end{figure}

\begin{figure}[t]
\begin{center}
\psfrag{x}{{$\ln \frac{H^2}{\Lambda^2}$}}
\psfrag{z}{{${\color{red} \hK_{1,0}^h}/{\color{blue} \hK_{1,0}^h}-1$}}
\psfrag{y}{{${\color{green} \hK_{1,0}^h}/{\color{blue} \hK_{1,0}^h}-1$}}
\includegraphics[width=2.6in]{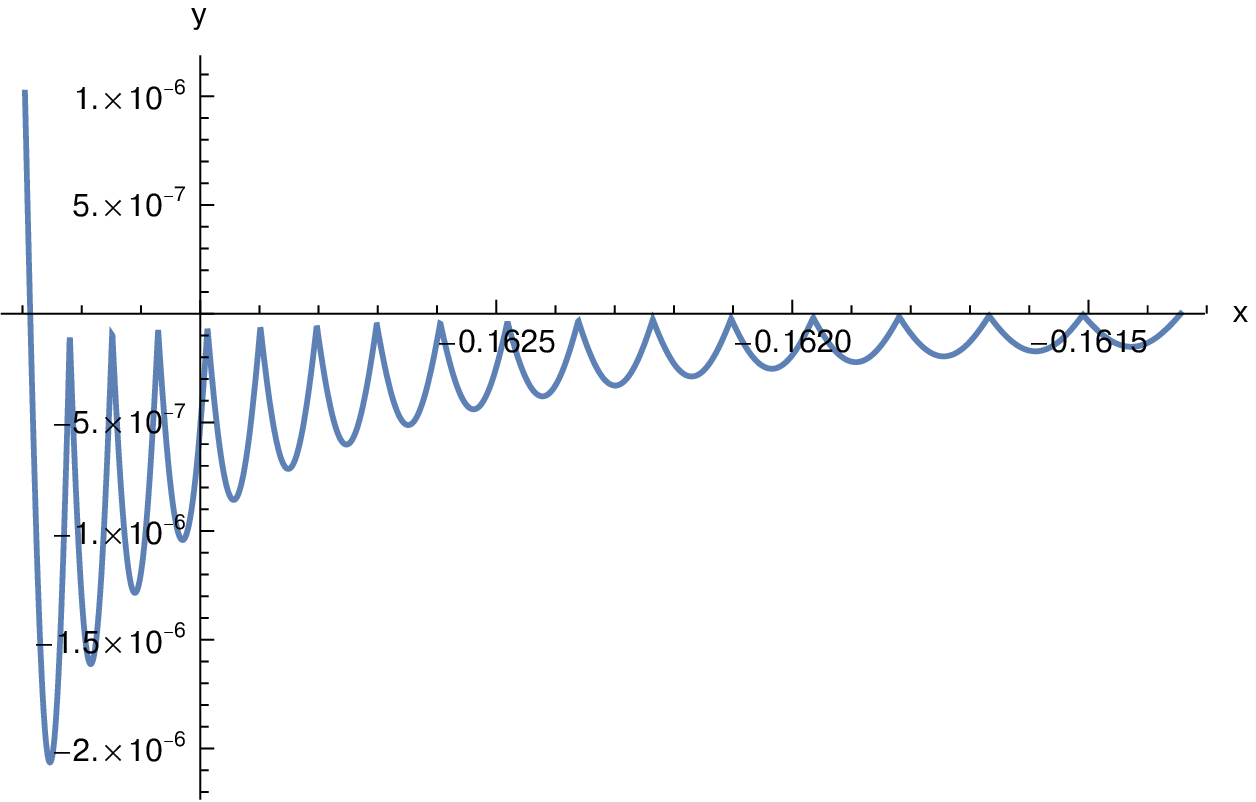}
\qquad \includegraphics[width=2.6in]{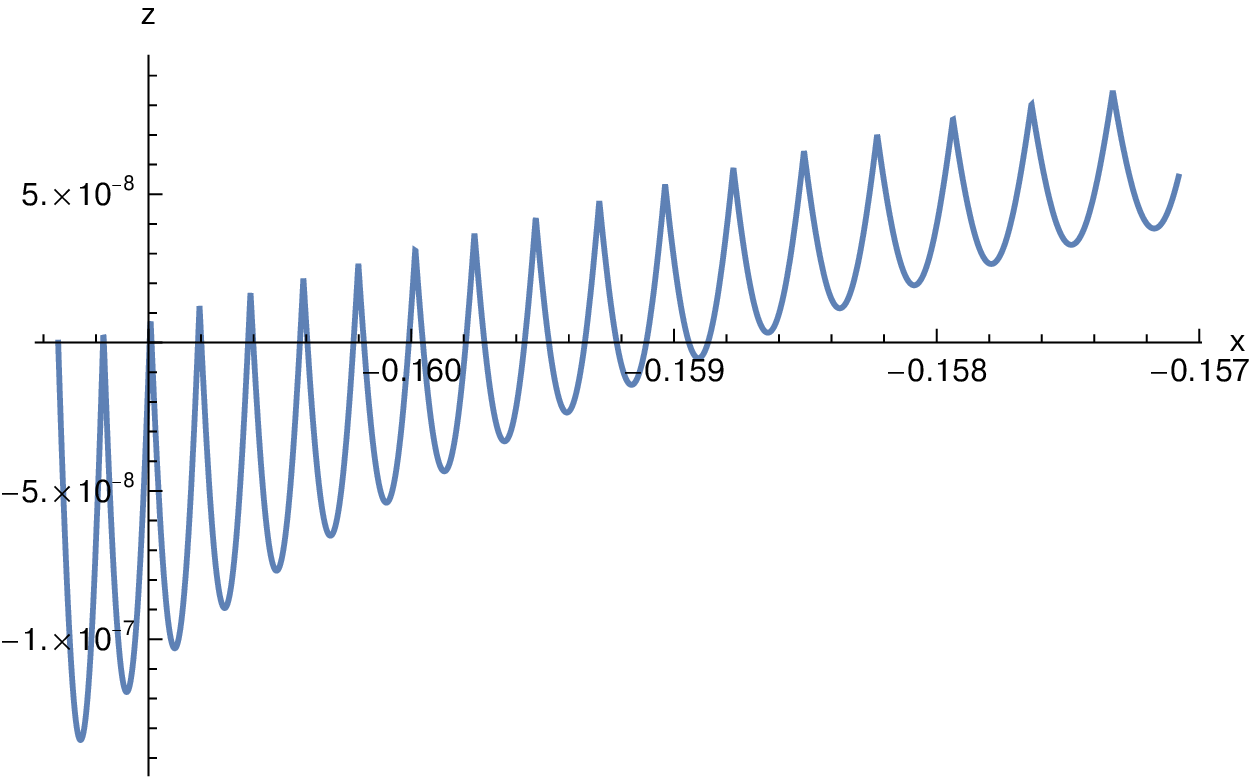}
\end{center}
  \caption{
Left panel: comparison of ${\color{green} \hK_{1,0}^h}$ (the computational scheme SchemeIII) with
${\color{blue} \hK_{1,0}^h}$ (the computational scheme SchemeI). Right panel:
comparison of ${\color{red} \hK_{1,0}^h}$ (the computational scheme SchemeII) with
${\color{blue} \hK_{1,0}^h}$ (the computational scheme SchemeI).  
} \label{plottypeabfg2}
\end{figure}

Following \eqref{collapsetypeab}, we collect
results of $\{\hf_{a,0}^h-\hf_{b,0}^h,\hK_{1,0}^h\}$ as  functions of  
$\ln \frac{H^2}{\Lambda^2}$ in different computational schemes in fig.~\ref{plottypeabfg1}:
SchemeI (blue curves), SchemeII (red curves) and Scheme III (green curves).
The accuracy of the collapsed results in different schemes is highlighted in fig.~\ref{plottypeabfg2}
for $\hK_{1,0}^h$ --- the remaining parameters follow the same trend. 
Notice that TypeA$_b$ vacua exist only for $H\ge H_{min}^b$ \eqref{defkcrit}; furthermore,
in the limit $H\to H_{min}^b+0$, all the chiral symmetry breaking
condensates \eqref{uvirparslin} vanish as $\propto (H-H_{min}^b)^{1/2}$, typical for
a spontaneous symmetry breaking with a mean-field exponent $\frac 12$.

Next, FG frame TypeA$_b$ de Sitter vacua have to be reinterpreted in EF frame, see appendix \ref{apb2}.
The diffeomorphism transformation is performed at the radial location as in \eqref{mapfgef}.
Details of numerical construction of EF frame vacua from FG frame vacua are collected in
appendix \ref{apcef}. An important quantity is the parameter $s_0^h$,  see \eqref{ef1i}, and
\eqref{sh0FGEF}. As with FG frame UV/IR parameters \eqref{uvirparstypeab}, results for $s_0^h$ should not
depend on the choice of the computational scheme, provided we compare properly
dimensionless and rescaled quantities, \ie $\ln \frac{H^2}{\Lambda^2}$ and
$\hat{s}_0^h$ \eqref{finalsh0},
\begin{equation}
\begin{split}
&{\rm SchemeI:}\qquad \ln\frac{H^2}{\Lambda^2}=k_s\,,\qquad  \hs_{0}^h=s_0^h\,;\\
&{\rm SchemeII:}\qquad \ln\frac{H^2}{\Lambda^2}=\frac {K_0^{\star}}{b}+\ln b\,,\qquad \hs_{0}^h=\frac{1}{b^{1/4}}s_0^h\,;
\\
&{\rm SchemeIII:}\qquad \ln\frac{H^2}{\Lambda^2}=K_0^{\star}+\ln\a\,,\qquad \hs_{0}^h=\frac{1}{\a^{1/2}}s_0^h\,.
\end{split}
\eqlabel{collapsetypeassigmatypeab}
\end{equation}

\begin{figure}[t]
\begin{center}
\psfrag{x}{{$\ln \frac{H^2}{\Lambda^2}$}}
\psfrag{s}{{$\hs_{0}^h$}}
\includegraphics[width=2.6in]{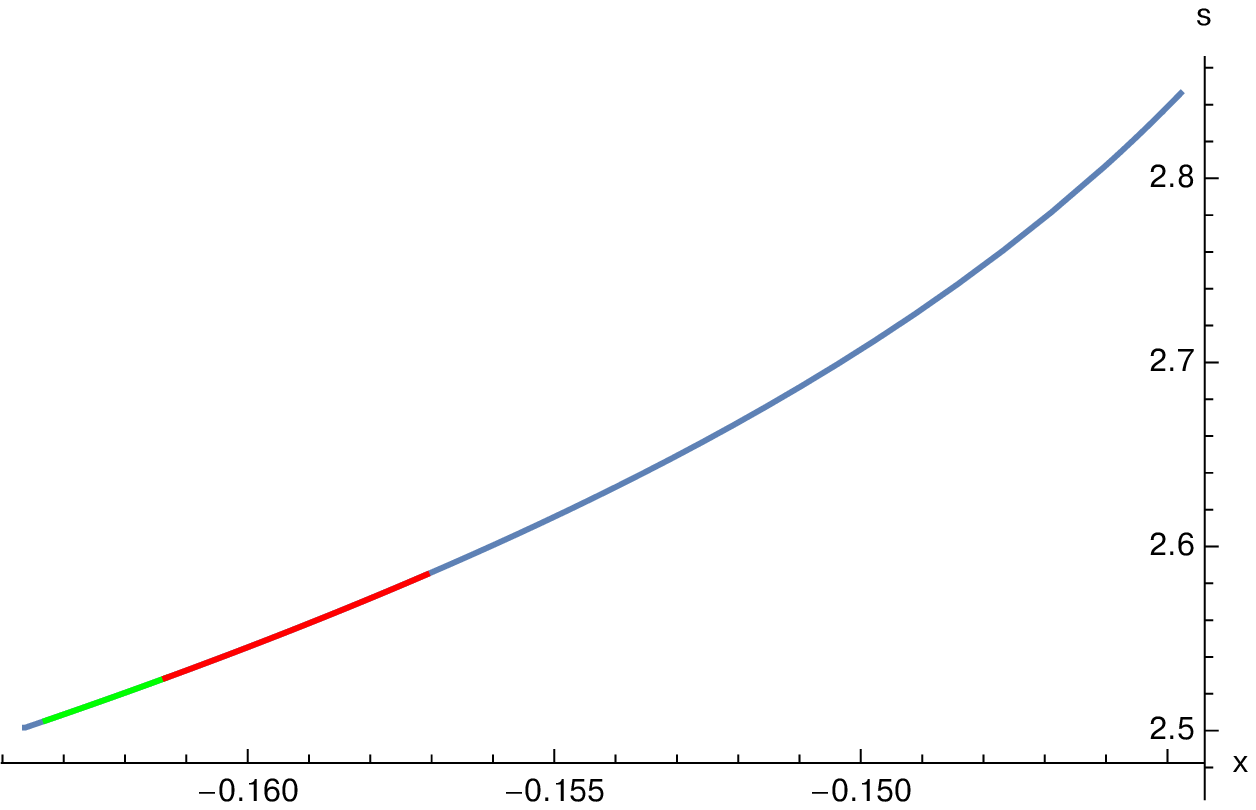}
\end{center}
  \caption{
Parameters $\hs_{0}^h$ of TypeA$_s$ de Sitter vacua of the cascading
gauge theory as functions of $\ln \frac{H^2}{\Lambda^2}$ in different computational
schemes \eqref{collapsetypeab}: SchemeI (blue), SchemeII (red) and Scheme III (green). 
} \label{sh0typeab1}
\end{figure}

\begin{figure}[t]
\begin{center}
\psfrag{x}{{$\ln \frac{H^2}{\Lambda^2}$}}
\psfrag{y}{{${\color{red} \hs_{0}^h}/{\color{blue} \hs_{0}^h}-1$}}
\psfrag{z}{{${\color{green} \hs_{0}^h}/{\color{blue} \hs_{0}^h}-1$}}
\includegraphics[width=2.6in]{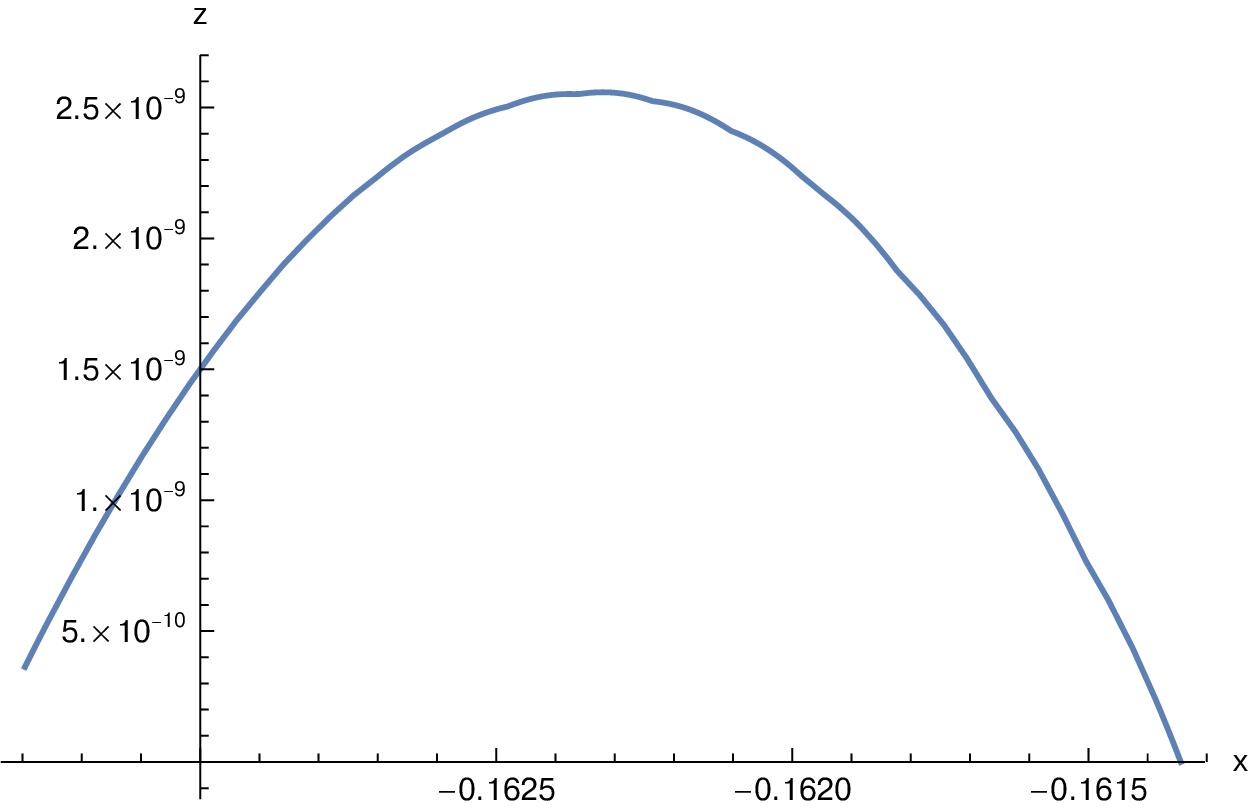}
\qquad \includegraphics[width=2.6in]{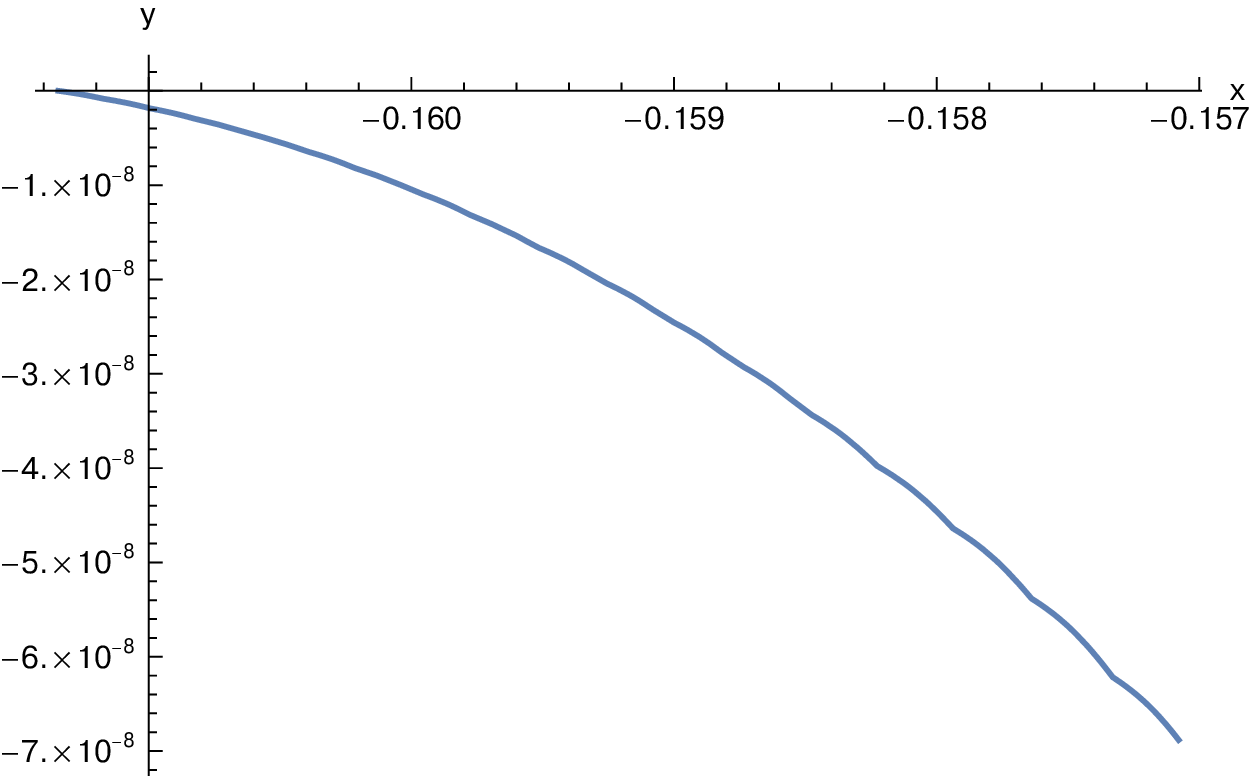}
\end{center}
  \caption{
Left panel: comparison of ${\color{green} \hs_{0}^h}$ (the computational scheme SchemeIII) with
${\color{blue} \hs_{0}^h}$ (the computational scheme SchemeI). Right panel:
comparison of ${\color{red} \hs_{0}^h}$ (the computational scheme SchemeII) with
${\color{blue} \hs_{0}^h}$ (the computational scheme SchemeI).  
} \label{sh0typeab2}
\end{figure}

Following \eqref{collapsetypeassigmatypeab}, we collect
(subset of the) results of $\hs_{0}^h$ as  functions of  
$\ln \frac{H^2}{\Lambda^2}$ in different computational schemes in fig.~\ref{sh0typeab1}:
SchemeI (blue curves), SchemeII (red curves) and Scheme III (green curves).
The accuracy of the collapsed results in different schemes is highlighted in fig.~\ref{sh0typeab2}.

EF frame equations of motion \eqref{efv2}-\eqref{efc1} are solved subject to the initial conditions
set by the asymptotic expansions \eqref{typebi1}-\eqref{typeabi9} at $z=0$. These equations have to be integrated
on the interval
\begin{equation}
z\in [0,z_{AH}]\,,
\eqlabel{rangez42}
\end{equation}
where $z_{AH}=-r_{AH}$ is the location of the apparent horizon at asymptotically late times, see \eqref{ahsvac}.
To determine the location of the apparent horizon, along with integrating the gravitational
background functions $\{a,\sigma,w_{a,b,c,2},K_{1,2,3},g\}$, we evaluate the AH location function
$\call_{AH}(z)$, see \eqref{defcall}. AH is located at the first zero of this function for $z>0$.
Once the AH is identified, TypeA$_b$ vacua entanglement entropy is computed following \eqref{ahsvac}:
\begin{equation}
s_{ent}=\frac{H^3P^4g_s^2}{4G_5}\ \biggl\{
\hat{\sigma}^3 \hat{w}_{c2}^{1/2}\hw_{a2}\hw_{b2}\biggr\}\bigg|_{\hat{z}=\hat{z}_{AH}}
=\frac{3^5M^4g_s^2}{2^5\pi^3}\ H^3\  \biggl\{
\hat{\sigma}^3 \hat{w}_{c2}^{1/2}\hw_{a2}\hw_{b2}\biggr\}\bigg|_{\hat{z}=\hat{z}_{AH}}\,,
\eqlabel{senttypeab}
\end{equation}
where following \eqref{scaleout} we introduced dimensionless and rescaled functions
and the radial coordinate:
\begin{equation}
\begin{split}
&\{z\,,\ a\,,\, \sigma\,,\, w_{a2,b2,c2}\,,\, K_{1,2,3}\,,\ g \}\qquad
\Longrightarrow\qquad \{\hat{z}\,,\ \ha\,,\, \hat{\sigma}\,,\, \hw_{a2,b2,c2}\,,\, \hK_{1,2,3}\,,\ \hat{g}\}\,;\\
&z={H Pg_s^{1/2}}\ \hat{z}\,,\qquad a=H^2P g_s^{1/2}\ \ha\,,\qquad \sigma=H P^{1/2}g_s^{1/4}\ \hat{\sigma}\,,\\
&w_{a2,b2,c2}=P g_s^{1/2}\ \hw_{a2,b2,c2}\,,\qquad  K_{1,3}=P^2g_s\ \hK_{1,3}\,,\qquad
  K_{2}=\hK_{2}\,,\qquad
g=g_s\ \hat{g}  \,.
\end{split}
\eqlabel{scaleouttypeab}
\end{equation}
In the last equality in \eqref{senttypeab} we used expressions for $G_5$ \eqref{g5deff} and $P$ \eqref{defpm}.
We compute entanglement entropy in different computational schemes; results must agree, provided
we compare dimensionless and rescaled quantities, see \eqref{defhatsent}.
Explicitly,
\begin{equation}
\begin{split}
&{\rm SchemeI:}\qquad \ln\frac{H^2}{\Lambda^2}=k_s\,,\qquad  \hs_{ent}=s_{ent}\,;\\
&{\rm SchemeII:}\qquad \ln\frac{H^2}{\Lambda^2}=\frac {K_0^\star}{b}+\ln b\,,\qquad \hs_{ent}=\frac{1}{b^2}s_{ent}\,;
\\
&{\rm SchemeIII:}\qquad \ln\frac{H^2}{\Lambda^2}=K_0^\star+\ln\a\,,\qquad \hs_{ent}=\frac{1}{\a^{3/2}}s_{ent}\,.
\end{split}
\eqlabel{collapsetypebsent}
\end{equation}

\begin{figure}[t]
\begin{center}
\psfrag{x}{{$\ln \frac{H^2}{\Lambda^2}$}}
\psfrag{y}{{$4 G_5\ \hat{s}_{ent}$}}
\includegraphics[width=5.2in]{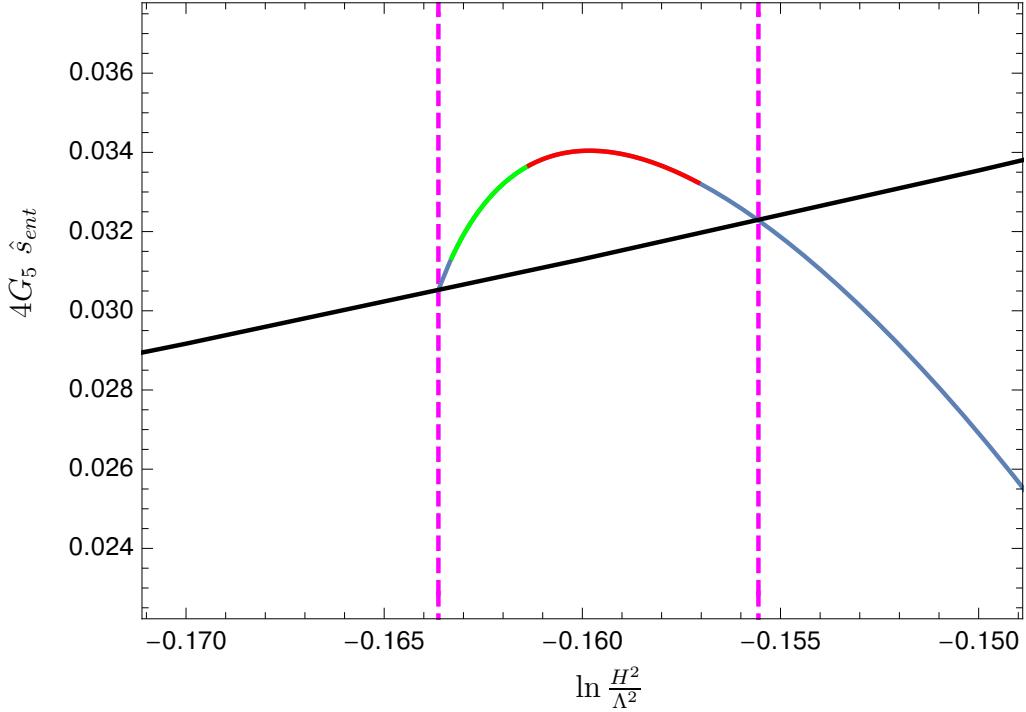}
\end{center}
  \caption{
Entanglement entropy $\hat{s}_{ent}$  \eqref{defhatsent}
of TypeA$_s$ (black curve) and TypeA$_b$ (different computational
schemes \eqref{collapsetypebsent}: SchemeI (blue), SchemeII (red) and Scheme III (green))
de Sitter vacua of the cascading
gauge theory as functions of $\ln \frac{H^2}{\Lambda^2}$. Dashed
vertical magenta lines indicate the range of the Hubble constant $H$
such that  $s_{ent}\bigg|_{{\rm TypeA}_b}\ \ge\ s_{ent}\bigg|_{{\rm TypeA}_s}$,
see \eqref{main}.
} \label{comparesentasab}
\end{figure}

\begin{figure}[t]
\begin{center}
\psfrag{x}{{$\ln \frac{H^2}{\Lambda^2}$}}
\psfrag{y}{{${\color{red} \hs_{ent}}/{\color{blue} \hs_{ent}}-1$}}
\psfrag{z}{{${\color{green} \hs_{ent}}/{\color{blue} \hs_{ent}}-1$}}
\includegraphics[width=2.6in]{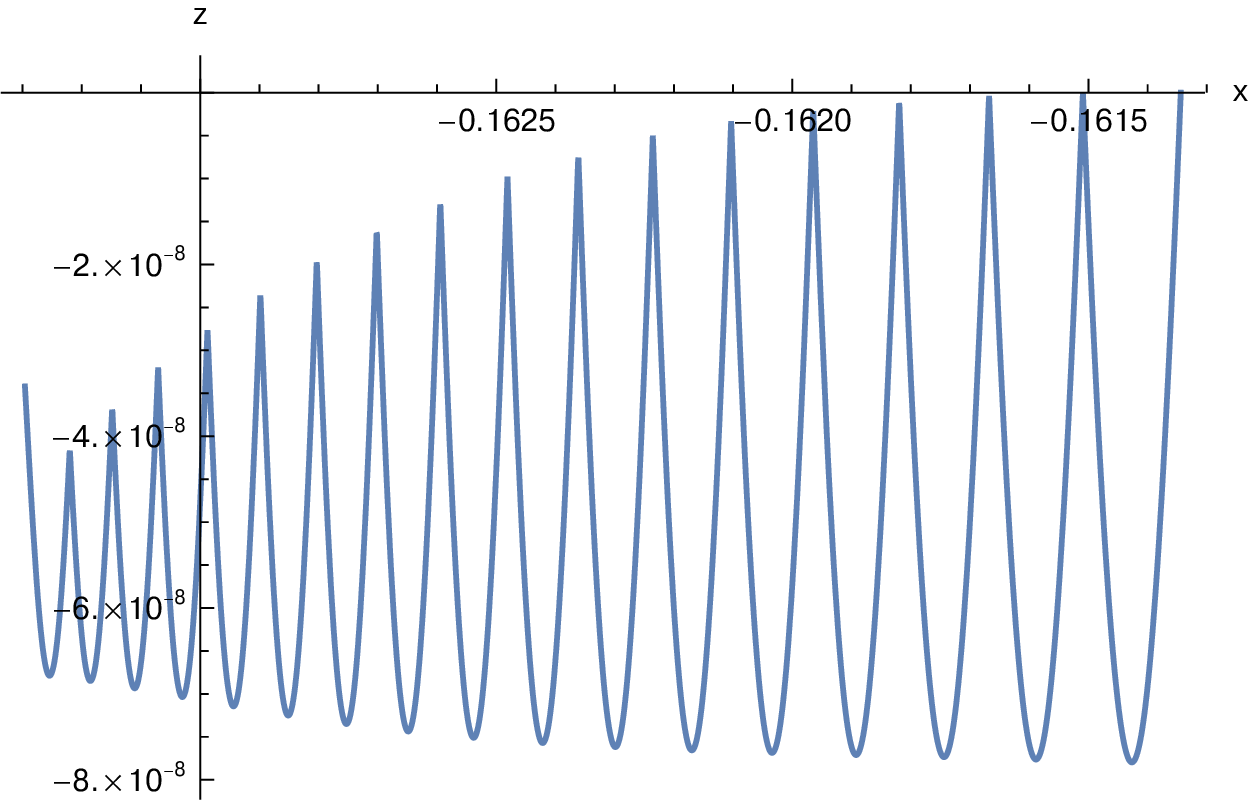}
\qquad \includegraphics[width=2.6in]{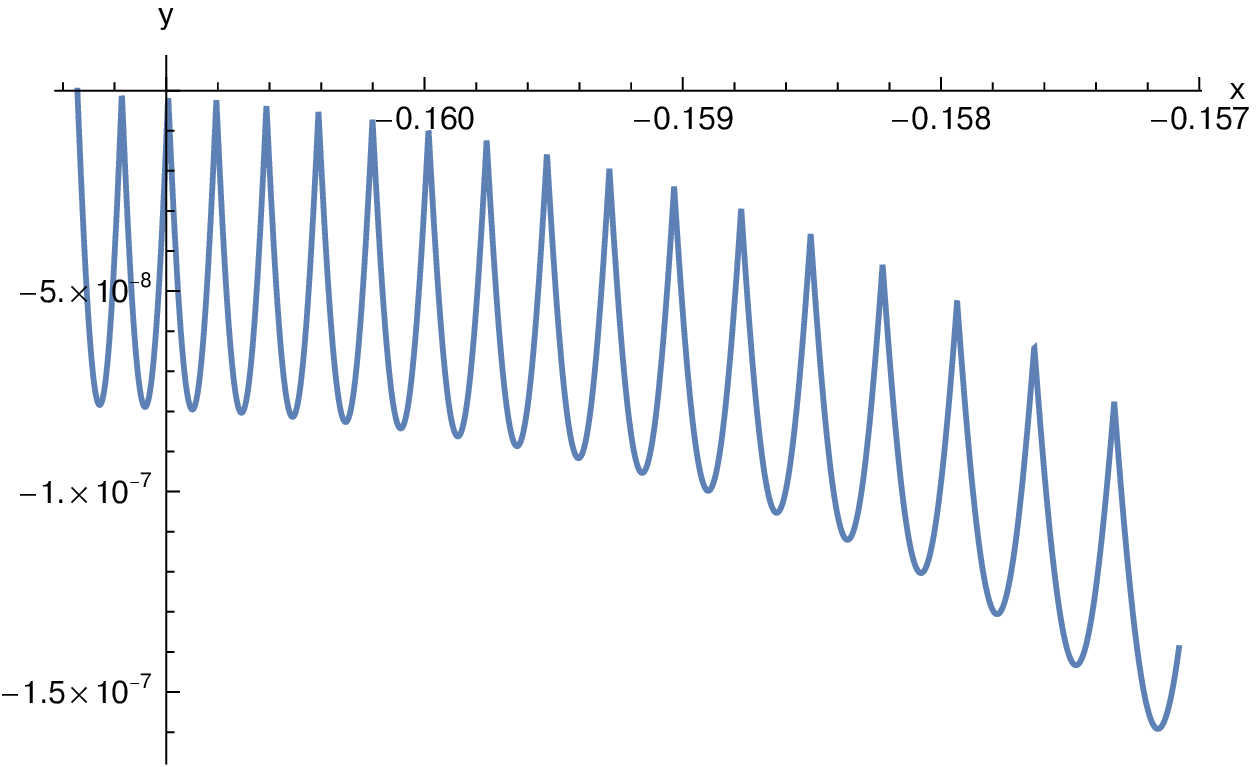}
\end{center}
  \caption{
Left panel: comparison of ${\color{green} \hs_{ent}}$ (the computational scheme SchemeIII) with
${\color{blue} \hs_{ent}}$ (the computational scheme SchemeI). Right panel:
comparison of ${\color{red} \hs_{ent}}$ (the computational scheme SchemeII) with
${\color{blue} \hs_{ent}}$ (the computational scheme SchemeI).  
} \label{errorsentb}
\end{figure}

Following \eqref{collapsetypebsent}, we collect
(subset of the) results of $(4G_5\ \hs_{ent})$ as  functions of  
$\ln \frac{H^2}{\Lambda^2}$ in different computational schemes in fig.~\ref{comparesentasab}:
SchemeI (blue curves), SchemeII (red curves) and Scheme III (green curves).
Additionally, we replot the results for the entanglement entropy of TypeA$_s$ vacua (black curve).
Fig.~\ref{comparesentasab} is the main result of the paper:
it demonstrates that the entanglement entropy of TypeA$_b$ vacua is larger than that of
TypeA$_s$ vacua provided (the values $H_{min}^b$ and $H_{max}$ are denoted by vertical
dashed magenta lines)
\begin{equation}
H_{min}^b\le H\le H_{max}\,,
\eqlabel{main}
\end{equation}
where
\begin{equation}
\frac{H_{min}^b}{\Lambda}=0.92(1)\,,\qquad \frac{H_{max}}{\Lambda}=0.92(5)\,.
\eqlabel{hminhmax}
\end{equation}
This is an unexpected result, as it implies that $SU(N)\times SU(N+M)$
cascading gauge theory with a strong coupling scale $\Lambda$ undergoes spontaneous
chiral symmetry breaking in de Sitter space time with a Hubble constant $H$
in the interval \eqref{main}.

The accuracy of the collapsed results for TypeA$_b$ vacua in different schemes is highlighted in
fig.~\ref{errorsentb}.

\subsection{Validity of supergravity approximation for TypeA$_b$ vacua}\label{typeabsugra}

\begin{figure}[t]
\begin{center}
\psfrag{x}{{$\ln \frac{H^2}{\Lambda^2}$}}
\psfrag{y}{{$\ln\hat{K}_{AH}$}}
\psfrag{t}{{$1/\hat{K}_{AH}$}}
\includegraphics[width=2.6in]{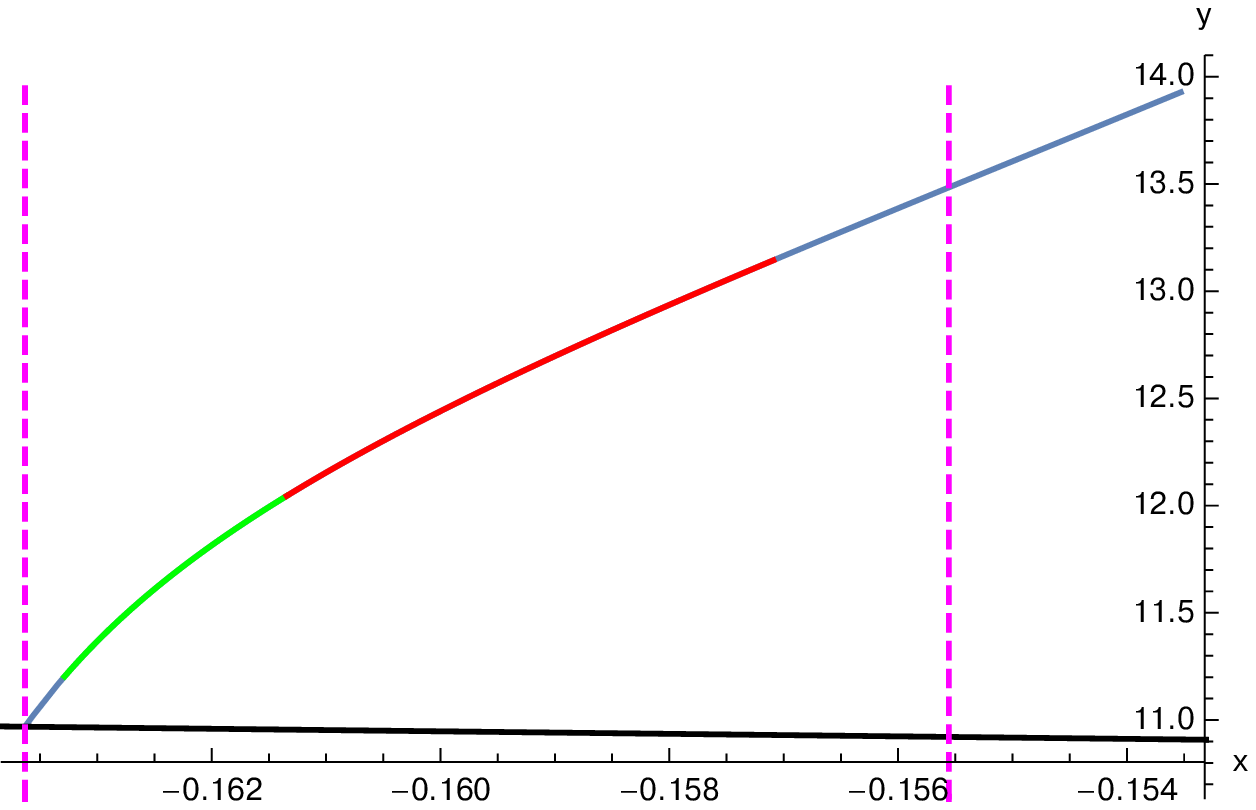}
\end{center}
  \caption{
Kretschmann scalar of  \eqref{ef1i} evaluated at the
apparent horizon as  functions of $\ln\frac{H^2}{\Lambda^2}$ for TypeA$_b$ vacua
in different computation schemes  \eqref{collapsekb}:
SchemeI (blue), SchemeII (red) and Scheme III
(green). The black curve
is the Kretschmann scalar of  \eqref{ef1i} evaluated at the
apparent horizon as a function of $\ln\frac{H^2}{\Lambda^2}$ for TypeA$_s$ vacua.
Vertical dashed magenta lines indicate the range of dominance of
TypeA$_b$ vacua over TypeA$_s$, see \eqref{main}. 
}\label{kb}
\end{figure}

In this section we briefly comment on the validity of the supergravity approximation in
construction of TypeA$_b$ vacua. 
In fig.~\ref{kb} we present the  Kretschmann scalar of  \eqref{ef1i} evaluated at the
apparent horizon in different computations schemes for the TypeA$_b$ vacua, see appendix \ref{kretschmann}:
\begin{equation}
\begin{split}
&{\rm SchemeI:}\qquad \ln\frac{H^2}{\Lambda^2}=k_s\,,\qquad \hat{K}=K\,;\\
&{\rm SchemeII:}\qquad \ln\frac{H^2}{\Lambda^2}=\frac {K_0^\star}{b}+\ln b\,,\qquad \hat{K}=b K\,;
\\
&{\rm SchemeIII:}\qquad \ln\frac{H^2}{\Lambda^2}=K_0^\star+\ln\a\,,\qquad \hat{K}=K\,.
\end{split}
\eqlabel{collapsekb}
\end{equation}
Vertical dashed magenta lines indicate the range of dominance of
TypeA$_b$ vacua over TypeA$_s$, see \eqref{main}. 
Additionally, we replot the Kretschmann scalar of  \eqref{ef1i} evaluated at the
apparent horizon for TypeA$_s$ vacua (black curve). $K_{AH}$ is the same for  TypeA$_b$ and  TypeA$_s$
vacua at $H=H_{min}^b$; the former is about 13 times larger for TypeA$_b$ vacuum at
$H=H_{max}$ and continues to increase as $\frac{H}{\Lambda}$ increases. We do not study the
breakdown of the supergravity approximation for TypeA$_b$ vacua for $H> H_{max}$, as these vacua
are irrelevant.

\section{TypeB de Sitter vacua}\label{typebv}

TypeB de Sitter vacua were studied previously in \cite{Buchel:2013dla}.
We showed in section \ref{sentb} that the entanglement entropy of these vacua
vanishes. Thus, these vacua can arise as late-time dynamical attractors of the cascading
gauge theory in de Sitter only when neither TypeA$_s$ nor TypeA$_b$ vacua exist
(for the corresponding values $\frac{H}{\Lambda}$). Recall that TypeA$_s$ vacua exist only
for $H\gtrsim H_{min}^s$ \eqref{typeasdne}, and TypeA$_b$ vacua exist only when
$H\ge  H_{min}^b$ \eqref{hminhmax}. In this section we establish that
TypeB vacua do exist for $H\lesssim H_{max}^B$ with $H_{max}^B > \{H_{min}^s,H_{min}^b\}$,
see \eqref{typebexists}.
In section \ref{typebnum} we present numerical results for
TypeB vacua for generic values of $\frac{H^2}{\Lambda^2}$.
In section \ref{typebsugra} we estimate $H_{max}^B$
above which TypeB vacua construction in type IIB supergravity
becomes unreliable/does not exist. We identify the source of breaking of the supergravity approximation.

\subsection{Numerical results: TypeB}\label{typebnum}

\begin{figure}[t]
\begin{center}
\psfrag{a}{{$\a$}}
\psfrag{f}{{$f_{a,0}^h$}}
\psfrag{h}{{$h_{0}^h$}}
\psfrag{x}{{$k_{1,3}^h$}}
\psfrag{y}{{$k_{2,2}^h$}}
\psfrag{z}{{$k_{2,4}^h$}}
\psfrag{t}{{$k_{3,1}^h$}}
\psfrag{u}{{$g_{0}^h$}}
\psfrag{i}{{$f_{a,3,0}$}}
\psfrag{o}{{$k_{2,3,0}$}}
\includegraphics[width=1.7in]{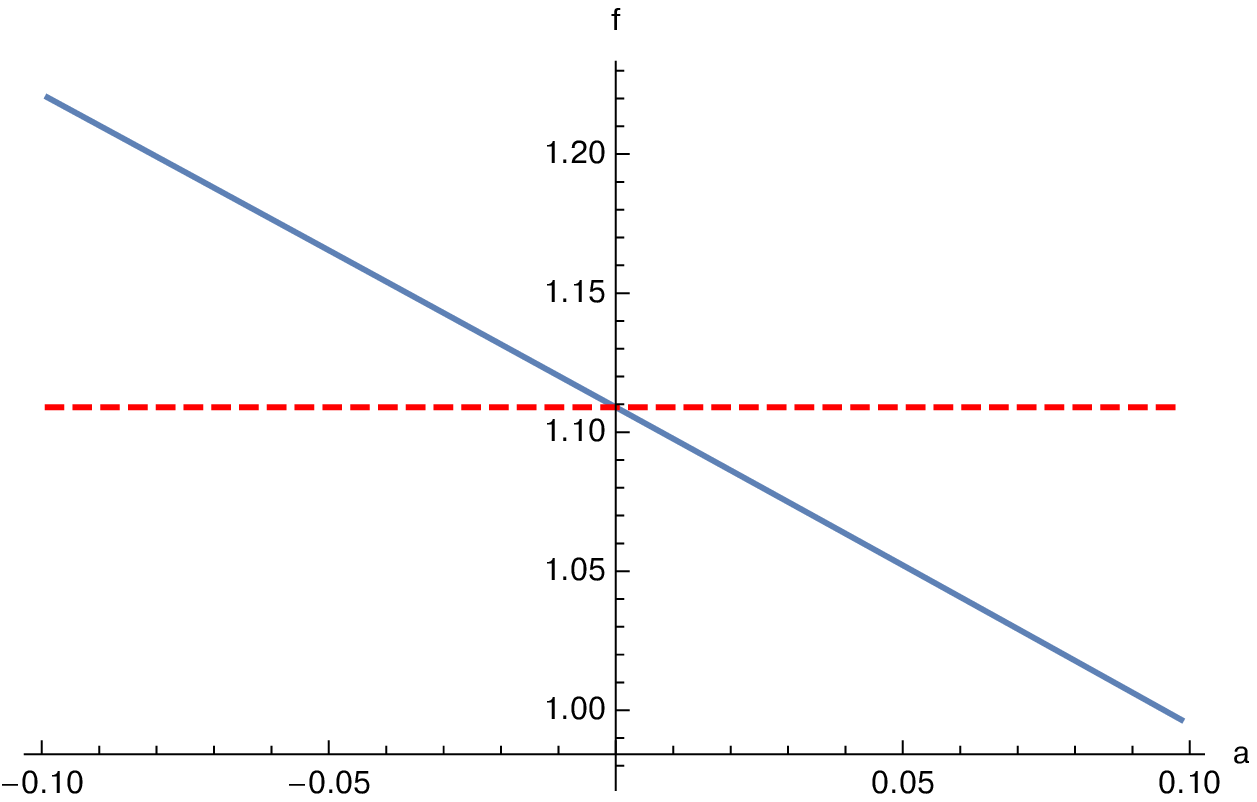}\qquad 
\includegraphics[width=1.7in]{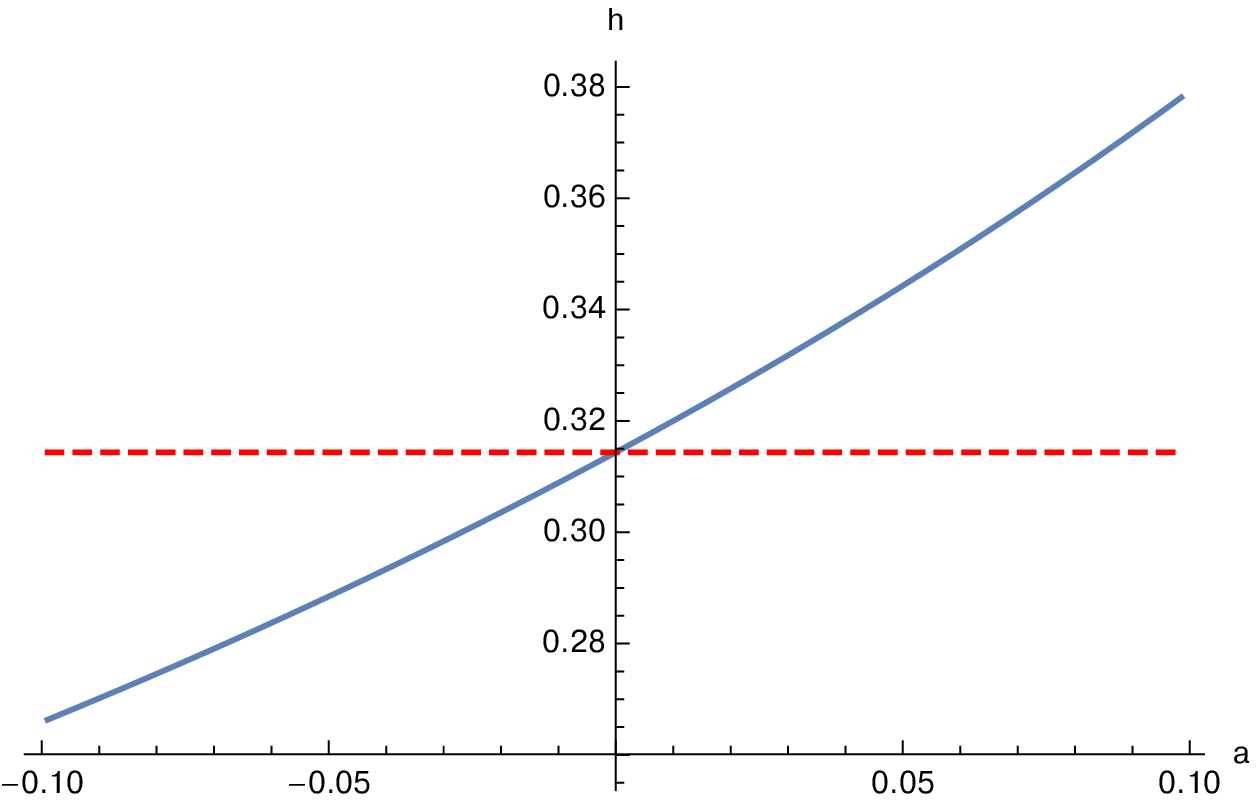}\qquad 
\includegraphics[width=1.7in]{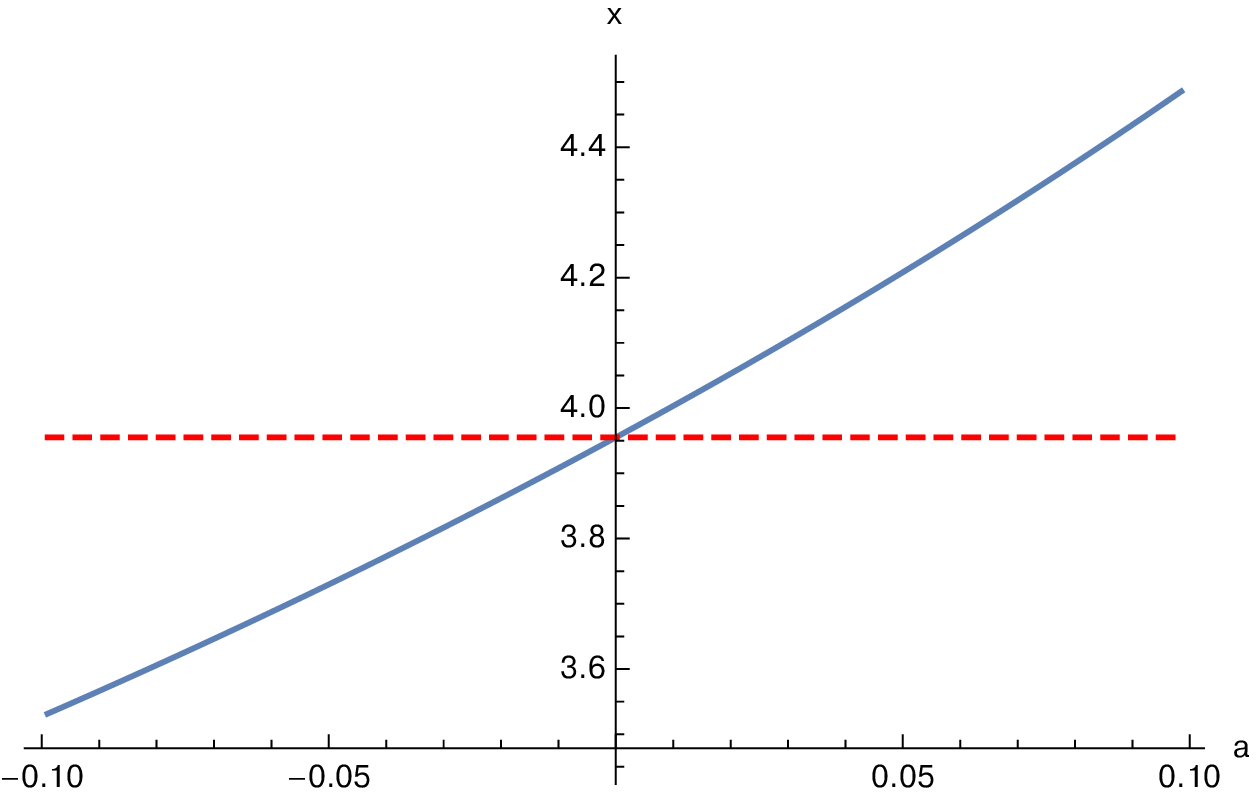}\\
\vspace{1cm}
\includegraphics[width=1.7in]{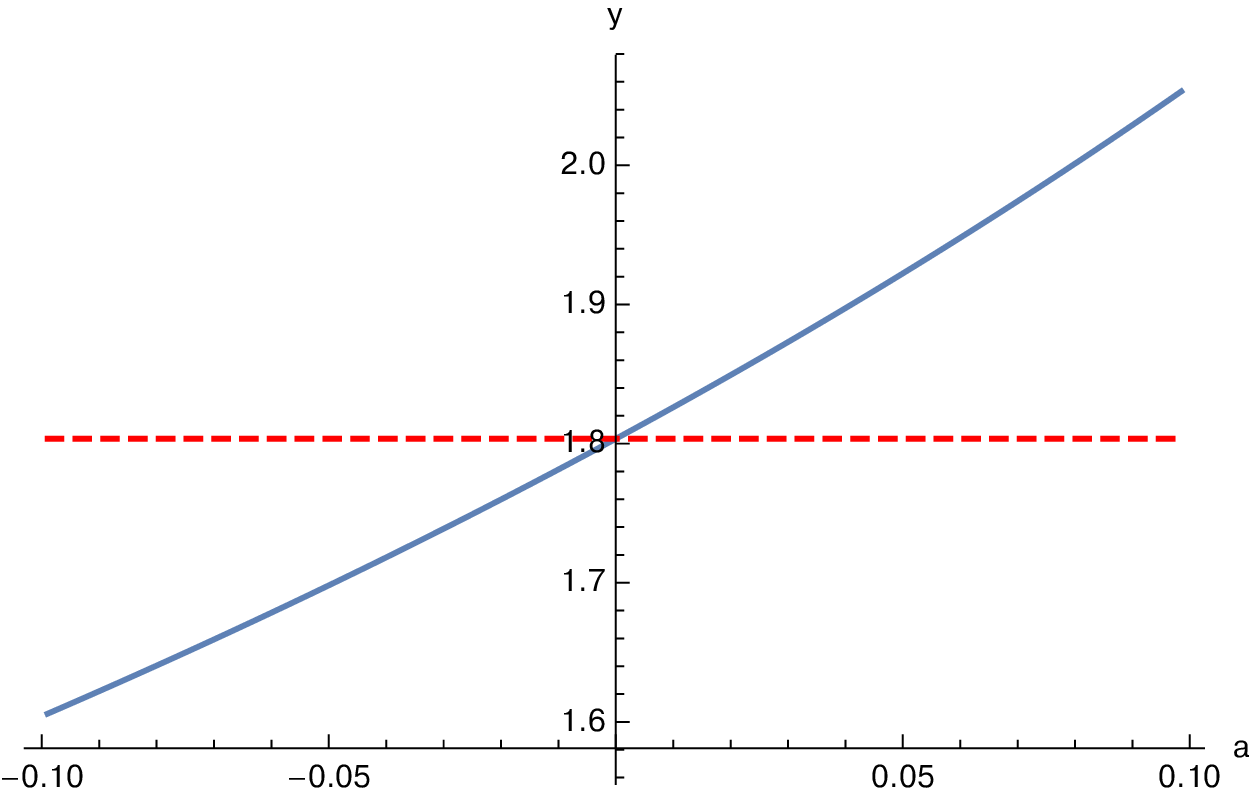}\qquad
\includegraphics[width=1.7in]{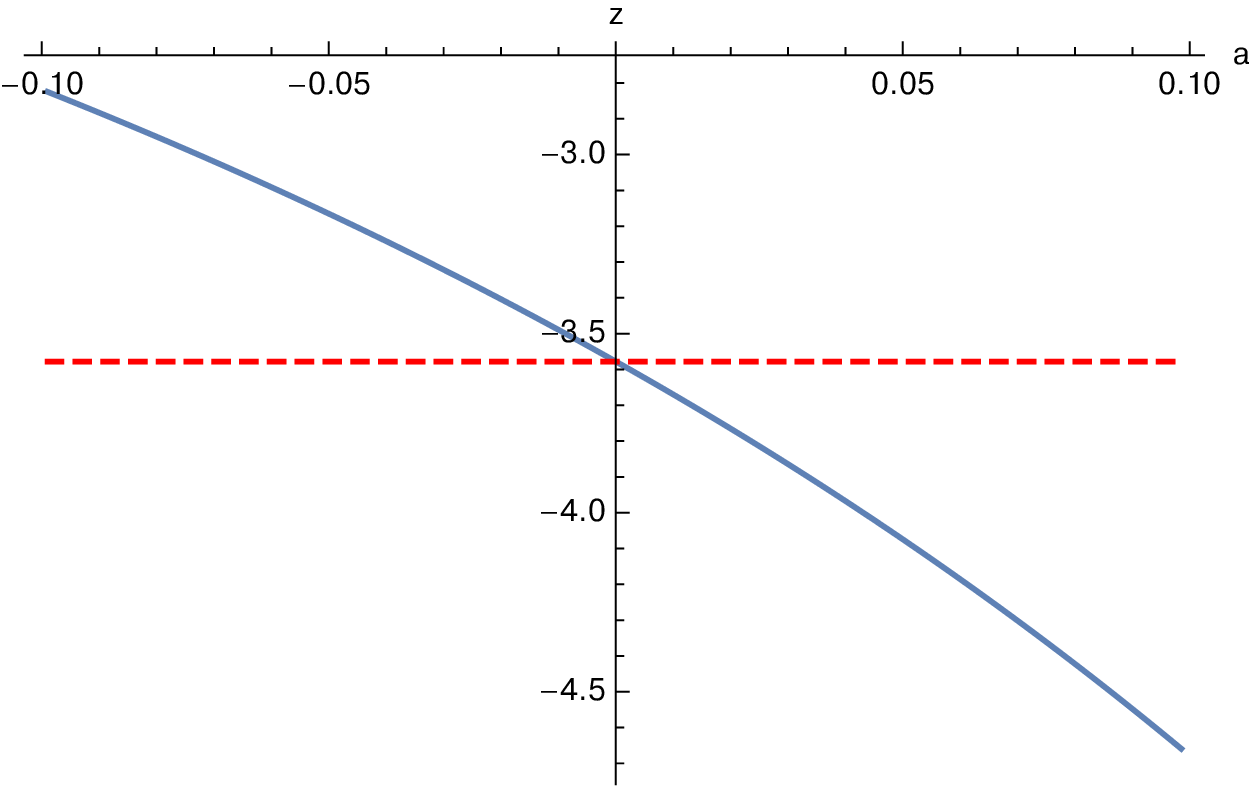}\qquad
\includegraphics[width=1.7in]{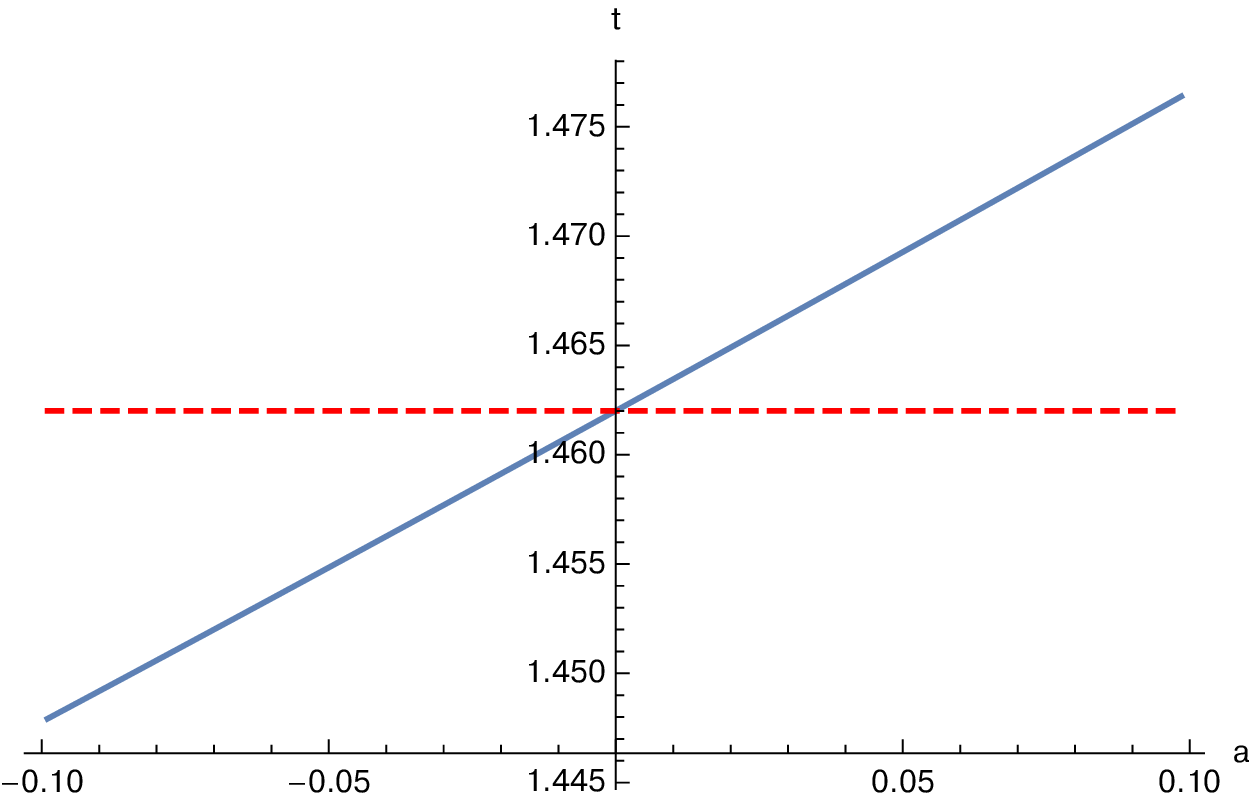}\\
\vspace{1cm}
\includegraphics[width=1.7in]{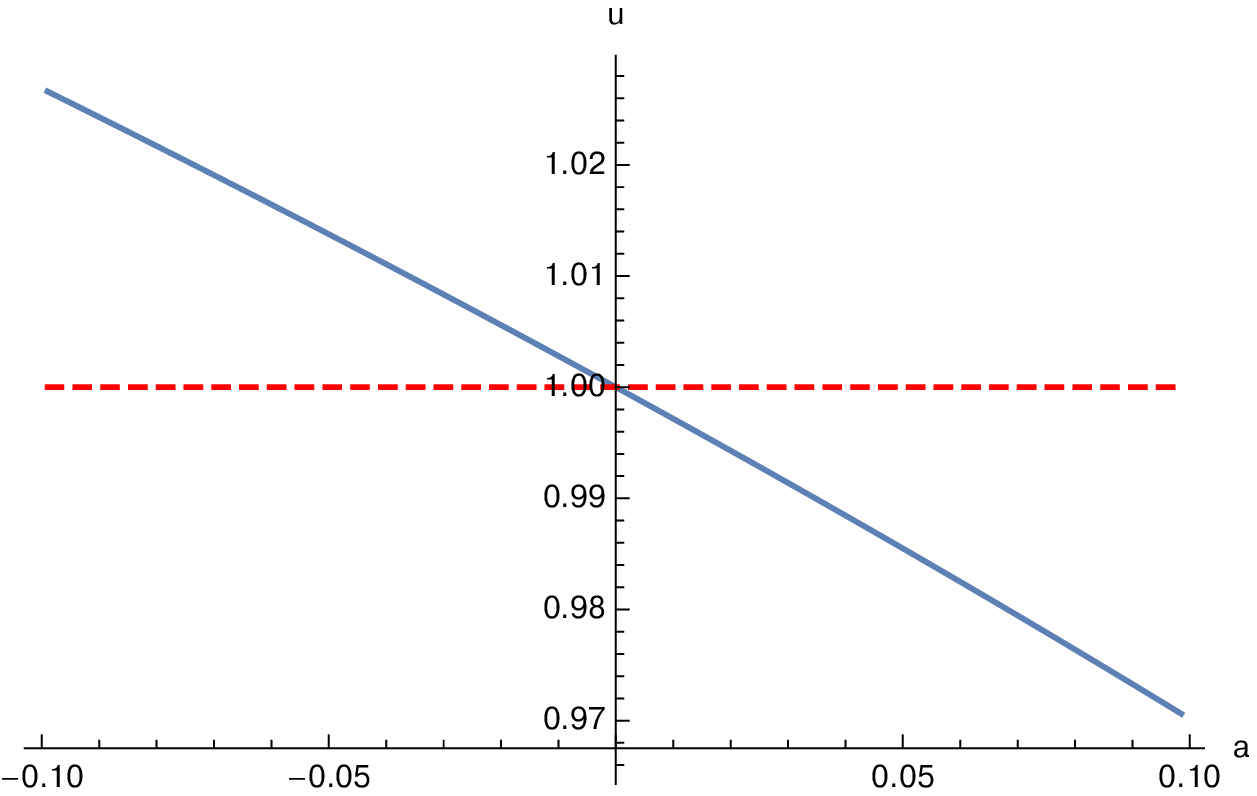}\qquad
\includegraphics[width=1.7in]{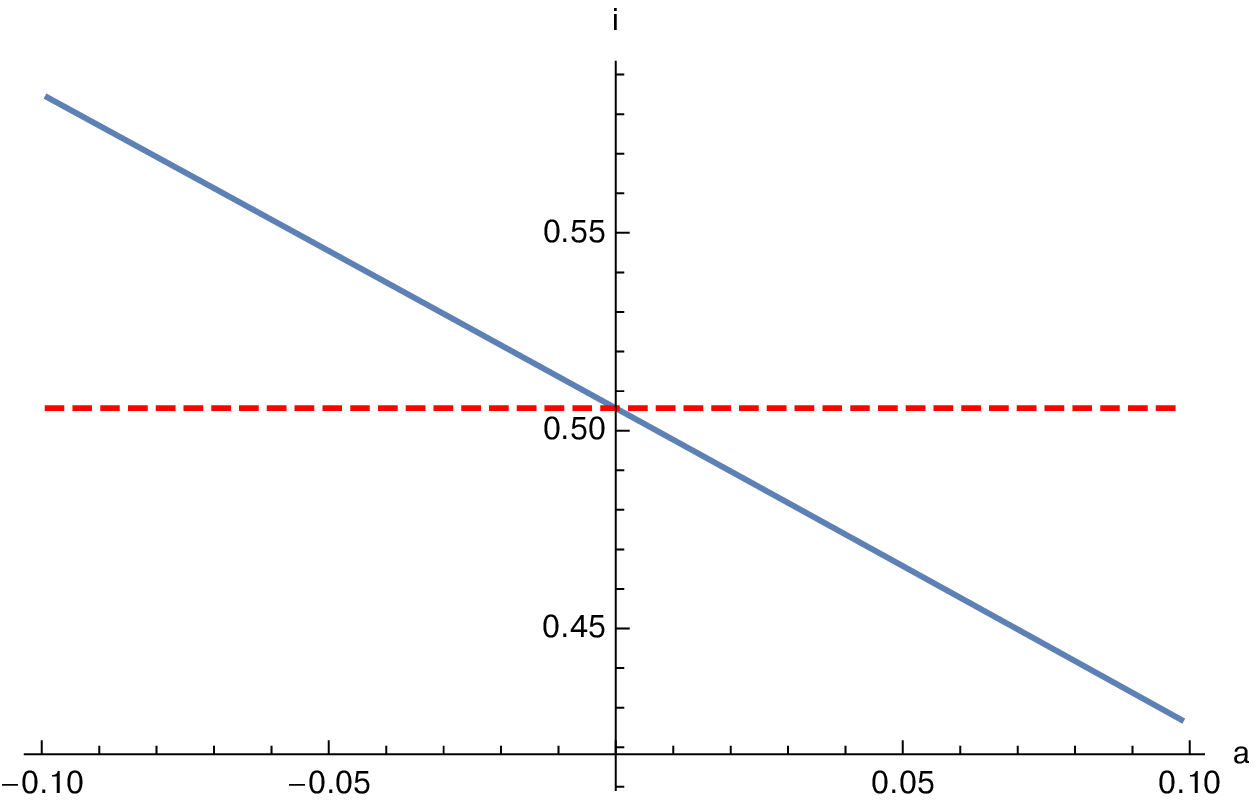}\qquad
\includegraphics[width=1.7in]{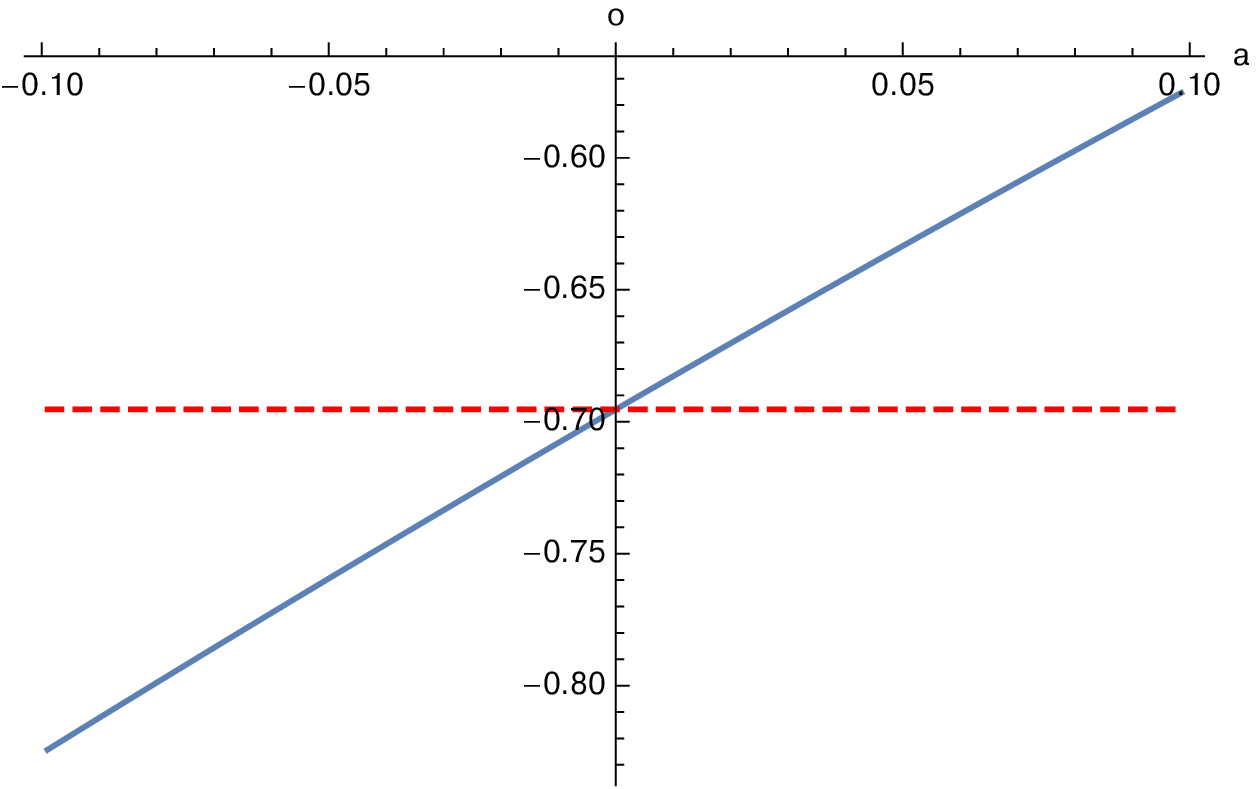}\qquad
\end{center}
  \caption{
  TypeB vacua IR parameters and select UV parameters \eqref{uvirparstypeb}
  in computational SchemeIII as  functions of $\a\equiv H^2$ (solid blue curves).
  Red dashed horizontal lines represent comparison with extremal KS solution, see \eqref{epiii},
  at $\a=0$.
}\label{typebsusycompare}
\end{figure}

To establish the existence of TypeB vacua it is sufficient to construct them
in FG frame \eqref{fg1i}. The construction follows the steps implemented
for TypeA$_s$ vacua in section \ref{typeasnum}.
There are 8 second order equations
\eqref{kseq2}-\eqref{kseq9} and 1 first order equation \eqref{kseq10}.
The first order equation
\eqref{kseq10} involves (linearly) $f_c'$ and can be used instead of one of the
second order equations
(namely, the one involving $f_c''$). Thus, altogether we have a coupled system of 7 second order
ODEs (linear in  $\{f_a'',f_b'',h'',K_1'',K_2'',K_3'',g''\}$) and a single first order equation
(linear in $f_c'$). As a result,
a unique solution must be characterized by $15=2\times 7+1$ parameters;
these are the UV/IR parameters 
\begin{equation}
\begin{split}
&{\rm UV:}\qquad \{f_{a,1,0}\,,\, f_{a,3,0}\,,\, k_{2,3,0}\,,\, g_{4,0}\,,\, f_{c,4,0}\,,\ f_{a,6,0}\,,\, f_{a,7,0}\,,\, f_{a,8,0}\}\,;\\
&{\rm IR:}\qquad \{f_{a,0}^h\,,\ h_{0}^h\,,\ k_{1,3}^h\,,\ k_{2,2}^h\,,\ k_{2,4}^h\,,\
k_{3,1}^h\,,\ g_{0}^h\}\,.
\end{split}
\eqlabel{uvirparstypeb}
\end{equation}
It is rather challenging to find the solutions of the corresponding system of ODEs
in 15-dimensional parameter space by brute force --- fortunately,
a special case of TypeB vacua, namely, the limit $H\to 0$, is the
supersymmetric Minkowski space-time Klebanov-Strassler solution
\cite{Klebanov:2000hb}, see appendix \ref{ksextremal}. Using this
extremal KS solution as a seed, we can construct TypeB vacua turning on the
deformation parameter $\a\equiv H^2$ in the ODEs \eqref{kseq2}-\eqref{kseq10}.

To validate our results, we use two different computation schemes: SchemeI and SchemeIII, 
see \eqref{compschemes}. Numerical results must not depend on which
computational scheme is adopted.
We illustrate now that this is indeed the case using a sample of  IR parameters in \eqref{uvirparstypeb} as
an example\footnote{The same is true for the rest of IR parameters and the UV parameters as well.}.
Comparison of the different computational schemes is done using dimensionless and rescaled quantities:
$\ln \frac {H^{2}}{\Lambda^2}$ (as a vacuum label) \eqref{defks}
and $\{\hf_{a}^h\,,\, \hh_0^h\,,\, \hk_{1,3}^h\,,\, \hk_{2,2}^h\,,\, \hk_{2,4}^h\,,\,
\hk_{3,1}^h\,,\, \hat{g}_0^h\}$
\eqref{irtypebscaled}.  Explicitly:
\begin{equation}
\begin{split}
&{\rm SchemeI:}\qquad \ln\frac{H^2}{\Lambda^2}=k_s\,,\ \hf_{a}^h=f_{a}^h\,,\ \hh_{0}^h=h_{0}^h\,,\
\hk_{1,3}^h=k_{1,3}^h\,,\ \hk_{2,2}^h=k_{2,2}^h\,,\ \hk_{2,4}^h=k_{2,4}^h\,,\\
&\qquad\qquad\qquad\qquad\qquad\qquad\qquad \hk_{3,1}^h=k_{3,1}^h\,,\
\hat{g}_0^h=g_0^h\,;\\
&{\rm SchemeIII:}\qquad \ln\frac{H^2}{\Lambda^2}=\frac14+\ln\a\,,\ \hf_{a}^h
=\frac{1}{\a}f_{a}^h\,,\ \hh_{0}^h={\a^2}h_{0}^h\,,\
\hk_{1,3}^h=\a^{3/2}k_{1,3}^h\,,\\
&\qquad\qquad\qquad\qquad\qquad \hk_{2,2}^h=\a k_{2,2}^h\,,\ \hk_{2,4}^h=\a^{2}k_{2,4}^h\,,\
\hk_{3,1}^h={\a^{1/2}}k_{3,1}^h\,,\ \hat{g}_0^h=g_0^h\,.
\end{split}
\eqlabel{collapsetypeb}
\end{equation}

Fig.~\ref{typebsusycompare} presents all the IR parameters and select UV parameters
($f_{a,3,0}$ and $k_{2,3,0}$), see \eqref{uvirparstypeb},
of TypeB vacua in computational SchemeIII as functions of $\a$.
Extremal KS parameters are represented by dashed horizontal red lines
and must agree with the corresponding TypeB parameters at $\a=0$.
While negative values of $\a$ are not physical, we run numerical codes
for $\a<0$ to extract more precisely this comparison at $\a=0$.
Extremal KS parameters in computational SchemeIII
can be determined from \eqref{susyuv} and \eqref{susyir} provided we set
\begin{equation}
\begin{split}
&K_0=P^2g_s\biggl(-\ln 3+\frac53\ \ln2-\frac43\ \ln\e-\frac23\biggr)\biggl|_{\rm SchemeIII}=\frac 14\\
\Longrightarrow\qquad&\\
&\e\bigg|_{\rm SchemeIII}=\frac23\ 6^{1/4}\ e^{-11/16}\,.
\end{split}
\eqlabel{epiii}
\end{equation}
We find remarkable agreements, \eg
\begin{equation}
\frac{f_{a,0}^h(\a=0)}{f_{a,0}^h(KS)}-1 \sim 5\times 10^{-10}\,,\qquad
\frac{k_{2,4}^h(\a=0)}{k_{2,4}^h(KS)}-1 \sim 2\times 10^{-10}\,.
\eqlabel{comparefah0bks}
\end{equation}
The remaining parameters are validated at $\sim 10^{-6}$ level or better.

\begin{figure}[t]
\begin{center}
\psfrag{x}{{$\ln \frac{H^2}{\Lambda^2}$}}
\psfrag{k}[r]{{$\hk_{1,3}^h$\qquad  }}
\psfrag{z}{{${\color{green} \hk_{1,3}^h}/{\color{blue} \hk_{1,3}^h}-1$}}
\includegraphics[width=2.6in]{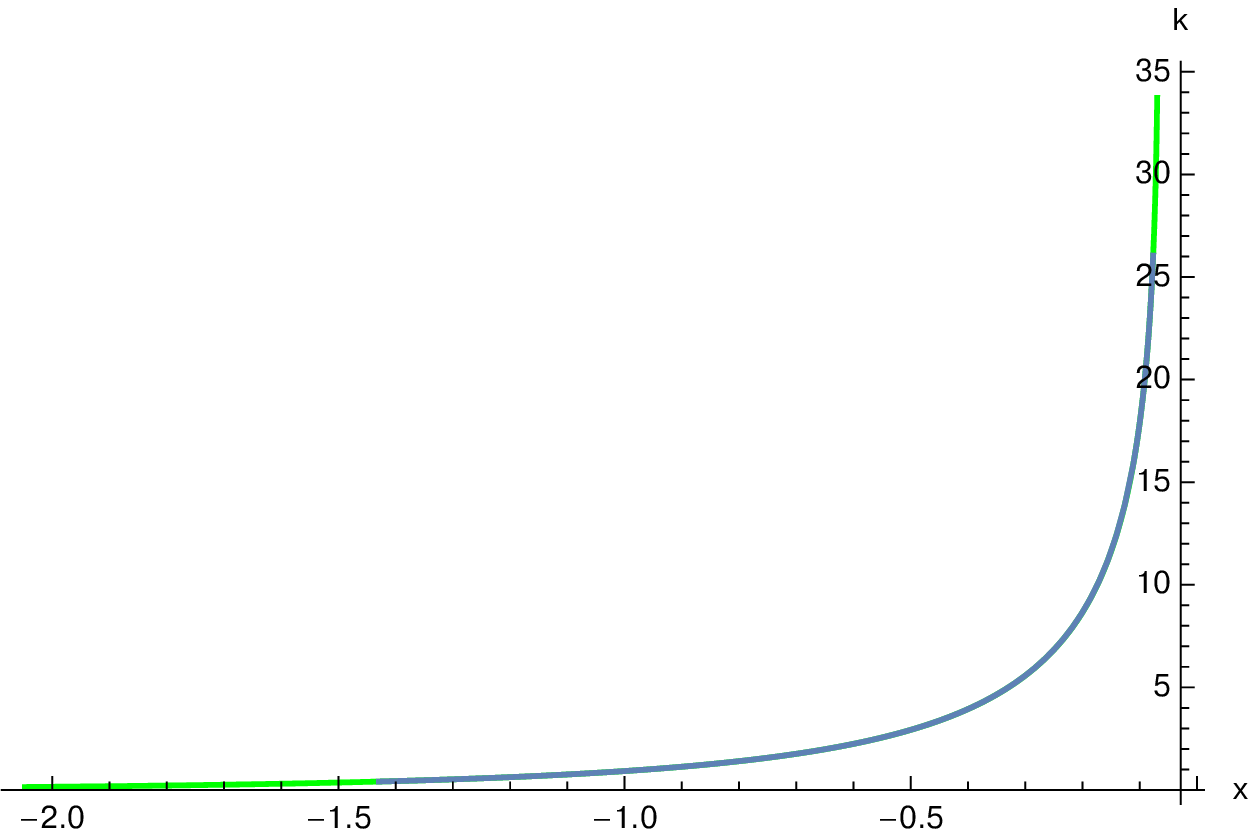}
\qquad \includegraphics[width=2.6in]{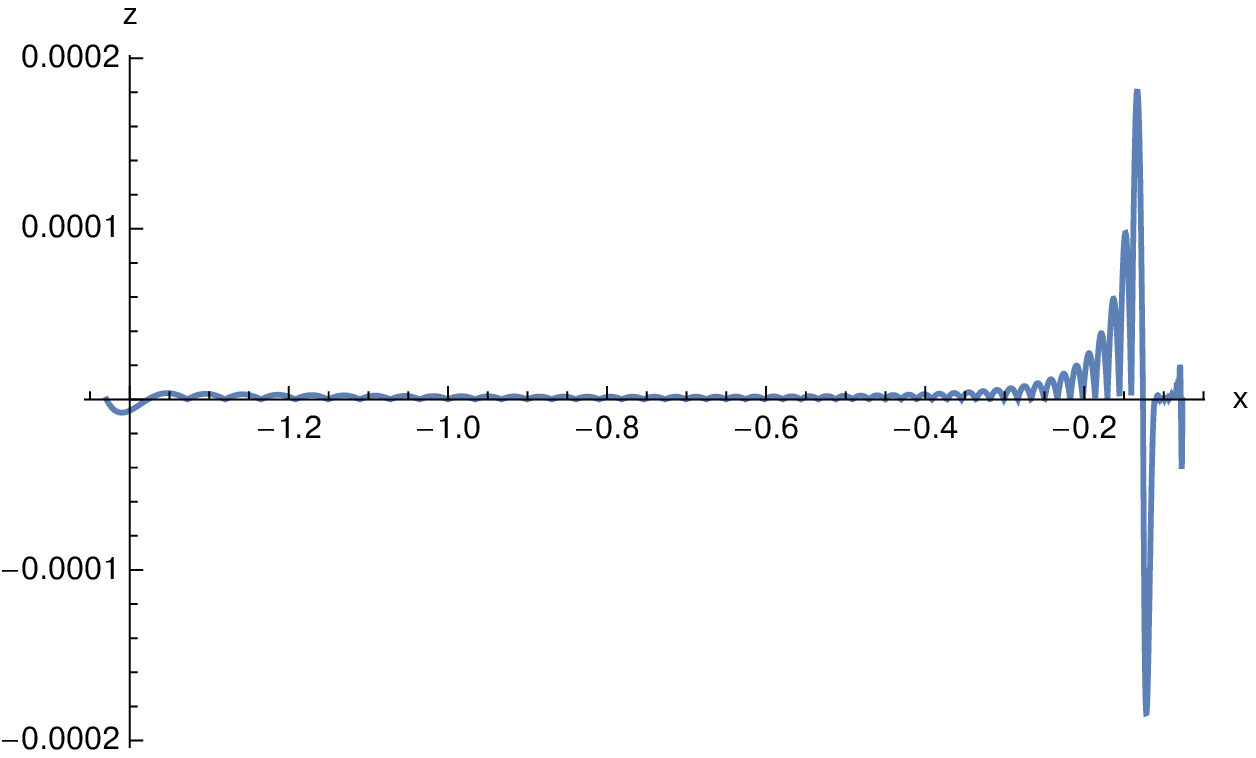}
\end{center}
  \caption{
Left panel: infrared parameter $\hk_{1,3}^h$  of the  
Fefferman-Graham coordinate frame of TypeB de Sitter vacua of the cascading
gauge theory as functions of $\ln \frac{H^2}{\Lambda^2}$ in different computational
schemes \eqref{collapsetypeb}: SchemeI (blue)  and Scheme III (green).
Right panel:
comparison of ${\color{green} \hk_{1,3}^h}$ (the computational scheme SchemeIII) with
${\color{blue} \hk_{1,3}^h}$ (the computational scheme SchemeI).  
} \label{plottypebfg1}
\end{figure}

Following \eqref{collapsetypeb}, we collect
results of $\hk_{1,3}^h$ as  functions of  
$\ln \frac{H^2}{\Lambda^2}$ in different computational schemes in fig.~\ref{plottypebfg1}:
SchemeI (blue curves) and Scheme III (green curves) (left panel);
the accuracy of the collapsed results in different schemes is highlighted in
right panel. Comparison of the remaining parameters follows the same trend.
Note the degradation in accuracy as $\frac{H}{\Lambda}$ increases --- in section
\ref{typebsugra} we relate this to the breakdown of the supergravity approximation.

\subsection{Validity of supergravity approximation for TypeB vacua}\label{typebsugra}

\begin{figure}[t]
\begin{center}
\psfrag{x}{{$\ln \frac{H^2}{\Lambda^2}$}}
\psfrag{k}[r]{{$1/\hat{K}_{AH}\qquad\qquad $}}
\psfrag{r}[r]{{$R^2_{S^3}/(Pg_s^{1/2})\qquad\qquad $}}
\includegraphics[width=2.6in]{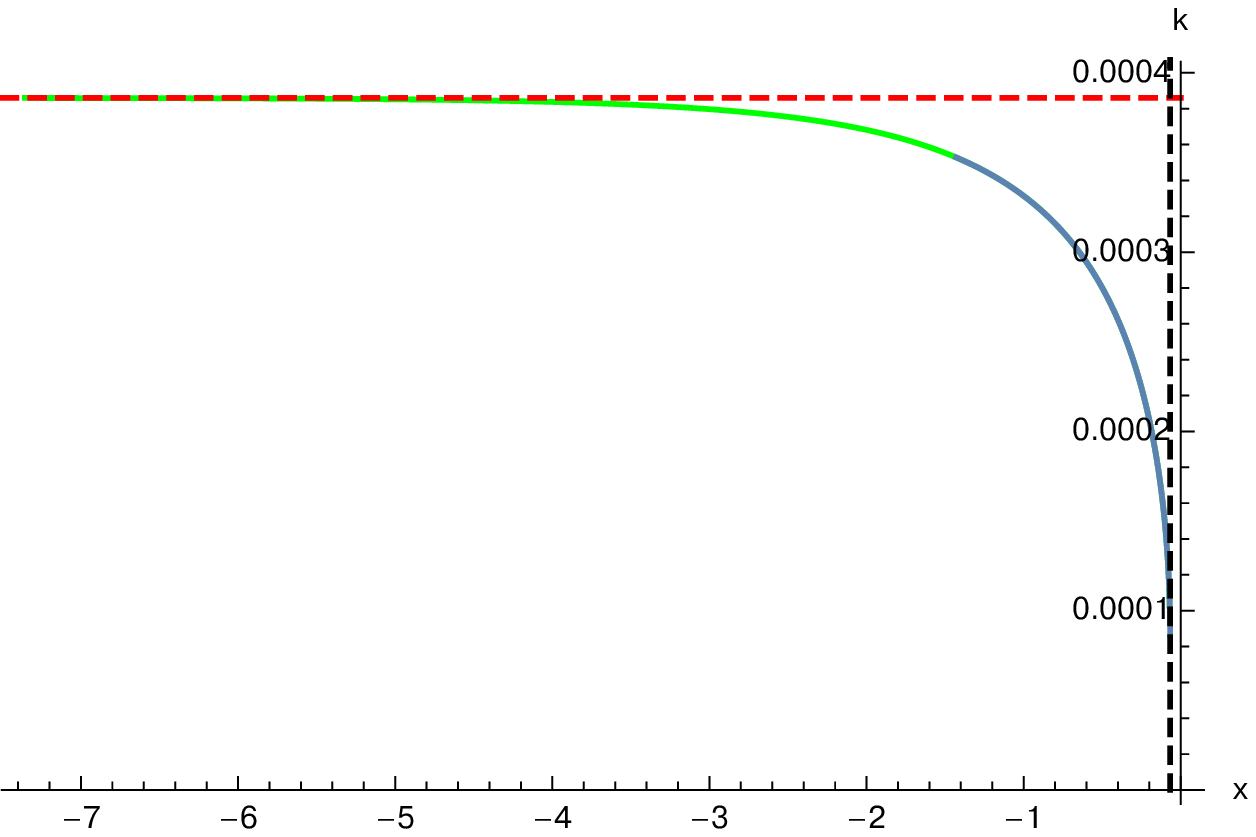}\qquad
\includegraphics[width=2.6in]{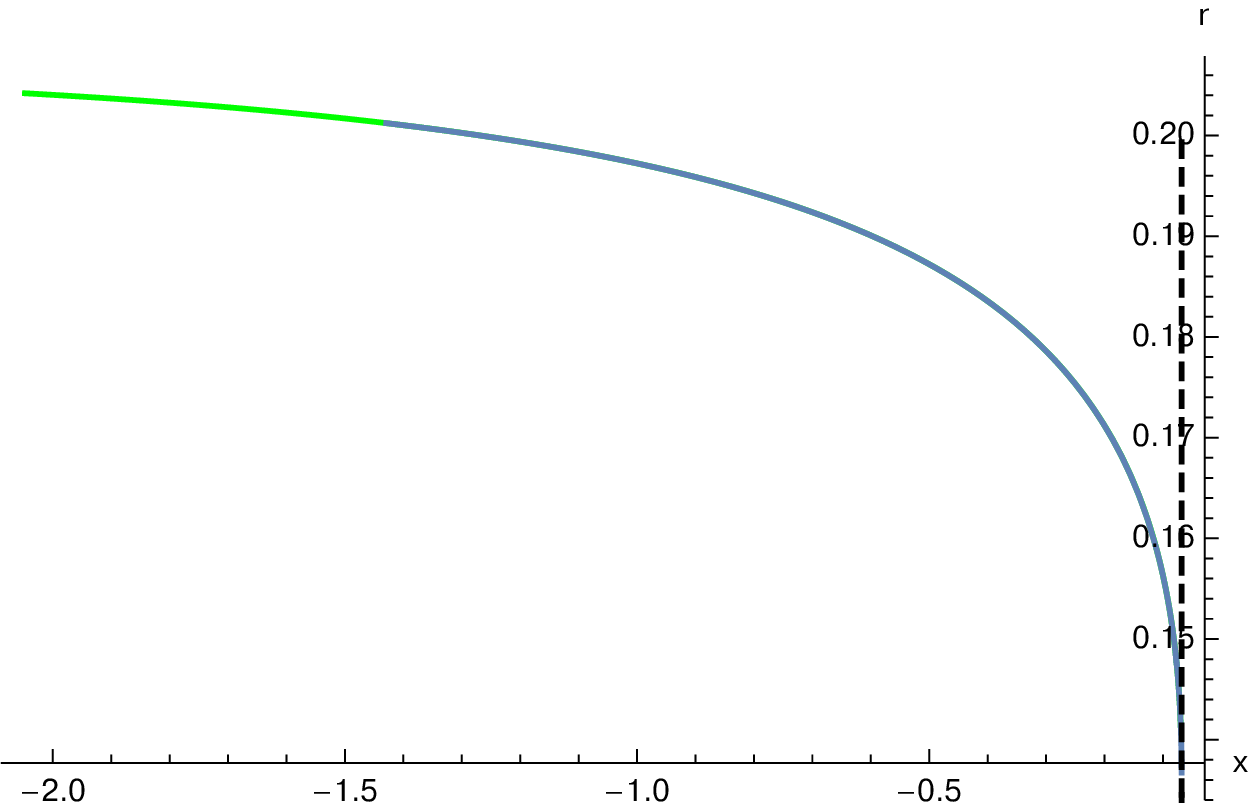}
\end{center}
  \caption{
Left panel: Inverse Kretschmann scalar of  \eqref{ef1i} evaluated at the
apparent horizon for TypeB vacua as  functions of $\ln\frac{H^2}{\Lambda^2}$
in different computation schemes  \eqref{compschemes}:
SchemeI (blue) and Scheme III
(green). Horizontal red dashed line represents $\frac{1}{\hK_{AH}}$ for the extremal KS solution,
which is recovered in the limit $\frac{H}{\Lambda}\to 0$.
Right panel: the divergence of the Kretschmann scalar as $H\to H_{max}^B$ is associated with the
collapse of the 3-cycle, see \eqref{3cycleb}.
Vertical black dashed lines represent $\frac{H_{max}^B}{\Lambda}$.
}\label{kTypeB}
\end{figure}

As clear from fig.~\ref{plottypebfg1} the accuracy in constructing TypeB vacua deteriorates
as $H$ increases; we have been able to construct TypeB vacua for
\begin{equation}
\ln\frac{H^2}{\Lambda^2}\le -0.06(8)\qquad \Longrightarrow\qquad H\le H_{max}^B=0.966(5)\Lambda\,.
\eqlabel{hmaxtypeb}
\end{equation}
Besides numerical (technical) difficulties associated with construction of these vacua, there are conceptual
ones, associated with the breakdown of the supergravity approximation --- the effective action \eqref{5action}
becomes less reliable as the background space-time curvature of \eqref{ef1i} grows.
In fig.~\ref{kTypeB} (left panel) we present the inverse Kretschmann scalar of
\eqref{ef1i} evaluated at the
apparent horizon in different computations schemes, see appendix \ref{kretschmann},
specifically \eqref{kahtypeb}:
\begin{equation}
\begin{split}
&{\rm SchemeI:}\qquad \ln\frac{H^2}{\Lambda^2}=k_s\,,\qquad \hat{K}=K\,;\\
&{\rm SchemeIII:}\qquad \ln\frac{H^2}{\Lambda^2}=\frac 14+\ln\a\,,\qquad \hat{K}=K\,.
\end{split}
\eqlabel{collapsektypeb}
\end{equation}
In the limit $\frac{H}{\Lambda}\to 0$ we recover the  inverse Kretschmann scalar
of the extremal KS solution \eqref{kks}, represented by a horizontal red dashed line.
As $H$ approached $H_{max}^B$, represented by vertical dashed black line,
the Kretschmann scalar at the AH of the holographic dual to
TypeB de Sitter vacua of the cascading gauge theory appears to grow faster than any polynomial
of $\Lambda/(H_{max}^B-H)$ --- we take $H_{max}^B$ in \eqref{hmaxtypeb} as the limiting value
for the existence of TypeB vacua. In the right panel of fig.~\ref{kTypeB}
we associate the growth of the Kretschmann scalar in the limit $H\to H_{max}^B$
with the collapse of the 3-cycle (the $S^{3}$ supporting the RR 3-form
flux \eqref{pquantization}) at the horizon, see \eqref{3cycle},
\begin{equation}
R_{S^3}^2=\frac{f_{a,0}^h(h_0^h)^{1/2}}{3}=Pg_s^{1/2}\ \frac{\hf_{a,0}^h(\hh_0^h)^{1/2}}{3}\,, 
\eqlabel{3cycleb}
\end{equation}
where in the second equality we used \eqref{irtypebscaled}.

\section{Conclusion}\label{conclude}

In this paper we presented a comprehensive analysis of the vacua structure of
the cascading gauge theory in de Sitter. The cascading gauge theory in Minkowski space-time
is characterized by a single modulus $g_s$ and the strong coupling scale $\Lambda$;
it confines with the spontaneous breaking of the chiral symmetry. de Sitter space-time
presents a new mass scale --- the Hubble constant $H$.
There are three distinct types of  de Sitter vacua of the theory --- Type$A_s$
(resembling the thermal deconfined states of KS theory with the unbroken chiral symmetry),
Type$A_b$ (resembling the thermal deconfined states of KS theory with the spontaneously
broken chiral symmetry) and TypeB (resembling the thermal confined states of KS theory with the spontaneously
broken chiral symmetry)---
with the different (Euclidean) topology, and the global symmetry. All three types play a role
of being an attractor of the late-time de Sitter dynamics, depending on the interplay
of the strong coupling scale $\Lambda$ and the Hubble constant $H$.
We discover an intriguing
pattern of the chiral symmetry breaking in the theory depending on the ratio
$\frac{H}{\Lambda}$. While it is natural to expect that the chiral symmetry is spontaneously broken
for sufficiently small $\frac{H}{\Lambda}$ (in fact, the extremal KS solution is a limiting
case $\frac{H}{\Lambda}\to 0$), we find that the chiral symmetry is spontaneously broken
as well when $H\in [H_{min}^b,H_{max}]$, with $\{H_{min}^b,H_{max}\}\sim \Lambda$. 
In the former case, TypeB de Sitter vacua, the vacuum entanglement entropy density
vanishes\footnote{More precisely it is order $\calo(N^0)$.}  
much like for the confining thermal states, while in the latter, TypeA$_b$ de Sitter vacua,
the vacuum entanglement entropy is finite, much like for the thermal deconfined states.
Since $H_{min}^s< H_{min}^b$, the chiral symmetry breaking and the confinement/deconfinement
are two separate transitions in the cascading gauge theory in de Sitter. This is in contrast
to thermal transitions in the cascading gauge theory in Minkowski space-time,
where the chiral symmetry breaking is always accompanied by the confinement
\cite{Buchel:2010wp,Buchel:2018bzp}.

There is a number of open questions and future directions:
\nxt We argued that vacua TypeA$_s$ do not exist for sufficiently
small $\frac{H}{\Lambda}$. It is important to rigorously establish this fact.
Indeed, TypeA$_s$ vacua, unlike TypeB vacua, are characterized by
the nonzero entanglement entropy density, and thus, when exist, will always dominate over TypeB vacua
as the late-time dynamical attractors. 
\nxt We mentioned that TypeA$_b$ vacua resemble the thermal states of the deconfined cascading gauge theory
with $\zet_2$ chiral symmetry. The holographic dual of these states is a Klebanov-Strassler black hole 
\cite{Buchel:2018bzp}, which is unstable to local energy density perturbations --- the sound waves
in the cascading gauge theory plasma.
It would be interesting to study the fate of spatial inhomogeneities in TypeA$_b$ de
Sitter vacua. 
\nxt Ideally, we would like to develop numerical simulations of the cascading gauge theory in
de Sitter, akin to the model studied in \cite{Buchel:2017lhu}. As a first step, it would be interesting
to compute the spectrum (the quasinormal modes) of the chiral symmetry breaking
fluctuations about TypeA$_s$ vacua for $H\in [H_{min}^b,H_{max}]$.
\nxt It is important to explore the spontaneous symmetry breaking and the role played by the
de Sitter vacuum entanglement entropy in other top-down examples of massive holography. 
\nxt In this paper we studied confinement/deconfinement and chiral symmetry breaking of
strongly coupled gauge theories in de Sitter. It would be extremely interesting
to pursue these questions in other curved background space-times, and specifically in
anti-de Sitter. There is an ample literature on the subject\footnote{See \eg \cite{Aharony:2012jf}
and references/citations there.
}, mostly from the field theory perspective. A natural starting point would be to understand the
dynamics of $\caln=2^*$ gauge theory in $AdS_4$, expanding on \cite{Buchel:2004rr}.

\section*{Acknowledgments}
Research at Perimeter
Institute is supported by the Government of Canada through Industry
Canada and by the Province of Ontario through the Ministry of
Research \& Innovation. This work was further supported by
NSERC through the Discovery Grants program.

\appendix
\section{EF frame equations of motion}\label{efeoms}

Within Eddington-Finkelstein metric ansatz (with spatially homogeneous and isotropic background
metric of the cascading gauge theory --- $ d\boldsymbol{x}^2$)
\begin{equation}
ds_{10}^2 = 2dt\ \left(dr-A\ dt\right)+\Sigma^2\ d\boldsymbol{x}^2 + \Omega_1^2\ g_5^2
+ \Omega_2^2\ (g_3^2 + g_4^2) + \Omega_3^2\ (g_1^2 + g_2^2)\,,
\eqlabel{ef1}
\end{equation}
with
\begin{equation}
\begin{split}
A=A(t,r)\,,\ \Sigma=\Sigma(t,r)\,,\ \Omega_i=\Omega_i(t,r)\,,\ K_i=K_i(t,r)\,,\
\Phi=\ln g(t,r)\,,
\end{split}
\eqlabel{ef2}
\end{equation}
we find from \eqref{5action} the following evolution ($'\equiv \del_r$ and
$d_+\equiv \del_t+ A\del_r$): 
\begin{equation}
\begin{split}
&0=\left(d_+\Sigma\right)'+\biggl(\frac{d_+\Omega_2}{\Omega_2}
+\frac{d_+\Omega_3}{\Omega_3}+\frac{d_+\Omega_1}{2\Omega_1}\biggr)\ \Sigma'
+\biggl(
\frac{\Omega_2'}{\Omega_2}+\frac{\Omega_3'}{\Omega_3}+2 \frac{\Sigma'}{\Sigma}+\frac{\Omega_1'}{2\Omega_1}
\biggr)\ d_+\Sigma\\
&-\frac{ P^2g \Sigma  K_2'}{1296\Omega_2^2 \Omega_3^2}\ d_+K_2
-\frac{\Sigma  K_1'}{1152\Omega_3^4 P^2g }\ d_+K_1
-\frac{\Sigma K_3'}{1152 \Omega_2^4 P^2g}\  d_+K_3-\frac{\Sigma (K_1-K_3)^2}
{4608\Omega_3^2 \Omega_2^2  \Omega_1^2 P^2g}\\
&-\frac{P^2g K_2^2 \Sigma  (\Omega_2^4+\Omega_3^4)}{5184\Omega_2^4 \Omega_1^2 \Omega_3^4}
+\frac{P^2 g K_2\Sigma }{1296\Omega_2^4 \Omega_1^2} -\frac{P^2 g \Sigma }{1296\Omega_2^4 \Omega_1^2}
-\frac{\Sigma (K_1 K_2-K_3 K_2-2 K_1)^2}{373248\Omega_1^2 \Omega_2^4 \Omega_3^4}\,,
\end{split}
\eqlabel{ev1}
\end{equation}
\begin{equation}
\begin{split}
&0=\left(d_+\Omega_1\right)'+\biggl(
\frac{3 \Sigma'}{2\Sigma}+\frac{\Omega_2'}{\Omega_2}+\frac{\Omega_3}{\Omega_3}
\biggr)\ d_+\Omega_1
+\biggl(\frac{d_+\Omega_2}{\Omega_2}+\frac{d_+\Omega_3}{\Omega_3}
+\frac{3d_+\Sigma}{2\Sigma}\biggr)\ \Omega_1'\\
&-\frac{K_1' \Omega_1}{1152 P^2g \Omega_3^4}\  d_+K_1
-\frac{\Omega_1 P^2 g K_2'}{1296\Omega_3^2 \Omega_2^2}\  d_+K_2
-\frac{\Omega_1  K_3'}{1152\Omega_2^4 P^2 g}\ d_+K_3
+\frac{(K_3-K_1)^2}{1536\Omega_3^2 \Omega_2^2 \Omega_1 P^2g}\\
&+\frac{(K_3 K_2-K_1 K_2+2 K_1)^2}{373248\Omega_3^4 \Omega_1 \Omega_2^4}
-\frac{(2 \Omega_1^2+\Omega_2^2-\Omega_3^2) (2 \Omega_1^2-\Omega_2^2+\Omega_3^2)}{8\Omega_3^2 \Omega_2^2 \Omega_1}
+\frac{P^2gK_2^2 (\Omega_2^4+\Omega_3^4)}{1728\Omega_3^4 \Omega_1 \Omega_2^4}\\
&-\frac{P^2gK_2}{432\Omega_1 \Omega_2^4} +\frac{P^2g}{432\Omega_1 \Omega_2^4}\,,
\end{split}
\eqlabel{ev2}
\end{equation}
\begin{equation}
\begin{split}
&0=\left(d_+\Omega_2\right)'+\biggl(
\frac{\Omega_2'}{\Omega_2}+\frac{\Omega_3'}{\Omega_3}+\frac{3\Sigma'}{2\Sigma}+\frac{\Omega_1'}{2\Omega_1}
\biggr)\ d_+\Omega_2
+\biggl(
\frac{d_+\Omega_3} {\Omega_3}+\frac{d_+\Omega_1}{2\Omega_1}
+\frac{3d_+\Sigma}{2\Sigma}\biggr)\ \Omega_2' \\
&+\frac{P^2g  K_2'}{1296\Omega_2 \Omega_3^2}\ d_+K_2
-\frac{\Omega_2  K_1'}{1152 \Omega_3^4 P^2g}\ d_+K_1
+\frac{K_3'}{384 \Omega_2^3 P^2g}\ d_+K_3
+\frac{(K_1-K_3)^2}{4608\Omega_3^2 \Omega_2  \Omega_1^2P^2g}
\\&-\frac{K_2^2 P^2 g (\Omega_2^4-3 \Omega_3^4)}{5184\Omega_2^3 \Omega_1^2 \Omega_3^4}
-\frac{K_2 P^2g}{432\Omega_2^3 \Omega_1^2} +\frac{P^2g}{432\Omega_2^3 \Omega_1^2}
+\frac{(K_1 K_2-K_3 K_2-2 K_1)^2}{373248\Omega_2^3 \Omega_1^2 \Omega_3^4}\\
&+\frac{4 \Omega_1^4-8 \Omega_1^2 \Omega_3^2-\Omega_2^4+\Omega_3^4}{16\Omega_3^2 \Omega_2 \Omega_1^2}\,,
\end{split}
\eqlabel{ev3}
\end{equation}
\begin{equation}
\begin{split}
&0=\left(d_+\Omega_3\right)'+\biggl(
\frac{\Omega_2'}{\Omega_2}+\frac{\Omega_3'}{\Omega_3}+\frac{3\Sigma'}{2\Sigma}
+\frac{\Omega_1'}{2\Omega_1}\biggr)\ d_+\Omega_3
+\biggl( \frac{d_+\Omega_2} {\Omega_2}+\frac{d_+\Omega_1}{2\Omega_1}
+\frac{3d_+\Sigma}{2\Sigma}\biggr)\ \Omega_3'\\
&+\frac{ P^2g  K_2'}{1296\Omega_2^2 \Omega_3}\ d_+K_2
+ \frac{K_1'}{384 \Omega_3^3 P^2g}\ d_+K_1
-\frac{\Omega_3  K_3'}{1152 \Omega_2^4 P^2g}\ d_+K_3
+\frac{(K_1-K_3)^2}{4608\Omega_3 \Omega_2^2  \Omega_1^2P^2g}\\
&+\frac{P^2 g K_2^2 (3 \Omega_2^4-\Omega_3^4)}{5184\Omega_2^4 \Omega_1^2 \Omega_3^3}
+\frac{P^2 gK_2 \Omega_3}{1296\Omega_2^4 \Omega_1^2}
-\frac{P^2g \Omega_3}{1296\Omega_2^4 \Omega_1^2}
+\frac{(K_1 K_2-K_3 K_2-2 K_1)^2}{373248\Omega_2^4 \Omega_1^2 \Omega_3^3}\\
&+\frac{4 \Omega_1^4-8 \Omega_1^2 \Omega_2^2+\Omega_2^4-\Omega_3^4}{16\Omega_3 \Omega_2^2 \Omega_1^2}\,,
\end{split}
\eqlabel{ev4}
\end{equation}
\begin{equation}
\begin{split}
&0=\left(d_+K_1\right)'+ \biggl(
\frac{d_+\Omega_2}{\Omega_2}- \frac{d_+\Omega_3}{\Omega_3}+ \frac{d_+\Omega_1}{2\Omega_1}
+\frac{3d_+\Sigma}{2\Sigma}-\frac{d_+g}{2g}\biggr)\ K_1'
+\biggl(
\frac{\Omega_2'}{\Omega_2}-\frac{\Omega_3'}{\Omega_3}+ \frac{3\Sigma'}{2\Sigma}
\\&+\frac{\Omega_1'}{2\Omega_1}-\frac{g'}{2g}\biggr)\ d_+K_1
-\frac{\Omega_3^2 (K_1-K_3)}{4\Omega_2^2 \Omega_1^2}-\frac{P^2g (K_2-2)
(K_1 K_2-K_3 K_2-2 K_1)}{648\Omega_2^4 \Omega_1^2}\,,
\end{split}
\eqlabel{ev5}
\end{equation}
\begin{equation}
\begin{split}
&0=\left(d_+K_2\right)'+\biggl(
\frac{d_+\Omega_1}{2\Omega_1}+\frac{3d_+\Sigma}{2\Sigma}+\frac{d_+g}{2g}
\biggr)\ K_2'
+\biggl(
\frac{\Omega_1'}{2\Omega_1}+\frac{g'}{2g}+\frac{3\Sigma'}{2\Sigma}\biggr)\ d_+K_2
\\& -\frac{(K_3-K_1) (K_3 K_2-K_1 K_2+2 K_1)}{576\Omega_1^2 \Omega_2^2 \Omega_3^2 P^2g}
-\frac{K_2(\Omega_2^4+\Omega_3^4)}{4\Omega_1^2 \Omega_2^2 \Omega_3^2}+\frac{\Omega_3^2}{2\Omega_1^2 \Omega_2^2}\,,
\end{split}
\eqlabel{ev6}
\end{equation}
\begin{equation}
\begin{split}
&0=\left(d_+K_3\right)'+\biggl(
\frac{3d_+\Sigma}{2\Sigma}- \frac{d_+g}{2g}
- \frac{d_+\Omega_2}{\Omega_2}+ \frac{d_+\Omega_3}{\Omega_3}+ \frac{d_+\Omega_1}{2\Omega_1}
\biggr)\ K_3'+\biggl(
\frac{\Omega_3'}{\Omega_3}-\frac{\Omega_2'}{\Omega_2}+\frac{\Omega_1'}{2\Omega_1}
\\&-\frac{g'}{2g}+\frac{3\Sigma'}{2\Sigma}\biggr)\ d_+K_3
-\frac{\Omega_2^2 (K_3-K_1)}{4\Omega_3^2 \Omega_1^2}-
 \frac{P^2 g K_2  (K_3 K_2-K_1 K_2+2 K_1)}{648\Omega_1^2 \Omega_3^4}\,,
\end{split}
\eqlabel{ev7}
\end{equation}
\begin{equation}
\begin{split}
&0=\left(d_+g\right)'+\biggl(
\frac{d_+\Omega_2}{\Omega_2}+ \frac{d_+\Omega_3}{\Omega_3}+ \frac{d_+\Omega_1}{2\Omega_1}
+\frac{3d_+\Sigma}{2\Sigma} \biggr)\ g'
-\frac{P^2 g^2K_2'  }{324\Omega_2^2 \Omega_3^2}\ d_+K_2\\&+\frac{K_3'}{288\Omega_2^4 P^2}\ d_+K_3
+\frac{K_1'}{288\Omega_3^4 P^2}\ d_+K_1+\biggl(
\frac{\Omega_2'}{\Omega_2}+\frac{\Omega_3'}{\Omega_3}
+\frac{\Omega_1'}{2\Omega_1}-\frac{g'}{g}+\frac{3\Sigma'}{2\Sigma}\biggr)\ d_+g
\\&+\frac{(K_3-K_1)^2}{1152\Omega_3^2 \Omega_2^2 \Omega_1^2P^2}
-\frac{P^2g^2K_2^2 (\Omega_2^4+\Omega_3^4)}{1296\Omega_1^2 \Omega_2^4 \Omega_3^4}
+\frac{P^2g^2K_2}{324\Omega_1^2 \Omega_2^4} -  \frac{P^2g^2}{324\Omega_1^2 \Omega_2^4}\,,
\end{split}
\eqlabel{ev8}
\end{equation}
\begin{equation}
\begin{split}
&0=A''-\biggl(
\frac{2 \Omega_2'}{\Omega_1 \Omega_2}+\frac{2\Omega_3'}{\Omega_1 \Omega_3}
+\frac{3\Sigma'}{\Sigma \Omega_1}\biggr)\ d_+\Omega_1
-\biggl(
\frac{2 \Omega_2'}{\Omega_2^2}+\frac{4 \Omega_3'}{\Omega_2 \Omega_3}
+\frac{6 \Sigma'}{\Sigma \Omega_2}+\frac{2 \Omega_1'}{\Omega_1 \Omega_2}
\biggr)\ d_+\Omega_2\\
&-\biggl(
\frac{4 \Omega_2'}{\Omega_2 \Omega_3}+\frac{2\Omega_3'}{\Omega_3^2}
+\frac{6\Sigma'}{\Sigma \Omega_3}
+\frac{2 \Omega_1'}{\Omega_1 \Omega_3}\biggr)\ d_+\Omega_3
-\biggl(
\frac{6 \Omega_2'}{\Omega_2 \Sigma}+\frac{6\Omega_3'}{\Omega_3 \Sigma}
+\frac{6 \Sigma'}{\Sigma^2}+\frac{3\Omega_1'}{\Omega_1 \Sigma}
\biggr)\ d_+\Sigma\\
&+\frac{g'}{2g^2}\ d_+g+\frac{ P^2g  K_2'}{648\Omega_2^2 \Omega_3^2}\ d_+K_2
+\frac{K_1'}{576 \Omega_3^4 P^2g}\ d_+K_1
+\frac{K_3'}{576 \Omega_2^4 P^2g}\  d_+K_3
\\&-\frac{(K_1-K_3)^2}{768\Omega_3^2 \Omega_2^2  \Omega_1^2P^2g}
-\frac{P^2g K_2^2 (\Omega_2^4+\Omega_3^4)}{864\Omega_2^4 \Omega_1^2 \Omega_3^4}
+\frac{P^2g K_2}{216\Omega_2^4 \Omega_1^2}-\frac{P^2g}{216\Omega_2^4 \Omega_1^2}
\\&-\frac{(K_1 K_2-K_3 K_2-2 K_1)^2}{93312\Omega_1^2 \Omega_2^4 \Omega_3^4}
-\frac{4 \Omega_1^4-8 \Omega_1^2 \Omega_2^2-8 \Omega_1^2 \Omega_3^2
+\Omega_2^4-2 \Omega_2^2 \Omega_3^2+\Omega_3^4}{8\Omega_2^2 \Omega_3^2 \Omega_1^2}\,;
\end{split}
\eqlabel{ev9}
\end{equation}
and the constraint equations
\begin{equation}
\begin{split}
&0=\Sigma''+\frac{\Sigma}{6} \biggl(
\frac{(g')^2}{g^2}+\frac{4 \Omega_3''}{\Omega_3}+\frac{4 \Omega_2''}{\Omega_2}
+\frac{2 \Omega_1''}{\Omega_1}+  \frac{P^2g (K_2')^2}{162\Omega_3^2 \Omega_2^2}
+\frac{(K_3')^2}{144P^2g \Omega_2^4}+\frac{(K_1')^2}{144P^2g \Omega_3^4}
\biggr)\,,
\end{split}
\eqlabel{con1}
\end{equation}
\begin{equation}
\begin{split}
&0=d_+^2\Sigma+\frac{\Sigma}{3\Omega_1}\ d_+^2\Omega_1
+\frac{2\Sigma}{3\Omega_2}\ d_+^2\Omega_2+\frac{2\Sigma}{3\Omega_3}\ d_+^2\Omega_3
-\biggl(\frac{\Sigma }{3\Omega_1}\ d_+\Omega_1+ \frac{2\Sigma}{3\Omega_2}\ d_+\Omega_2
\\&+\frac{2\Sigma}{3\Omega_3}\ d_+\Omega_3+ d_+\Sigma\biggr)\ A'
+\frac{\Sigma  P^2g }{972\Omega_2^2 \Omega_3^2}\ (d_+K_2)^2
+\frac{\Sigma }{864 \Omega_2^4 P^2g}\ (d_+K_3)^2\\&
+\frac{\Sigma }{864 \Omega_3^4 P^2g}\ (d_+K_1)^2
+\frac{\Sigma}{6g^2}\ (d_+g)^2\,.
\end{split}
\eqlabel{con2}
\end{equation}

To derive the late-time geometry dual to the cascading gauge theory vacuum in de Sitter, we introduce 
following \cite{Buchel:2017pto}
\begin{equation}
\lim_{t\to \infty}\biggl\{A(t,r)\,,\,\frac{\Sigma(t,r)}{e^{H t}}\,,\, 
K_{i}(t,r)\,,\, g(t,r)\biggr\}=\left\{a(r)\,,\,\sigma(r)\,,\, K_i(r)\,,\, g(r)\right\}\,;
\eqlabel{ltlimit1}
\end{equation}
furthermore,
\begin{equation}
\lim_{t\to \infty}\biggl\{\Omega_1^2(t,r)\,,\, \Omega_2^2(t,r)\,,\, \Omega_3^2(t,r)\biggr\}=
\left\{
\frac{1}{9}w_{c2}(r)\,,\, \frac{1}{6}w_{a2}(r)\,,\, \frac{1}{6}w_{b2}(r)
\right\}  \,.
\eqlabel{ltlimit2}
\end{equation}
We find from \eqref{ev1}-\eqref{con2} in the $t\to \infty$ limit 9
second order ODEs:
\begin{equation}
\begin{split}
&0=\s''+\frac{5(\s')^2}{4\s}+\frac{5a' \s'}{8a}+\frac{\s}{16} \biggl(
\frac{2 \s'}{\s}-\frac{a'}{a}\biggr) \biggl(
\frac{w_{c2}'}{w_{c2}}+\frac{2 w_{a2}'}{w_{a2}}+\frac{2 w_{b2}'}{w_{b2}}\biggr)
+\frac{H \s}{16a} \biggl(
\frac{30 \s'}{\s}+\frac{w_{c2}'}{w_{c2}}\\
&+\frac{2 w_{a2}'}{w_{a2}}
+\frac{2 w_{b2}'}{w_{b2}}\biggr)
-\frac{\s}{8} \biggl(
\frac 12\left(\frac{w_{a2}'}{w_{a2}}+\frac{w_{b2}'}{w_{b2}}\right)^2+\frac{w_{a2}' w_{b2}'}{w_{a2} w_{b2}}
+\frac{w_{a2}' w_{c2}'}{w_{a2} w_{c2}}+\frac{w_{c2}' w_{b2}'}{w_{c2} w_{b2}}\biggr)
+\frac{\s (g')^2}{16g^2}\\
&-\frac{2g P^2 (K_2')^2}{9w_{b2} w_{a2}}-\frac{(K_3')^2}{4 g P^2 w_{a2}^2}-\frac{(K_1')^2}{4g P^2w_{b2}^2}
-\frac{27\s (K_3-K_1)^2}{256w_{b2} w_{a2} a  w_{c2} gP^2}-\frac{\s}{128a w_{b2}^2 w_{a2}^2 w_{c2}} \biggl(
\\&
5 K_2^2 (K_3-K_1)^2+2 w_{b2} w_{a2} (9 w_{b2}^2-18 w_{b2} w_{a2}-48 w_{b2} w_{c2}
+9 w_{a2}^2-48 w_{a2} w_{c2}+16 w_{c2}^2)\\&+20 K_1 (K_3 K_2-K_1 K_2+K_1)\biggr)
-\frac{3\s g P^2(K_2 (w_{b2}^2 K_2+w_{a2}^2 K_2-4 w_{b2}^2)+4 w_{b2}^2)}{32a w_{b2}^2 w_{a2}^2 w_{c2}}\,;
\end{split}
\eqlabel{efv1}
\end{equation}
\begin{equation}
\begin{split}
&0=a''+\frac{21a' \s'}{4\s}+\frac{a}{8} \left( \frac{18\s'}{\s}+\frac{7 a'}{a}\right)
\biggl(\frac{w_{c2}'}{w_{c2}}+\frac{2w_{a2}'}{w_{a2}}+\frac{2 w_{b2}'}{w_{b2}}\biggr)
+\frac{3 H}{8} \biggl(\frac{26\s'}{\s}+\frac{3 w_{c2}'}{w_{c2}}\\
&+\frac{6 w_{a2}'}{w_{a2}}+\frac{6 w_{b2}'}{w_{b2}}\biggr)
+\frac{3a}{8} \biggl(
\frac{12 (\s')^2}{\s^2}+\frac{2 w_{a2}' w_{c2}'}{w_{a2} w_{c2}}
+\frac{2 w_{c2}' w_{b2}'}{w_{c2} w_{b2}}+\frac{2 w_{a2}' w_{b2}'}{w_{a2} w_{b2}}
+\left(\frac{w_{a2}}{w_{a2}}+\frac{w_{b2}'}{w_{b2}}\right)^2\ \biggr)\\
&-\frac{3(g')^2 a}{8g^2}-\frac{5 a}{32} \biggl(
\frac{8g P^2 (K_2')^2}{9w_{b2} w_{a2}}+\frac{(K_3')^2}{gP^2 w_{a2}^2}+\frac{(K_1')^2}{g P^2 w_{b2}^2}
\biggr)
+\frac{9 (K_3-K_1)^2}{128 w_{b2} w_{a2}  w_{c2}g P^2}\\
&-\frac{1}{64w_{c2} w_{a2}^2 w_{b2}^2} \biggl(K_2^2 (K_3-K_1)^2-6 w_{b2} w_{a2} (9 w_{b2}^2-18 w_{b2} w_{a2}
-48 w_{b2} w_{c2}+9 w_{a2}^2\\&-48 w_{a2} w_{c2}+16 w_{c2}^2)+4 K_1 ((K_3 -K_1) K_2+K_1)\biggr)+\frac{g P^2}{16w_{c2} w_{b2}^2 w_{a2}^2}
\biggl((w_{a2}^2+w_{b2}^2) K_2^2\\&+4 (1-K_2) w_{b2}^2\biggr)\,;
\end{split}
\eqlabel{efv2}
\end{equation}
\begin{equation}
\begin{split}
&0=w_{a2}''+\frac{w_{a2} a'}{8 a} \biggl(
\frac{6 w_{a2}'}{w_{a2}}-\frac{6 \s'}{\s}-\frac{2 w_{b2}'}{w_{b2}}-\frac{w_{c2}'}{w_{c2}}\biggr)
-\frac{w_{a2}}{8} \biggl(
\frac{(w_{a2}')^2}{w_{a2}^2}-\frac{4 w_{a2}' w_{b2}'}{w_{a2} w_{b2}}+\frac{(w_{b2}')^2}{w_{b2}^2}
\\&-\frac{12 w_{a2}' \s'}{w_{a2} \s}+\frac{12 w_{b2}' \s'}{\s w_{b2}}
+\frac{12 (\s')^2}{\s^2}-\frac{2 w_{a2}' w_{c2}'}{w_{a2} w_{c2})}+\frac{2 w_{c2}' w_{b2}'}{w_{c2} w_{b2}}
+\frac{6 w_{c2}' \s'}{\s w_{c2}}\biggr)
+\frac{(g')^2 w_{a2}}{8g^2}\\
&+\frac{g P^2 (K_2')^2}{12w_{b2}}-\frac{w_{a2} (K_1')^2}{ 32w_{b2}^2 gP^2}+\frac{7 (K_3')^2}{32 w_{a2}g P^2}
+\frac{3 H w_{a2}}{8 a} \biggl(\frac{2 w_{a2}'}{ w_{a2}}-\frac{2 w_{b2}'}{w_{b2}}-\frac{6 \s'}{\s}-\frac{w_{c2}'}{w_{c2}}\biggr)
\\&+\frac{9 (K_3-K_1)^2}{128w_{b2}  w_{c2} a gP^2}+\frac{3}{64w_{b2}^2 w_{c2} w_{a2} a}
\biggl(
K_2^2 (K_3-K_1)^2+2 w_{b2} w_{a2} (16 w_{c2}^2-48 w_{c2} w_{b2}\\&+16 w_{a2} w_{c2}+9 w_{b2}^2
+6 w_{a2} w_{b2}-15 w_{a2}^2)+4 K_1 (K_3 K_2-K_1 K_2+K_1)\biggr)
\\&+\frac{g P^2}{16w_{b2}^2 w_{c2} w_{a2} a} \biggl(K_2^2 (5 w_{b2}^2-3 w_{a2}^2)+20 (1-K_2) w_{b2}^2\biggr)\,;
\end{split}
\eqlabel{efv3}
\end{equation}
\begin{equation}
\begin{split}
&0=w_{b2}''+\frac{a' w_{b2}}{8 a} \biggl(\frac{6 w_{b2}'}{w_{b2}}-\frac{6 \s'}{\s}-\frac{2 w_{a2}'}{w_{a2}}
-\frac{w_{c2}'}{w_{c2}}\biggr)
-\frac{w_{b2}}{8} \biggl(
\frac{(w_{a2}')^2}{w_{a2}^2}-\frac{4 w_{a2}'w_{b2}'}{w_{a2} w_{b2}}
+\frac{(w_{b2}')^2}{w_{b2}^2}\\&+\frac{12 w_{a2}'\s'}{w_{a2} \s}
-\frac{12 w_{b2}'\s'}{\s w_{b2}}+\frac{12 (\s')^2}{\s^2}
+\frac{2 w_{a2}' w_{c2}'}{w_{a2} w_{c2}}
-\frac{2w_{c2}'w_{b2}'}{w_{c2} w_{b2}}
+\frac{6 w_{c2}'\s'}{\s w_{c2}}\biggr)
+\frac{(g')^2 w_{b2}}{8g^2}\\
&+\frac{g P^2 (K_2')^2}{12w_{a2}}+\frac{7(K_1')^2}{32 w_{b2}g P^2}-\frac{w_{b2} (K_3')^2}{32 w_{a2}^2 gP^2}
-\frac{3 H w_{b2}}{8a} \biggl(\frac{2 w_{a2}'}{w_{a2} }-\frac{2 w_{b2}'}{w_{b2} }+\frac{6 \s'}{\s }+\frac{w_{c2}'}{w_{c2} }
\biggr)
\\&+\frac{9 (K_3-K_1)^2}{128w_{a2} g P^2 w_{c2} a}+\frac{3}{64w_{b2} w_{a2}^2 w_{c2} a} \biggl(
K_2^2 (K_3-K_1)^2+2 w_{a2} w_{b2} (16 w_{c2}^2+16 w_{c2} w_{b2}\\&-48 w_{a2} w_{c2}
-15 w_{b2}^2+6 w_{a2} w_{b2}+9 w_{a2}^2)+4 K_1 (K_3 K_2-K_1 K_2+K_1)\biggr)
\\&-\frac{g P^2}{16w_{b2} w_{a2}^2 w_{c2} a} \biggl(K_2^2 (3 w_{b2}^2-5 w_{a2}^2)+12 (1-K_2) w_{b2}^2\biggr)\,;
\end{split}
\eqlabel{efv4}
\end{equation}
\begin{equation}
\begin{split}
&0=w_{c2}''-\frac{w_{c2} a'}{8 a} \biggl(
\frac{2 w_{b2}'}{w_{b2}}+\frac{6 \s'}{\s}-\frac{7w_{c2}'}{w_{c2}}+\frac{2w_{a2}'}{w_{a2}}\biggr)
-\frac{w_{c2}}{8} \biggl(
\frac{(w_{a2}')^2}{w_{a2}^2}+\frac{4 w_{a2}'w_{b2}'}{w_{a2} w_{b2}}
+\frac{(w_{b2}')^2}{w_{b2}^2}\\&+\frac{12 w_{a2}'\s'}{w_{a2} \s}
+\frac{12 w_{b2}'\s'}{\s w_{b2}}+\frac{12 (\s')^2}{\s^2}
-\frac{6 w_{a2}'w_{c2}'}{w_{a2} w_{c2}}-\frac{6 w_{c2}'w_{b2}'}{w_{c2} w_{b2}}
-\frac{18 w_{c2}'\s'}{\s w_{c2}}+\frac{4 (w_{c2}')^2}{w_{c2}^2}\biggr)
\\&+\frac{(g')^2 w_{c2}}{8g^2}-\frac{w_{c2} g P^2 (K_2')^2}{36w_{b2} w_{a2}}-\frac{w_{c2} (K_1')^2}{32 w_{b2}^2 gP^2}
-\frac{w_{c2} (K_3')^2}{32 w_{a2}^2 gP^2}-\frac{3 H w_{c2}}{8 a}
\biggl(\frac{2 w_{a2}'}{w_{a2}}+\frac{2w_{b2}'}{w_{b2}}+\frac{6 \s'}{\s}
\\&-\frac{3 w_{c2}'}{w_{c2}}\biggr)
+\frac{45(K_3-K_1)^2}{128w_{a2} w_{b2} g P^2 a}+\frac{3}{64w_{b2}^2 w_{a2}^2 a} \biggl(
K_2^2 (K_3-K_1)^2-2 w_{a2} w_{b2} (48 w_{c2}^2\\&-16 w_{c2} w_{b2}-16 w_{a2} w_{c2}-21 w_{b2}^2
+42 w_{a2} w_{b2}-21 w_{a2}^2)+4 K_1 (K_3 K_2-K_1 K_2+K_1)\biggr)
\\&+\frac{5g P^2}{16w_{b2}^2 w_{a2}^2 a} \biggl(K_2^2 (w_{b2}^2+w_{a2}^2)+4 (1-K_2) w_{b2}^2\biggr)\,;
\end{split}
\eqlabel{efv5}
\end{equation}
\begin{equation}
\begin{split}
&0=K_1''+\biggl(
\frac{3H}{2a} +\frac{w_{a2}'}{w_{a2}}+\frac{w_{c2}'}{2w_{c2}}-\frac{w_{b2}'}{w_{b2}}
+\frac{3 \s'}{\s}
+\frac{a'}{a}-\frac{g'}{g}\biggr) K_1'-\frac{9w_{b2} (K_1-K_3)}{4w_{a2} a w_{c2}}
\\&-\frac{(K_2-2) (K_2 K_1-K_2 K_3-2 K_1) gP^2}{2a w_{c2} w_{a2}^2}\,;
\end{split}
\eqlabel{efv6}
\end{equation}
\begin{equation}
\begin{split}
&0=K_2''+\biggl(
\frac{a'}{a}+\frac{3 H}{2a}+\frac{w_{c2}'}{2w_{c2}}+\frac{g'}{g}
+\frac{3 \s'}{\s}\biggr) K_2'
-\frac{9(K_1-K_3) (K_2 (K_1-K_3)-2 K_1)}{16w_{a2} w_{b2} g P^2 w_{c2} a}
\\&-\frac{9((w_{a2}^2+w_{b2}^2) K_2-2 w_{b2}^2)}{4w_{a2} w_{b2} w_{c2} a}\,;
\end{split}
\eqlabel{efv7}
\end{equation}
\begin{equation}
\begin{split}
&0=K_3''+\biggl(
\frac{3H}{2a}-\frac{w_{a2}'}{w_{a2}}+\frac{3 \s'}{\s}+\frac{w_{c2}'}{2w_{c2}}
+\frac{w_{b2}'}{w_{b2}}+\frac{a'}{a}
-\frac{g'}{g}\biggr) K_3'+\frac{9w_{a2} (K_1-K_3)}{4w_{b2} a w_{c2}}\\&
+\frac{K_2  (K_2 (K_1-K_3)-2 K_1) gP^2}{2a w_{b2}^2 w_{c2}}\,;
\end{split}
\eqlabel{efv8}
\end{equation}
\begin{equation}
\begin{split}
&0=g''+\biggl(
\frac{3H}{2a}+\frac{a'}{a}+\frac{w_{c2}'}{2w_{c2}}-\frac{g'}{g}
+\frac{w_{b2}'}{w_{b2}}+\frac{w_{a2}'}{w_{a2}}+\frac{3 \s'}{\s}\biggr) g'
-\frac{g^2 P^2 (K_2')^2}{9w_{b2} w_{a2}}+\frac{(K_1')^2}{8P^2 w_{b2}^2}
\\&+\frac{(K_3')^2}{8P^2 w_{a2}^2}+\frac{9(K_1-K_3)^2}{32a w_{b2} w_{c2} w_{a2} P^2}
-\frac{g^2 P^2(K_2^2 (w_{a2}^2+w_{b2}^2)+4 (1-K_2) w_{b2}^2)}{4w_{b2}^2 a w_{c2} w_{a2}^2}\,,
\end{split}
\eqlabel{efv9}
\end{equation}
and 2 first order ODEs:
\begin{equation}
\begin{split}
&0=\s'+\frac{\s}{2a} \biggl(H-a'\biggr)\,;
\end{split}
\eqlabel{efc1}
\end{equation}
\begin{equation}
\begin{split}
&0=\frac{(g')^2}{g^2}-\frac{3 H}{a} \biggl(
\frac{2 w_{b2}'}{w_{b2}}+\frac{4 \s'}{\s}+\frac{2w_{a2}'}{w_{a2}}+\frac{w_{c2}'}{w_{c2}}
+\frac{a'}{a}-\frac{H}{a}\biggr)
-\frac{2 w_{a2}'a'}{a w_{a2}}-\frac{6 \s'a'}{\s a}-\frac{4w_{b2}'w_{a2}'}{w_{b2} w_{a2}}
\\&-\frac{12 w_{b2}'\s'}{\s w_{b2}}-\frac{12 w_{a2}'\s'}{\s w_{a2}}
-\frac{2 w_{b2}'a'}{a w_{b2}}-\frac{6 w_{c2}'\s'}{\s w_{c2}}-\frac{w_{c2}'a'}{a w_{c2}}
-\frac{2 w_{b2}'w_{c2}'}{w_{b2} w_{c2}}-\frac{2 w_{c2}'w_{a2}'}{w_{c2} w_{a2}}-\frac{(w_{b2}')^2}{w_{b2}^2}
\\&-\frac{12 (\s')^2}{\s^2}-\frac{(w_{a2}')^2}{w_{a2}^2}
+\frac{2P^2 g (K_2')^2}{9w_{b2} w_{a2}}+\frac{(K_1')^2}{4w_{b2}^2 P^2 g}+\frac{(K_3')^2}{4w_{a2}^2 P^2 g}
-\frac{9(K_1-K_3)^2}{16a w_{c2} g P^2 w_{b2} w_{a2}}
\\&-\frac{1}{8a w_{b2}^2 w_{c2} w_{a2}^2} \biggl(
K_2^2 (K_1-K_3)^2+2 w_{a2} w_{b2} (9 w_{a2}^2-18 w_{b2} w_{a2}
-48 w_{c2} w_{a2}+9 w_{b2}^2\\&-48 w_{b2} w_{c2}+16 w_{c2}^2)
-4 K_1 ((K_1-K_3) K_2-K_1)\biggr)-\frac{P^2 g}{2w_{b2}^2 a w_{c2} w_{a2}^2} \biggl(K_2^2 (w_{a2}^2+w_{b2}^2)\\&+4 (1-K_2) w_{b2}^2\biggr)\,.
\end{split}
\eqlabel{efc2}
\end{equation}
It is straightforward to verify the \eqref{efv1}-\eqref{efv9} are consistent with \eqref{efc1}-\eqref{efc2};
thus the latter ODEs can be used for drop \eqref{efv1} and \eqref{efv5} and eliminate $\s'$ and $w_{c2}'$ in the
remaining second order ODEs.

The cascading gauge theory de Sitter vacuum equations of motion \eqref{efv1}-\eqref{efc2} are invariant under the following
symmetries ($\l\equiv {\rm const}$),
\nxt symmetry SEF1:
\begin{equation}
r\ \to\ r+\l\,,\qquad \left\{H,P,a,\s,w_{a2,b2,c2},K_{1,2,3},g\right\}\ \to\ \left\{H,P,a,\s,w_{a2,b2,c2},K_{1,2,3},g\right\}\,;
\eqlabel{efsym1}
\end{equation}
\nxt symmetry SEF2:
\begin{equation}
P\ \to \l P\,,\qquad g\ \to\ \frac g\l \,,\ \ \left\{r,H,a,\s,w_{a2,b2,c2},K_{1,2,3}\right\}\ \to\ \left\{r,H,a,\s,w_{a2,b2,c2},
K_{1,2,3}\right\}\,;
\eqlabel{efsym2}
\end{equation}
\nxt symmetry SEF3:
\begin{equation}
\begin{split}
&\{P,r,a,w_{a2,b2,c2}\}\ \to\ \l\{P,r,a,w_{a2,b2,c2}\}\,,\qquad \s\ \to\ \l^{1/2}\s\,,\qquad
\{K_{1,3}\}\ \to\ \l^2\{K_{1,3}\}\,,\\
&\left\{H,K_{2},g\right\}\ \to\ \left\{H,K_{2},g\right\}\,; 
\end{split}
\eqlabel{efsym3}
\end{equation}
\nxt symmetry SEF4:
\begin{equation}
\begin{split}
&\{r,H\}\ \to\ \l\{r,H\}\,,\qquad
\left\{P,\s,w_{a2,b2,c2},K_{1,2,3},g\right\}\ \to\ \left\{P,\s,w_{a2,b2,c2},K_{1,2,3},g\right\} \\
&a\ \to \l^2 a\,.
\end{split}
\eqlabel{efsym4}
\end{equation}

\section{FG frame equations of motion,  asymptotics, relation to EF frame and
extremal Klebanov-Strassler solution}\label{apb}
Fefferman-Graham frame can be used to describe only (the patch of) the gravitational dual to
the cascading gauge theory de Sitter vacua. It is useful to setup the asymptotic boundary conditions,
analytical continuation to Euclidean (Bunch-Davies) vacua, and study the $H\to 0$ limit in which
one recovers the KS solution \cite{Klebanov:2000hb}.

Within the metric ansatz
\begin{equation}
\begin{split}
&ds_{10}^2 = \frac{1}{h^{1/2}\r^{2}} \left(d\calm_4^{f,c}\right)^2+
\frac{h^{1/2}}{\r^{2}} \left(d\r\right)^2+ \frac {f_c h^{1/2}}{9}  g_5^2
+ \frac{f_a h^{1/2}}{6} (g_3^2 + g_4^2) + \frac{f_b h^{1/2}}{6} (g_1^2 + g_2^2)\\
&\left(d\calm_4^f\right)^2=-d\t^2 + e^{2 H \t}d\boldsymbol{x}^2\,,\qquad
\left(d\calm_4^c\right)^2=-d\t^2 + \frac{1}{H^2}\cosh^2(H\t)\left(dS^3\right)^2\,,
\end{split}
\eqlabel{fg1}
\end{equation}
where we used the FG frame time $\t$ and the radial coordinate $\r$
to distinguish them from the EF frame time $t$ and the radial coordinate $r$ 
in \eqref{ef1},
\begin{equation}
f_{a,b,c}=f_{a,b,c}(\r)\,,\qquad h=h(\r)\,,\qquad K_{1,2,3}=K_{1,2,3}(\r)\,,\qquad
g=g(\r)\,,
\eqlabel{fg2}
\end{equation}
we find the following equations of
motion (independent of whether we use the flat boundary spatial slicing
$\left(d\calm_4^f\right)^2$ or the closed boundary spatial slicing
$\left(d\calm_4^c\right)^2$) describing de Sitter
vacuum of the cascading gauge theory \cite{Buchel:2013dla}:
\begin{equation}
\begin{split}
&0=f_c''- \frac{3f_c'}{\r}-3 h f_c H^2
-\frac{(f_c')^2}{2f_c}+\frac{5 f_c}{\r^2}+\frac{f_c (g')^2}{8g^2}+\frac{3f'_b f'_c}{4f_b}
+\frac{63 f_a}{16f_b \r^2}+\frac{63 f_b}{16f_a \r^2}+\frac{3 f_c}{f_a \r^2}\\
&-\frac{f_c (f_a')^2}{8f_a^2}
+\frac{3 f_a' f'_c}{4f_a}+\frac{f_c (h')^2}{8h^2}- \frac{f_c (f_b')^2}{8f_b^2}+\frac{3 f_c}{f_b \r^2}-\frac{63}{8 \r^2}
-\frac{K_1^2}{8f_a^2 h^2 f_b^2 \r^2}+ \frac{3g P^2}{2f_a^2 h \r^2}\\
&-\frac{f_c f_a' f'_b}{2f_a f_b}- \frac{27K_1 K_3}{32f_a h f_b g P^2 \r^2}
-\frac{K_2^2 K_1^2}{32f_a^2 h^2 f_b^2 \r^2}+\frac{K_2 K_1^2}{8f_a^2 h^2 f_b^2 \r^2}
- \frac{K_2^2 K_3^2}{32f_a^2 h^2 f_b^2 \r^2}- \frac{3f_c (K_1')^2}{32h f_b^2 g P^2}
\\
&- \frac{3f_c (K_3')^2}{32f_a^2 h g P^2}+\frac{3g P^2 K_2^2}{8h f_b^2 \r^2}+\frac{3 g P^2 K_2^2}{8f_a^2 h \r^2}
-\frac{3 g P^2 K_2}{2f_a^2 h \r^2}-\frac{9 f_c^2}{f_a f_b \r^2}+\frac{f_c h'}{h \r}+\frac{K_2^2 K_1 K_3}{16f_a^2 h^2 f_b^2 \r^2}
\\
&-\frac{K_2 K_1 K_3}{8f_a^2 h^2 f_b^2 \r^2}-\frac{g P^2 f_c (K_2')^2}{12f_a h f_b}
+\frac{27 K_1^2}{64f_a h f_b g P^2 \r^2}+\frac{27 K_3^2}{64f_a h f_b g P^2 \r^2}\,,
\end{split}
\eqlabel{kseq2}
\end{equation}
\begin{equation}
\begin{split}
&0=f_a''-\frac{45 f_a^2}{16f_c f_b \r^2}+\frac{f_a h'}{h \r}
+\frac{g P^2 (K_2')^2}{36h f_b}+\frac{5(K_3')^2}{32f_a h g P^2}
- \frac{f_a f'_b f'_c}{4f_c f_b}-\frac{(f_a')^2}{8f_a}+\frac{5 f_a}{\r^2}
-\frac{3 f_a'}{\r}\\
&-\frac{K_2^2 K_1^2}{32f_c f_a h^2 f_b^2 \r^2}+ \frac{K_2 K_1^2}{8f_c f_a h^2 f_b^2 \r^2}
-\frac{K_2^2 K_3^2}{32f_c f_a h^2 f_b^2 \r^2}-\frac{3 g P^2 K_2}{2f_c f_a h \r^2}+\frac{3 g P^2 K_2^2}{8f_c f_a h \r^2}
\\
&-\frac{9 K_3^2}{64f_c h f_b g P^2 \r^2}-\frac{9 K_1^2}{64f_c h f_b g P^2 \r^2}
+\frac{3 f_a}{f_b \r^2}+\frac{3 f_c}{f_b \r^2}+\frac{9 K_1 K_3}{32f_c h f_b g P^2 \r^2}
+\frac{K_2^2 K_1 K_3}{16f_c f_a h^2 f_b^2 \r^2}
\\&- \frac{K_2 K_1 K_3}{8f_c f_a h^2 f_b^2 \r^2}-\frac{5 f_a g P^2 K_2^2}{8f_c h f_b^2 \r^2}
-\frac{K_1^2}{8f_c f_a h^2 f_b^2 \r^2}+\frac{3g P^2}{2f_c f_a h \r^2}
-\frac{3 f_a (K_1')^2}{32h f_b^2 g P^2}-\frac{9}{\r^2}+\frac{f_a (g')^2}{8g^2}\\
&-3 f_a h H^2
+\frac{f_a' f'_b}{2f_b}+\frac{f'_c f_a'}{4f_c}-\frac{f_a (f_b')^2}{8f_b^2}
+\frac{9f_a}{8f_c \r^2}+ \frac{f_a (h')^2}{8h^2}+\frac{27f_b}{16f_c \r^2}\,,
\end{split}
\eqlabel{kseq3}
\end{equation}
\begin{equation}
\begin{split}
&0=f_b''-\frac{3 f'_b}{\r}-\frac{(f_b')^2}{8f_b}
+\frac{5 f_b}{\r^2}-\frac{45 f_b^2}{16f_c f_a \r^2}+\frac{f_b h'}{h \r}
-\frac{K_1^2}{8f_c h^2 f_a^2 f_b \r^2}-\frac{3 f_b (K_3')^2}{32h g f_a^2 P^2}
\\
&-\frac{K_2^2 K_1^2}{32f_c h^2 f_a^2 f_b \r^2}
+ \frac{K_2 K_1^2}{8f_c h^2 f_a^2 f_b \r^2}-\frac{K_2^2 K_3^2}{32f_c h^2 f_a^2 f_b \r^2}
-\frac{9 K_1^2}{64f_c h g f_a P^2 \r^2}+\frac{3 g P^2 K_2^2}{f_c h f_b \r^2}
\\
&-\frac{9K_3^2}{64f_c h g f_a P^2 \r^2}
-\frac{5 g f_b P^2}{2f_c h f_a^2 \r^2}+\frac{3 f_b}{f_a \r^2}+\frac{3 f_c}{f_a \r^2}
-\frac{f_b f'_c f_a'}{4f_c f_a}
+\frac{5 (K_1')^2}{32h g f_b P^2}+ \frac{g P^2 (K_2')^2}{36h f_a}\\
&-\frac{9}{\r^2}
+\frac{27 f_a}{16f_c \r^2}+\frac{9 f_b}{8f_c \r^2}+ \frac{K_2^2 K_1 K_3}{16f_c h^2 f_a^2 f_b \r^2}
-\frac{K_2 K_1 K_3}{8f_c h^2 f_a^2 f_b \r^2}+\frac{5 g f_b P^2 K_2}{2f_c h f_a^2 \r^2}
-\frac{5 g f_b P^2 K_2^2}{8f_c h f_a^2 \r^2}\\
&+\frac{9 K_1 K_3}{32f_c h g f_a P^2 \r^2}
+ \frac{f_b (g')^2}{8g^2}-3 h f_b H^2+ \frac{f_a' f'_b}{2f_a}
-\frac{f_b (f_a')^2}{8f_a^2}+\frac{f'_b f'_c}{4f_c}+ \frac{f_b (h')^2}{8h^2}\,,
\end{split}
\eqlabel{kseq4}
\end{equation}
\begin{equation}
\begin{split}
&0=h''+\frac{ K_2^2 K_1^2}{4f_c f_a^2 f_b^2 h \r^2}
-\frac{K_2 K_1^2}{f_c f_a^2 f_b^2 h \r^2}+ \frac{K_2^2 K_3^2}{4f_c f_a^2 f_b^2 h \r^2}
+\frac{9 K_1^2}{16f_c f_a f_b \r^2 g P^2}+\frac{9 K_3^2}{16f_c f_a f_b \r^2 g P^2}
\\
&+\frac{2 h f'_c}{f_c \r}+\frac{4 h f'_b}{f_b \r}+\frac{4 h f_a'}{f_a \r}+ \frac{(K_1')^2}{8f_b^2 g P^2}
+\frac{(K_3')^2}{8f_a^2 g P^2}+\frac{g P^2 K_2^2}{2f_c f_b^2 \r^2}+\frac{ g P^2 K_2^2}{2f_c f_a^2 \r^2}
-\frac{2 g P^2 K_2}{f_c f_a^2 \r^2}+ \frac{f'_c h'}{2f_c}
\\
&+\frac{h' f'_b}{f_b}+\frac{h' f_a'}{f_a}-\frac{16 h}{\r^2}-\frac{(h')^2}{h}+12 h^2 H^2
-\frac{K_2^2 K_1 K_3}{2f_c f_a^2 f_b^2 h \r^2}+\frac{K_2 K_1 K_3}{f_c f_a^2 f_b^2 h \r^2}+\frac{K_1^2}{f_c f_a^2 f_b^2 h \r^2}
\\
&+\frac{2 g P^2}{f_c f_a^2 \r^2}+ \frac{g P^2 (K_2')^2}{9f_a f_b)}
-\frac{9 K_1 K_3}{8f_c f_a f_b \r^2 g P^2}-\frac{3 h'}{\r}\,,
\end{split}
\eqlabel{kseq5}
\end{equation}
\begin{equation}
\begin{split}
&0=K_1''-\frac{g K_2^2 K_1 P^2}{f_c f_a^2 h \r^2}+\frac{g K_2^2 K_3 P^2}{f_c f_a^2 h \r^2}
+\frac{4 g K_2 K_1 P^2}{f_c f_a^2 h \r^2}-\frac{2 g K_2 K_3 P^2}{f_c f_a^2 h \r^2}- \frac{9f_b K_1}{2f_c f_a \r^2}
+ \frac{9f_b K_3}{2f_c f_a \r^2}\\
&-\frac{4 g K_1 P^2}{f_c f_a^2 h \r^2}+\frac{K'_1 f'_c}{2f_c}
-\frac{K'_1 g'}{g}-\frac{K'_1 h'}{h}+\frac{f_a' K'_1}{f_a}-\frac{3 K'_1}{\r}-\frac{K'_1 f'_b}{f_b}\,,
\end{split}
\eqlabel{kseq6}
\end{equation}
\begin{equation}
\begin{split}
&0=K_3''+\frac{g K_2^2 K_1 P^2}{f_c f_b^2 h \r^2}-\frac{g K_2^2 K_3 P^2}{f_c f_b^2 h \r^2}
-\frac{2 g K_2 K_1 P^2}{f_c f_b^2 h \r^2}+ \frac{9f_a K_1}{2f_c f_b \r^2}
- \frac{9f_a K_3}{2f_c f_b \r^2}+ \frac{K'_3 f'_c}{2f_c}\\
&-\frac{K'_3 g'}{g}
+\frac{f'_b K'_3}{f_b}-\frac{K'_3 h'}{h}
-\frac{3 K'_3}{\r}-\frac{K'_3 f_a'}{f_a}\,,
\end{split}
\eqlabel{kseq7}
\end{equation}
\begin{equation}
\begin{split}
&0=K_2''-\frac{9f_b K_2}{2f_c f_a \r^2}
- \frac{9f_a K_2}{2f_c f_b \r^2)}+\frac{9 f_b}{f_c f_a \r^2}-\frac{9 K_2 K_1^2}{8f_c g P^2 h f_b f_a \r^2}
+\frac{9 K_2 K_1 K_3}{4f_c g P^2 h f_b f_a \r^2}\\
&-\frac{9 K_2 K_3^2}{8f_c g P^2 h f_b f_a \r^2}
+\frac{9 K_1^2}{4f_c g P^2 h f_b f_a \r^2}- \frac{9K_1 K_3}{4f_c g P^2 h f_b f_a \r^2}
+ \frac{K'_2 f'_c}{2f_c}+\frac{K'_2 g'}{g}-\frac{K'_2 h'}{h}\\
&-\frac{3 K'_2}{\r}\,,
\end{split}
\eqlabel{kseq8}
\end{equation}
\begin{equation}
\begin{split}
&0=g''- \frac{g^2 P^2 K_2^2}{2f_c f_a^2 h \r^2}-\frac{g^2 P^2 K_2^2}{2f_c f_b^2 h \r^2}
+\frac{2 g^2 P^2 K_2}{f_c f_a^2 h \r^2}+\frac{9K_1^2}{16f_c f_a f_b h \r^2 P^2}+\frac{9 K_3^2}{16f_c f_a f_b h \r^2 P^2}
\\&-\frac{(g')^2}{g}- \frac{9K_1 K_3}{8f_c f_a f_b h \r^2 P^2}
+\frac{(K_3')^2}{8f_a^2 h P^2}+\frac{(K_1')^2}{8f_b^2 h P^2}
-\frac{2 g^2 P^2}{f_c f_a^2 h \r^2}-\frac{g^2 P^2 (K_2')^2}{9f_a f_b h}
+ \frac{g' f'_c}{2f_c}\\
&+\frac{g' f_a'}{f_a}+\frac{g' f'_b}{f_b}-\frac{3 g'}{\r}\,.
\end{split}
\eqlabel{kseq9}
\end{equation}
Additionally, we have the first order constraint
\begin{equation}
\begin{split}
&0=\frac89 g^2 (K_2')^2 f_b f_a P^4+(K_3')^2 f_b^2+(K_1')^2 f_a^2-\frac{4 g^2 K_2^2 f_a^2 P^4}{f_c \r^2}
+\frac{4 g f_a^2 f_b^2 P^2 (h')^2}{h}\\
&+\frac{4 h (g')^2 f_a^2 f_b^2 P^2}{g}+\frac{96 h g f_a^2 f_b P^2}{\r^2}+\frac{96 h g f_a f_b^2 P^2}{\r^2}
-\frac{96 h g f_a^2 f_b^2 P^2}{\r^2}-\frac{4 g K_1^2 P^2}{f_c h \r^2}
\\
&+96 h^2 g f_a^2 f_b^2 P^2 H^2+\frac{9 K_1 K_3 f_b f_a}{f_c \r^2}
+\frac{32 g f_a^2 f_b^2 P^2 h'}{\r}+\frac{16 g^2 K_2 f_b^2 P^4}{f_c \r^2}
-\frac{4 g^2 K_2^2 f_b^2 P^4}{f_c \r^2}\\
&-\frac{g K_2^2 K_1^2 P^2}{f_c h \r^2}
+\frac{4 g K_2 K_1^2 P^2}{f_c h \r^2}-\frac{g K_2^2 K_3^2 P^2}{f_c h \r^2}
+\frac{64 h g f_a^2 f_b P^2 f'_b}{\r}+\frac{64 h g f_a f_b^2 P^2 f_a'}{\r}
\\&-16 h g f_a f_b P^2 f_a' f'_b-\frac{32 f_c h g f_a f_b P^2}{\r^2}
-\frac{18 h g f_a f_b^3 P^2}{f_c \r^2}-\frac{18 h g f_a^3 f_b P^2}{f_c \r^2}
+\frac{36 h g f_a^2 f_b^2 P^2}{f_c \r^2}\\
&-\frac{9 K_3^2 f_b f_a}{f_c \r^2}
-4 h g f_b^2 P^2 (f_a')^2-4 h g f_a^2 P^2 (f_b')^2-\frac{16 g^2 f_b^2 P^4}{f_c \r^2}
+\frac{2 g K_2^2 K_1 K_3 P^2}{f_c h \r^2}\\
&-\frac{4 g K_2 K_1 K_3 P^2}{f_c h \r^2}
-\frac{8 h g f_b^2 f_a P^2 f'_c f_a'}{f_c}+\frac{32 h g f_a^2 f_b^2 P^2 f'_c}{f_c \r}
-\frac{8 h g f_a^2 f_b P^2 f'_b f'_c}{f_c}-\frac{9 K_1^2 f_b f_a}{2f_c \rho^2}\,.
\end{split}
\eqlabel{kseq10}
\end{equation}

The cascading gauge theory de Sitter vacuum equations of motion \eqref{kseq2}-\eqref{kseq10} are
invariant under the following symmetries ($\l\equiv {\rm const}$) (compare with \eqref{efsym1}-\eqref{efsym4}):
\nxt symmetry SFG1:
\begin{equation}
\left( \begin{array}{c}
\r  \\
H\\
P\\
h  \\
f_{a,b,c}\\
K_{1,2,3}\\
g   
\end{array} \right)\
\qquad \longrightarrow\qquad  
\left( \begin{array}{c}
{\r}/{(1+\l\ \r)}  \\
H\\
P\\
(1+\l\ \r)^4\ h \\
(1+\l\ \r)^{-2}\ f_{a,b,c}\\
K_{1,2,3}\\
{g}   \end{array} \right)\,;
\eqlabel{kssym4}
\end{equation}
\nxt symmetry SFG2:
\begin{equation}
P\to \lambda P\,,\ g\to \frac g\l \,,\ \{\r,H,f_{a,b,c},h,K_{1,2,3}\}\to \{\r,H,f_{a,b,c},h,K_{1,2,3}\}\,;
\eqlabel{kssym1}
\end{equation} 
\nxt symmetry SFG3:
\begin{equation}
P\to \lambda P\,,\ \r\to \frac \r\l\,,\ \{h,K_{1,3}\}\to \l^2\{h,K_{1,3}\}\,,\ 
\{H,f_{a,b,c},K_{2},g\}\to \{H,f_{a,b,c},K_{2},g\}\,;
\eqlabel{kssym2}
\end{equation} 
\nxt symmetry SFG4:
\begin{equation}
\r\to \lambda \r\,,\ H\to \frac H\l \,,\ 
\{P,f_{a,b,c},h,K_{1,2,3},g\}\to \{P,f_{a,b,c},h,K_{1,2,3},g\}\,.
\eqlabel{kssym3}
\end{equation}

FG frame makes analytical continuation to Euclidean Bunch-Davies vacuum
obvious:
\begin{equation}
\left(d\calm_4^c\right)^2\qquad \underbrace{\longrightarrow}_
{\t\to i\frac{\theta+\pi/2}{H}}\qquad \frac{1}{H^2}\biggl((d\theta)^2
+\sin^2(\theta)\ \left(dS^3\right)^2\biggr)=\frac{1}{H^2}\
\left(dS^4\right)^2\,.
\eqlabel{fgn1}
\end{equation}

\subsection{Asymptotics}\label{apb1}
The general UV (as $\r\to 0$) asymptotic solution of \eqref{kseq2}-\eqref{kseq10} describing the phase of the cascading
 gauge theory with spontaneously broken chiral symmetry takes the form
\begin{equation}
\begin{split}
&f_c=1+f_{a,1,0} \r+\biggl(-\frac38 P^2g_s H^2-\frac14 K_0 H^2+\frac14 f_{a,1,0}^2+\frac12 P^2 g_s H^2\ln\r\biggr) \r^2
\\&-\frac14  P^2g_s H^2f_{a,1,0}  \r^3
+\sum_{n=4}^\infty\sum_k f_{c,n,k}\ {\r^n}\ln^k\r\,,
\end{split}
\eqlabel{ksfc}
\end{equation}
\begin{equation}
\begin{split}
&f_a=1+f_{a,1,0}\r+\biggl(-\frac12 P^2g_s H^2-\frac 14 K_0H^2+\frac 14 f_{a,1,0}^2+\frac12 P^2 g_s H^2\ln\r
\biggr) \r^2+ f_{a,3,0}\r^3\\
&+\sum_{n=4}^\infty\sum_k f_{a,n,k}\ {\r^n}\ln^k\r\,,
\end{split}
\eqlabel{ksfa}
\end{equation}
\begin{equation}
\begin{split}
&f_b=1+f_{a,1,0}\r+\biggl(-\frac12 P^2g_s H^2-\frac14 K_0H^2+\frac14 f_{a,1,0}^2+\frac12 P^2 g_s H^2\ln\r\biggr)
 \r^2\\&-\biggl(\frac12 P^2 g_s H^2 f_{a,1,0} +f_{a,3,0}\biggr) \r^3
+\sum_{n=4}^\infty\sum_k f_{b,n,k}\ {\r^n}\ln^k\r\,,
\end{split}
\eqlabel{ksfb}
\end{equation}
\begin{equation}
\begin{split}
&h=\frac18 P^2g_s +\frac14 K_0-\frac12 P^2 g_s \ln\r+\biggl(P^2 g_s \ln\r-\frac12 K_0\biggr) f_{a,1,0} \r
+\biggl(\biggl(
-\frac14  P^2 g_s\\&-\frac54 P^2 g_s \ln\r
+\frac58 K_0\biggr) f_{a,1,0}^2+\frac{119}{576} P^4 g_s^2 H^2+\frac{31}{96} P^2 g_s H^2 K_0+\frac18 H^2 K_0^2
+\frac12 P^4 g_s^2 H^2\ln\r^2
\\&
-\frac{31}{48} P^4 g_s^2 H^2\ln\r-\frac12 P^2 g_s H^2K_0 \ln\r\biggr)\r^2+\biggl(\biggl(\frac54 P^2 g_s \ln\r
+\frac{11}{24}  P^2g_s-\frac58 K_0\biggr) f_{a,1,0}^3
\\&+\biggl(-\frac32 P^4 g_s^2 \ln\r^2+\frac{23}{16} P^4 g_s^2 \ln\r-\frac{19}{64} P^4 g_s^2+\frac32 P^2 g_s K_0\ln\r
-\frac{23}{32} P^2 g_s K_0\\
&-\frac38 K_0^2\biggr) H^2 f_{a,1,0}\biggr)\r^3+\sum_{n=4}^\infty\sum_k h_{n,k}\ {\r^n}\ln^k\r\,,
\end{split}
\eqlabel{ksh}
\end{equation}
\begin{equation}
\begin{split}
&K_1=K_0-2 P^2 g_s \ln\r+P^2g_s f_{a,1,0}  \r+\biggl(-\frac14 P^2 f_{a,1,0}^2 g_s-\frac14 P^4 g_s^2 H^2\ln\r
+\frac{9}{16} P^4 g_s^2H^2\\
&+\frac18 P^2 g_s H^2 K_0\biggr)\r^2
+\biggl(\frac{1}{12} f_{a,1,0}^3 P^2g_s+\frac{1}{48}  P^2g_s \biggl(36 P^2 g_s \ln\r-13 P^2 g_s\\
&-6 K_0
\biggr) H^2 f_{a,1,0}+\frac{2}{3}  P^2g_s \biggl(3 f_{a,3,0} \ln\r+ f_{a,3,0}+ k_{2,3,0}
\biggr)\biggr)\r^3\\
&+\sum_{n=4}^\infty\sum_k k_{1,n,k}\ {\r^n}\ln^k\r\,,
\end{split}
\eqlabel{ksK1}
\end{equation}
\begin{equation}
\begin{split}
&K_2=1+\left(k_{2,3,0}+\frac34 H^2 f_{a,1,0} P^2 g_s \ln\r+3 f_{a,3,0} \ln\r\right) \r^3
+\sum_{n=4}^\infty\sum_k k_{2,n,k}\ {\r^n}\ln^k\r\,,
\end{split}
\eqlabel{ksK2}
\end{equation}
\begin{equation}
\begin{split}
&K_3=K_0-2 P^2 g_s \ln\r+P^2 g_sf_{a,1,0}  \r+\biggl(-\frac14 P^2 g_sf_{a,1,0}^2 -\frac14 P^4 g_s^2 H^2\ln\r
+\frac{9}{16}
 P^4 g_s^2H^2\\&+\frac18 P^2 g_s H^2 K_0\biggr)\r^2
+\biggl(\frac{1}{12} f_{a,1,0}^3  P^2g_s-\frac{1}{48}  P^2 g_s\biggl(12 P^2 g_s \ln\r+29 P^2 g_s\\
&+6 K_0
\biggr) H^2f_{a,1,0}-\frac{2}{3}  P^2 g_s\biggl(3 f_{a,3,0} \ln\r+ f_{a,3,0}
+ k_{2,3,0}\biggr)\biggr)\r^3\\&+\sum_{n=4}^\infty\sum_k k_{3,n,k}\ {\r^n}\ln^k\r\,,
\end{split}
\eqlabel{ksK3}
\end{equation}
\begin{equation}
\begin{split}
&g=g_s\biggl(1-\frac12 P^2 g_sH^2\r^2+\frac12 f_{a,1,0} P^2 g_s H^2\r^3
+\sum_{n=4}^\infty\sum_k g_{n,k}\ {\r^n}\ln^k\r\biggr)\,.
\end{split}
\eqlabel{ksg}
\end{equation}
It is characterized by 11 parameters:
\begin{equation}
\{K_0\,,\ H\,,\ g_s\,,\  f_{a,1,0}\,,\ \underbrace{f_{a,3,0}\,,\ k_{2,3,0}}_{\calo_3^\a}\,,\
\underbrace{g_{4,0}\,,\ f_{c,4,0}}_{\calo_4^\b}\,,\ \underbrace{f_{a,6,0}}_{\calo_6}\,,\
\underbrace{f_{a,7,0}}_{\calo_7}\,,\ \underbrace{f_{a,8,0}}_{\calo_8}\}\,,
\eqlabel{uvparks}
\end{equation}
where we indicated the dual cascading gauge theory operators which expectation values
these parameters characterize. $g_s$ is the asymptotic string coupling, and
$K_0$ is related to strong coupling scale $\Lambda$ of the cascading gauge theory
(see appendix \ref{ksextremal}) as \cite{Buchel:2013dla}
\begin{equation}
\Lambda^2=\frac{1}{P^2g_s}\ e^{-\frac{K_0}{P^2g_s}}\,.
\eqlabel{deflambda}
\end{equation}
Finally, $f_{a,1,0}$ corresponds to a diffeomorphism parameter $(-2\l)$ in symmetry transformation
SFG1, see \eqref{kssym4}.

To understand IR asymptotics of the FG frame solutions it is convenient to consider Euclidean continuation
of the background geometry \eqref{fg1}. For a fixed radial coordinate $\r$
the resulting Euclidean space is topologically $S^4\times S^2\times S^3$, where $S^4$ is an
analytical continuation of $\calm^c$ \eqref{fgn1}, and $S^2\times S^3$ is a  compact part of the warped
deformed conifold\footnote{See \cite{Herzog:2001xk} for a nice review.}. Without loss
of generality we assume that the radial coordinate
\begin{equation}
\r\ \in\ [0,+\infty)\,,
\eqlabel{rangerho}
\end{equation}
so that $y\equiv \frac 1\r$ corresponds to the IR asymptotic. The range \eqref{rangerho} can always be
enforced with an appropriate symmetry transformation SFG1 \eqref{kssym4}. Ten dimensional
Euclidean manifold is geodesically complete if one of the compact factors $S^4$ or $S^2$
smoothly shrinks to zero size as $y\to 0$. Note that $S^3$ can not shrink to zero size without
causing a naked singularity since it supports nonzero (when $P\ne 0$) RR 3-form flux
\eqref{fluxes}. Thus, from purely topological considerations we expect several
inequivalent de Sitter vacua of the cascading gauge theory: TypeA (shrinking $S^4$)
and TypeB (shrinking $S^2$).
\begin{itemize}
\item TypeA de Sitter vacua of the cascading gauge theory.
To identify smooth Euclidean FG frame geometries with vanishing $S^4$ as $y\to 0$
we introduce\footnote{Other holographic models in this class were discussed earlier in
\cite{Buchel:2001iu,Buchel:2002wf,Buchel:2002kj,Buchel:2003qm,Buchel:2004qg}}
\begin{equation}
h^h\equiv y^{-2}\ h\,,\qquad f^{h}_{a,b,c}\equiv y\ f_{a,b,c}\,.
\eqlabel{defhors4}
\end{equation}
The IR asymptotic expansion 
\begin{equation}
\begin{split}
&f_{a,b,c}^h=\sum_{n=0} f_{a,b,c,n}^h y^n\,,\qquad h^h=\frac {1}{4 H^2}+\sum_{n=1} h^h_{n} y^n\,,\\
&K_{1,2,3}=\sum_{n=0} K_{1,2,3,n}^h y^n\,,\qquad g=\sum_{n=0} g_{n}^h y^n\,,
\end{split}
\eqlabel{phase1ir1}
\end{equation}
is characterized by 7 parameters:
\begin{equation}
\{f_{a,0}^h\,,\ f_{b,0}^h\,,\ f_{c,0}^h\,,\ K_{1,0}^h\,,\ K_{2,0}^h\,,\ K_{3,0}^h\,,\ g_{0}^h\}\,.
\eqlabel{irph1par}
\end{equation}
Note that given \eqref{phase1ir1},
\begin{equation}
\begin{split}
&\frac{1}{h^{1/2}\r^{2}} \left(d\calm_4^{c}\right)^2+
\frac{h^{1/2}}{\r^{2}} \left(d\r\right)^2\qquad
\underbrace{\longrightarrow}_{\t\to i \frac{\theta+\pi/2}{H}}\qquad 
\frac{1}{h^{1/2}\r^{2}}\ \frac{1}{H^2} (dS^4)^2+
\frac{h^{1/2}}{\r^{2}} \left(d\r\right)^2\\
&=\frac{y}{(h^h)^{1/2}}\ \frac{1}{H^2} (dS^4)^2+
\frac{(h^h)^{1/2}}{y} \left(dy\right)^2\qquad \underbrace{\longrightarrow}_{y\equiv z^2\to 0}
\qquad \frac{2}{H}\ \biggl(z^2 (dS^4)^2+(dz)^2\biggr) \,,
\end{split}
\eqlabel{smoothph1}
\end{equation}
\ie $S^4$ indeed smoothly shrinks to zero size as $y\to 0$. It is important to emphasize that
TypeA vacua defined by \eqref{phase1ir1}  have either $U(1)$ or $\zet_2$ chiral symmetry --- chiral
symmetry is unbroken in the former ($\as$), and spontaneously broken in the latter ($\ab$). 
\item TypeB de Sitter vacua of the cascading gauge theory.
To identify smooth Euclidean FG frame geometries with vanishing $S^2$ as $y\to 0$
we introduce \cite{Buchel:2013dla}
\begin{equation}
h^h\equiv y^{-4}\ h\,,\qquad f^{h}_{a,b,c}\equiv y^2\ f_{a,b,c}\,.
\eqlabel{defhors2}
\end{equation}
The IR asymptotic expansion 
\begin{equation}
\begin{split}
&f_{a}^h=f_{a,0}^h+\sum_{n=1} f_{a,n}^h y^{2n}\,,\ \ f_b^h=3 y^2+\sum_{n=2} f_{b,n}^h y^{2n}\,,\ \
f_{c}^h=\frac 34 f_{a,0}^h+\sum_{n=1} f_{c,n}^h y^{2n}\,,\\
&K_{1}=k_{1,3}^h y^3+\sum_{n=2} k_{1,n}^h y^{2n+1}\,,\qquad K_{2}=k_{2,2}^h y^2+k_{2,4}^h y^4
+\sum_{n=3} k_{2,n}^h y^{2n}\,,\\
&K_{3}=k_{3,1}^hy+\sum_{n=1} k_{3,n}^h y^{2n+1}\,,\qquad 
h^h=h_0^h+\sum_{n=1} h_{n}^h y^{2n}\,,\qquad g=g_0^h+\sum_{n=1} g_{n}^h y^{2n}\,,
\end{split}
\eqlabel{phase1ir2}
\end{equation}
is characterized by 7 parameters:
\begin{equation}
\{f_{a,0}^h\,,\ h_{0}^h\,,\ k_{1,3}^h\,,\ k_{2,2}^h\,,\ k_{2,4}^h\,,\ k_{3,1}^h\,,\ g_{0}^h\}\,.
\eqlabel{irph2par}
\end{equation}
Note that given \eqref{phase1ir2},
\begin{equation}
\begin{split}
&\frac{h^{1/2}}{\r^{2}} \left(d\r\right)^2+\frac{f_bh^{1/2}}{6}(g_1^2+g_2^2)
={(h^h)^{1/2}} \left(dy\right)^2+{(h^h)^{1/2}} y^2\ \frac12(g_1^2+g_2^2)\bigg|_{\rm 2-cycle} \\
&\underbrace{\longrightarrow}_{y\to 0}
\qquad \left(h_0^h\right)^{1/2}\ \biggl(y^2 (dS^2)^2+(dy)^2\biggr) \,,
\end{split}
\eqlabel{smoothph2}
\end{equation}
where $\bigg|_{S^2}$ means restriction to
a 2-cycle. Following \cite{Herzog:2001xk}, this means setting $\psi=0$,
$\phi_2=-\phi_1$, $\theta_2=-\theta_1$ in one-forms $\{g_i\}$ on $T^{1,1}$:
\begin{equation}
\left(g_1^2+g_2^2\right)\bigg|_{\rm 2-cycle}=2 \left(\left(d\theta_1\right)^2+\sin^2\theta_1 \left(d\phi_1\right)^2\right)
=2 \left(dS^2\right)^2\,.
\eqlabel{2cycle}
\end{equation}
On the other hand, the 3-cycle supporting RR flux remains finite, provided $f_{a,0}^h h_0^h\ne 0$:
\begin{equation}
\begin{split}
&\frac{f_ch^{1/2}}{9}g_5^2+\frac{f_ah^{1/2}}{6}(g_3^2+g_4^2)
=\frac{f_c^h(h^h)^{1/2}}{9}g_5^2+\frac{f_a^h(h^h)^{1/2}}{6}(g_3^2+g_4^2)\\
&\underbrace{\longrightarrow}_{y\to 0}\qquad \frac{f_{a,0}^h(h_0^h)^{1/2}}{6}\
\left(\frac 12 g_5^2+g_3^2+g_4^2\right)\biggl|_{{\rm3-cycle}:\ \theta_2=\phi_2=0,\theta_1=2\eta, \psi=\xi_1+\xi_2,\phi_1=\xi_1-\xi_2}\\
&=\frac{f_{a,0}^h(h_0^h)^{1/2}}{6}\ 2\left((d\eta)^2+\cos^2\eta (d\xi_1)^2+\sin^2\eta (d\xi_2)^2\right)
=\frac{f_{a,0}^h(h_0^h)^{1/2}}{3}\ \left(dS^3\right)^2\,.
\end{split}
\eqlabel{3cycle}
\end{equation}
From \eqref{smoothph2}, $S^2$ indeed smoothly shrinks to zero size as $y\to 0$. Because
$f_a\ne f_b$ as $y\to 0$,  
TypeB vacua defined by \eqref{phase1ir2}  have $\zet_2$ chiral symmetry --- chiral
symmetry is spontaneously broken. 
\end{itemize}

\subsubsection{$\as$ vacua asymptotics}\label{apb11}
We provide here connection with the extensive earlier studies   of
$\as$ vacua in  \cite{Buchel:2013dla}.

Chirally symmetric de Sitter vacua of the cascading gauge theory ($\as$) correspond to a consistent truncation
\begin{equation}
f_c\equiv f_2\,,\qquad f_a=f_b\equiv f_3\,,\qquad K_1=K_3\equiv K\,,\qquad  K_2=1\,.
\eqlabel{ktks1}
\end{equation}
We find:
\nxt in the UV, \ie as $\r\to 0$,
\begin{equation}
\begin{split}
&f_2=1+f_{2,1,0}\ \r+ \biggl(-\frac38 H^2 P^2 g_s-\frac14 H^2 K_0+\frac14 f_{2,1,0}^2+\frac12 H^2 P^2 g_s \ln\r\biggr)\ \r^2
\\&-\frac14 H^2 P^2g_s f_{2,1,0}\ \r^3+\sum_{n=4}^\infty\sum_k f_{2,n,k}\ \r^n\ln^k\r\,,
\end{split}
\eqlabel{ktks2}
\end{equation}
\begin{equation}
\begin{split}
&f_3=1+f_{2,1,0}\ \r+ \biggl(-\frac12 H^2 P^2 g_s-\frac14 H^2 K_0+\frac14 f_{2,1,0}^2+\frac12 H^2 P^2 g_s \ln\r\biggr)\ \r^2
\\&-\frac14  H^2 P^2 g_s f_{2,1,0}\ \r^3 +\sum_{n=4}^\infty\sum_k f_{3,n,k}\ \r^n\ln^k\r\,,
\end{split}
\eqlabel{ktks3}
\end{equation}
\begin{equation}
\begin{split}
&h=\frac18 P^2 g_s+\frac14 K_0-\frac12 P^2 g_s \ln\r-\frac12 f_{2,1,0} (-2 P^2 g_s \ln\r+K_0)\ \r
+ \biggl(
\frac{119}{576} H^2 P^4 g_s^2\\
&+\frac{31}{96} H^2 K_0 P^2 g_s-\frac14 P^2 g_sf_{2,1,0}^2 +\frac18 H^2 K_0^2
+\frac58 f_{2,1,0}^2 K_0-\frac{1}{96} P^2 g_s (62 H^2 P^2 g_s\\
&+48 H^2 K_0+120 f_{2,1,0}^2) \ln\r
+\frac12 H^2 P^4 g_s^2 \ln^2\r\biggr)\ \r^2
-\frac{1}{192} f_{2,1,0} \biggl(288 H^2 P^4 g_s^2 \ln^2\r\\
&+\biggl(-276 H^2 P^4 g_s^2 
-288 H^2 K_0  P^2 g_s-240  P^2 g_sf_{2,1,0}^2\biggr)\ln\r+57 H^2 P^4 g_s^2\\&+138 H^2 K_0 P^2 g_s
-88 P^2 f_{2,1,0}^2 g_s+72 H^2 K_0^2+120 f_{2,1,0}^2 K_0\biggr)\ \r^3+\sum_{n=4}^\infty\sum_k h_{n,k}\ \r^n\ln^k\r\,,
\end{split}
\eqlabel{ktks4}
\end{equation}
\begin{equation}
\begin{split}
&K=K_0-2 P^2 g_s \ln\r+ P^2 g_sf_{2,1,0}\ \r+\frac{1}{16} P^2 g_s (-4 H^2 P^2 g_s \ln\r+9 H^2 P^2 g_s+2 H^2 K_0\\
&-4 f_{2,1,0}^2)\ \r^2
-\frac{1}{48} P^2 g_s f_{2,1,0} (-12 H^2 P^2 g_s \ln\r+21 H^2 P^2 g_s+6 H^2 K_0-4 f_{2,1,0}^2)\ \r^3
\\&+\sum_{n=4}^\infty\sum_k k_{n,k}\ \r^n\ln^k\r\,,
\end{split}
\eqlabel{ktks5}
\end{equation}
\begin{equation}
\begin{split}
&g=g_s\biggl(1-\frac12  H^2 P^2g_s\ \r^2+\frac12 H^2 P^2 g_sf_{2,1,0}\ \r^3+\sum_{n=4}^\infty\sum_k g_{n,k}\r^n\ln^k\r\biggr)\,,
\end{split}
\eqlabel{ktks6}
\end{equation}
characterized by 8 parameters:
\begin{equation}
\{K_0\,,\ H\,,\ g_s\,,\  f_{2,1,0}\,,\ g_{4,0}\,,\ f_{2,4,0}\,,\ f_{2,6,0}\,,\
f_{2,8,0}\}\,;
\eqlabel{uvparkt}
\end{equation} 
\nxt in the IR, \ie as $y\equiv \frac 1\r\to 0$,
\begin{equation}
\begin{split}
&f_{2,3}^h=\sum_{n=0} f_{2,3,n}^h y^n\,,\qquad h^h=\frac {1}{4 H^2}+\sum_{n=1} h^h_{n} y^n\,,\\
&K=\sum_{n=0} K_n^h y^n\,,\qquad g=\sum_{n=0} g_{n}^h y^n\,,
\end{split}
\eqlabel{phase2ir1}
\end{equation}
characterized by 4 parameters:
\begin{equation}
\{f_{2,0}^h\,,\ f_{3,0}^h\,,\ K_0^h\,,\ g_{0}^h\}\,.
\eqlabel{irph2par2}
\end{equation}

Comparing \eqref{ksfc}-\eqref{ksg} with \eqref{ktks2}-\eqref{ktks6} to $\calo(\r^8)$ we identify
\begin{equation}
\begin{split}
&f_{a,1,0}=f_{2,1,0}\,,\qquad f_{a,3,0}=-\frac14 H^2P^2g_sf_{2,1,0}\,,\qquad k_{2,3,0}=0\,,\qquad f_{c,4,0}=f_{2,4,0}\,,\\
\end{split}
\eqlabel{map1}
\end{equation}
\begin{equation}
\begin{split}
&f_{a,6,0}=\biggl(-\frac{96087}{4096000} K_0 P^4 g_s^2-\frac{3409}{409600} K_0^2 P^2 g_s-\frac{11056513}{245760000}P^6 g_s^3\biggr) H^6
+\biggl(\\&\frac{1171}{20480} P^4 g_s^2f_{2,1,0}^2 -\frac{13}{10240} K_0 P^2 g_sf_{2,1,0}^2 \biggr) H^4
+\biggl(-\frac{1}{512} P^2 g_sf_{2,1,0}^4 -\frac{307}{1280}  P^2g_s f_{2,4,0}
\\&+\frac{31}{320} P^2 g_s g_{4,0}-\frac{87}{640} K_0 f_{2,4,0}\biggr) H^2-\frac14 f_{2,6,0}+\frac{3}{16} f_{2,1,0}^2 f_{2,4,0}\,,\\
\end{split}
\eqlabel{map2}
\end{equation}
\begin{equation}
\begin{split}
&f_{a,7,0}=\biggl(
\frac{13331}{196608} P^6 g_s^3 f_{2,1,0}+\frac{753}{16384} P^4 g_s^2f_{2,1,0}  K_0+\frac{547}{24576} K_0^2 f_{2,1,0}  P^2g_s\biggr) H^6
+\biggl(\\&-\frac{2077}{18432} P^4 g_s^2 f_{2,1,0}^3-\frac{77}{3072} K_0 P^2 g_sf_{2,1,0}^3 \biggr) H^4
+\biggl(\frac{21}{1280} P^2 g_s f_{2,1,0}^5+\frac{19}{64} K_0 f_{2,1,0} f_{2,4,0}\\
&+\frac{61}{128} f_{2,1,0} P^2 g_s f_{2,4,0}
-\frac{7}{32} f_{2,1,0} P^2 g_s g_{4,0}\biggr) H^2-\frac{3}{8} f_{2,1,0}^3 f_{2,4,0}+\frac12 f_{2,1,0} f_{2,6,0}\,,\\
\end{split}
\eqlabel{map3}
\end{equation}
\begin{equation}
\begin{split}
&f_{a,8,0}=\frac{1}{70K_0-141 P^2 g_s}
\biggl[
\biggl(-\frac{40244584228943}{5689958400000} K_0 P^8g_s^4-\frac{12213914790101}{3034644480000} K_0^2 P^6 g_s^3
\\&-\frac{931679}{4915200} K_0^4 P^2 g_s-\frac{9161577517}{7225344000} K_0^3 P^4 g_s^2+\frac{25292565670124671}{19118260224000000}
P^{10} g_s^5\biggr) H^8
+\biggl(\\&-\frac{173957}{81920} P^2 g_s K_0^3 f_{2,1,0}^2-\frac{5131309293043}{303464448000} P^8 g_s^4f_{2,1,0}^2
-\frac{12991428547}{1032192000} K_0 P^6 g_s^3f_{2,1,0}^2 \\
&+\frac{504197}{1433600} P^4 g_s^2f_{2,1,0}^2  K_0^2\biggr) H^6
+\biggl(
\frac{1892623}{92160} P^4 g_s^2 K_0 f_{2,1,0}^4+\frac{63}{8} P^2 g_s K_0^2 f_{2,1,0}^4
\\&+\frac{2093}{768} P^2 g_s K_0^2 g_{4,0}-\frac{11179}{2560} K_0^3 f_{2,4,0}+\frac{176710639657}{4741632000} P^6 g_s^3f_{2,4,0}
-\frac{2470057}{1290240} P^6 g_s^3f_{2,1,0}^4 \\&-\frac{259362731}{33868800} P^6 g_s^3 g_{4,0}
-\frac{132413627}{16128000} K_0 P^4g_s^2 f_{2,4,0} -\frac{6266917}{537600} K_0^2 P^2 g_sf_{2,4,0}
\\&+\frac{698651}{80640} P^4 g_s^2 K_0 g_{4,0}\biggr) H^4
+\biggl(-\frac{15365}{3072} P^4 g_s^2 f_{2,1,0}^6-\frac{69139}{960} P^4 g_s^2 f_{2,6,0}\\&
-\frac{2751}{128} K_0^2 f_{2,1,0}^2 f_{2,4,0}
-\frac{3675}{512} P^2 g_s K_0 f_{2,1,0}^6-\frac{1215}{128} P^4 g_s^2 f_{2,1,0}^2 g_{4,0}-\frac{1699}{16} P^2 g_s K_0 f_{2,6,0}
\\&-\frac{14827}{320} P^2 g_s K_0 f_{2,1,0}^2 f_{2,4,0}+\frac{2177}{64} P^2 g_s K_0 f_{2,1,0}^2 g_{4,0}-\frac{385}{16} K_0^2 f_{2,6,0}
\\&-\frac{1540367}{17920} P^4g_s^2 f_{2,1,0}^2  f_{2,4,0}\biggr) H^2
+\frac{3085}{32} P^2 g_s f_{2,1,0}^4 f_{2,4,0}-\frac{1375}{8} P^2 g_s f_{2,1,0}^2 f_{2,6,0}\\
&+21 P^2 g_s f_{2,4,0} g_{4,0}
-\frac{2527}{10} K_0 f_{2,4,0}^2+70 K_0 f_{2,8,0}-\frac{63}{2P^2g_s} K_0^2 f_{2,4,0}^2+\frac{2275}{16} K_0 f_{2,1,0}^4 f_{2,4,0}
\\&-\frac{875}{4} K_0 f_{2,1,0}^2 f_{2,6,0}+42 K_0 f_{2,4,0} g_{4,0}+14 P^2 g_s g_{4,0}^2+104 P^2 g_s f_{2,8,0}
-\frac{45539}{280} P^2 g_sf_{2,4,0}^2 
\biggr]\,.
\end{split}
\eqlabel{map4}
\end{equation}
Comparing \eqref{phase1ir1} with \eqref{phase2ir1}  we identify
\begin{equation}
f_{c,0}^h=f_{2,0}^h\,,\qquad f_{a,0}^h=f_{b,0}^h=f_{3,0}^h\,,\qquad K_{1,0}^h=K_{3,0}^h=K_0^h\,,\qquad K_{2,0}^h=1\,.
\eqlabel{map5}
\end{equation}

\subsection{From FG to EF frame}\label{apb2}

A general map between the FG and EF frame de Sitter vacua of the holographic duals
was worked out in \cite{Buchel:2017pto}. Specifically, given
\begin{equation}
\begin{split}
&ds^2\bigg|_{EF}=2dt\ \left(dr- a(r)\ dt\right)+\s(r)^2e^{2Ht}\ d\boldsymbol{x}^2+\cdots\,,\\
&ds^2\bigg|_{FG}=c_1(\r)^2\ \left(-d\t^2+e^{2H\t}\ d\boldsymbol{x}^2\right)+c_2(\r)^2\ (d\r)^2+\cdots\,,\\
\end{split}
\eqlabel{fgtoef}
\end{equation}
where $\cdots$ are metric components along the compact directions,
\begin{equation}
\begin{split}
&r=-\int^\r ds\ c_1(s)c_2(s)+{\rm const}\,,\qquad t=\t-\int_0^\r ds\ \frac{c_2(s)}{c_1(s)}\,,\\
&a(r)=\frac{1}{2}c_1(\r)^2\,,\qquad \s(r)=c_1(\r)\ \exp\biggl[H\int_0^\r ds\ \frac{c_2(s)}{c_1(s)}\biggr] \,.
\end{split}
\eqlabel{mapfin}
\end{equation}

Using \eqref{fg1}, we find from \eqref{mapfin}:
\begin{equation}
\begin{split}
r=\frac 1\r+{\rm const}=y+{\rm const}\,,\qquad a=\frac{1}{2h^{1/2}\r^2}\,,\qquad t=\t-\int_0^\r ds\ h(s)^{1/2}\,.
\end{split}
\eqlabel{resmap}
\end{equation}
Note that asymptotically in UV, \ie as   $\r\to 0$, the EF and the FG times coincide:
\begin{equation}
t-\t\qquad \sim\qquad -\int^\r_0 ds\ \left(-\frac 12 P^2g_s\ln s\right)^{1/2}\qquad \longrightarrow\qquad  0\,.
\end{equation}

Without loss of generality we fix ${\rm const}$ in \eqref{resmap}
so that $r=0\Longleftrightarrow \frac 1\r\equiv y=0$. 
Introducing
\begin{equation}
z\equiv -r\,,
\eqlabel{defzef}
\end{equation}
we find from \eqref{defhors4}-\eqref{phase1ir1}, \eqref{defhors2}-\eqref{phase1ir2}, and \eqref{phase1ir2}
the following asymptotic expansions for the EF frame vacua:
\nxt TypeA$_s$ vacua:
\begin{equation}
\begin{split}
&a=-H z+
\frac{H((f_{2,0}^h)^2 (f_{3,0}^h)^2-6 f_{2,0}^h (f_{3,0}^h)^3+3 H^2P^2 (f_{3,0}^h)^2 g_0^h+
10 H^4 (K_0^h)^2)}{5(f_{3,0}^h)^4 f_{2,0}^h} z^2+\calo(z^3)\,,\\
&\s=s_0^h \left(1-\frac{
(f_{2,0}^h)^2 (f_{3,0}^h)^2-6 f_{2,0}^h (f_{3,0}^h)^3+3 H^2 P^2(f_{3,0}^h)^2 g_0^h+10 H^4 (K_0^h)^2}
{5(f_{3,0}^h)^4 f_{2,0}^h}z+\calo(z^2)
\right)\,,\\
&w_{c2}\equiv f_{2}^h \left(h^h\right)^{1/2}=\frac{f_{2,0}^h}{2H}
-\frac{2}{5H} \frac{4 (f_{2,0}^h)^2 (f_{3,0}^h)^2-3 H^2 P^2 (f_{3,0}^h)^2 g_0^h-4 H^4
(K_0^h)^2}{(f_{3,0}^h)^4}z+\calo(z^2)\,,\\
&w_{a2}=w_{b2}\equiv f_{3}^h \left(h^h\right)^{1/2}=\frac{f_{3,0}^h}{2H}\\
&\qquad\qquad +\frac{2}{5H}
\frac{2 (f_{2,0}^h)^2 (f_{3,0}^h)^2-6 f_{2,0}^h (f_{3,0}^h)^3+H^2 P^2 (f_{3,0}^h)^2 g_0^h+4H^4 (K_0^h)^2}{(f_{3,0}^h)^3 f_{2,0}^h}z+\calo(z^2)\,,\\
&K_1=K_3\equiv K=K_0^h-\frac{16}{5} \frac{H^2 P^2 K_0^h g_0^h}{(f_{3,0}^h)^2 f_{2,0}^h}z+\calo(z^2)\,,\qquad K_2=1\,,\\
&g=g_0^h-\frac85 \frac{H^2P^2 (g_0^h)^2}{(f_{3,0}^h)^2 f_{2,0}^h}z+\calo(z^2)\,;
\end{split}
\eqlabel{typeasi}
\end{equation}
\nxt TypeA$_b$ vacua:
\begin{equation}
\begin{split}
&a=-H z+\frac{H}{80 (f_{a,0}^h)^2 (f_{b,0}^h)^2 g_0^h f_{c,0}^h P^2}
\biggl(
24 H^2 (g_0^h)^2 ((K_{2,0}^h)^2 (f_{a,0}^h)^2+(K_{2,0}^h)^2 (f_{b,0}^h)^2\\
&-4 K_{2,0}^h (f_{b,0}^h)^2
+4 (f_{b,0}^h)^2) P^4+g_0^h (40 H^4 (K_{1,0}^h)^2 (K_{2,0}^h)^2-80 H^4 K_{1,0}^h (K_{2,0}^h)^2 K_{3,0}^h\\
&+40 H^4 (K_{2,0}^h)^2
(K_{3,0}^h)^2-160 H^4 (K_{1,0}^h)^2 K_{2,0}^h+160 H^4 K_{1,0}^h K_{2,0}^h K_{3,0}^h\\
&+160 H^4 (K_{1,0}^h)^2
+9 (f_{a,0}^h)^3 f_{b,0}^h-18 (f_{a,0}^h)^2 (f_{b,0}^h)^2-48 (f_{a,0}^h)^2 f_{b,0}^h f_{c,0}^h
+9 f_{a,0}^h (f_{b,0}^h)^3\\
&-48 f_{a,0}^h (f_{b,0}^h)^2 f_{c,0}^h+16 f_{a,0}^h f_{b,0}^h (f_{c,0}^h)^2) P^2
+27 H^2 f_{a,0}^h f_{b,0}^h (K_{1,0}^h-K_{3,0}^h)^2
\biggr) z^2\\&+\calo(z^3)\,,
\end{split}
\eqlabel{typebi1}
\end{equation}
\begin{equation}
\begin{split}
&\sigma=s^h_0\biggl(1-\frac{1}{80 (f_{a,0}^h)^2 (f_{b,0}^h)^2 g_0^h f_{c,0}^h P^2)} \biggl(
24 H^2 (g_0^h)^2 ((K_{2,0}^h)^2 (f_{a,0}^h)^2+(K_{2,0}^h)^2 (f_{b,0}^h)^2\\
&-4 K_{2,0}^h (f_{b,0}^h)^2
+4 (f_{b,0}^h)^2) P^4+g_0^h (40 H^4 (K_{1,0}^h)^2 (K_{2,0}^h)^2-80 H^4 K_{1,0}^h (K_{2,0}^h)^2 K_{3,0}^h
\\&+40 H^4 (K_{2,0}^h)^2 (K_{3,0}^h)^2-160 H^4 (K_{1,0}^h)^2 K_{2,0}^h+160 H^4 K_{1,0}^h K_{2,0}^h K_{3,0}^h
\\&+160 H^4 (K_{1,0}^h)^2+9 (f_{a,0}^h)^3 f_{b,0}^h-18 (f_{a,0}^h)^2 (f_{b,0}^h)^2
-48 (f_{a,0}^h)^2 f_{b,0}^h f_{c,0}^h+9 f_{a,0}^h (f_{b,0}^h)^3\\
&-48 f_{a,0}^h (f_{b,0}^h)^2 f_{c,0}^h
+16 f_{a,0}^h f_{b,0}^h (f_{c,0}^h)^2) P^2+27 H^2 f_{a,0}^h f_{b,0}^h (K_{1,0}^h-K_{3,0}^h)^2
\biggr) z\\&+\calo(z^2)\biggr)\,,
\end{split}
\eqlabel{typebi2}
\end{equation}
\begin{equation}
\begin{split}
&w_{c2}\equiv f_c^h (h^h)^{1/2}=\frac{f_{c,0}^h}{2H}+\frac{1}{H (f_{a,0}^h)^2 (f_{b,0}^h)^2 g_0^h P^2}
\biggl(
\frac35 H^2 (g_0^h)^2 ((K_{2,0}^h)^2 (f_{a,0}^h)^2\\
&+(K_{2,0}^h)^2 (f_{b,0}^h)^2-4 K_{2,0}^h (f_{b,0}^h)^2+4 (f_{b,0}^h)^2)
P^4+\frac{1}{10} g_0^h (4 H^4 (K_{1,0}^h)^2 (K_{2,0}^h)^2\\
&-8 H^4 K_{1,0}^h (K_{2,0}^h)^2 K_{3,0}^h+4 H^4 (K_{2,0}^h)^2
(K_{3,0}^h)^2-16 H^4 (K_{1,0}^h)^2 K_{2,0}^h\\
&+16 H^4 K_{1,0}^h K_{2,0}^h K_{3,0}^h+16 H^4 (K_{1,0}^h)^2+9 (f_{a,0}^h)^3
f_{b,0}^h-18 (f_{a,0}^h)^2 (f_{b,0}^h)^2\\&+9 f_{a,0}^h (f_{b,0}^h)^3-16 f_{a,0}^h f_{b,0}^h (f_{c,0}^h)^2) P^2
+\frac{27}{40} H^2 f_{a,0}^h f_{b,0}^h (K_{1,0}^h-K_{3,0}^h)^2\biggr)
z+\calo(z^2)\,,
\end{split}
\eqlabel{typebi3}
\end{equation}
\begin{equation}
\begin{split}
&w_{a2}\equiv f_a^h (h^h)^{1/2}=\frac{f_{a,0}^h}{2H}+\frac{1}{f_{a,0}^h H (f_{b,0}^h)^2 g_0^h f_{c,0}^h P^2} \biggl(
-\frac15 H^2 (g_0^h)^2 ((K_{2,0}^h)^2 (f_{a,0}^h)^2\\
&-3 (K_{2,0}^h)^2 (f_{b,0}^h)^2+12 K_{2,0}^h (f_{b,0}^h)^2
-12 (f_{b,0}^h)^2) P^4+\frac{1}{20} g_0^h (8 H^4 (K_{1,0}^h)^2 (K_{2,0}^h)^2\\
&-16 H^4 K_{1,0}^h (K_{2,0}^h)^2
K_{3,0}^h+8 H^4 (K_{2,0}^h)^2 (K_{3,0}^h)^2-32 H^4 (K_{1,0}^h)^2 K_{2,0}^h+32 H^4 K_{1,0}^h K_{2,0}^h K_{3,0}^h
\\&+32 H^4 (K_{1,0}^h)^2-9 (f_{a,0}^h)^3 f_{b,0}^h+9 f_{a,0}^h (f_{b,0}^h)^3-48 f_{a,0}^h (f_{b,0}^h)^2 f_{c,0}^h
+16 f_{a,0}^h f_{b,0}^h (f_{c,0}^h)^2) P^2\\
&+\frac{9}{40} H^2 f_{a,0}^h f_{b,0}^h (K_{1,0}^h-K_{3,0}^h)^2\biggr) z+\calo(z^2)\,,
\end{split}
\eqlabel{typebi4}
\end{equation}
\begin{equation}
\begin{split}
&w_{b2}\equiv f_b^h (h^h)^{1/2}=\frac{f_{b,0}^h}{2H}+\frac{1}{f_{b,0}^h H (f_{a,0}^h)^2 g_0^h f_{c,0}^h P^2} \biggl(
\frac15 H^2 (g_0^h)^2 (3 (K_{2,0}^h)^2 (f_{a,0}^h)^2\\
&-(K_{2,0}^h)^2 (f_{b,0}^h)^2+4 K_{2,0}^h (f_{b,0}^h)^2-4 (f_{b,0}^h)^2)
P^4+\frac{1}{20} g_0^h (8 H^4 (K_{1,0}^h)^2 (K_{2,0}^h)^2\\&
-16 H^4 K_{1,0}^h (K_{2,0}^h)^2 K_{3,0}^h+8 H^4 (K_{2,0}^h)^2
(K_{3,0}^h)^2-32 H^4 (K_{1,0}^h)^2 K_{2,0}^h+32 H^4 K_{1,0}^h K_{2,0}^h K_{3,0}^h\\
&+32 H^4 (K_{1,0}^h)^2+9 (f_{a,0}^h)^3
f_{b,0}^h-48 (f_{a,0}^h)^2 f_{b,0}^h f_{c,0}^h-9 f_{a,0}^h (f_{b,0}^h)^3+16 f_{a,0}^h f_{b,0}^h (f_{c,0}^h)^2) P^2
\\&+\frac{9}{40} H^2 f_{a,0}^h f_{b,0}^h (K_{1,0}^h-K_{3,0}^h)^2\biggr) z+\calo(z^2)\,,
\end{split}
\eqlabel{typebi5}
\end{equation}
\begin{equation}
\begin{split}
&K_1=K_{1,0}^h-\frac{1}{5 (f_{a,0}^h)^2 f_{c,0}^h} \biggl(
8 H^2 g_0^h (K_{2,0}^h-2) (K_{1,0}^h K_{2,0}^h-K_{2,0}^h K_{3,0}^h-2 K_{1,0}^h) P^2\\
&+9 f_{a,0}^h f_{b,0}^h (K_{1,0}^h-K_{3,0}^h)
\biggr) z+\calo(z^2)\,,
\end{split}
\eqlabel{typebi6}
\end{equation}
\begin{equation}
\begin{split}
&K_2=K_{2,0}^h-\frac{9}{5 f_{a,0}^h f_{b,0}^h g_0^h f_{c,0}^h P^2} \biggl(
g_0^h (K_{2,0}^h (f_{a,0}^h)^2+K_{2,0}^h (f_{b,0}^h)^2-2 (f_{b,0}^h)^2) P^2\\
&+H^2 (K_{1,0}^h-K_{3,0}^h)
(K_{1,0}^h K_{2,0}^h-K_{2,0}^h K_{3,0}^h-2 K_{1,0}^h)\biggr) z+\calo(z^2)\,,
\end{split}
\eqlabel{typebi7}
\end{equation}
\begin{equation}
\begin{split}
&K_3=K_{3,0}^h+\frac{1}{5 (f_{b,0}^h)^2 f_{c,0}^h} \biggl(
8 H^2 K_{2,0}^h g_0^h (K_{1,0}^h K_{2,0}^h-K_{2,0}^h K_{3,0}^h-2 K_{1,0}^h) P^2\\&+9 f_{a,0}^h f_{b,0}^h (K_{1,0}^h-K_{3,0}^h)
\biggr) z+\calo(z^2)\,,
\end{split}
\eqlabel{typebi8}
\end{equation}
\begin{equation}
\begin{split}
&g=g_0^h-\frac{H^2}{10 (f_{a,0}^h)^2 (f_{b,0}^h)^2 f_{c,0}^h P^2} \biggl(
8 (g_0^h)^2 ((K_{2,0}^h)^2 (f_{a,0}^h)^2+(K_{2,0}^h)^2 (f_{b,0}^h)^2-4 K_{2,0}^h (f_{b,0}^h)^2\\
&+4 (f_{b,0}^h)^2)
P^4-9 f_{a,0}^h (K_{1,0}^h-K_{3,0}^h)^2 f_{b,0}^h) z+\calo(z^2)\,;
\end{split}
\eqlabel{typeabi9}
\end{equation}
\nxt TypeB vacua:
\begin{equation}
\begin{split}
&a=\frac{1}{2(h^h_0)^{1/2}}+\calo(z^2)\,,\qquad \sigma=s_0^h\biggl(1+(h^h_0)^{1/2} H z+\calo(z^2)\biggr)\,,\\
&w_{c2}\equiv f_c^h (h^h)^{1/2}=\frac34 f_{a,0}^h (h^h_0)^{1/2}+\calo(z^2)\,,\qquad
w_{a2}\equiv f_a^h (h^h)^{1/2}= f_{a,0}^h (h^h_0)^{1/2}+\calo(z^2)\,,\\
&w_{b2}\equiv f_b^h (h^h)^{1/2}=3 (h^h_0)^{1/2} z^2+\calo(z^4)\,,\qquad K_1=-k_{1,3}^h z^3+\calo(z^5)\,,\\
&K_2=k_{2,2}^h z^2+\calo(z^4)\,,\qquad K_3= -k_{3,1}^h z+\calo(z^3)\,,\qquad g=g_0^h+\calo(z^2)\,,
\end{split}
\eqlabel{typebi}
\end{equation}
where
\begin{equation}
s_0^h=\sigma\bigg|_{y=0}^{{\rm FG\ frame}}\,.
\eqlabel{defsh0}
\end{equation}

\subsection{Extremal KS solution limit $H\to 0$}\label{ksextremal}
We review here extremal KS solution \cite{Klebanov:2000hb} following \cite{Buchel:2013dla}
and identify the relation of the strong coupling scale $\Lambda$ \eqref{deflambda} to the
conifold deformation parameter $\e$ \eqref{ksks}.

We use the radial coordinate ${\hat r}\in [0,\infty)$  to describe KS solution:
\begin{equation}
\begin{split}
&ds_5^2=H_{KS}^{-1/2}\ \left(-dt^2+d\boldsymbol{x}^2\right)+H_{KS}^{1/2}\ \w_{1,KS}^2\ d{\hat r}^2\,,\\
&\om_i=\w_{i,KS}\ H^{1/4}_{KS}\,,\qquad K_i=K_{i,KS}\,,
\end{split}
\eqlabel{ks1}
\end{equation}
\begin{equation}
\begin{split}
&K_{1,KS}=\frac23P^2 g_s\ \frac{\cosh {\hat r}-1}{\sinh {\hat r}}
\left(\frac{{\hat r}\cosh {\hat r}}{\sinh {\hat r}}-1\right)\,,\qquad 
K_{2,KS}=1-\frac {{\hat r}}{\sinh {\hat r}}\,,\\
&K_{3,KS}=\frac 23 P^2 g_s\ \frac{\cosh {\hat r}+1}{\sinh {\hat r}}
\left(\frac{{\hat r}\cosh {\hat r}}{\sinh {\hat r}}-1\right)\,,
\qquad g=g_s\,,\\
&\w_{1,KS}=\frac{\epsilon^{2/3}}{\sqrt{6}{\hat K_{KS}}}\,,\qquad 
\w_{2,KS}=\frac{\epsilon^{2/3}{\hat K_{KS}}^{1/2}}{\sqrt{2}}\cosh\frac {\hat r}2\,,\qquad  \w_{3,KS}=\frac{\epsilon^{2/3}
{\hat K_{KS}}^{1/2}}{\sqrt{2}}
\sinh\frac {\hat r}2\,,
\end{split}
\eqlabel{ksks}
\end{equation}
with 
\begin{equation}
{\hat K_{KS}}=\frac{(\sinh (2{\hat r})-2{\hat r})^{1/3} }{2^{1/3}\sinh {\hat r}}\,,\ H'_{KS}=\frac{2}{27}\ \frac{(K_{1,KS}-K_{3,KS})K_{2,KS}-2 K_{1,KS}}
{\epsilon^{8/3}{\hat K_{KS}}^2\sinh^2 {\hat r} }\,,
\eqlabel{kk}
\end{equation}
where now ${\hat r}\to \infty$ is the boundary and ${\hat r}\to 0$ is the IR.

Comparing the metric ansatz in \eqref{ks1} and \eqref{fg1} we identify 
\begin{equation}
\frac{(d\r)^2}{\r^4}=(w_{1,KS}({\hat r}))^2 (d{\hat r})^2\,.
\eqlabel{rrho}
\end{equation}
Introducing 
\begin{equation}
z\equiv e^{-{\hat r}/3}\,,
\eqlabel{defz}
\end{equation}
we find from \eqref{rrho}
\begin{equation}
\frac1\r=\frac {\sqrt{6}\ (2\e)^{2/3}}{4}\ \int_1^z\ du\  \frac{u^6-1}{u^2(1-u^{12}+12u^6 \ln u)^{1/3}}\,.
\eqlabel{solverho}
\end{equation}
In the UV, ${\hat r}\to \infty$, $z\to 0$ and $\r\to 0$ we have
\begin{equation}
\begin{split}
&e^{-{\hat r}/3}\equiv z=
\frac{\sqrt{6}\ (2\e)^{2/3}}{4} \r \biggl(1+\calq \r+\calq^2 \r^2+\calq^3 \r^3+\calq^4 \r^4+\calq^5 \r^5
+\biggl(\frac{27}{80} \e^4 \ln 3+\calq^6\\
&+\frac{27}{800} \e^4-\frac{9}{16} \e^4 \ln 2+\frac{9}{20}
 \e^4 \ln\e+\frac{27}{40} \e^4 \ln\r\biggr) \r^6+\biggl(
-\frac{63}{16} \e^4 \calq \ln 2+\frac{189}{80} \e^4 \calq \ln 3+\calq^7\\
&+\frac{729}{800} \calq \e^4+\frac{63}{20} \e^4 \calq \ln\e
+\frac{189}{40} \calq \e^4 \ln\r\biggr) \r^7+\biggl(\frac{2403}{400} \e^4 \calq^2
-\frac{63}{4} \e^4 \calq^2 \ln 2+\frac{189}{20} \e^4 \calq^2 \ln 3\\
&+\frac{63}{5} \e^4 \calq^2 \ln\e+\calq^8
+\frac{189}{10} \e^4 \calq^2 \ln\r\biggr) \r^8+\biggl(\frac{189}{5} \e^4 \calq^3 \ln\e+\frac{9729}{400}
 \e^4 \calq^3-\frac{189}{4} \e^4 \calq^3 \ln 2\\
&+\frac{567}{20} \e^4 \calq^3 \ln 3+\calq^9+\frac{567}{10} \e^4 \calq^3 
\ln\r\biggr) 
\r^9+\calo(\r^{10}\ln\r)\biggr)\,,
\end{split}
\eqlabel{rrhouv}
\end{equation} 
where 
\begin{equation}
\begin{split}
\calq=&\frac{\sqrt{6}\ (2\e)^{2/3}}{4}\ \biggl\{
\int_0^1\ du\  \biggl(\frac{1-u^6}{u^2(1-u^{12}+12u^6 \ln u)^{1/3}}-\frac{1}{u^2}\biggr)-1\biggr\}\\
=&-\frac{\sqrt{6}\ (2\e)^{2/3}}{4}\ \times\ 0.839917(9)\,.
\end{split}
\eqlabel{qdef}
\end{equation}
In the IR, ${\hat r}\to 0$, $z\to 1_-$ and $\frac1\r\to 0$ we have
\begin{equation}
\begin{split}
{\hat r}=\frac{\sqrt 6\ 2^{1/3}}{3^{1/3}\ \e^{2/3}}\ y\ \biggl(1-\frac{2^{2/3}\ 3^{1/3}}{15\ \e^{4/3}}\  y^2+
\frac{71\ 3^{2/3}\ 2^{1/3}}{2625\ \e^{8/3}}\  y^4
+\calo(y^6)\biggr)\,.
\end{split}
\eqlabel{rrhoir}
\end{equation}
Using \eqref{rrhouv} and \eqref{rrhoir}, and the exact analytic solution describing the Klebanov-Strassler Minkowski vacuum of 
the cascading gauge theory \eqref{ksks}, \eqref{kk}  we can identify parameters \eqref{uvparks}: 
\begin{equation}
\begin{split}
&K_0=P^2g_s\biggl(-\ln 3+\frac53\ \ln2-\frac43\ \ln\e-\frac23\biggr)\,,\qquad f_{a,1,0}=-2 \calq\,,\\
&k_{2,3,0}=\frac{3 \sqrt{6}}{8} \e^2 (3 \ln3-5 \ln2+4 \ln\e)\,,\qquad f_{c,4,0}=0\,,\qquad f_{a,3,0}=\frac{3\sqrt{6}}{4}\ \e^2\,,\\
&f_{a,6,0}=\biggl(-\frac{27}{16} \ln2
+\frac{81}{50}+\frac{81}{80} \ln3+\frac{27}{20} \ln\e\biggr) \e^4+\frac{3\sqrt{6}}{4} \calq^3  \e^2\,,
\\ &f_{a,7,0}=\left(\frac{27}{5}\ln\e-\frac{27}{4}\ln 2+\frac{81}{20}\ln 3+\frac{1701}{200}\right) \e^4\calq
+\frac{3\sqrt{6}}{4}  \e^2 \calq^4\,,\\
&f_{a,8,0}=\left(\frac{27}{2}\ln\e-\frac{135}{8}\ln2+\frac{81}{8}\ln 3+\frac{405}{16}\right) \calq^2\e^4
+\frac{3\sqrt{6}}{4} \calq^5  \e^2\,,\qquad g_{4,0}=0\,,
\end{split}
\eqlabel{susyuv}
\end{equation} 
in the UV, and parameters \eqref{irph2par}:
\begin{equation}
\begin{split}
&f_{a,0}^h=2^{1/3}\ 3^{2/3}\ \e^{4/3}\,,\qquad h_0^h=P^2g_s\ \e^{-8/3}\ \times\ 0.056288(0)\,,\\
&k_{1,3}^h=\frac{4\sqrt{6}}{9\ \e^2}\ P^2g_s\,,\qquad k_{2,2}^h=\frac{2^{2/3}}{3^{2/3}\ \e^{4/3}}\,,\qquad 
k_{2,4}^h=-\frac{11\ 2^{1/3}\ 3^{2/3}}{45\ \e^{8/3}}\,,\\
&k_{3,1}^h=\frac{4\sqrt{6}\ 2^{1/3}\ 3^{2/3}}{27\e^{2/3}}\ P^2g_s\,,\qquad g_0^h=g_s\,,
\end{split}
\eqlabel{susyir}
\end{equation}
in the IR.

Given \eqref{deflambda}, we identify from \eqref{susyuv}
\begin{equation}
\Lambda=\frac{3^{1/2}e^{1/3}\e^{2/3}}{2^{5/6}(P^2g_s)^{1/2}}=\frac{2^{1/6}e^{1/3}\e^{2/3}}{3^{3/2}M\a' g_s^{1/2}}
=\frac{2^{1/6}e^{1/3}g_s^{1/2}}{3^{3/2}}\ m_{glueball} \ \approx\ 0.3 g_s^{1/2} m_{glueball} \,,
\eqlabel{deflambdaf}
\end{equation}
where in the second equality we used \eqref{defpm};  the glueball mass scale is defined as in \eqref{defmglue}.

\section{Numerical procedure}\label{apc}

\subsection{FG frame de Sitter vacua}\label{apcfg}

Equations of motion for the FG frame de Sitter vacua of the cascading gauge theory, along with the asymptotics and the symmetries
of the dual holographic formulation, are presented in appendix \ref{apb}. Generically, we have 
eight functions of the radial coordinate $\r$, see \eqref{fg2}. When the chiral symmetry is unbroken,
there are only five functions, see \eqref{ktks1}. The solution to the equations of motion is
unique\footnote{Apart from the discrete choices associated with the IR boundary conditions
leading to classification of topologically distinct holographic vacua: TypeA$_{s,b}$ or TypeB, see
appendix \ref{apb1}.}
once we fix the Hubble constant $H$, the asymptotic string coupling $g_s$, the 3-form flux $P$ (alternatively the
rank difference of gauge group factors $M$ in the cascading theory), see \eqref{pquantization} and \eqref{defpm},
and the strong coupling scale $\Lambda$ of the cascading gauge theory (alternatively $K_0$, see \eqref{deflambda}, or
the conifold deformation parameter $\epsilon$, see \eqref{deflambdaf}).
Of these, parameters $H,\Lambda,P$ are dimensionful. The radial coordinate
$\r$ is dimensionful as well, albeit in units of 'mass'. 
As a result, UV/IR parameters of the solutions, see \eqref{uvparks}, \eqref{irph1par} and \eqref{irph2par},
have complicated dimensional dependence.
It is possible to completely eliminate the dimensional dependence (and the $g_s$ dependence) from all the equations of motion and the
asymptotic expansions with appropriate rescaling:
\begin{equation}
\begin{split}
&\{\r\,,\ f_{a,b,c}\,,\  h\,,\ K_{1,2,3}\,,\ g \}\qquad \Longrightarrow\qquad \{\hr\,,\ \hf_{a,b,c}\,,\ \hh\,,\ \hK_{1,2,3}\,,\
\hat{g}\}\,;\\
&\r=\frac{1}{H Pg_s^{1/2}}\ \hr\,,\qquad f_{a,b,c}=\hf_{a,b,c}\,,\qquad h=P^2g_s\ \hh\,,\\
& K_{1,3}=P^2g_s\ \hK_{1,3}\,,\qquad K_2=\hK_2\,,\qquad g=g_s\ \hat{g}  \,.
\end{split}
\eqlabel{scaleout}
\end{equation}
Additionally we introduce a dimensionless  parameter $k_s$ as
\begin{equation}
k_s\equiv \frac{K_0}{P^2 g_s}+\ln\left(H^2 P^2 g_s\right)\,,
\eqlabel{defks}
\end{equation}
leading from \eqref{deflambda} to the identification
\begin{equation}
k_s=\ln\frac{H^2}{\Lambda^2}\,.
\eqlabel{ks2}
\end{equation}
Notice that the conformal limit in the cascading gauge theory, \ie $H\gg \Lambda$, corresponds to $k_s\to \infty$.

We do not present the relations between all the UV/IR parameters stemming from \eqref{scaleout} and \eqref{defks}
--- they are straightforward to work out, but
too long to be illuminating --- and instead focus on the few ones for which we are reporting the numerical results: 
\nxt TypeA$_{s,b}$ vacua,
\begin{equation}
f^h_{a,b,c,0}=HPg_s^{1/2}\ \hf^h_{a,b,c,0}\,,\qquad K^h_{1,3,0}=P^2g_s\ \hK^h_{1,3,0}\,,\qquad K^h_{2,0}=\hK^h_{2,0}\,,\qquad
g^h_{0}=g_s\ \hat{g}^h_0\,;
\eqlabel{irtypeascaled}
\end{equation}
\nxt TypeB vacua,
\begin{equation}
\begin{split}
&f_{a,0}^h=H^2P^2g_s\ \hf_{a,0}^h\,,\qquad h_0^h =\frac{\hh_0^h}{H^4P^2 g_s}\,,\qquad
k_{1,3}^h=\frac{\hk_{1,3}^h}{H^3Pg_s^{1/2}}\,,\qquad k_{2,2}^h=\frac{\hk_{2,2}^h}{H^2P^2g_s}\\
&k_{2,4}^h=\frac{\hk_{2,4}^h}{H^4P^4g_s^2}\,,\qquad k_{3,1}^h=\frac{Pg_s^{1/2}}{H}\ \hk_{3,1}^h\,,
\qquad g^h_{0}=g_s\ \hat{g}^h_0\,.
\end{split}
\eqlabel{irtypebscaled}
\end{equation}

Numerical analysis of the bulk differential equations describing de Sitter vacua are rather involved.
To trust them, we would like to have various consistency checks. Here, the symmetry transformations
SFG2-SFG4 \eqref{kssym1}-\eqref{kssym3} are very useful: we can produce different data sets
fixing three of the four parameters $\{H,P,g_s,K_0\}$. As we demonstrate,
with appropriate rescaling, the distinct data sets must collapse. We find it useful to implement
three different computational schemes:
\begin{equation}
\begin{split}
{\rm SchemeI:}&\qquad H=P=g_s=1\,,\qquad k_s\qquad {\rm is\ varied}\,;\\
{\rm SchemeII:}&\qquad H=g_s=K_0=1\,,\qquad b\equiv P^2\qquad {\rm is\ varied}\,;\\
{\rm SchemeIII:}&\qquad P=g_s=1\,,\ K_0=\frac 14\,,\qquad \a\equiv H^2\qquad {\rm is\ varied}\,.\\
\end{split}
\eqlabel{compschemes}
\end{equation}
Note that:
\nxt SchemeI is equivalent to performing computations in the hatted variables in \eqref{scaleout},
with \eqref{defks};
\nxt SchemeII is convenient to take a conformal limit to Klebanov-Witten solution \cite{Klebanov:1998hh}
in TypeA$_s$ vacua: $b\to 0$;
\nxt SchemeIII is convenient to study the extremal KS \cite{Klebanov:2000hb} limit in TypeB vacua: $\a\to 0$.

Numerical computations are done adopting the algorithms developed in \cite{Aharony:2007vg}.
Altogether, there are 8 second order differential equations \eqref{kseq2}-\eqref{kseq9} and
a single first order constraint \eqref{kseq10} for 8 functions
$\{f_{a},f_b,f_c,h,K_1,K_2,K_2,g\}$. Notice that the constraint \eqref{kseq10} involves
$f_c'$ linearly. Thus, we can use the latter equation and eliminate the redundant equation \eqref{kseq2}.
The final set of ODEs --- 7 second order equations and 1 first order equation --- necessitates
$15=2\times 7+1$ parameters.
\nxt TypeA$_{s,b}$ vacua:\\
The result of the numerical computations are the data files with entries for the 8 UV parameters
$\{f_{a,1,0},f_{a,3,0},k_{2,3,0},g_{4,0},f_{c,4,0},f_{a,6,0},f_{a,7,0},f_{a,8,0}\}$
and the 7 IR parameters $\{f_{a,0}^h,f_{b,0}^h,f_{c,0}^h,K_{1,0}^h,K_{2,0}^h,K_{3,0}^h,g_0^h\}$
(see appendix \ref{apb1}) labeled by $k_s$ (for the computational scheme SchemeI), $b$
(for the computational scheme SchemeII) or $\a$ (for the computational scheme SchemeIII).
The number of parameters are reduced to 5 (in the UV) and 4 (in the IR) when chiral symmetry is
unbroken (see appendix \ref{apb11}).
\nxt TypeB vacua:\\
The result of the numerical computations are the data files with entries for the 8 UV parameters
$\{f_{a,1,0},f_{a,3,0},k_{2,3,0},g_{4,0},f_{c,4,0},f_{a,6,0},f_{a,7,0},f_{a,8,0}\}$
and the 7 IR parameters $\{f_{a,0}^h,h_{0}^h,k_{1,3}^h,k_{2,2}^h,k_{2,4}^h,k_{3,1}^h,g_0^h\}$
(see appendix \ref{apb1}) labeled\footnote{We will not use
the computation scheme SchemeII here.} by $k_s$ (for the computational scheme SchemeI),
or $\a$ (for the computational scheme SchemeIII).

\subsection{EF frame de Sitter vacua}\label{apcef}

In total, there are 11 (8 with unbroken chiral symmetry) coupled ODEs \eqref{efv1}-\eqref{efc2}
describing EF frame de Sitter vacua involving 5 metric warp factors
$\{a,\sigma,w_{a2},w_{b2},w_{c2}\}$ (see \eqref{ef1i}), 3 flux functions $\{K_1,K_2,K_3\}$ (see \eqref{redef})
and the string coupling $g$ as a function of a radial coordinate $z\equiv -r$, see \eqref{defzef}.
The full set of ODEs is redundant, and in practice we use 9 equations \eqref{efv2}-\eqref{efc1}:
we drop \eqref{efv1} in favor of \eqref{efc1}, and we use \eqref{efv5} (it involves $w_{c2}''$ linearly)
instead of \eqref{efc2} (though it involves $w_{c2}'$ linearly). The reason for this is to reduce the
complexity of the system of ODEs --- unlike construction of de Sitter vacua in FG frame which is
a boundary value problem, representation of de Sitter vacua in EF frame is an initial value problem,
and thus we can get away with using a higher order system of ODEs.

The initial
conditions for these equations are set at $z\to 0_+$ with asymptotic expansions \eqref{typeasi} for
TypeA$_s$ de Sitter vacua, and with asymptotic expansions \eqref{typebi1}-\eqref{typeabi9} for Type$A_b$
de Sitter vacua. The EF frame equations of motion are integrated on the interval
\begin{equation}
z\in [0,z_{AH}]\,,
\eqlabel{rangez}
\end{equation}
where $z_{AH}$ is the first zero of the AH location function $\call_{AH}$ (see \eqref{ahsvac}):
\begin{equation}
\call_{AH}(z)\equiv 3H\ \sigma^3 \w_{c2}^{1/2}\w_{a2}\w_{b2}-a\frac{d}{dz}\biggl\{
\sigma^3 \w_{c2}^{1/2}\w_{a2}\w_{b2}\biggr\}\,.
\eqlabel{defcall}
\end{equation}
Using \eqref{typeasi}-\eqref{typeabi9},
\begin{equation}
\begin{split}
&{\rm TypeA}_s:\qquad \call_{AH}=\frac{3\sqrt{2}}{8H^{3/2}} (s_0^h)^3 (f_{2,0}^h)^{1/2} (f_{3,0}^h)^2+\calo(z)\,;
\\
&{\rm TypeA}_b:\qquad \call_{AH}=\frac{3\sqrt{2}}{8H^{3/2}} (s_0^h)^3 (f_{c,0}^h)^{1/2} f_{a,0}^h f_{b,0}^h+\calo(z)\,,
\end{split}
\eqlabel{callzsmall}
\end{equation}
\ie both for TypeA$_s$ and TypeA$_b$ vacua
\begin{equation}
\call_{AH}(z=0)>0\,,\qquad \frac{d}{dz}\call_{AH}(z=0)\ <\ 0\,,
\eqlabel{callzsmall1}
\end{equation}
where the second inequality is a numerical observation. 
Notice that to set-up the initial conditions for  \eqref{efv2}-\eqref{efc1},
besides the FG frame IR data \eqref{irph1par} (or \eqref{irph2par2} when the chiral symmetry
is unbroken), one needs parameter $s_0^h$, see \eqref{defsh0}, 
\begin{equation}
\begin{split}
s_0^h=\lim_{z\to 0_+} \sigma(z)=&\lim_{\rho\to +\infty} \biggl\{c_1(\r)\ \exp\left[H \int_0^\r ds\
\frac{c_2(s)}{c_1(s)}\right]\biggr\}\\
=&\lim_{\rho\to +\infty} \biggl\{\frac{1}{(h(\r))^{1/4}\r}\ \exp\left[H \int_0^\r ds\
(h(s))^{1/2}\right]\biggr\}\,,
\end{split}
\eqlabel{sh0}
\end{equation}
where we used \eqref{mapfin} and explicit expressions 
\begin{equation}
c_1=\frac{1}{h^{1/4}\r}\,,\qquad c_2=\frac{h^{1/4}}{\r}
\eqlabel{c1c2}
\end{equation}
from comparing \eqref{fgtoef} and \eqref{fg1i}.
The limit in \eqref{sh0} must be taken carefully, as the integral is divergent
at the upper limit of integration: using the asymptotic expression for $h$ as
$y\equiv \frac 1\r\to 0$ \eqref{defhors4} and \eqref{phase1ir1} we can regulate it as follows,
\begin{equation}
\int_0^\r ds\
(h(s))^{1/2}=\int_0^\r ds\
\biggl((h(s))^{1/2}-\frac{Pg_s^{1/2}}{2(H Pg_s^{1/2}s+1)}\biggr)+\frac{1}{2H}\ \ln\left(1+HPg_s^{1/2}\r\right) \,,
\eqlabel{regint}
\end{equation}
or in dimensionless/rescaled quantities \eqref{scaleout}
\begin{equation}
\frac{1}{H}\int_0^{\hat\r} d\hat{s}\
(\hh(\hat{s}))^{1/2}=\frac 1H \int_0^{\hat{\r}} d\hat{s}\
\biggl((\hh(\hat{s}))^{1/2}-\frac{1}{2(\hat{s}+1)}\biggr)+\frac{1}{2H}\ \ln\left(1+\hat{\r}\right) \,,
\eqlabel{regintres}
\end{equation}
leading to
\begin{equation}
s_0^h=2^{1/2}\ H P^{1/2}g_s^{1/4}\  \exp\biggl[ \int_0^{\hat{\r}} d\hat{s}\
\biggl((\hh(\hat{s}))^{1/2}-\frac{1}{2(\hat{s}+1)}\biggr)\biggr]\equiv  H P^{1/2}g_s^{1/4}\ \hat{s}_0^h\,,
\eqlabel{finalsh0}
\end{equation}
where the last equality defines dimensionless/rescaled $\hat{s}_0^h$.

\section{$b\to 0$ of TypeA$_s$ vacua}\label{apd}

\subsection{FG frame}\label{apdfg}

The conformal, \ie $H\gg \Lambda$, limit of TypeA$_s$ vacua is best described
in computational SchemeII \eqref{compschemes}. Using perturbative
expansions \eqref{ktschb1} we find ($'=\frac{d}{d\r}$),
\nxt for $n=1$:
\begin{equation}
\begin{split}
&0=k_1''-\frac{\r+6}{2\r(1+\r)} k_1'-\frac{8}{(1+\r) \r^2}\,,
\end{split}
\eqlabel{n11}
\end{equation}
\begin{equation}
\begin{split}
&0=g_1''-\frac{\r+6}{2\r(1+\r)}g_1'+(k_1')^2-\frac{4}{(1+\r) \r^2}\,,
\end{split}
\eqlabel{n12}
\end{equation}
\begin{equation}
\begin{split}
&0=f_{21}'+h_1'+4 f_{31}'+\frac{(1+\r) \r}{2(\r+2)} (k_1')^2
+\frac{2}{\r (\r+2)} \left(f_{21}+4 f_{31}-4 k_1-1\right)
\\&+\frac{(\r+4) (3 \r+4)}{2(\r+2) (1+\r) \r} h_1\,,
\end{split}
\eqlabel{n13}
\end{equation}
\begin{equation}
\begin{split}
&0=f_{31}''+\frac 14 (k_1')^2+\frac{(\r+2)}{2\r (1+\r)} h_1'
-\frac{(\r+6)}{2\r (1+\r)} f_{31}'+\frac{1}{(1+\r) \r^2}
\left(5 f_{21}+8 f_{31}-4 k_1-1\right)
\\&- \frac{3 \r^2-16 \r-16}{4(1+\r)^2 \r^2} h_1\,,
\end{split}
\eqlabel{n14}
\end{equation}
\begin{equation}
\begin{split}
&0=h_1''+\frac12(k_1')^2-\frac{(3\r+10)}{2\r(1+\r)}
h_1+\frac{2}{(1+\r)\r^2}\left(3+20k_1-9f_{21}-36f_{31}\right)
\\&+\frac{(3\r^2-80\r-80}{2(1+\r)^2\r^2} h_1\,;
\end{split}
\eqlabel{n15}
\end{equation}
\nxt for $n=2$:
\begin{equation}
\begin{split}
&0=k_2''-\frac{\r+6}{2\r (1+\r)} k_2'-\frac{1}{4\r (1+\r) (\r+2)} \biggl(
(4 g_1'+6 h_1' +8 f_{31}' )\r^3
+(12 g_1' +18 h_1' +3  h_1
\\&+24 f_{31}' )\r^2 +(8 g_1' +12 h_1' -16 k_1 
+4 f_{21} +16 h_1 +16 f_{31}' +16 f_{31})\r -16 k_1+4 f_{21}
\\&+16 h_1+16 f_{31}-4 \r-4\biggr) k_1'
-\frac{(1+\r) \r}{4(\r+2)} (k_1')^3-\frac{8 (k_1+g_1-f_{21}-h_1-2 f_{31})}{(1+\r) \r^2}\,,
\end{split}
\eqlabel{n21}
\end{equation}
\begin{equation}
\begin{split}
&0=g_2''-\frac{\r+6}{2\r (1+\r)} g_2'-(g_1')^2+2 k_1' k_2'
-\frac{1}{4\r (\r+2) (1+\r)} \biggl(
(k_1')^2 \r^4+(2 (k_1')^2 \r^3\\&+2 h_1') \r^3
+( (k_1')^2 +3  h_1+6 h_1') \r^2
+(16 f_{31} +4 f_{21} +16 h_1 +4 h_1' -16 k_1 )\r
+16 f_{31}\\&+4 f_{21}+16 h_1-16 k_1-4 \r-4\biggr) g_1'-(2 f_{31}+h_1) (k_1')^2
+\frac{4 (2 f_{31}-2 g_1+f_{21}+h_1)}{(1+\r) \r^2}\,,
\end{split}
\eqlabel{n22}
\end{equation}
\begin{equation}
\begin{split}
&0=f_{22}'+4 f_{32}'+h_2'+\frac{1}{\r+2}\biggl(
f_{31}' \r^2+(f_{21} -h_1 +f_{31}') \r
+2 f_{21}-2 h_1\biggr) h_1'\\&+\frac{5\r (1+\r)}{2(\r+2)}
(f_{31}')^2+ \frac{\r (1+\r)}{4(\r+2)} (g_1')^2
+\frac{\r (1+\r)}{4(\r+2)} (h_1')^2+\frac{(1+\r)\r}{2(\r+2)^2} \biggl(
f_{31}' \r^2+(f_{21} -g_1 \\&-h_1 -2 f_{31} +f_{31}') \r
+2 f_{21}-2 g_1-2 h_1-4 f_{31}\biggr)  (k_1')^2
+\frac{\r  (1+\r)}{\r+2}k_1' k_2'+\frac{1}{2(\r+2)^2} \biggl(
\\&(8  f_{21}+3 h_1 -8 f_{31})\r^2 +(36 f_{21} -16 k_1 +16 h_1 -16 f_{31})\r
+36 f_{21}-16 k_1+16 h_1-16 f_{31}\\&-4 \r-4\biggr) f_{31}'
+\frac{1}{2(\r+2) \r (1+\r)} \biggl(
3 f_{21} h_1 \r^2+(-8 f_{21} f_{31} +32 k_1 h_1 +64 k_1 f_{31} +4 f_{22} 
\\&-16 k_2) \r-8 f_{21} f_{31}+32 k_1 h_1+64 k_1 f_{31}
-8 k_1^2 -4 f_{21}^2+(16 f_{32} -4 f_{21}^2 -8 k_1^2-64 h_1 f_{31} \\&-4 g_1 +4 h_1
-24 h_1^2 +16 h_2 +8 f_{31} -68 f_{31}^2 )\r-64 h_1 f_{31}+-4 g_1-16 k_2+4 f_{22}+8 f_{31}\\&+16 f_{32}+4 h_1+16 h_2-24 h_1^2
-68 f_{31}^2+3 h_2 \r^2\biggr)\,,
\end{split}
\eqlabel{n23}
\end{equation}
\begin{equation}
\begin{split}
&0=f_{32}''-\frac{\r+6}{2\r (1+\r)} f_{32}'
+\frac{\r+2}{2\r (1+\r)} h_2'+\frac12 k_1' k_2'
+\frac18 (g_1')^2+\frac18(h_1')^2+\frac14(f_{31}')^2
-\frac14 (g_1\\&+ h_1+ f_{31}) (k_1')^2
-\frac{\r+2}{2\r (1+\r)} (h_1-f_{31}) h_1'
+\frac{1}{4(1+\r)^2 \r^2} \biggl(
(4 f_{21} +16 f_{21} k_1 -16 f_{21} h_1 \\&-36 f_{21} f_{31} +32 k_1 h_1
+48 k_1 f_{31} +20 f_{22} 
-16 k_2 -8 f_{21}^2 
-8 k_1^2 -48 h_1 f_{31} -4 g_1 
+4 h_1 \\&-24 h_1^2 +16 h_2 +4 f_{31} -36 f_{31}^2 +32 f_{32} )\r+16 f_{21} k_1
-36 f_{21} f_{31}+32 k_1 h_1+48 k_1 f_{31}\\&-8 f_{21}^2-8 k_1^2-3 h_1 f_{31} \r^2
-48 h_1 f_{31}
-4 g_1-16 k_2+4 f_{21}+20 f_{22}+4 f_{31}+32 f_{32}
+4 h_1\\&+16 h_2-24 h_1^2-36 f_{31}^2-16 f_{21} h_1-3 h_2 \r^2\biggr)\,,
\end{split}
\eqlabel{n24}
\end{equation}
\begin{equation}
\begin{split}
&0=h_2''-\frac{3 \r+10}{2\r (1+\r)} h_2'-\frac74 (h_1')^2
-\frac52 (f_{31}')^2-\frac{1}{4(\r+2)} \biggl(
(h_1' +2 f_{31}') \r^2+(h_1'
+2 g_1 \\&+4 f_{31} +2 f_{31}') \r
+4 g_1+8 f_{31}\biggr) (k_1')^2-\frac14 (g_1')^2+k_1' k_2'
-\frac{1}{4\r (1+\r) (\r+2)} \biggl(
4 f_{31}' \r^3+(3 h_1 \\&+12 f_{31}') \r^2
+(4 f_{21} 
-16 k_1 +16 h_1 +16 f_{31} +8 f_{31}') \r
+4 f_{21}-16 k_1+16 h_1+16 f_{31}-4 \r\\&-4\biggr) h_1'
-\frac{1}{2\r (1+\r) (\r+2)} \biggl(
3 h_1 \r^2+(4 f_{21} -16 k_1 +16 h_1 +16 f_{31}) \r+4 f_{21}-16 k_1
\\&+16 h_1+16 f_{31}-4 \r-4\biggr) f_{31}'
-\frac{1}{2(1+\r)^2 \r^2} \biggl(
(80 f_{21} k_1 -44 f_{21} h_1-152 f_{21} f_{31}
+80 k_1 h_1\\& +320 k_1 f_{31} +36 f_{22} 
-80 k_2 -40 f_{21}^2 
-40 k_1^2-176 h_1 f_{31} -12 g_1 -40 h_1^2 
+80 h_2 +24 f_{31} \\&-388 f_{31}^2  +144 f_{32}+12 f_{21}  )\r +80 f_{21} k_1
-152 f_{21} f_{31}+80 k_1 h_1+320 k_1 f_{31}
-40 f_{21}^2\\&-40 k_1^2-176 h_1 f_{31}
-44 f_{21} h_1-12 g_1-80 k_2+12 f_{21}+36 f_{22}+24 f_{31}+144 f_{32}+80 h_2
\\&-40 h_1^2-388 f_{31}^2-3 (h_1^2 + h_2) \r^2\biggr)\,.
\end{split}
\eqlabel{n25}
\end{equation}
The UV ($\rho\to 0$) and the IR ($y\equiv \frac 1\r\to 0$) asymptotic expansions can be obtained from
\eqref{ktks2}-\eqref{ktks6} and \eqref{phase2ir1} correspondingly, using the SchemeII parameters
\eqref{compschemes}, where
\begin{equation}
\begin{split}
&f_{2,1,0}=1+f_{2,1,0;1}\ b+f_{2,1,0;2}\ b^2+\calo(b^3)\,,\qquad g_{4,0}=g_{4,0;1}\ b+g_{4,0;2}\ b^2+\calo(b^3)\,,\\
&f_{2,4,0}=\left(-\frac{1}{12}+\frac43\ k_{4,0;1}\right)\ b
+ \biggl(-\frac{139}{1152}+\frac{1}{24}\ f_{2,1,0;1}-\frac{22}{9} k_{4,0;1}+\frac23 g_{4,0;1}\\&
+\frac43 k_{4,0;2}\biggr)
\ b^2+\calo(b^3)\,,\\
&f_{2,6,0}=f_{2,6,0;1}\ b+f_{2,6,0;2}\ b^2+\calo(b^3)\,,\qquad f_{2,8,0}=f_{2,8,0;1}\ b+f_{2,8,0;2}\ b^2+\calo(b^3)\,,\\
&f_{2,0}^h=1+f_{2,0;1}^h\ b+f_{2,0;2}^h\ b^2+\calo(b^3)\,,\qquad f_{3,0}^h=1+f_{3,0;1}^h\ b+f_{3,0;2}^h\ b^2+\calo(b^3)\,,\\
&K_{0}^h=1+K_{0;1}^h\ b+K_{0;2}^h\ b^2+\calo(b^3)\,,\qquad g_{0}^h=1+g_{0;1}^h\ b+g_{0;2}^h\ b^2+\calo(b^3)\,.
\end{split}
\eqlabel{pertcond}
\end{equation}
Note that in lieu of $f_{2,4,0;1}$ and $f_{2,4,0;2}$ in \eqref{pertcond} we used $k_{4,0;1}$ and $k_{4,0;2}$:
\begin{equation}
\begin{split}
&k_1= -2 \ln\r+\r-\frac18 \r^2-\frac{1}{24} \r^3+\left(\frac{3}{64} \ln\r+k_{4,0;1}\right) \r^4+\calo(\r^5)\,,\\
&k_2= f_{2,1,0;1}\ \r+\left(-\frac14 \ln\r+\frac{9}{16}-\frac12 f_{2,1,0;1}\right) \r^2
+\biggl( \frac14 \ln\r-\frac{7}{16}+\frac18 f_{2,1,0;1}\biggr) \r^3\\
&+\left(-\frac{3}{16} \ln^2\r+\left(\frac{11}{64}-4 k_{4,0;1}\right) \ln\r+k_{4,0;1}\right) \r^4
+\calo(\r^5)\,.
\end{split}
\eqlabel{defk40s}
\end{equation}
This is done for computational convenience --- the equations for $k_1$ (see \eqref{n11}) and $k_2$ (see \eqref{n21})
decouple from all the other equations at the corresponding order.

We are able to solve analytically only the equation for $k_1$ \eqref{n11},
\begin{equation}
k_1=\frac\r4+\frac{1}{4+4\r}-\frac14-4\ln2+\frac{\r^3-6\r^2-24\r-16}{8 (1+\r)^{3/2}}\
\ln\frac{\sqrt{1+\r}-1}{\sqrt{1+\r}+1}\,,
\eqlabel{k1anal}
\end{equation}
resulting in
\begin{equation}
k_{4,0;1}=\frac{29}{259}-\frac{3}{32}\ln 2\,,\qquad K_{0;1}^h=\frac 53-4\ln 2\,,
\eqlabel{analk40s}
\end{equation}
and the equation for $g_1$ \eqref{n12},
\begin{equation}
\begin{split}
&g_1=-\frac{\r^2}{32} -\frac{7\r}{16}-\frac{1}{32 (1+\r)^2}
-\frac{3}{8+8 \r}+\frac{13}{32}-\frac{\r^4 (\r+2)}{32(1+\r)^{5/2}}\
\ln\frac{\sqrt{1+\r}-1}{\sqrt{1+\r}+1}\\
&+\biggl(\frac{23}{64}-\frac{\r^3}{128}+\frac{15\r^2}{128}-\frac{15\r}{64}
+\frac{9}{64 (1+\r)^2}-\frac{63}{128+128 \r}-\frac{1}{128 (1+\r)^3}\biggr)
\\&\times \ln^2 \frac{\sqrt{1+\r}-1}{\sqrt{1+\r}+1}\,;
\end{split}
\eqlabel{analg1}
\end{equation}
\begin{equation}
g_{4,0;1}=-\frac{17}{32}+\frac38\ \ln^22+\frac18\ \ln2\,,\qquad g_{0;1}^h=-\frac {13}{18} \,.
\eqlabel{analg40s}
\end{equation}
All the remaining equations are solved numerically, using the shooting algorithm developed in
\cite{Aharony:2007vg}. We find:
\begin{equation}
\begin{split}
&f_{2,1,0;1}=0.434278\,,\qquad  f_{2,1,0;2}=0.357298\,,\\
&g_{4,0;1}=-0.264437\,,\qquad  g_{4,0;2}=-0.64466\,,\\
&k_{4,0;1}=0.0482987\,,\qquad  k_{4,0;2}=0.184174\,,\\
&f_{2,6,0;1}=-0.407036\,,\qquad  f_{2,6,0;2}=-0.489017\,,\\
&f_{2,8,0;1}=-0.427022\,,\qquad  f_{2,8,0;2}=-0.609369\,,\\
&f_{2,0;1}^h=-0.156614\,,\qquad f_{2,0;2}^h=0.54009\,,\\
&f_{3,0;1}^h=-0.378836\,,\qquad f_{3,0;2}^h=0.638051\,,\\
&K_{0;1}^h=-1.10592\,,\qquad K_{0;2}^h=1.65245\,,\\
&g_{0;1}^h=-0.722222\,,\qquad  g_{0;2}^h=0.311658\,,
\end{split}
\eqlabel{numresults}
\end{equation}
where we used the same numerical methods to solve \eqref{n11} and $\eqref{n12}$. Comparing the numerical
results for $\{k_{4,0;1},\ g_{4,0;1},\ K_{0;1}^h,\ g_{0;1}^h\}$ from \eqref{numresults}
with the analytic predictions \eqref{analk40s} and \eqref{analg40s} we find agreement
at the fractional level of $\sim 10^{-10}$ or better.

\subsection{EF frame}\label{apdef}

Using perturbative expansions \eqref{ktefschb1} and $w_{c2n}\equiv v_n-w_{a2n}$,
we find from \eqref{efv2}-\eqref{efc1} $\left('=\frac{d}{dz}\right)$,
\nxt for $n=1$:
\begin{equation}
\begin{split}
0=&k_1''+\frac{5(2 z-1)}{2(z-1) z}\ k_1' -\frac{8}{(z-1) z}\,,
\end{split}
\eqlabel{efn11}
\end{equation}
\begin{equation}
\begin{split}
0=&v_1''-\frac{27}{(z-1) z}\ v_1-\frac{15(2 z-1)}{2(z-1) z}\ a_1'+\frac{11}{4}\ (k_1')^2
+\frac{60}{(z-1) z}\ k_1-\frac{15(2 z^2-2 z+1)}{2z^2 (z-1)^2}\ a_1\\
&+\frac{9}{(z-1) z}\,,
\end{split}
\eqlabel{efn12}
\end{equation}
\begin{equation}
\begin{split}
0=&a_1''+\frac{7(2 z-1)}{2(z-1) z}\ a_1'+\frac{10 z^2-10 z+3}{2z^2 (z-1)^2}\ a_1-\frac14 (k_1')^2
+\frac{9}{(z-1) z}\ v_1-\frac{20}{(z-1) z}\ k_1\\
&-\frac{3}{(z-1) z}\,,
\end{split}
\eqlabel{efn13}
\end{equation}
\begin{equation}
\begin{split}
0=&w_{a21}''+\frac{5(2 z-1)}{2(z-1) z}\ w_{a21}'-\frac{12}{(z-1) z}\ w_{a21}
-\frac{(2 z-1)}{2(z-1) z}\ v_1'-\frac{3(2 z-1)}{2(z-1) z}\ a_1'+\frac 34\ (k_1')^2
\\&-\frac{3}{(z-1) z}\ v_1-\frac{3(2 z^2-2 z+1)}{2z^2 (z-1)^2}\ a_1+\frac{12}{(z-1) z}\ k_1
+\frac{1}{(z-1) z}\,,
\end{split}
\eqlabel{efn14}
\end{equation}
\begin{equation}
\begin{split}
0=&s_1'-\frac12 a_1'+\frac{1}{2(z-1) z}\ a_1\,,
\end{split}
\eqlabel{efn15}
\end{equation}
\begin{equation}
\begin{split}
0=&g_1''+\frac{5(1-2 z)}{2z (1-z)}\ g_1'+(k_1')^2+\frac{4}{z (1-z)}\,;
\end{split}
\eqlabel{efn16}
\end{equation}
\nxt for $n=2$:
\begin{equation}
\begin{split}
0=&k_2''+\frac{5(2 z-1)}{2(z-1) z}\ k_2'+\left(
\frac 12 v_1'+\frac52 a_1'-2 w_{a21}'-g_1'
\right) k_1'-\frac{8 (k_1-a_1-v_1+g_1+2 w_{a21})}{(z-1) z}\,,
\end{split}
\eqlabel{efn21}
\end{equation}
\begin{equation}
\begin{split}
&0=v_2''-\frac{27}{(z-1) z}\ v_2-\frac{15(2 z-1)}{2(z-1) z}\ a_2'
+\frac{11}{2}\ k_1' k_2'+\frac{60}{(z-1) z}\ k_2-\frac{15(2 z^2-2 z+1)}{2z^2 (z-1)^2}\ a_2
\\&-\frac12 (v_1')^2+
\left(\frac{11}{2} w_{a21}'-\frac{10 (2 z-1)}{(z-1) z}\ w_{a21}+\frac{2 (2 z-1)}{(z-1) z}\ v_1\right) v_1'
+\frac{3(2 z-1) (5 a_1-v_1)}{2(z-1) z}\ a_1'\\&
-\frac{15}{4} (a_1')^2-\left(
\frac{11}{4} g_1+\frac14 v_1+\frac32 w_{a21}\right) (k_1')^2
-\frac{55}{4} (w_{a21}')^2-\frac{10 (2 z-1) (v_1-5 w_{a21})}{(z-1) z}\ w_{a21}'\\
&+\frac58 (g_1')^2+\frac{15 v_1^2}{(z-1) z}+\frac{30 k_1^2}{(z-1) z}+\frac{75 w_{a21}^2}{(z-1) z}
+\frac{15(4 z^2-4 z+3) a_1^2}{4z^2 (z-1)^2}-\frac{12 (5 a_1+4 v_1) k_1}{(z-1) z}
\\&+\left(\frac{3(16 z^2-16 z-1) v_1}{2z^2 (z-1)^2}-\frac{9}{(z-1) z}\right)a_1
-\frac{2 (15 v_1-1) w_{a21}}{(z-1) z}+\frac{9 g_1-4 v_1}{(z-1) z}\,,
\end{split}
\eqlabel{efn22}
\end{equation}
\begin{equation}
\begin{split}
&0=a_2''+\frac{7(2 z-1)}{2(z-1) z}\ a_2'+\frac{10 z^2-10 z+3}{2z^2 (z-1)^2}\ a_2
+\frac{9}{(z-1) z}\ v_2-\frac 12 k_1' k_2'-\frac{20}{(z-1) z}\ k_2
\\&+\left(-\frac14 a_1+\frac14 g_1+\frac12 w_{a21}\right) (k_1')^2+\frac18 (g_1')^2
+\frac54 (w_{a21}')^2+\frac34 (a_1')^2-\frac12 w_{a21}' v_1'-\frac{10 k_1^2}{(z-1) z}
\\&-\frac{105 w_{a21}^2}{(z-1) z}+\frac{20 v_1 k_1}{(z-1) z}-\frac{10 v_1^2}{(z-1) z}
-\frac{3a_1^2}{4z^2 (z-1)^2}+\frac{6 (7 v_1-1) w_{a21}}{(z-1) z}+\frac{3 (v_1-g_1)}{(z-1) z}\,,
\end{split}
\eqlabel{efn23}
\end{equation}
\begin{equation}
\begin{split}
&0=w_{a22}''+\frac{5(2 z-1)}{2z (z-1)}\ w_{a22}'-\frac{12}{(z-1) z}\ w_{a22}
-\frac{3(2 z-1)}{2z (z-1)}\ a_2'-\frac{2 z-1}{2z (z-1)}\ v_2'
\\&-\frac{3(2 z^2-2 z+1)}{2z^2 (z-1)^2}\ a_2-\frac{3}{(z-1) z}\ v_2+\frac{12}{(z-1) z}\ k_2
+\frac32 k_1' k_2'+\frac 14 (w_{a21}')^2-\frac34 (a_1')^2\\
&-\frac34 (g_1+w_{a21}) (k_1')^2+\frac 18 (g_1')^2+\left(\frac52 a_1'
-\frac{2 (2 z-1) (v_1-5 w_{a21})}{(z-1) z}\right) w_{a21}'
\\&+\left(\frac{(2 z-1) (v_1-5 w_{a21})}{2(z-1) z}-\frac 12 a_1'\right) v_1'
+\frac{3(2 z-1) (a_1-w_{a21})}{2(z-1) z} a_1'+\frac{6v_1^2}{(z-1) z}
\\&+\frac{3(4 z^2-4 z+3)a_1^2}{4z^2 (z-1)^2} +\frac{75w_{a21}^2}{(z-1) z} +\frac{6k_1^2}{(z-1) z}
-\frac{12 (v_1+a_1-w_{a21})k_1}{z (z-1)} \\
&+\frac{(3 a_1-33 w_{a21}-1)v_1}{z (z-1)}
+\frac{g_1}{(z-1) z}+\frac{3(6 z^2-6 z-1)w_{a21}a_1}{2z^2 (z-1)^2}-\frac{a_1-3 w_{a21}}{(z-1) z}\,,
\end{split}
\eqlabel{efn24}
\end{equation}
\begin{equation}
\begin{split}
0=&s_2'-\frac12 a_2'+\frac{1}{2(z-1) z}\ a_2+\frac12 (a_1-s_1) a_1'+\frac{a_1 (a_1-s_1)}{2(1-z) z}\,,
\end{split}
\eqlabel{efn25}
\end{equation}
\begin{equation}
\begin{split}
0=&g_2'+\frac{5(2 z-1)}{2(z-1) z}\ g_2'+2 k_1' k_2'-2 w_{a21} (k_1')^2-(g_1')^2+\left(\frac12 v_1'
+5 a_1'\right) g_1'\\
&+\frac{4 (v_1-2 g_1+a_1-2 w_{a21})}{(z-1) z}\,.
\end{split}
\eqlabel{efn26}
\end{equation}

Initial conditions for \eqref{efn11}-\eqref{efn26} can be deduced from \eqref{typeasi} using \eqref{pertcond}
and \eqref{numsigma}:
\begin{equation}
\begin{split}
&k_1=K_{0;1}^h-\frac{16}{5} z+\calo(z^2)\,,\\
&v_1= f_{2,0;1}^h+4 f_{3,0;1}^h
+\left(32 K_{0;1}^h-\frac{64}{5} f_{2,0;1}^h-\frac{256}{5} f_{3,0;1}^h+\frac{28}{5}\right) z+\calo(z^2)\,,\\
&a_1=\left(-4 K_{0;1}^h+\frac95 f_{2,0;1}^h+\frac{36}{5} f_{3,0;1}^h-\frac35\right) z+\calo(z^2)\,,\\
&w_{a21}=f_{3,0;1}^h+\left(\frac{32}{5} K_{0;1}^h-\frac85 f_{2,0;1}^h-\frac{56}{5} f_{3,0;1}^h+\frac45\right) z+\calo(z^2)\,,\\
&s_1=s_{0;1}^h+\calo(z)\,,\qquad g_1=g_{0;1}^h-\frac85 z+\calo(z^2)\,,
\end{split}
\eqlabel{efincond1}
\end{equation}
for $n=1$, and
\begin{equation}
\begin{split}
&k_2=K_{0;2}^h+\left(\frac{32}{5} f_{3,0;1}^h-\frac{16}{5} K_{0;1}^h-\frac{16}{5} g_{0;1}^h+\frac{16}{5} f_{2,0;1}^h\right)
z+\calo(z^2)\,,\\
&v_2=f_{2,0;2}^h+4 f_{3,0;2}^h+\biggl(\frac{224}{5} f_{2,0;1}^h f_{3,0;1}^h-\frac{512}{5} K_{0;1}^h
f_{3,0;1}^h+16 (K_{0;1}^h)^2+32 K_{0;2}^h-\frac{64}{5} f_{2,0;2}^h\\
&-\frac{256}{5} f_{3,0;2}^h
+\frac{28}{5} g_{0;1}^h+\frac{48}{5} (f_{2,0;1}^h)^2+\frac{528}{5} (f_{3,0;1}^h)^2-8 f_{3,0;1}^h
-\frac{128}{5} K_{0;1}^h f_{2,0;1}^h-\frac{16}{5} f_{2,0;1}^h\biggr)
z\\
&+\calo(z^2)\,,\\
&a_2=\biggl(-2 (K_{0;1}^h)^2+4 K_{0;1}^h f_{2,0;1}^h+16 K_{0;1}^h f_{3,0;1}^h-2 (f_{2,0;1}^h)^2
-\frac{38}{5} f_{2,0;1}^h f_{3,0;1}^h
-\frac{97}{5} (f_{3,0;1}^h)^2\\
&-4 K_{0;2}^h+\frac35 f_{2,0;1}^h+\frac95 f_{2,0;2}^h+\frac65 f_{3,0;1}^h+\frac{36}{5} f_{3,0;2}^h
-\frac35 g_{0;1}^h\biggr) z+\calo(z^2)\,,\\
&w_{a22}=f_{3,0;2}^h+\biggl(8 f_{2,0;1}^h f_{3,0;1}^h-\frac{96}{5} K_{0;1}^h f_{3,0;1}^h-\frac{32}{5}
K_{0;1}^h f_{2,0;1}^h+\frac{16}{5} (K_{0;1}^h)^2+\frac{32}{5} K_{0;2}^h
\\
&-\frac85 f_{2,0;2}^h-\frac{56}{5} f_{3,0;2}^h
+\frac45 g_{0;1}^h+\frac{16}{5} (f_{2,0;1}^h)^2+\frac{104}{5} (f_{3,0;1}^h)^2
-\frac45 f_{3,0;1}^h-\frac45 f_{2,0;1}^h\biggr) z+\calo(z^2)\,,\\
&s_2=s_{0;2}^h+\calo(z)\,,\\
&g_2=g_{0;2}^h+\left(\frac85 f_{2,0;1}^h+\frac{16}{5} f_{3,0;1}^h-\frac{16}{5} g_{0;1}^h\right) z+\calo(z^2)\,,
\end{split}
\eqlabel{efincond2}
\end{equation}
for $n=2$.

\section{Kretschmann scalar of EF frame background geometry}\label{kretschmann}

We collect here  the expression for the Kretschmann scalar $K$
\begin{equation}
K\equiv \calr_{\a\b\g\dd}\calr^{\a\b\g\dd}
\eqlabel{defK}
\end{equation}
of gravitational bulk geometries \eqref{ef1i}
dual to de Sitter vacua of the cascading gauge theories. Growth of $K$ evaluated
at the apparent horizon as $\frac{H^2}{\Lambda^2}$ varies signals the breakdown of the
supergravity approximation. Explicitly evaluating \eqref{defK} we find, $'=\frac{d}{dr}=
-\frac{d}{dz}$,
\begin{equation}
\begin{split}
&K=4 a^2 \biggl(\frac{12(\s'')^2}{\s^2}+\frac{(a'')^2}{a^2}
+\frac{2 (w_{a2}'')^2}{w_{a2}^2}+\frac{2 (w_{b2}'')^2}{w_{b2}^2}
+\frac{(w_{c2}'')^2}{w_{c2}^2}\biggr)+\frac{24(\s'')}{\s} \biggl(
H^2 -H  a'\\
&+\frac{4 H a\s'}{\s}+\frac{2 a \s' a'}{s}\biggr)
+\frac{4 a^2w_{c2}'w_{c2}''}{w_{c2}^2}\left(\frac{a'}{a}- \frac{w_{c2}'}{w_{c2}}\right)
+\frac{8 a^2w_{a2}'w_{a2}''}{w_{a2}^2}\left(\frac{a'}{a}-\frac{w_{a2}'}{w_{a2}}\right)
+\frac{8 a^2w_{b2}'w_{b2}''}{w_{b2}^2}\\
&\times\left(\frac{a'}{a}-\frac{w_{b2}'}{w_{b2}}\right)
+3 H^2 \left(\frac{24 (\s')^2}{\s^2}
+\frac{(w_{c2}')^2}{w_{c2}^2}+\frac{2 (w_{b2}')^2}{w_{b2}^2}+\frac{2 (w_{a2}')^2}{w_{a2}^2}\right)
+2(a')^2 \biggl(
\frac{12 (\s')^2}{\s^2}\\&
+\frac{(w_{c2}')^2}{w_{c2}^2}+\frac{2 (w_{b2}')^2}{w_{b2}^2}
+\frac{2 (w_{a2}')^2}{w_{a2}^2}\biggr)
+\frac{12 H \s'a}{\s} \biggl(
\frac{4 \s' a'}{a \s}+\frac{8 (\s')^2}{\s^2}+\frac{(w_{c2}')^2}{w_{c2}^2}
+\frac{2 (w_{b2}')^2}{w_{b2}^2}\\
&+\frac{2 (w_{a2}')^2}{w_{a2}^2}
\biggr)
-2 a a' \biggl(
\frac{(w_{c2}')^3}{w_{c2}^3}+\frac{2 (w_{b2}')^3}{w_{b2}^3}+\frac{2 (w_{a2}')^3}{w_{a2}^3}
\biggr)
+\frac{12 a^2 (\s')^2}{\s^2} \biggl(
\frac{(w_{c2}')^2}{w_{c2}^2}+\frac{2 (w_{b2}')^2}{w_{b2}^2}\\
&+\frac{2 (w_{a2}')^2}{w_{a2}^2}\biggr)
+\frac{48 a^2(\s')^4}{\s^4}+a^2 \biggl(
\frac{2 (w_{b2}')^2 (w_{c2}')^2}{w_{b2}^2 w_{c2}^2}
+\frac{2 (w_{a2}')^2 (w_{c2}')^2}{w_{a2}^2 w_{c2}^2}
+\frac{4 (w_{a2}')^2 (w_{b2}')^2}{w_{a2}^2 w_{b2}^2}
\\&+\frac{(w_{c2}')^4}{w_{c2}^4}+\frac{3 (w_{b2}')^4}{w_{b2}^4}
+\frac{3 (w_{a2}')^4}{w_{a2}^4}\biggr)
+\frac{a(w_{a2}')^2}{w_{b2} w_{c2} w_{a2}^3}
(27 w_{a2}^2+9 w_{b2}^2-36 w_{b2} w_{c2}+16 w_{c2}^2)
\\&+\frac{a(w_{b2}')^2}{w_{a2} w_{c2} w_{b2}^3} (9 w_{a2}^2-36 w_{a2} w_{c2}+27 w_{b2}^2+16 w_{c2}^2)
+\frac{3 a (w_{c2}')^2}{w_{a2} w_{b2} w_{c2}^3} (3 w_{a2}^2-6 w_{a2} w_{b2}\\
&+3 w_{b2}^2+16 w_{c2}^2)
-\frac{a w_{a2}'w_{c2}'}{2w_{b2} w_{a2}^2 w_{c2}^2} (63 w_{a2}^2-18 w_{a2} w_{b2}+24 w_{a2} w_{c2}-45 w_{b2}^2-24 w_{b2} w_{c2}
\\&+112 w_{c2}^2)
+\frac{a w_{b2}'w_{c2}'}{2w_{a2} w_{c2}^2 w_{b2}^2}
(45 w_{a2}^2+18 w_{a2} w_{b2}+24 w_{a2} w_{c2}-63 w_{b2}^2
-24 w_{b2} w_{c2}\\
&-112 w_{c2}^2)
-\frac{a w_{a2}' w_{b2}'}{2w_{c2} w_{b2}^2 w_{a2}^2}
(63 w_{a2}^2+18 w_{a2} w_{b2}-24 w_{a2} w_{c2}+63 w_{b2}^2
-24 w_{b2} w_{c2}-80 w_{c2}^2)\\
&+\frac{136 w_{c2}^2}{w_{a2}^2 w_{b2}^2}-\frac{144 (w_{a2}+w_{b2}) w_{c2}}{
w_{a2}^2 w_{b2}^2}+\frac{81(13 w_{a2}^2+6 w_{a2} w_{b2}+13 w_{b2}^2)(w_{a2}-w_{b2})^2}
{32w_{c2}^2 w_{b2}^2 w_{a2}^2}\\
&-\frac{54 (w_{a2}^2-w_{b2}^2) (w_{a2}-w_{b2})}{w_{a2}^2 w_{b2}^2 w_{c2}}
+\frac{9 (9 w_{a2}^2+14 w_{a2} w_{b2}+9 w_{b2}^2)}{w_{a2}^2 w_{b2}^2}\,.
\end{split}
\eqlabel{kres}
\end{equation}
Introducing the dimensionless and rescaled functions and the radial coordinate
$\hat{r}\equiv -\hat{z}$ as in \eqref{scaleouttypeas},
\begin{equation}
K=\frac{1}{P^2g_s}\ \hat{K}\,.
\eqlabel{defhatk}
\end{equation}

\subsection{Kretschmann scalar at AH of TypeB de Sitter vacua}\label{ktypeb}

In section \ref{sentb} we showed that the AH horizon of the bulk gravitational dual to
TypeB de Sitter vacua of the cascading gauge theory is located at $r_{AH}=-z_{AH}=0$, see
\eqref{ahsvactypeb}. Using \eqref{typebi} we find from \eqref{kres}:
\begin{equation}
\begin{split}
&K_{AH}\bigg|_{\rm TypeB}=300 h_0^h H^4+H^2 \biggl(
\frac{16P^2g_0^h (3 (k_{2,2}^h)^2 (f_{a,0}^h)^2+20)}{3(f_{a,0}^h)^3 h_0^h}
+\frac{72}{(f_{a,0}^h)^2 k_{2,2}^h} \biggl(5 k_{2,4}^h (f_{a,0}^h)^2\\
&+3 k_{2,2}^h f_{a,0}^h+18\biggr)
+\frac{5 (k_{1,3}^h)^2 k_{2,2}^h (f_{a,0}^h)^2
-15 k_{2,2}^h (k_{3,1}^h)^2-36 k_{1,3}^h k_{3,1}^h}{k_{2,2}^h P^2 (f_{a,0}^h)^2 g_0^h h_0^h}\biggr)
\\&+\frac{1}{3840(f_{a,0}^h)^4 (h_0^h)^3 P^4 (g_0^h)^2 (k_{2,2}^h)^2}
\biggl(
355 (k_{1,3}^h)^4 (k_{2,2}^h)^2 (f_{a,0}^h)^4-30 (k_{1,3}^h)^2 (k_{2,2}^h)^2 (k_{3,1}^h)^2 (f_{a,0}^h)^2
\\&+2283 (k_{2,2}^h)^2 (k_{3,1}^h)^4+6912 k_{1,3}^h k_{2,2}^h (k_{3,1}^h)^3
+6912 (k_{1,3}^h)^2 (k_{3,1}^h)^2
\biggr)
\\&+\frac{3}{10(f_{a,0}^h)^4 P^2 g_0^h (h_0^h)^2 (k_{2,2}^h)^2} \biggl(
25 (k_{1,3}^h)^2 (k_{2,2}^h)^2 (f_{a,0}^h)^3-60 k_{2,2}^h k_{2,4}^h (k_{3,1}^h)^2 (f_{a,0}^h)^2
\\&-120 k_{1,3}^h k_{2,4}^h k_{3,1}^h (f_{a,0}^h)^2-37 (k_{2,2}^h)^2 (k_{3,1}^h)^2 f_{a,0}^h
-24 k_{1,3}^h k_{2,2}^h k_{3,1}^h f_{a,0}^h-216 k_{2,2}^h (k_{3,1}^h)^2\\&-432 k_{1,3}^h k_{3,1}^h\biggr)
+\frac{1}{1080(f_{a,0}^h)^5 (h_0^h)^3 (k_{2,2}^h)^2} \biggl(
175 (k_{1,3}^h)^2 (k_{2,2}^h)^4 (f_{a,0}^h)^4\\&
+194400 (k_{2,4}^h)^2 (f_{a,0}^h)^5 (h_0^h)^2
-491 (k_{2,2}^h)^4 (k_{3,1}^h)^2 (f_{a,0}^h)^2+77760 k_{2,2}^h k_{2,4}^h (f_{a,0}^h)^4 (h_0^h)^2
\\&-1152 k_{1,3}^h (k_{2,2}^h)^3 k_{3,1}^h (f_{a,0}^h)^2+746496 (k_{2,2}^h)^2 (f_{a,0}^h)^3 (h_0^h)^2
-2220 (k_{1,3}^h)^2 (k_{2,2}^h)^2 (f_{a,0}^h)^2\\
&+1399680 k_{2,4}^h (f_{a,0}^h)^3 (h_0^h)^2
+279936 k_{2,2}^h (f_{a,0}^h)^2 (h_0^h)^2-3492 (k_{2,2}^h)^2 (k_{3,1}^h)^2
\\&-13824 k_{1,3}^h k_{2,2}^h k_{3,1}^h+2519424 f_{a,0}^h (h_0^h)^2\biggr)
+\frac{8P^2 g_0^h}{45k_{2,2}^h (f_{a,0}^h)^5 (h_0^h)^2} \biggl(
60 (k_{2,2}^h)^2 k_{2,4}^h (f_{a,0}^h)^4\\&+37 (k_{2,2}^h)^3 (f_{a,0}^h)^3
+216 (k_{2,2}^h)^2 (f_{a,0}^h)^2+720 k_{2,4}^h (f_{a,0}^h)^2-756 k_{2,2}^h f_{a,0}^h
+2592\biggr)
\\&+\frac{P^4 (g_0^h)^2}{3645(h_0^h)^3
(f_{a,0}^h)^6}  \biggl(881 (k_{2,2}^h)^4 (f_{a,0}^h)^4+10584 (k_{2,2}^h)^2 (f_{a,0}^h)^2+184464\biggr)\,.
\end{split}
\eqlabel{kahtypeb}
\end{equation}

A special case of \eqref{kahtypeb} is the Kretschmann scalar at the ``AH'' of the extremal
KS solution, see section \ref{ksextremal}: setting $H=0$ and using \eqref{susyir} we find
\begin{equation}
\lim_{H\to 0}\ K_{AH}\bigg|_{{\rm TypeB}}=\frac{1}{P^2g_s}\ \frac{32\cdot 12^{2/3} (110\cdot 12^{1/3}+177147\delta^2)}{295245\dd^3}\,,
\eqlabel{kks}
\end{equation}
where we denoted, see $h_0^h$ in \eqref{susyir},
\begin{equation}
\dd\equiv 0.056288(0)\,.
\eqlabel{defdelta}
\end{equation}

\section{Static linearized $\csb$ fluctuations about TypeA$_s$ vacua}\label{lincsb}

Static linearized $\csb$ fluctuations about TypeA$_s$ vacua in FG frame are parameterized as in \eqref{fldef}.
From \eqref{kseq3}-\eqref{kseq4} and \eqref{kseq6}-\eqref{kseq8} we find,
($'=\frac{d}{d\r}$ and $P=H=g_s=1$):
\begin{equation}
\begin{split}
&0=\dd f''-\frac{1}{16\r g^2 f_2 h^2 f_3^3 (f_3' \r
-2 f_3)} \biggl(
-48 f_3^4 h^3 f_2 g^2 \r^2-2 (h')^2 g^2 f_3^4 f_2 \r^2
-2 (g')^2 h^2 f_2 f_3^4 \r^2\\
&+12 g^2 h^2 (f_3')^2 f_3^2 f_2 \r^2
-16 h (h') g^2 f_3^4 f_2 \r-16 h^2 g^2 f_3^3 (f_3') f_2 \r
-48 f_3^4 h^2 g^2 f_2\\
&-f_2 \r^2 (K')^2 h f_3^2 g+16 f_3^2 h^2 f_2^2 g^2
-96 f_3^3 h^2 f_2 g^2+4 g^3 f_3^2 h+2 g^2 K^2
\biggr)\ \dd f'
-\frac{K'}{2g h f_3}\ {\dd k_1'}\\
&-\frac{2 g}{f_2 f_3 h \r^2}\ {\dd k_2}
+\frac{1}{8g^2 f_2 h^2 \r^2 f_3^3 (f_3' \r-2 f_3)}
\biggl(
-48 g^2 f_3' f_2 h^3 f_3^3 \r^3+8 g^2 (f_3')^3 f_2 h^2 f_3 \r^3
\\&+48 f_3^4 h^3 f_2 g^2 \r^2-2 (h')^2 g^2 f_3^4 f_2 \r^2
-2 (g')^2 h^2 f_2 f_3^4 \r^2-36 g^2 h^2 (f_3')^2 f_3^2 f_2 \r^2
\\&-16 h h' g^2 f_3^4 f_2 \r+64 h^2 g^2 f_3^3 f_3' f_2 \r
-4 g f_3' (K')^2 f_2 h f_3 \r^3
+32 g^2 f_3' f_2^2 h^2 f_3 \r-72 g^2 f_3' h^2 f_3^3 \r
\\
&-80 f_3^4 h^2 g^2 f_2+7 f_2 \r^2 (K')^2 h f_3^2 g
-48 f_3^2 h^2 f_2^2 g^2-96 f_3^3 h^2 f_2 g^2+144 f_3^4 h^2 g^2
\\
&-16 g^3 f_3' h f_3 \r+36 g^3 f_3^2 h
+2 g^2 K^2\biggr)\  \dd f\,,
\end{split}
\eqlabel{fl1}
\end{equation}
\begin{equation}
\begin{split}
&0={\dd k_1}''-\frac{1}{16\r g^2 f_2 h^2 f_3^3 (f_3' \r-2 f_3)}
\biggl(
-48 f_3^4 h^3 f_2 g^2 \r^2-2 (h')^2 g^2 f_3^4 f_2 \r^2
\\&+16 f_3^3 h f_2 g^2 \r^2 f_3' h'
-2 (g')^2 h^2 f_2 f_3^4 \r^2
+16 f_3^3 f_2 f_3' h^2 g g' \r^2
+12 g^2 h^2 (f_3')^2 f_3^2 f_2 \r^2
\\&-48 h h' g^2 f_3^4 f_2 \r-32 f_3^4 f_2 h^2 g g' \r
-16 h^2 g^2 f_3^3 f_3' f_2 \r-48 f_3^4 h^2 g^2 f_2
-f_2 \r^2 (K')^2 h f_3^2 g\\
&+16 f_3^2 h^2 f_2^2 g^2
-96 f_3^3 h^2 f_2 g^2+4 g^3 f_3^2 h+2 g^2 K^2\biggr)\
\dd k_1'+\frac{2 K'}{f_3}\ \dd f'-\frac{9}{\r^2 f_2}\ {\dd k_1}
\\&+\frac{2 g K}{\r^2 f_2 h f_3^2}\ {\dd k_2}
+\frac{2 (-f_3' K' f_2 h f_3 \r^2+2 g K)}{f_3^3 \r^2 f_2 h}\ \dd f\,,
\end{split}
\eqlabel{fl2}
\end{equation}
\begin{equation}
\begin{split}
&0={\dd k_2}''-\frac{1}{16\r g^2 f_2 h^2 f_3^3 (f_3' \r-2 f_3)}
\biggl(
-48 f_3^4 h^3 f_2 g^2 \r^2-2 (h')^2 g^2 f_3^4 f_2 \r^2
\\&+16 f_3^3 h f_2 g^2 \r^2 f_3' h'
-2 (g')^2 h^2 f_2 f_3^4 \r^2
-16 f_3^3 f_2 f_3' h^2 g g' \r^2
+12 g^2 h^2 (f_3')^2 f_3^2 f_2 \r^2
\\
&-48 h h' g^2 f_3^4 f_2 \r+32 f_3^4 f_2 h^2 g g' \r
-16 h^2 g^2 f_3^3 f_3' f_2 \r-48 f_3^4 h^2 g^2 f_2
-f_2 \r^2 (K')^2 h f_3^2 g\\
&+16 f_3^2 h^2 f_2^2 g^2
-96 f_3^3 h^2 f_2 g^2+4 g^3 f_3^2 h+2 g^2 K^2\biggr)\
{\dd k_2}'-\frac{9}{\r^2 f_2}\ {\dd k_2}+\frac{9K}{2\r^2 f_2 f_3^2 h g}\ {\dd k_1}\\
&-\frac{18}{f_3 \r^2 f_2}\ \dd f\,.
\end{split}
\eqlabel{fl3}
\end{equation}
Performing the asymptotic expansions, we determine:
\nxt  in the UV, \ie as $\r\to 0$, using \eqref{ktks1}-\eqref{ktks6},
\begin{equation}
\begin{split}
&\dd f=\dd f_{1,0}\ \r+\frac12 f_{2,1,0}\  \dd f_{1,0}\ \r^2+\biggl(\dd f_{3,0}+\left(\frac14 \dd f_{1,0}\ k_s-\frac{11}{8}
\dd f_{1,0}\right) \ln\r
\\&-\frac14 \dd f_{1,0} \ln^2\r\biggr)\
\r^3+\sum_{n=4}\sum_k \dd f_{n,k}\ \r^n\ln^k\r\,,
\end{split}
\eqlabel{uvdf}
\end{equation}
\begin{equation}
\begin{split}
&\dd k_1=-\frac12 \dd f_{1,0}\ \r+\frac14 f_{2,1,0}\ \dd f_{1,0}\ \r^2+\biggl(\dd k_{1,3,0}
+\left(-\frac{1}{24} \dd f_{1,0}\ k_s-\frac{47}{144} \dd f_{1,0}+2 \dd f_{3,0}\right) \ln\r
\\&+\left(-\frac43 \dd f_{1,0}+\frac14 \dd f_{1,0}\ k_s\right) \ln^2\r-\frac16 \dd  f_{1,0}\ \ln^3\r\biggr) \r^3+
+\sum_{n=4}\sum_k \dd k_{1,n,k}\ \r^n\ln^k\r\,,
\end{split}
\eqlabel{uvdk1}
\end{equation}
\begin{equation}
\begin{split}
&\dd k_2=-\frac94 \dd f_{1,0}\ \r+\frac98 f_{2,1,0}\ \dd f_{1,0}\ \r^2
+\biggl(-\frac{13}{48}\dd f_{1,0}\ k_s-\frac38\dd f_{1,0}\ f_{2,1,0}^2+\frac32\dd k_{1,3,0}\\
&-\frac{163}{144}\ \dd f_{1,0}
-\dd f_{3,0}+\left(-\frac{5}{16}\ \dd f_{1,0}\ k_s+\frac{137}{96} \dd f_{1,0}+3\dd f_{3,0}\right) \ln\r
+\biggl(-\frac74 \dd f_{1,0}\\
&+\frac38\dd f_{1,0}\ k_s\biggr) \ln^2\r-\frac14\dd f_{1,0}\ \ln^3\r\biggr)
\r^3+\sum_{n=4}\sum_k \dd k_{2,n,k}\ \r^n\ln^k\r\,,
\end{split}
\eqlabel{uvdk2}
\end{equation}
characterized by 4 parameters (compare with \eqref{uvparks}):
\begin{equation}
\{\dd f_{1,0}\,,\, \underbrace{\dd f_{3,0}\,,\, \dd k_{1,3,0}}_{\calo_3^\a}\,,\, \underbrace{\dd f_{7,0}}_{\calo_7} \}\,,
\eqlabel{uvlinpar}
\end{equation}
where $\dd f_{1,0}$ is an explicit chiral symmetry breaking scale ($\propto$ the gaugino mass term), and the remaining
parameters are the expectation values of the chiral symmetry breaking operators in the cascading gauge theory;
\nxt  in the IR, \ie as $\frac 1\r=y\to 0$, using \eqref{phase2ir1},
\begin{equation}
\begin{split}
\dd f=\frac1y\ \sum_{n=0}\ \dd f_{n}^h\ y^n\,,\qquad \dd k_{1,2}=\sum_{n=0}\ \dd k_{1,2,n}^n\ y^n\,,
\end{split}
\eqlabel{irdfk1k2}
\end{equation}
characterized by 3 parameters:
\begin{equation}
\{\dd f_0^h\,,\, \dd k_{1,0}^h\,,\, \dd k_{2,0}^h\}\,.
\eqlabel{irlinpar}
\end{equation}

\bibliographystyle{JHEP}
\bibliography{ksdesitter}

\end{document}